\documentclass{JHEP3}
\usepackage{amsmath,amssymb,url}
\usepackage{epsfig,graphicx,tabularx}
\usepackage{psfrag}

\graphicspath{{images/}}

\newcommand{\D}{\displaystyle}

\newcommand{\etal}{\textit{et al.}}
\newcommand{\ie}{{\sl i.e.\ }}
\newcommand{\eg}{{\sl e.g. }}

\newcommand{\mev}{{\ensuremath\rm MeV}}
\newcommand{\gev}{{\ensuremath\rm GeV}}

\newcommand{\br}{{\ensuremath\rm BR}}

\newcommand{\ifb}{{\ensuremath\rm fb^{-1}}}

\psfrag{HWW}[c][][1][0]{ $\Delta_{WWH}$}
\psfrag{HZZ}[c][][1][0]{ $\Delta_{ZZH}$}
\psfrag{Htt}[c][][1][0]{ $\Delta_{ttH}$}
\psfrag{Hbb}[c][][1][0]{ $\Delta_{bbH}$}
\psfrag{Htautau}[c][][1][0]{ $\Delta_{\tau\tau H}$}
\psfrag{Hgg}[c][][1][0]{ $\Delta_{ggH}$}
\psfrag{Hgamgam}[c][][1][0]{ $\Delta_{\gamma\gamma H}$}
\psfrag{Hii}[c][][1][0]{ $\Delta_\Gamma$}
\psfrag{-9}[c][c][1][0]{-9}
\psfrag{-5}[c][c][1][0]{-5}
\psfrag{-4}[c][c][1][0]{-4}
\psfrag{-3}[c][c][1][0]{-3}
\psfrag{-2}[c][c][1][0]{-2}
\psfrag{-1}[c][c][1][0]{-1}
\psfrag{ 0}[c][c][1][0]{0}
\psfrag{ 0.0005}[c][c][1][0]{0.0005 }
\psfrag{ 0.001}[c][c][1][0]{0.001 }
\psfrag{ 0.002}[c][c][1][0]{0.002 }
\psfrag{ 0.003}[c][c][1][0]{0.003 }
\psfrag{ 0.004}[c][c][1][0]{0.004 }
\psfrag{ 0.006}[c][c][1][0]{0.006 }
\psfrag{ 0.008}[c][c][1][0]{0.008 }
\psfrag{ 0.02}[c][c][1][0]{0.02 }
\psfrag{ 0.04}[c][c][1][0]{0.04 }
\psfrag{ 0.06}[c][c][1][0]{0.06 }
\psfrag{ 0.2}[c][c][1][0]{0.2 }
\psfrag{ 0.4}[c][c][1][0]{0.4 }
\psfrag{ 0.6}[c][c][1][0]{0.6 }
\psfrag{ 0.8}[c][c][1][0]{0.8 }
\psfrag{ 1}[c][c][1][0]{1}
\psfrag{ 2}[c][c][1][0]{2}
\psfrag{ 3}[c][c][1][0]{3}
\psfrag{ 4}[c][c][1][0]{4}
\psfrag{ 5}[c][c][1][0]{5}
\psfrag{ 6}[c][c][1][0]{6}
\psfrag{ 7}[c][c][1][0]{7}
\psfrag{ 8}[c][c][1][0]{8}
\psfrag{ 9}[c][c][1][0]{9}
\psfrag{ 10}[c][c][1][0]{10}
\psfrag{ 15}[c][c][1][0]{15}
\psfrag{ 20}[c][c][1][0]{20}
\psfrag{ 160}[c][c][1][0]{160 }
\psfrag{ 180}[c][c][1][0]{180 }
\psfrag{ 200}[c][c][1][0]{200 }
\psfrag{TOPm}[c][c][1][0]{$m_t$}
\psfrag{TOPt}[c][c][1][0]{}

\preprint{KA-TP-01-2009 \\ SFB/CPP-09-28}

\date{April 24, 2009}

\title{Measuring the Higgs Sector}

\author{R\'emi Lafaye \\
        LAPP, Universit\'e Savoie, IN2P3/CNRS, Annecy, France}

\author{Tilman Plehn \\
        Institut f\"ur Theoretische Physik, Universit\"at Heidelberg, Germany}

\author{Michael Rauch \\
        Institut f\"ur Theoretische Physik, Universit\"at Karlsruhe, Germany}

\author{Dirk Zerwas \\
        LAL, IN2P3/CNRS, Orsay, France}

\author{Michael D\"uhrssen \\
        Physikalisches Institut, Universit\"at Freiburg, Germany}

\abstract{
  If we find a light Higgs boson at the LHC, there should be many
  observable channels which we can exploit to measure the relevant
  parameters in the Higgs sector. We use the SFitter framework to map
  these measurements on the parameter space of a general weak-scale
  effective theory with a light Higgs state of mass 120~GeV. Our analysis benefits
  from the parameter determination tools and the error treatment used
  in new--physics searches, to study individual parameters and their
  error bars as well as parameter correlations.
}

\begin{document}


\newpage

\section{Higgs physics at the LHC}

Over the coming years, the main goal of LHC physics is to understand
the breaking of the electroweak gauge symmetry. This symmetry we
usually assume to be spontaneously broken, \ie hidden by the presence
of a non-symmetric vacuum. Such a vacuum can be induced through a
gauge invariant Higgs potential involving powers of $H^\dagger H$,
where $H$ is a $SU(2)$ Higgs doublet~\cite{higgs,reviews}. One of the
degrees of freedom of this doublet becomes a physical Higgs scalar,
while the other three are absorbed as longitudinal degrees of freedom
into the massive $W,Z$ gauge bosons. In this model all particle masses
are proportional to the vacuum expectation value $v=246$~GeV, and the
Higgs couplings to other particles as well as the Higgs itself are
proportional to that particle's mass.

Gauge boson masses $m_{W,Z}$ arise from dimension--four $WWHH$ and
$ZZHH$ terms. As the pseudo--Goldstone modes of the Higgs doublet
are parts of the $W,Z$ fields there is no additional free parameter in
this construction. Both masses $m_{W,Z}$ are given by $v$ times
the gauge couplings $g$ times a rotation by the electroweak mixing
angle. This relation between the (measured) gauge boson masses and $v$
allows us to determine $v$ without observing the Higgs scalar.

Fermion masses arise from the dimension--four operators $H \Psi
\overline{\Psi}$.  This operator can be accompanied by any real
dimensionless number, the Yukawa coupling $y_f$. Once the Higgs field
develops a vacuum expectation value the combination $y_f \times v$
becomes the fermion mass. Note that in contrast to the gauge sector we
have not yet measured any Higgs couplings to fermions, with the
possible exception of $y_t$ in the electroweak precision 
analysis~\cite{Alcaraz:2007ri,Collaboration:2008ub}. The
link between the known fermion masses and the largely unknown Yukawa
couplings can be offset for example by non-standard Higgs
sectors. \bigskip

A crucial question at the LHC will be: if we observe a light scalar
particle, how can we tell that this state is the physical Higgs boson
and not something else or a mixture between a Higgs boson and another
new state. A good example for a state which looks like a Higgs scalar
and mixes with a Higgs scalar is the radion in theories with large
warped extra dimensions~\cite{radion}. A more conservative
modification of the Higgs sector in the Standard Model could be an
extension, for example by another Higgs doublet. A theory which
requires such an extended Higgs sector is supersymmetry, where we find
two CP-even scalar Higgs bosons which mix, leading to a light and a
heavy mass eigenstate~\cite{susy,m_h}. In the decoupling limit the
lighter of these mass eigenstates becomes the Standard Model Higgs
boson, but for finite supersymmetric and heavy--Higgs mass scales its
properties deviate from the Standard Model case.

Our strategy for studying the Higgs sector is analogous to our studies
of new--physics parameter spaces~\cite{sfitter}: there are many pieces
of information on the Higgs sector which are not continuous. These
include the spin of the Higgs boson, its CP quantum numbers, or the
structure of a CP-even $WWH$ vertex~\cite{wbf_vertex}. In this paper
we study the continuous Higgs parameters, \ie we assume that these
quantum numbers are known, or in other words, we assume that we know
the operator basis we need to include in the effective Lagrangian
describing Higgs physics at the LHC. However, for each of these
operators we have to determine its strength, to then compare the
result with the Standard Model or alternative weak-scale theories.

Such an analysis will test all accessible Higgs couplings to known
particles for a given Higgs
mass~\cite{couplings_first,duehrssen,maxim}, which we assume to be 120~GeV. This will first of all
include the gauge boson couplings, where as mentioned above we have
little room to alter the relations of the gauge boson masses and the
gauge couplings. In addition, we will measure the Yukawa couplings and
compare them to the known fermion masses. The observable Higgs
channels at the LHC involve gluon fusion as well as weak boson fusion
combined with a large number of branching ratios. Such an analysis will
be statistically challenging. Just as the high-dimensional parameter
space analyses studying new physics in the LHC era, the analysis
needs to focus on the proper treatment of all
errors~\cite{sfitter,fittino}.\bigskip

Some Higgs couplings will most likely only be measured at a later stage,
involving integrated luminosities of at least $300~\ifb$. One of them is
the only second-generation Higgs decay we might observe at the LHC, the
decay into muons~\cite{wbf_mu,tth_mu}. Owed to the outstanding $m_{\mu
\mu}$ resolution in particular of CMS the weak-boson-fusion channel
might come close to a $5\sigma$ signal using modern analysis tools.
However, because of the lack of experimental studies and due to this
small event rate we neglect all light-flavor Yukawa couplings in this
analysis.

Extracting the one-loop Higgs coupling to $Z\gamma$ is similar to the
two--photon channel in suffering from a reduced branching ratio, plus
an additional leptonic branching ratio of the $Z$ boson. We do not
include it in our analysis, because of its small event rate for
low-luminosity running.

Involving physics beyond the Standard Model a light Higgs can decay to
invisible particles. The issue in searching for this decay is
triggering the events, which should be possible for weak boson fusion
production~\cite{wbf_inv}. Then, we can extract an invisible Higgs
signal from an invisible $Z$ decay using kinematic distributions and a
careful analysis of signal-free control regions. Note that because of
the limited energy it is unlikely that such a signal would be mimicked
by the production of for example two neutralinos or charginos, because
phase space essentially only allows for the production of one particle
in this channel at the LHC~\cite{wbf_inos}.

Last but not least, the arguably hardest-to-measure Higgs coupling can
be regarded as the most interesting: if the electroweak symmetry is
really spontaneously broken, this means there exists a Higgs
potential. Such a potential can only exist if there is a Higgs self
coupling. While the $W_L W_L H$ coupling is induced by such a self
coupling, it would be most desirable if we could measure
the explicit self coupling $\lambda_{HHH}$ at the LHC~\cite{selfcoup}.
Finally, a measurement of the quartic self coupling $\lambda_{HHHH}$
could confirm the actual form of the Higgs potential~\cite{higgs_pot},
but it is unlikely that we will measure this last and final Higgs
parameter at any collider currently discussed, including the VLHC or
CLIC~\cite{quartic}. We skip such high-luminosity channels which
should not feed back into our leading parameter set. The analysis
presented in this paper can easily be extended. \bigskip

The final assumption we make for the structure of the Higgs sector is
CP symmetry. While in many new--physics models, like supersymmetry, it
is actually hard to avoid additional CP violation we currently have no
experimental hints for such complex phases in the weak-scale
Lagrangian~\cite{cp_phases}. In other words, the apparent absence of
additional CP violation from the TeV scale could be considered one of
the worst problems of TeV-scale model building; it serves as a benefit for
our Higgs parameter analysis.\bigskip

Any parameter study at the LHC or elsewhere needs to focus on the error
analysis. First of all, this means that we have to include statistical
and systematic errors including a full correlation matrix as well as
theory errors. The latter are numerically challenging, because they
are not Gaussian. From these errors we need to construct a likelihood
map over the entire parameter space, for example using Markov Chains
or simulated annealing. The techniques we employ to scan the
high-dimensional parameter space can have a large impact on the
numerical efficiency, but they should be equivalent in the final
result. One major technical difference between these methods is the
study of secondary maxima in the likelihood map.

An exclusive likelihood map is well suited for example to study
correlations between different model parameters, \ie different Higgs
couplings, as well as local properties of the likelihood map. However,
there are good reasons to ask more specific questions. The most
obvious are about the probability or likelihood distribution of one Higgs
coupling (\eg $g_{WWH}$) or the correlation of two of them (\eg
$g_{WWH}$ vs $g_{\tau\tau H}$). The answer to this question depends
on how we ask this questions, and what methodology we want to employ.

Bayesian probability distributions and frequentist profile likelihoods
are two ways to study a parameter space, where some model parameters
might be very well determined, others heavily correlated, and even
others basically unconstrained. Both of them rely on an exclusive
likelihood map as a starting point.  Following the SFitter strategy we
carefully compare the benefits and disadvantages of the frequentist
and the Bayesian approaches for each model parameter.\bigskip

The approach of mapping measurements onto a high--dimensional
parameter space for example realized in SFitter~\cite{sfitter} (or in
the largely similar Fittino tool~\cite{fittino}) are completely
general: model parameters as well as measurements are included in the
form of model and measurement files and can simply be replaced. SFitter
serves as a general tool to map typically up to 20--dimensional highly
complex parameter spaces onto a large sample of highly correlated
measurements of different quality including a proper error treatment.
While this approach is obviously mandatory for high-precision analyses,
it is also best suited to extract the maximum information from any kind
of correlated measurements.

\section{LHC measurements}
\label{sec:measurements}

\subsection{Production modes}
\label{sec:production}

At the LHC we rely on two main production processes for a
Standard--Model type Higgs boson: through a top loop the Higgs couples
to two gluons, which leads to cross sections of the order of $10
- 100$~pb (37~pb for a 120~GeV Higgs). The production cross section is known at
next-to-leading order including all top-mass
effects~\cite{gf_nlo,spirix}. At NNLO the gluon--fusion production
rate is known for an effective $ggH$ coupling~\cite{gf_nnlo}, where
the description of 
gluon radiation off the top loop might not reproduce all distributions correctly. Due to unknown
higher orders we assume a theory uncertainty of 13\% on the production
rate after acceptance and background rejection
cuts~\cite{gf_error}. This error includes the uncertainty on a jet
veto for $H \to WW$ decays~\cite{gf_w}. For a realistic analysis
including actual LHC data we would have to check in detail that cuts
or analysis strategies do not lead to large additional errors, for
example by inducing large logarithms~\cite{gf_mc,grazzini}.\bigskip

The second largest production cross section is mediated by weak boson
fusion, involving two forward jets with transverse momentum around
$m_W/2$ and typical cross sections of $1 -
10$~pb (4.5~pb for a 120~GeV Higgs). From a QCD perspective this production
channel is particularly clean, because there is no color flow between
the two DIS sides of the process. This simple fact we can make use of
by vetoing central jets between the two forward tagging
jets~\cite{mjv}. The total rate as well as the key distributions for
this signal are known to next-to-leading order~\cite{wbf_nlo},
including (as it turns out: negligible) interference
effects~\cite{wbf_int}. An interesting aspect is that due to the
suppression of higher--order QCD effects, strong and electroweak NLO
corrections are of the same size but opposite sign. For this
production process we estimate the theory error to range around
7\%~\cite{wbf_nlo}.

For weak boson fusion there remains a caveat: Higgs rate measurements
include the survival probability of a central minijet
veto~\cite{mjv}. It is one of the reasons why weak boson fusion
signatures have such a spectacular signal-over-background ratio. While
we will be able to measure this veto probability in $Z+$jets
production~\cite{mjv_kyle} for the signal we need to predict
it~\cite{manchester} --- including the underlying
event. While surely a minijet veto will work in LHC
analyses, in Section~\ref{sec:mjv} we will check what happens to the
Higgs sector analysis with an increased error.

The production channels of gluon fusion and weak boson fusion are not
independent. Higgs boson production with two jets will have
contributions from both gluon fusion and weak boson fusion. With
appropriate cuts, we can enhance one channel over the
other~\cite{wbf_gf}, but in our rate analysis we nevertheless need to
include both. We do not introduce an additional error for this
separation, assuming it to be already covered by the theory errors and
the systematic errors.\bigskip

A third production mechanism is the radiation of a Higgs boson off top
pairs. This signature, in spite of its sizable rate of up to 600~fb (450~fb for a 120~GeV Higgs),
suffers from its high--multiplicity final state and hence large
combinatorial errors. Essentially, any Higgs decay product will also
be present in the top decays. Its rate is known to
next-to-leading order~\cite{tth_nlo}, with a remaining theory error
around 13\%.\bigskip

Last, but not least, there might be hope to see a Higgs boson produced
in association with a $W$ or $Z$ gauge boson. Its reasonable rate of
$0.3 - 3$~pb (2.2~pb for a 120~GeV Higgs) suffers from the additional leptonic branching ratio
of the gauge bosons. Still, in particular in the context of modern
jet--structure driven searches this production channel could well turn
out useful if we combine it with a Higgs decay with a large branching
ratio. For this channel we assume a theory error of 7\%.\bigskip

Throughout this analysis we will skip the analysis of Higgs pair
production~\cite{pairs_basics}. In the Standard Model this channel has
a one-to-one correlation with the Higgs self coupling~\cite{selfcoup}
and is strongly suppressed. We note, however, that in models with
additional Higgs resonances the pair production of light Higgs bosons
might well be visible~\cite{self_rare}.\bigskip

\subsection{Signatures}

The challenge for Higgs searches at the LHC lies in triggering and the
extraction of the signal from backgrounds. In our analysis we use the
following list of most promising signatures for models not too different from the Standard Model~\cite{reaches}. This list
should not be considered the final work for the LHC, but simply reflects
our understanding at the current stage of Monte-Carlo studies:
\begin{itemize}
\item[] $H \to b\bar{b}$ --- while for small Higgs masses this is the
  dominant decay channel, with branching ratios up to 90\%, it is
  particularly hard to extract from QCD backgrounds.  Recent CMS and
  Atlas studies show that combinatorial backgrounds make it very hard
  to observe this decay in $t\bar{t}H$ production, where it might
  never reach the $5\sigma$ level~\cite{tth_b}. The best chance for
  this decay channel will probably be the $WH/ZH$ production
  mode~\cite{wh_b} with a subjet analysis~\cite{subjet}, currently under study by both collaborations.

\item[] $H \to WW$ --- even though two leptonic $W$ decays only allow
  us to reconstruct a transverse mass of the $WW$ pair this channel is
  arguably the most powerful search channel for a light Higgs
  boson. In the gluon fusion process, where the most important
  observables to cut on are correlations between the two leptons this
  decay signature can be extracted using the opening angle of the two
  leptons (being small if the two $W$ bosons come from a resonant
  scalar)~\cite{gf_w}. In weak boson fusion, this decay can be used
  even if one of the $W$s is far off-shell~\cite{wbf_w} with a
  spectacular $S/B \sim \mathcal{O}(1)$. Combined with the $t\bar{t}H$
  production mode this decay might turn out useful for on-shell
  decays~\cite{tth_w}, so we include all three production
  mechanisms.  
\item[] $H \to ZZ$ --- due to its spectacular fully reconstructable
  four--lepton final state this signature is usually referred to as
  the `golden channel'. It works in combination with the gluon
  fusion~\cite{gf_z} as well as with the weak boson
  fusion~\cite{wbf_z} production process, but is somewhat
  statistically limited to larger Higgs masses.
\item[] $H \to \tau\tau$ --- for this channel to be useful we need to
  reconstruct the invariant mass of the two taus. It is well known how
  to do this in the collinear approximation, where the taus are
  fast-moving~\cite{coll_taus}. Unfortunately, because of the steep
  gluon densities, gluon fusion produces Higgs bosons too close to
  threshold, which means we cannot use this production mechanism. In
  contrast, in weak boson fusion the Higgs recoils against the two
  tagging jets, so the taus are boosted. That way we can even measure
  the Higgs mass to $\sim 5$~GeV~\cite{wbf_tau}. Further improvement
  we expect from an improved statistical distinction of the signal
  from the $Z \to \tau \tau$ background~\cite{tautau_kyle}.  For the
  $t\bar{t}H$ production channel with a subsequent decay to taus we
  are not aware of experimental studies, so we leave it out for
  now~\cite{tth_tau}.
\item[] $H \to \gamma\gamma$ --- the decay to two photons has the
  advantage of being the only fully reconstructable channel for a very
  light Higgs boson, as preferred by electroweak precision data. The
  $\gamma \gamma$ mass resolution of 2~GeV or better compensates for
  the small branching ratio and makes this decay channel a promising
  candidate for an observation in gluon fusion
  production~\cite{gf_gamma}, weak boson fusion
  production~\cite{wbf_gamma} and even $t \bar{t} H$ associated
  production~\cite{tth_gamma}. Moreover, this channels allows us to
  measure the mass of a light Higgs boson with a precision 
  $\mathcal{O}(100~\mev)$, depending on the Higgs mass.
\item[] $H \to$~invisible --- the best chance to observe a Higgs decay
  to invisible particles~\cite{wh_inv} is in combination with weak
  boson fusion, namely looking for two tagging jets recoiling against
  missing energy~\cite{wbf_inv}. Of course, the determination of the
  Higgs branching ratio to invisible new states relies on our
  understanding of the production rate and therefore on the proper
  combination with all other Higgs signatures. Due to the lack of a
  full--simulation analysis, we do not include this signal as a
  measurement. On the other hand, because this decay channel can be of
  crucial importance for new-physics searches, we include it as an
  additional free parameter in part of our analysis.
\end{itemize}

Because of the recent developments in the $H \to b\bar{b}$ channels we
briefly mention other strategies to see this Higgs decay at the LHC:
first, there might be a faint chance to observe the $b \bar{b}$ decay
directly in weak boson fusion. Producing a single Higgs boson might
lead to a $2.9-5.2\sigma$ signal with $600~\ifb$ of triggered
data~\cite{wbf_b}. Secondly, we can look for a Higgs boson produced in
association with a $W$ boson in weak boson fusion. This signature has
been estimated to lead to a $4.4\sigma$ signal with $200~\ifb$ of data
including a functioning central jet veto~\cite{wbf_wh}. The channel has been
reanalysed recently leading to a significance $S/\sqrt{B}$ 
of about 1.8~\cite{wbf_whnew} for $100~\ifb$. Both of these
analyses we do not include due to their limited potential during low
luminosity running and a lack of full simulation. Last but not least,
we can look for a hard photon and a Higgs in weak boson
fusion~\cite{wbf_gammah}. For this channel we might expect 
a $1.2\sigma$ significance for an integrated luminosity of $30~\ifb$
with a signal-to-background ratio around 1/20. The inclusion of a mini--jet veto,
studied qualitatively in Ref.~\cite{wbf_gammah},
would lead to an improvement by a factor~$2$~\cite{wbf_gammah}.
When discussing the
$t\bar{t}H$ production channel with a Higgs decay to bottoms we should
therefore keep in mind that its impact could be improved significantly
after including $H \gamma$ production or similar channels in weak
boson fusion.\bigskip

Note that including signal and background contributions to counting
experiments makes it obvious that the rate measurements discussed
above only rarely fulfill the ideal condition $S \gg B$. Instead, for
gluon-fusion signatures we find that the number of signal events tends
to be significantly smaller than the number of background events, with
the positive aspect that for example in the $H \to \gamma\gamma$
channel a side bin analysis allows us to very precisely subtract the
background events. For weak boson fusion signatures the ratio $S/B$
typically looks better, but on the down side the event numbers are
comparably small, which in some channels forces us into the Poisson
region.

In our numerical analysis we use predictions for the production rates
as described in Ref.~\cite{spira_hqq}. These we table the points and
interpolate between them. For the branching ratios we use
Hdecay~\cite{hdecay}, modified to allow for arbitrary Higgs coupling
values. To estimate the effect of cuts, we compute an efficiency
factor from our rates and the values given in
Ref.~\cite{duehrssennote}. Compared to this study we reduce the signal
rate for $t\bar{t}H, H\to b\bar{b}$ by 50\%, in agreement with recent
Atlas~\cite{tth_b,atlas_tdr} and CMS~\cite{cms_tdr} studies.  For the $WH/ZH,
H\to b\bar{b}$ channel including a subjet analysis we use the rates
given in the original paper~\cite{subjet} and check for an effect of a
degradation through detector effects in Section~\ref{sec:bottom}.

\TABLE[b]{
\begin{small} 
\begin{tabular}{l|l||r|r@{ }l|r||r|r}
 production & decay & 
 $S+B$ & 
 \multicolumn{2}{|r|}{$B$ }& 
 $S$ & 
 $\Delta S^\text{(exp)}$ &  
 $\Delta S^\text{(theo)}$ \\ \hline
 $gg \to H$ & $ZZ$ & 
  13.4 & 6.6 & ($\times$ 5) & 6.8 & 3.9 & 0.8 \\
 $qqH$ & $ZZ$ & 
  1.0  & 0.2 & ($\times$ 5) & 0.8 & 1.0 & 0.1 \\
 $gg \to H$ & $WW$ & 
  1019.5 & 882.8 & ($\times$ 1) & 136.7 & 63.4 & 18.2 \\
 $qqH$ & $WW$ & 
  59.4 & 37.5 & ($\times$ 1) & 21.9 & 10.2 & 1.7 \\
 $t\bar{t}H$ & $WW (3 \ell)$ & 
  23.9 & 21.2 & ($\times$ 1) & 2.7 & 6.8 & 0.4 \\
 $t\bar{t}H$ & $WW (2 \ell)$ & 
  24.0 & 19.6 & ($\times$ 1) & 4.4 & 6.7 & 0.6 \\
 inclusive & $\gamma\gamma$ & 
  12205.0 & 11820.0 & ($\times$ 10) & 385.0 & 164.9 & 44.5 \\
 $qqH$ & $\gamma\gamma$ & 
  38.7 & 26.7 & ($\times$ 10) & 12.0 & 6.5 & 0.9 \\
 $t\bar{t}H$ & $\gamma\gamma$ & 
  2.1 & 0.4 & ($\times$ 10) & 1.7 & 1.5 & 0.2 \\
 $WH$ & $\gamma\gamma$ & 
  2.4 & 0.4 & ($\times$ 10) & 2.0 & 1.6 & 0.1 \\
 $ZH$ & $\gamma\gamma$ & 
  1.1 & 0.7 & ($\times$ 10) & 0.4 & 1.1 & 0.1 \\
 $qqH$ & $\tau\tau (2 \ell)$ & 
  26.3 & 10.2 & ($\times$ 2) & 16.1 & 5.8 & 1.2 \\
 $qqH$ & $\tau\tau (1 \ell)$ & 
  29.6 & 11.6 & ($\times$ 2) & 18.0 & 6.6 & 1.3 \\
 $t\bar{t}H$ & $b\bar{b}$ & 
  244.5 & 219.0 & ($\times$ 1) & 25.5 & 31.2 & 3.6 \\
 $WH/ZH$ & $b\bar{b}$ & 
 228.6 & 180.0 & ($\times$ 1) & 48.6 & 20.7 & 4.0 
\end{tabular}
\end{small} \vspace*{0mm}
\caption[]{Signatures included in our analysis for a Higgs mass of
  120~GeV. The Standard Model event numbers for $30~\ifb$ include
  cuts~\cite{duehrssennote}. The factor after the background rates
  describes how many events are used to extrapolate into the signal
  region. The last two columns give the one-sigma experimental and
  theory error bars on the signal.}
\label{tab:channels}
}

\subsection{Errors}
\label{sec:errors}

In order to obtain reliable error estimates for the Higgs sector
parameters, a proper treatment of experimental and theory errors
depending on their origin is mandatory. We follow the
CKMfitter prescription or Rfit scheme~\cite{ckmfitter}.
The complete set of errors includes statistical experimental errors,
systematic experimental errors, and theory errors. The statistical
experimental errors are treated as uncorrelated in the measured
observables. In contrast, the systematic experimental errors for
example from the luminosity or from tagging efficiencies are fully
correlated.

As efficiency factors are usually determined from other observed
channels, they really behave like statistical errors and should
therefore be Gaussian. The numerical values for systematic errors we
take from Ref.~\cite{duehrssennote}. For convenience, we have
included them in Table~\ref{tab:syst_error}.\bigskip

\TABLE[b]{
\begin{small}
\begin{tabular}{l|r}
luminosity measurement & 5 \% \\\hline
detector efficiency & 2 \% \\\hline
lepton reconstruction efficiency & 2 \% \\\hline
photon reconstruction efficiency & 2 \% \\\hline
WBF tag-jets / jet-veto efficiency & 5 \% \\\hline
$b$-tagging efficiency & 3 \% \\\hline
$\tau$-tagging efficiency (hadronic decay) & 3 \% \\\hline
lepton isolation efficiency ($H \rightarrow 4\ell$) & 3 \% 
\end{tabular}
\hspace*{12ex}
\begin{tabular}{l|r|l}
 & 
 $\Delta B^\text{(syst)}$ &  
 corr. \\ \hline
 $H \to ZZ$ & 
  $1 \%$ & yes \\
 $H \to WW$ & 
  $5 \%$ & no \\
 $H \to \gamma\gamma$ & 
  $0.1 \%$ & yes \\
 $H \to \tau\tau$ & 
  $5 \%$ & yes \\
 $H \to b\bar{b}$ & 
  $10 \%$ & no \\
\end{tabular}
\end{small}
\caption[]{Systematic errors used in our analysis. Left: systematic
  errors applying to both signal and background. Reconstruction and
  tagging efficiencies are defined per particle, \eg $H \to
  \gamma\gamma $ has a $4 \%$ error on the photon
  reconstruction. Right: systematic background errors, either fully
  correlated or independent between channels.  Tables same as
  Ref.~\cite{duehrssennote}.}
\label{tab:syst_error}
}

In contrast, theory errors reflecting unknown higher orders in
perturbation theory should not be Gaussian but flat box--shaped within
a certain range of validity of perturbation theory. In other words,
the probability assigned to any measurement does not depend on its
actual value, as long as it is within the interval covered by the
theory error. There could be a tail attached to these theory--error
distributions, but higher--order corrections are definitely not
allowed to become arbitrarily large. Confronted with a perturbatively
unstable observable one would instead have to rethink the perturbative
description of the underlying theory. The numerical size of the theory
errors we give in Table~\ref{tab:theo_error}. The branching ratios to
$c\bar{c}$ and $gg$ do not enter in any measurements but indirectly
via the total width of the Higgs boson. To be consistent with our
approach of flat theory errors the propagation of errors is
linear and not quadratic.

\TABLE[b]{
\begin{small}
\begin{tabular}{l|r}
$\sigma$ (gluon fusion)                   & 13 \% \\\hline
$\sigma$ (weak boson fusion)              & 7 \% \\\hline
$\sigma$ ($VH$-associated)                & 7 \% \\\hline
$\sigma$ ($t\bar{t}$-associated)          & 13 \% 
\end{tabular}
\hspace*{30ex}
\begin{tabular}{l|r}
$\text{BR}(H \rightarrow ZZ)$             & 1 \% \\\hline
$\text{BR}(H \rightarrow WW)$             & 1 \% \\\hline
$\text{BR}(H \rightarrow \tau\bar{\tau})$ & 1 \% \\\hline
$\text{BR}(H \rightarrow c\bar{c})$       & 4 \% \\\hline
$\text{BR}(H \rightarrow b\bar{b})$       & 4 \% \\\hline
$\text{BR}(H \rightarrow \gamma\gamma)$   & 1 \% \\\hline
$\text{BR}(H \rightarrow Z\gamma)$        & 1 \% \\\hline
$\text{BR}(H \rightarrow gg)$             & 2 \% 
\end{tabular}
\end{small}
\caption[]{Theory errors used in our analysis for a 120~GeV Higgs.}
\label{tab:theo_error}
}

The error due to a mini-jet veto survival probability in background
processes behaves like any other efficiency, \ie Atlas and CMS measure
it in signal-free phase space regions and assign a Gaussian
error. However, for the signal rate we have to rely on theory to
predict the survival probability. Hence, it is a fully correlated flat
error which we can add to the error on the production cross section in
weak boson fusion.\bigskip

Defining a scheme for flat theory errors includes their combination
with the (Gaussian) experimental errors. A simple (Bayesian)
convolution leads to the difference of two one--sided error functions
with a clear maximum, so the convolution knows about the central value
of theory prediction.  A better solution is a distribution which is
flat as long as the measured value is within the theoretically
acceptable interval and then drops off like a Gaussian with the width
of the experimental error, the Rfit profile likelihood
construction~\cite{ckmfitter}. The log--likelihood $\chi^2 = -2 \log
\mathcal{L}$ given a set of measurements $\vec d$ and in the presence
of a general correlation matrix then $C$ reads
\begin{alignat}{7}
\chi^2     &= {\vec{\chi}_d}^T \; C^{-1} \; \vec{\chi}_d  \notag \\ 
\chi_{d,i} &=
  \begin{cases}
  0  
          &|d_i-\bar{d}_i | <   \sigma^{\text{(theo)}}_i \\
  \frac{\D d_i-\bar{d}_i+ \sigma^{\text{(theo)}}_i}{\D \sigma^{\text{(exp)}}_i}
          & \; d_i-\bar{d}_i\; < - \sigma^{\text{(theo)}}_i \\
  \frac{\D d_i-\bar{d}_i- \sigma^{\text{(theo)}}_i}{\D \sigma^{\text{(exp)}}_i}
  \qquad  & \; d_i-\bar{d}_i\; >   \sigma^{\text{(theo)}}_i \; ,
  \end{cases}
\label{eq:flat_errors}
\end{alignat}
where $\bar{d}_i$ is the $i$-th data point predicted by the model
parameters and $d_i$ the actual measurement. 
Flat errors lead to
a technical complication with hill--climbing algorithms. Functions
describing box--shaped error distribution will have a discontinuities
of higher derivatives. The Rfit scheme has a step in the second
derivative which we need to accommodate. As a second complication the
log-likelihood becomes constant in the central region, so some
parameters vanish from the counting of degrees of
freedom~\cite{sfitter}.
If all experimental errors are Gaussian, $\sigma^{\text{(exp)}}$ 
of eq.~(\ref{eq:flat_errors}) is the convolution of the errors, 
\ie the square-root of the sum of the squares of the experimental errors.
The off-diagonal elements of the correlation matrix $C$ are constructed 
by a generalization of eq.~(2) of~\cite{sfitter}.
\bigskip

The uncertainty on the mass of top and bottom quarks is another source
of errors which we need to take into account. For these measurements we
use a single experimental error each which comprises all different
sources and is uncorrelated to any other measurement. Its numerical size
is $1.0$ and $0.07\gev$ for top and bottom quarks, respectively.
\bigskip

For the experimental errors the systematic parts are inferred from
large data samples, and it is safe to assume them as Gaussian.  On the
other hand, in some of the Higgs channels we will only observe a few
events, so we need to use a Poisson distribution to model the behavior
correctly. Therefore, we use Poisson statistics for all statistical
errors.  Our method of combining Gaussian and Poisson errors affecting
the same measurement we describe in detail in
Appendix~\ref{app:combine}.  Unless explicitly stated otherwise,
SFitter uses smeared toy measurements to evaluate the errors.  For
each toy measurement we determine the best-fit value, which means the
width of the distribution of the best-fit values gives the error on
the model parameters.

\section{General Higgs sector}
\label{sec:higgs}

As sketched earlier, we allow for a general Higgs sector including
couplings to all Standard--Model particles and an invisible final
state.  We assume that we observe a narrow 120~GeV Higgs candidate as a peak
either in the $\gamma \gamma$ or $ZZ$ invariant masses, in the
collinearly reconstructed invariant $\tau \tau$ mass, or in the
transverse mass of $WW$ pairs.  This most notably means that we are
searching for two-particle decays of the Higgs boson, which has an
important consequence: if the Higgs boson were to couple strongly to
light-flavor jets (which are swamped by QCD backgrounds) this would
reduce the fraction of Higgs bosons decaying into visible channels,
but it would also enhance the production cross section at the LHC. We
discuss such a scenario as well as invisible Higgs decays in
Section~\ref{sec:cch}.  As mentioned in the introduction, we will not
take into account Higgs couplings which might be visible in
high-luminosity running in a single channel and/or are not expected to
feed back into the determination of the leading Higgs parameters. This
includes the Higgs-muon Yukawa coupling, the Higgs self coupling, or
invisible Higgs decays. Such channels can be very useful to further
understand the Higgs sector, but their results can be considered
purely additional to our analysis.\bigskip


For the tree-level Standard Model Higgs couplings to any particle $j$
we allow for a deviation
\begin{equation}
 g_{jjH} \longrightarrow g_{jjH}^{\text{SM}} \; \left( 1 + \Delta_{jjH}
                                   \right)
\label{eq:Coupl_SM_Change}
\end{equation}
where the $\Delta_{jjH}$ are independent of each other. If necessary
redefining our Higgs field we can without loss of generality assume
that one of the tree-level Higgs couplings is positive, \ie
$g_{WWH}>0$ or $\Delta_{WWH}>-1$.  Note that these couplings are of
mass dimension three for gauge bosons and four for fermions, so they
are part of the renormalizable Lagrangian. In the minimal one-doublet
Higgs sector such corrections spoil gauge invariance. This can be
cured for example by additional heavy Higgs states which we integrate
out to define an effective one-Higgs-doublet Standard Model with free
couplings.

In addition, there are three relevant loop induced couplings in the
Higgs sector: $g_{ggH}, g_{\gamma\gamma H}$ and $g_{\gamma ZH}$. Such
couplings are sensitive to new particles in the spectrum, like for
example supersymmetry or a chiral fourth generation. In the latter
case the Tevatron limits on $gg \to H \to WW$ strongly constrain the
allowed Higgs masses in the presence of such a heavy generation which
leads to an enhancement of the production rate roughly by a
factor~9~\cite{fourth_gen}. Including a general new-physics scale
$\Lambda$~\cite{survey,higgs_pot} such couplings arise in the
effective Lagrangian as
\begin{equation}
 \mathcal{L} \supset 
  \frac{g_{ggH}}{\Lambda} \; (G^a)^{\mu \nu}  (G^a)_{\mu \nu} 
+ \frac{g_{\gamma\gamma H}}{\Lambda} \; F^{\mu \nu}  F_{\mu \nu} 
\label{eq:coupl_dim5}
\end{equation}
Note that in general new-physics models these couplings do not have to
be perturbatively suppressed, as they are in the Standard Model or in
the MSSM. In our weak-scale analysis we consider two different sources
of higher-dimensional couplings: first, there are known Standard Model
particles propagating in loops and inducing them, so any change in the
tree-level couplings eq.(\ref{eq:Coupl_SM_Change}) propagates into the
one-loop effective couplings. Secondly, we allow for manifestly
dimension-five operators from new physics
\begin{equation}
 g_{jjH} \longrightarrow 
  g_{jjH}^{\text{SM}} \; \left(
   1 + \Delta_{jjH}^{\text{SM}} + \Delta_{jjH} \right)
\label{eq:Coupl_SMeff_Change}
\end{equation}
where $g_{jjH}^{\text{SM}}$ is the loop-induced coupling in the
Standard Model, $\Delta_{jjH}^{\text{SM}}$ is the contribution
from modified tree-level couplings to Standard-Model particles, and
$\Delta_{jjH}$ is an additional dimension-five contribution, for
example from new heavy states.\bigskip

The ansatz described above describes our general effective theory of
the Higgs sector, allowing for deviation from the Standard Model
values. Such deviations will be loop induced even in the Standard
Model. If we allow for new physics in electroweak symmetry breaking
such loop effects can be enhanced by large loop corrections, a
strongly interacting sector mimicking a scalar Higgs boson, or an
extended Higgs sector. We can illustrate this using $g_{WWH}$: if we
were to measure a smaller value than the Standard Model prediction
this could point to additional Higgs fields sharing the burden of
creating the correct gauge boson masses $v^2 = v_1^2 + v_2^2 + \cdots$
This kind of behavior we know from the type-II two-Higgs-doublet model
of the MSSM where the couplings of the two Higgs scalars to the weak
gauge bosons are split according to $\sin^2 (\beta - \alpha) + \cos^2
(\beta - \alpha) =1$.

A too large value of $g_{WWH}$, in contrast, could point towards the
existence of an additional higher-dimensional operator. The way to
check such a hypothesis would involve an analysis of angular
correlations which can distinguish different tensor structures of the
$WWH$ coupling in weak boson fusion, independent of the decay
channel~\cite{wbf_vertex}.\bigskip

One basic assumption we have to make for the Higgs sector analysis is
the treatment of the total width, which can be varied just like the
couplings
\begin{equation} 
\Gamma_\text{tot} = \Gamma^\text{SM}_\text{tot} \; ( 1+\Delta_\Gamma )
\quad .
\label{eq:deltagamma}
\end{equation} 
Allowing for a simultaneous scaling of all couplings and the total
width we see that as long as we keep $g^4/\Gamma_\text{tot} = C$
finite and constant, none of our $(\sigma \cdot \br)$ measurements
will be affected and the shift in the couplings will not be
observable. Keeping this ratio and hence all measurements unaffected
implies that for example a theory-motivated upper limit on any of the
relevant couplings translates into an upper limit on the total Higgs
width, and vice versa.

Limiting ourselves to Higgs singlets and doublets we can derive such
an upper limit on $g_{WWH}$ from the perturbative unitarization of $WW
\to WW$ scattering. As seen in the last paragraph, the unitarization
of this process can be shared by several Higgs states, but since their
couplings to gauge bosons enter twice per amplitude it leads to
strictly a sum of the individual contributions. In other words, no
individual $WWH_j$ coupling of any of the Higgs states $j$
contributing can be larger than the value which alone unitarizes this
process, \ie $g_{WWH}^\text{SM}$. This upper limit on $g_{WWH}$ can be
translated into an upper limit on the total width.  As mentioned in
the last paragraph, this argument is limited to dimension-three $WWH$
operators and for example breaks down in the limit of strong
dimension-five interactions as shown in
eq.(\ref{eq:Coupl_SMeff_Change}).

In a similar manner, we can obtain a lower limit on the Higgs width:
obviously, the total width is at least as large as the sum of all
observable partial widths. However, these partial widths are computed
from the extracted couplings, and the couplings can be dialed to zero
as long as the total width vanishes simultaneously. 
We can insert the scaling behavior $\Gamma_\text{vis} \propto g^2$
into the expression for $C$, add an additional positive contribution
$\Gamma_x(g)>0$ with an arbitrary $g$ dependence and compute the
small-coupling limit
\begin{equation}
C = 
\lim_{g^2 \to 0} \; 
\frac{g^4}{\Gamma_\text{tot}}
= 
\lim_{g^2 \to 0} \; 
\frac{g^4}{g^2 (\Gamma_\text{vis}/g^2) + \Gamma_x} 
= 0 \quad .
\end{equation}
However, by definition this ratio $C$ has to be finite to protect our
rate measurements, which means that we cannot actually dial down all
partial widths simultaneously.  Instead, we have to consider a
quadratic equation in $g^2$ which indeed returns a lower limit on the
couplings and hence the total width.  It is obvious for any such
argument that the knowledge of as many as possible partial widths and
therefore the observation of $H \to b\bar{b}$ decays is crucial for
studies like this.\bigskip

For a light Standard-Model Higgs boson, the difference between the
observable partial width at the LHC and the Standard Model width is of
the order of $\lesssim 10\%$. Identifying the total width with the sum
of the observed partial widths is practically equivalent to including
an upper limit. Therefore, throughout this analysis we always simply
identify the total width with the sum of the observed partial
widths. Unless explicitly mentioned this includes all Standard Model
particles listed in Section~\ref{sec:measurements}. An example
scenario how this might mislead us we discuss in
Section~\ref{sec:cch}.

\section{Higgs likelihood map}
\label{sec:likelihood}

Due to the number of possible observable channels for a light Higgs
boson, the LHC should be able to not just discover a narrow scalar
state which might or might not be the Standard--Model Higgs boson, but
tell if such a state has anything to do with the Standard Model Higgs
boson. This is a statistical statement, in which all errors and the
resulting confidence level for a given hypothesis play a crucial
role. Therefore, it is vital to control the mapping of the
high--dimensional space of correlated measurements onto the
high--dimensional Higgs parameter space. Just like for any other
parameter study, the starting point of such a study is a completely
exclusive likelihood map.\bigskip

Such a likelihood map is a large list or array of points in model
space with their corresponding probability to agree with the given
data.  We can construct it using Markov chains where the points in the
Markov chain represent the entire parameter space in their probability
to agree with a data set. In principle, we could evaluate any kind of
function on this Markov chain, similar to general mass scales or dark
matter properties in analyses of physics beyond the Standard
Model~\cite{ted,ben,leszek}. In this analysis we are only interested in
the probability to agree with data itself which in the case of
supersymmetric parameters led us to consider weighted Markov
chains~\cite{sfitter}. 

The problem with such an exclusive likelihood map is that it is (a)
hard to graphically represent and (b) hard to interpret once we for
example want to know if one individual coupling agrees with the
Standard Model or does not. However, to reduce the dimensionality of
the parameter space we have to decide on a scheme, namely Bayesian
probabilities or profile likelihoods. 

If we want to maintain the mathematical properties of a probability
distribution and compare the height and size of different peaks we
need to integrate over the unwanted directions, which means we need to
define an integration measure in model space. Such a measure is always
a free choice, a prior, and the resulting Bayesian probability
distribution will depend on it. 
The
model space integration also leads to volume effects, just like the
usual example --- is a water molecule more likely to be found in a
highly concentrated spoon of water or in a huge cloud? Moreover, the
integration will produce noise effects which can obscure narrow
structures in the parameter space.

Alternatively, we can construct a likelihood without introducing any
measure, for example by simply picking the best point in the unwanted
direction and identify it with the value in the reduced likelihood
map. The result is referred to as a profile likelihood, has no noise
effects and is designed to best represent the structures of the
original likelihood. However, it does not allow us to compare the
integrated size of different peaks of alternative solutions in the
parameter space.\bigskip

In our new--physics parameter analyses~\cite{sfitter} we have learned
that we can extract the most information by showing both, a Bayesian
probability as well as a profile likelihood, side by side, accepting 
that they do not have to give the same answers
because they are not the same question. In this analysis we will find
that correlations around a well-defined central fit value are more
important than distinct equally good solutions, which implies that we
will rely largely on profile likelihoods.

Due to the probability distribution function which chooses their next
point Markov chains are intrinsically Bayesian. Correspondingly, we
find that profile likelihoods are harder to extract. To improve this
behavior we can cool down the Markov chains to improve the resolution
of local structures, provided we ensure that we do not miss structures
in parameter space.  If we slowly dial down the random number which
decides whether a worse point is still accepted, we start scanning the
parameter space similar to a standard Markov chain and then slowly
concentrate on one structure. This cooling significantly improves the
resolution of local structures around a peak and thereby yields a much
better resolution for profile likelihoods.  More details of this
approach we give in Appendix~\ref{app:coolmc}.


\subsection{Parameters and correlations}
\label{sec:corr}

Given the set of measurements described in
Section~\ref{sec:measurements} it is obvious that most of the Standard
Model couplings should be accessible to a full analysis. Nevertheless,
we start with a minimal set of Higgs sector parameters in which we
only allow for tree-level couplings to all Standard Model
particles. This implies that there are no new particles contributing
to the effective $ggH$ and $\gamma\gamma H$ couplings. 
Since we compute the Higgs width as the sum of all visible
partial widths, a measurement of the bottom Yukawa constitutes the
main fraction of the Higgs width.\bigskip

\begin{figure}[t]
\begin{center}
 \includegraphics[width=0.24\textwidth]{%
     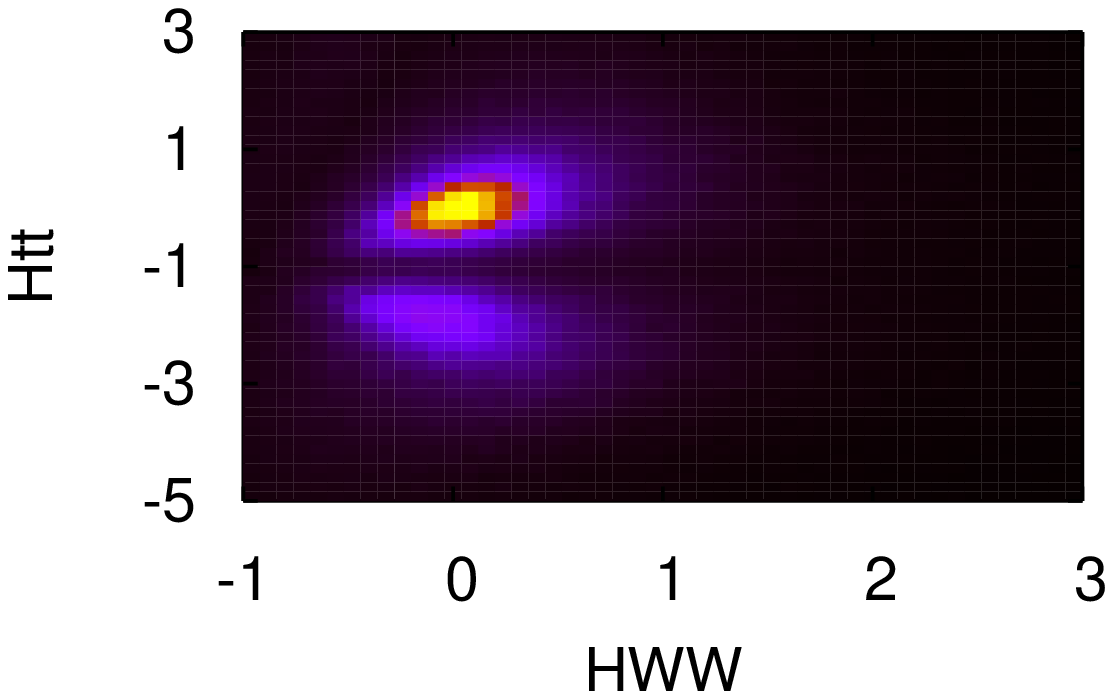} \hspace*{-2ex}
 \includegraphics[width=0.24\textwidth]{%
     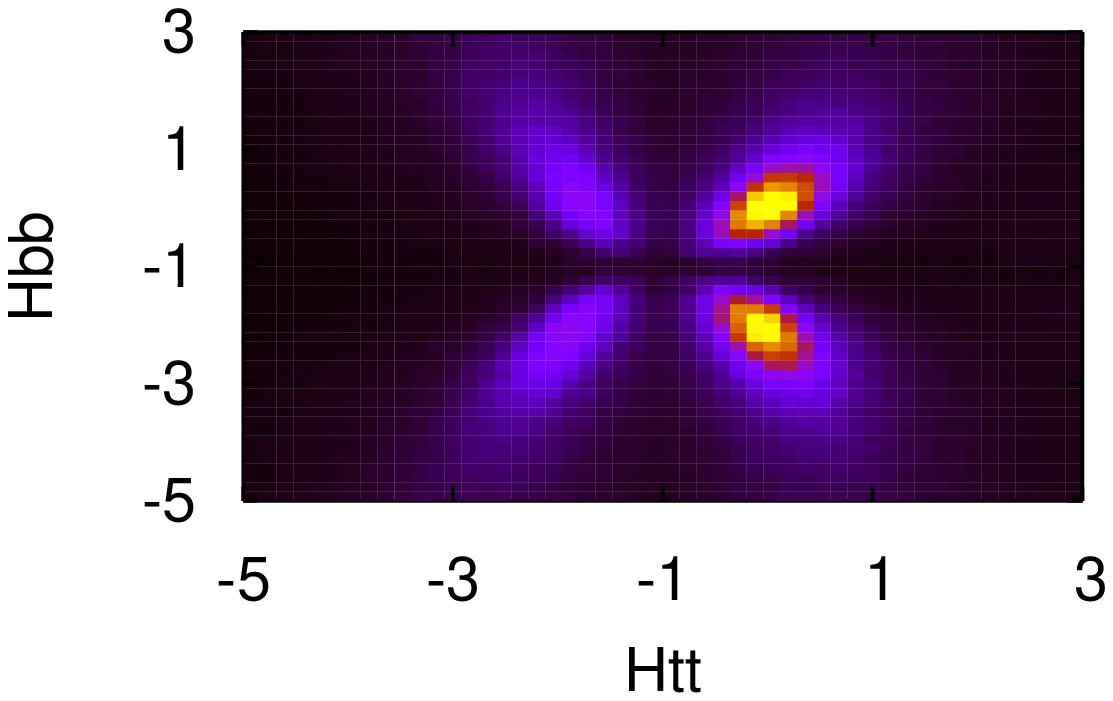} \hspace*{-2ex}
 \includegraphics[width=0.24\textwidth]{%
     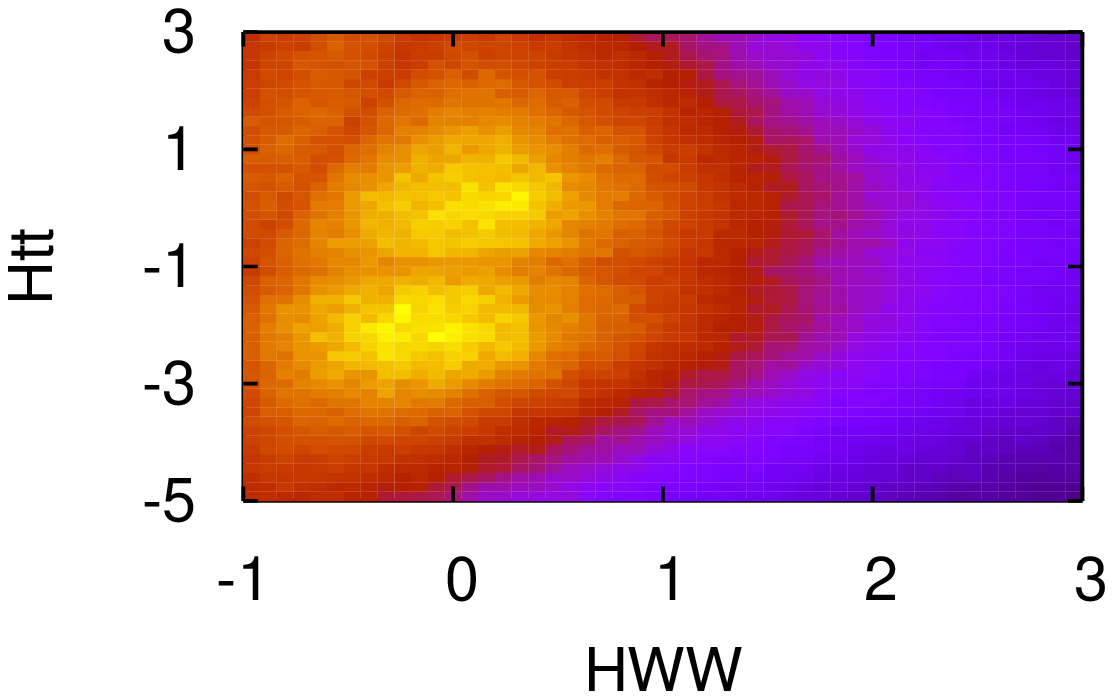} \hspace*{-2ex}
 \includegraphics[width=0.24\textwidth]{%
     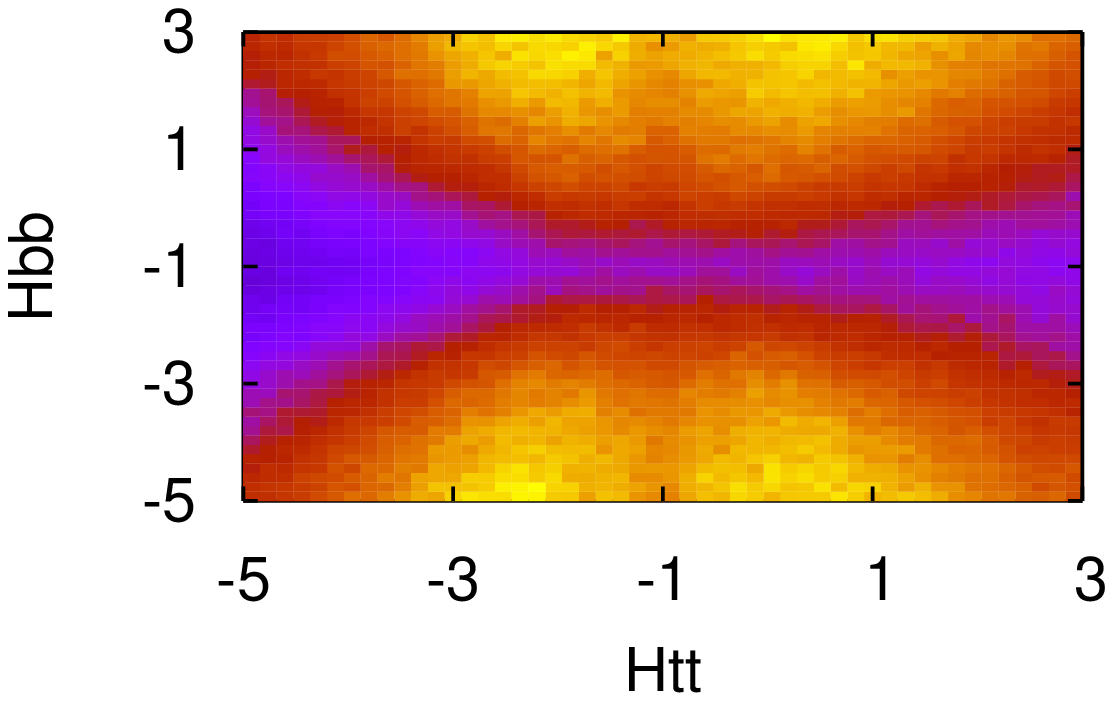} \\
 \includegraphics[width=0.24\textwidth]{%
     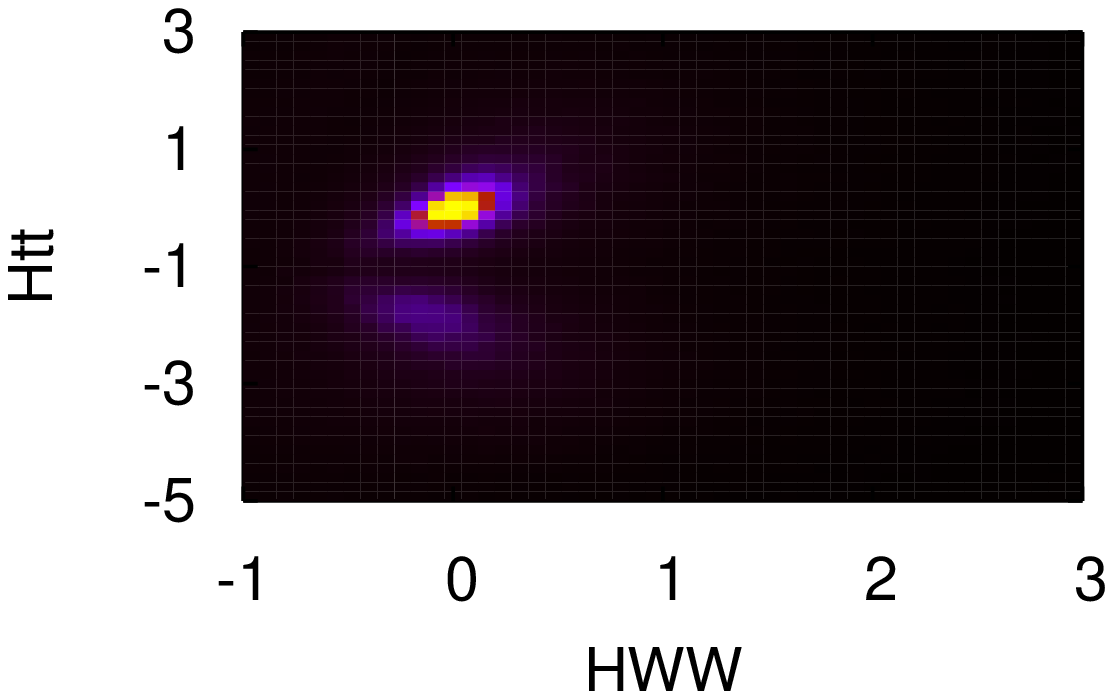} \hspace*{-2ex}
 \includegraphics[width=0.24\textwidth]{%
     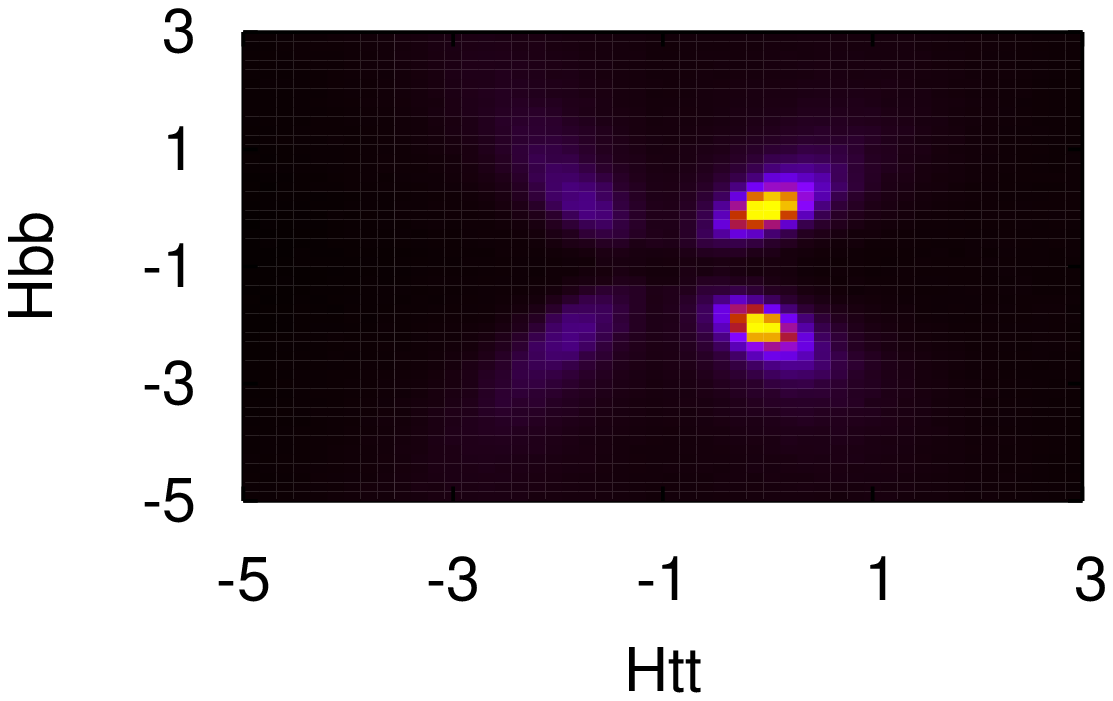} \hspace*{-2ex}
 \includegraphics[width=0.24\textwidth]{%
     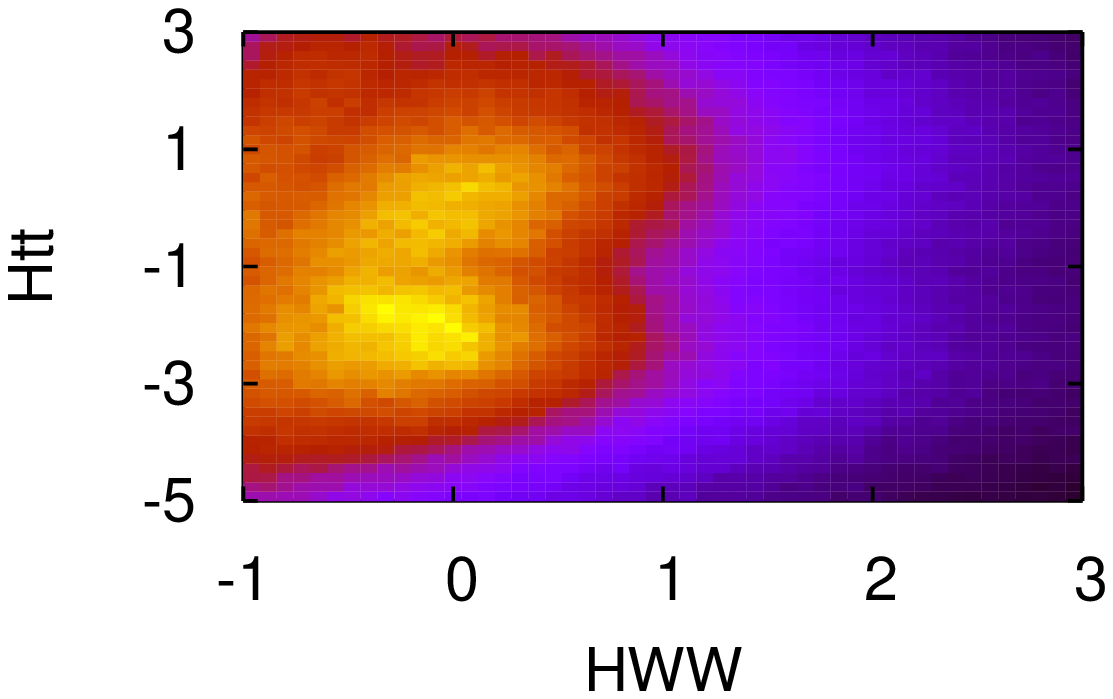} \hspace*{-2ex}
 \includegraphics[width=0.24\textwidth]{%
     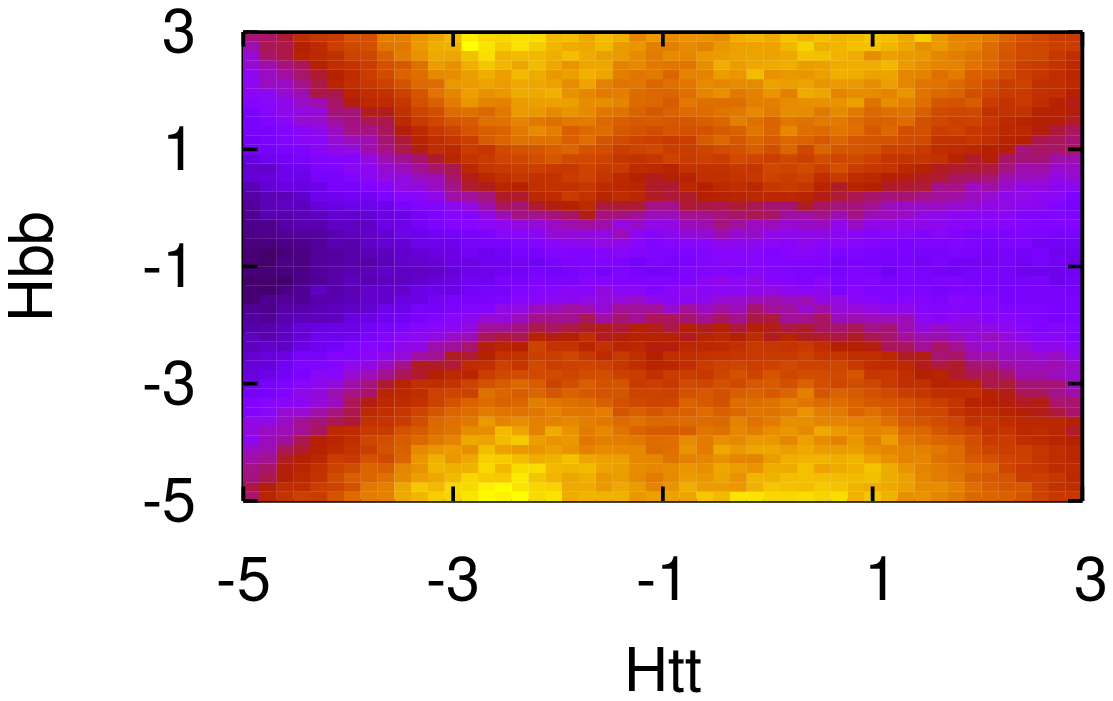} \\
 {\raggedright $1/\Delta \chi^2$\\[2ex]}
 \includegraphics[width=0.135\textwidth]{%
     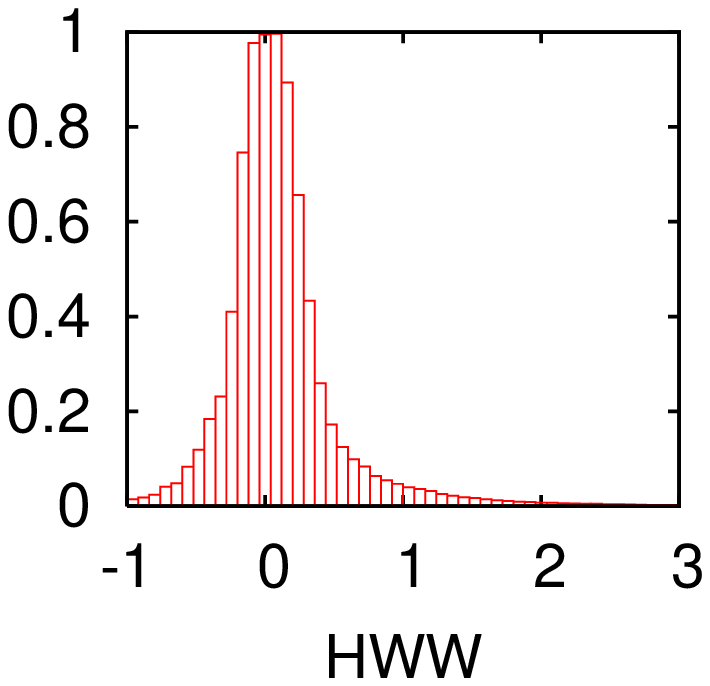} 
 \includegraphics[width=0.135\textwidth]{%
     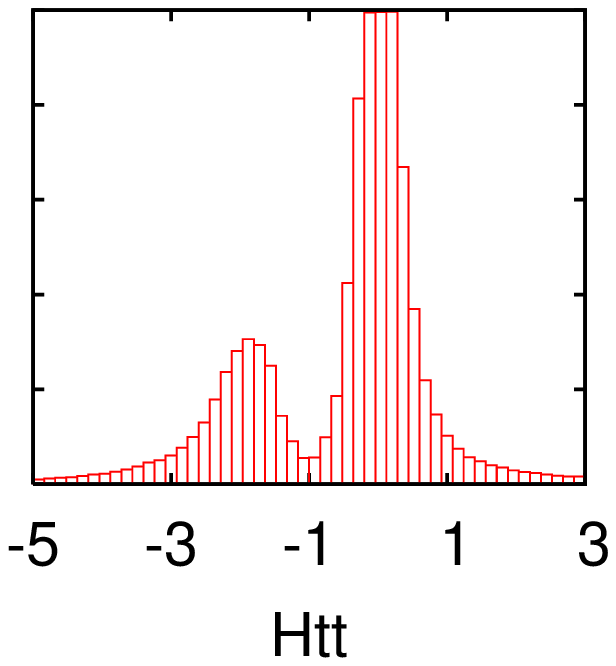} 
 \includegraphics[width=0.135\textwidth]{%
     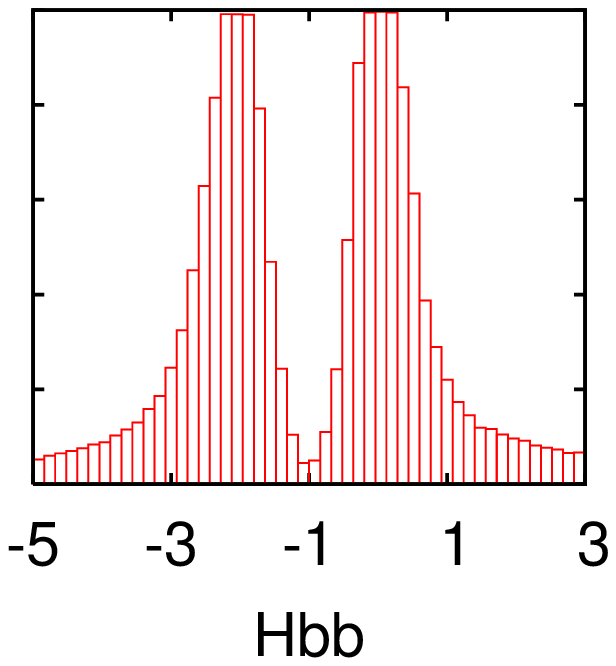} \hspace*{5ex}
 \includegraphics[width=0.135\textwidth]{%
     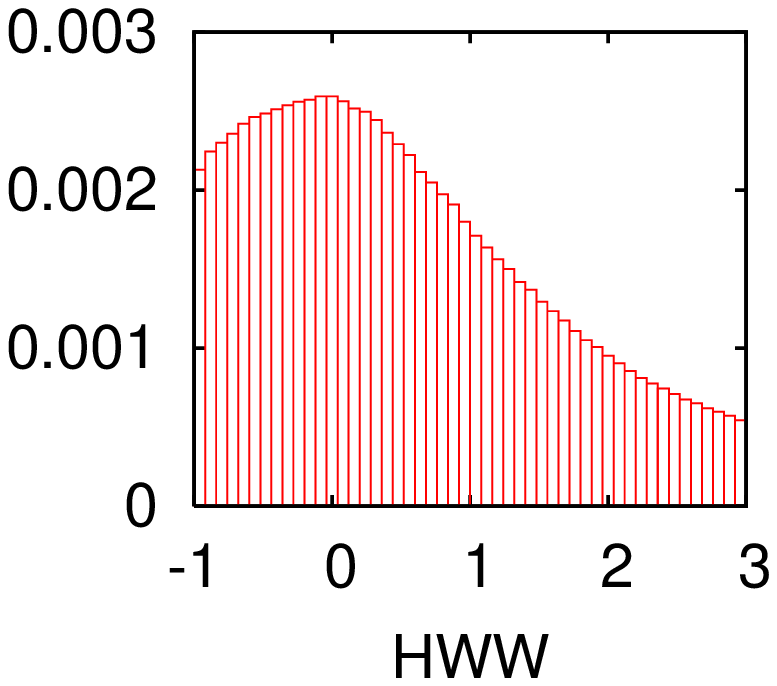} 
 \includegraphics[width=0.135\textwidth]{%
     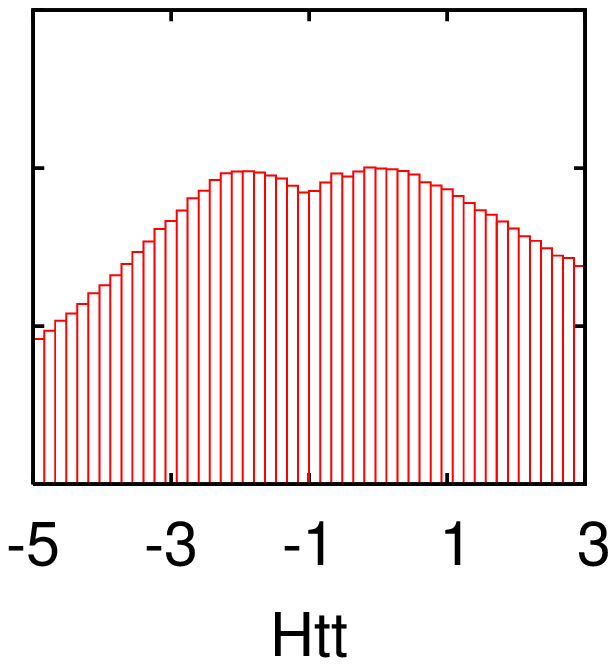} 
 \includegraphics[width=0.135\textwidth]{%
     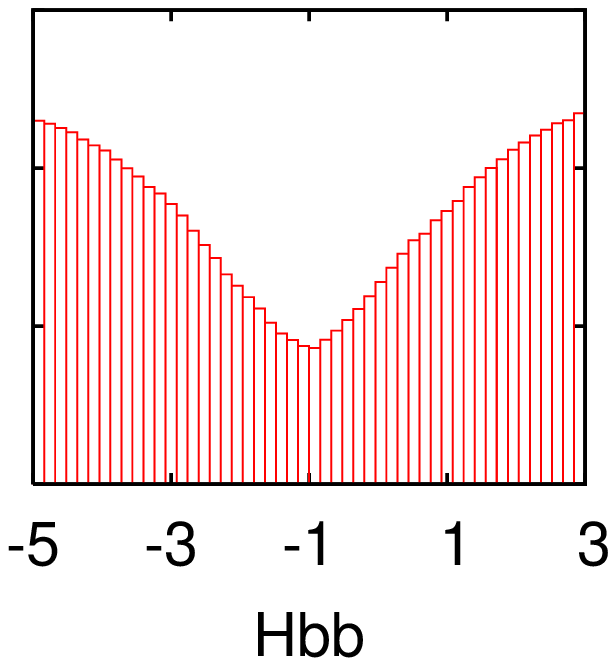} \\[2ex]
 \includegraphics[width=0.135\textwidth]{%
     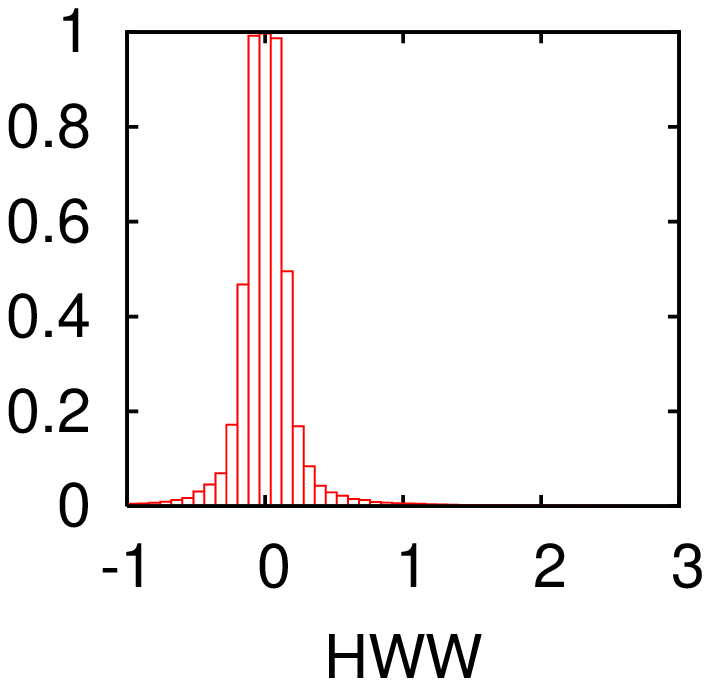} 
 \includegraphics[width=0.135\textwidth]{%
     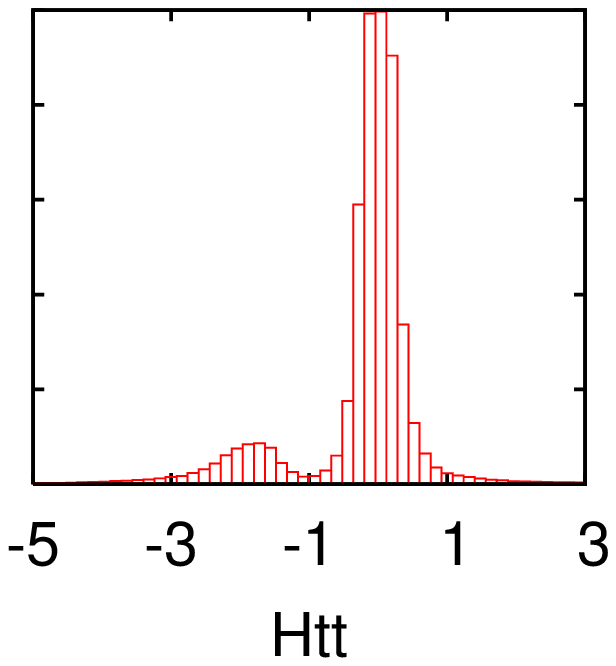} 
 \includegraphics[width=0.135\textwidth]{%
     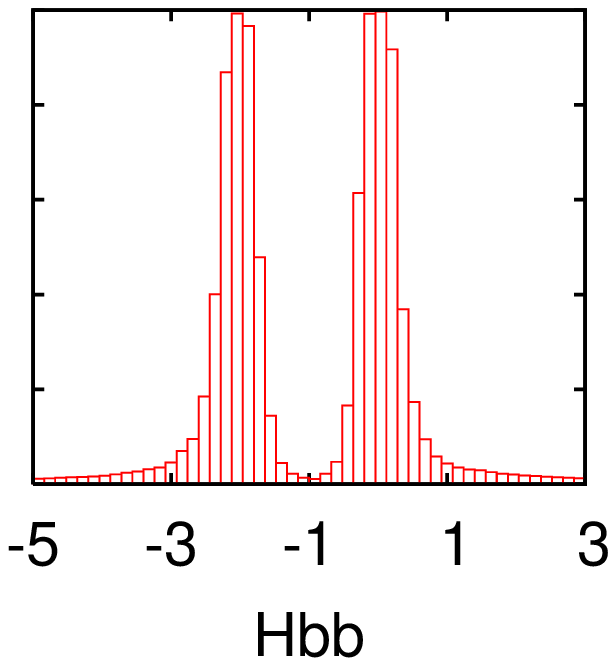} \hspace*{5ex}
 \includegraphics[width=0.135\textwidth]{%
     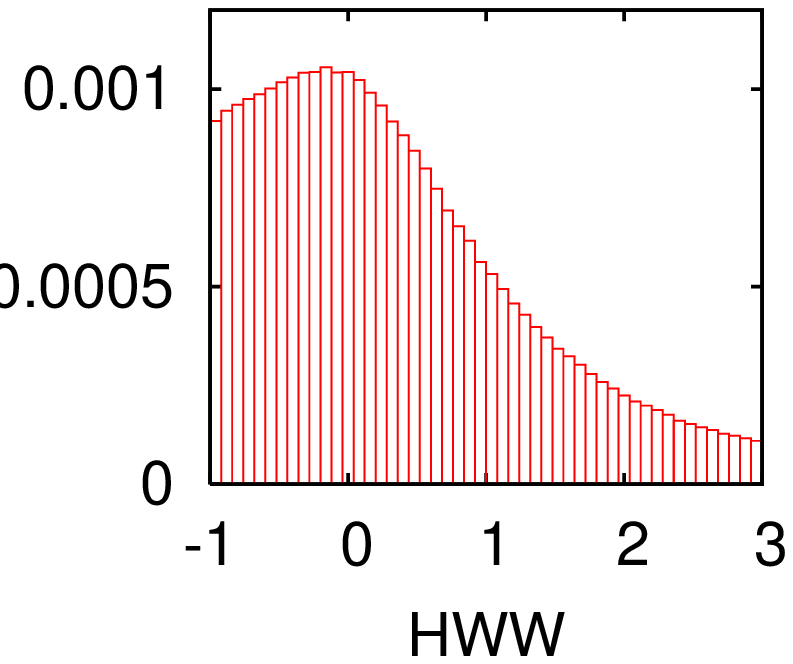} 
 \includegraphics[width=0.135\textwidth]{%
     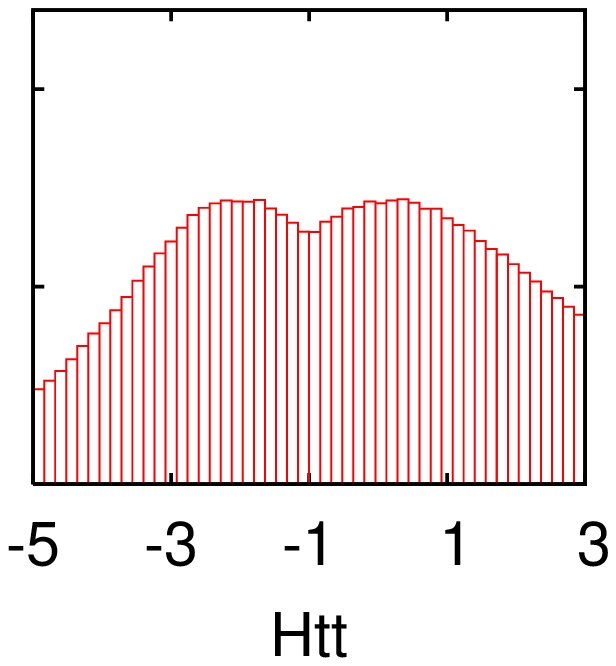} 
 \includegraphics[width=0.135\textwidth]{%
     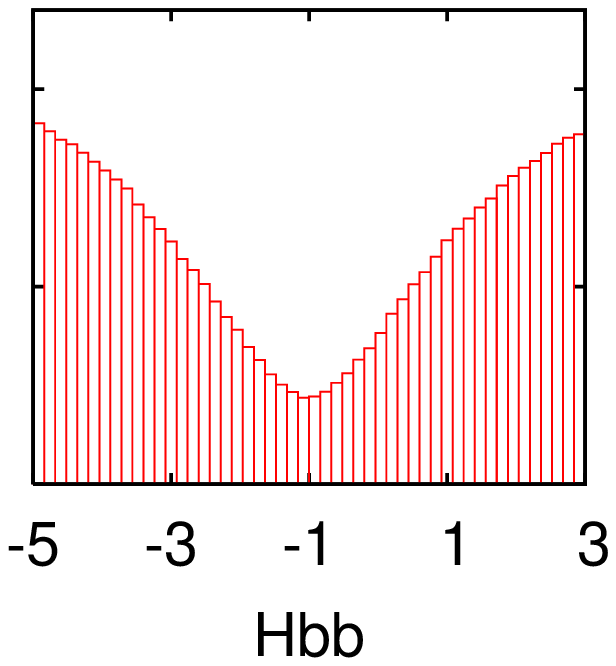} 
\end{center}
\caption[]{Profile likelihoods (left) and Bayesian probabilities
  (right) for the $WWH$, $ttH$, and $bbH$ couplings. Not allowing for
  additional $ggH$ or $\gamma\gamma H$ couplings we show results for
  $30~\ifb$ and for $300~\ifb$ in the upper and lower rows. The
  Higgs mass is chosen as 120~GeV. All
  experimental and theory errors are included. Here and in all other
  Figures we assume the $WWH$ coupling to be positive, \ie
  $\Delta_{WWH}>-1$.}
\label{fig:corr_1}
\end{figure}

Based on the studies of weak boson fusion we limit our study to
low-luminosity running and a conservative integrated luminosity of
$30~\ifb$. We can easily test how constraining this assumption is for
our analysis: without including any effective higher-dimensional
couplings we estimate the effect of an increase in luminosity from the
standard $30~\ifb$ to $300~\ifb$. Fig.~\ref{fig:corr_1} shows that the
existence of correlations is independent of the integrated luminosity,
\ie these correlations are physical, and even the relative weights of
alternative solutions hardly depend on the luminosity. 
Note that just like the entire analysis the error on the Higgs
rates does not decrease like $1/\sqrt{S}$ but that we always include
backgrounds, which makes $S+B$ the relevant number of events.\bigskip

Moving on, one striking feature in Fig.~\ref{fig:corr_1} is the
alternative minimum for a negative top Yukawa coupling. In the
comparison between profile likelihood and Bayesian probability we see
that while the best-fit point clearly prefers the correct sign of the
top Yukawa, volume effects wash out this distinction in the Bayesian
probability. The only term which breaks this sign degeneracy is the
loop-induced $g_{\gamma\gamma H}$, which only allows for a
synchronized sign flip of $g_{ttH}$ and $g_{WWH}$. The positive
correlation between these two couplings shown in the two-dimensional
panel of Fig.~\ref{fig:corr_1} arises on the one hand because of the
strong destructive interference of the two loops and on the other hand
due to a possible approximate scaling of all tree-level couplings
compensating a shift in the less well measured bottom Yukawa coupling.
A $g_{WWH}-g_{bbH}$ correlation shows the same kind of
positive correlations for the same two reasons.  As the $g_{bbH}$
Yukawa coupling is small, a small shift in $g_{WWH}$ or $g_{ttH}$
compensates a major change in $g_{bbH}$. However, such a large shift
in the bottom Yukawa coupling is not completely out of the world ---
in the supersymmetric two-Higgs-doublet model the additional parameter
$\tan\beta$ plays exactly this role.

From the one-dimensional histograms in Fig.~\ref{fig:corr_1} we see
that the LHC provides a very reasonable measurement of the top Yukawa
coupling via the effective $ggH$ coupling. The bottom Yukawa coupling
is better constrained than in previous analyses using the $t\bar{t}H$
production channel, if we take the subjet analysis at face
value~\cite{subjet}. The measurement of $g_{WWH}$ is a little less
precise than naively expected, given the weak-boson-fusion signatures
including ample information on this coupling. Comparing the
results for the two luminosities we see that the limiting factor for
the $WWH$ coupling is the statistics of the weak-boson-fusion
channels. For the gluon-fusion production process combined with a
decay $H \to WW$ we know that the kind of off-shell decay appearing
for light Higgs masses is essentially impossible to extract from the
sizable backgrounds.

Comparing the left and right set of panels, we see that in the
Bayesian probabilities volume effects completely hide any kind of sign
preference which might come out of the interference in the effective
$g_{\gamma\gamma H}$ and $g_{ggH}$.\bigskip

\begin{figure}[t]
 \begin{center}
 \includegraphics[width=0.24\textwidth]{%
     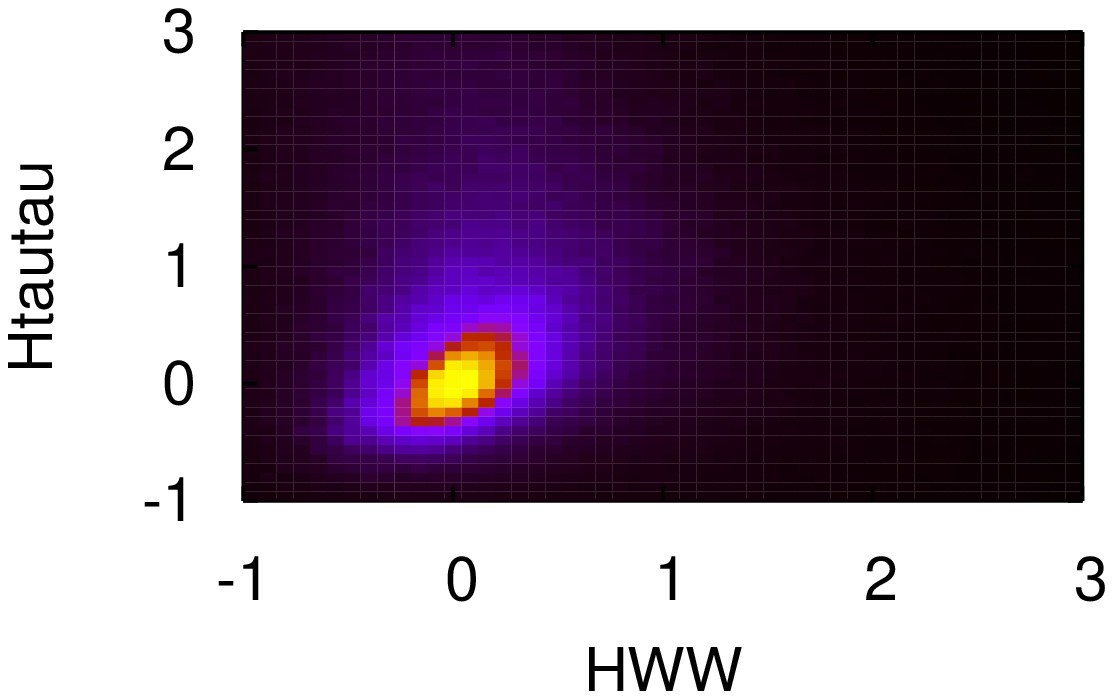} \hspace*{-2ex}
 \includegraphics[width=0.24\textwidth]{%
     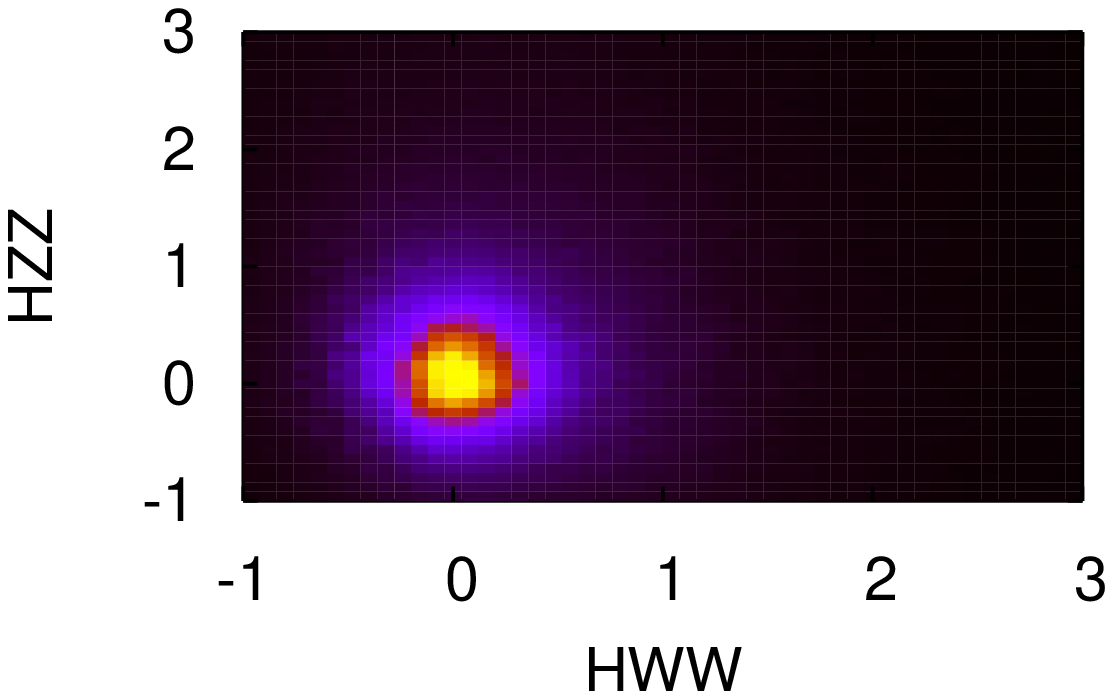} \hspace*{-2ex}
 \includegraphics[width=0.24\textwidth]{%
     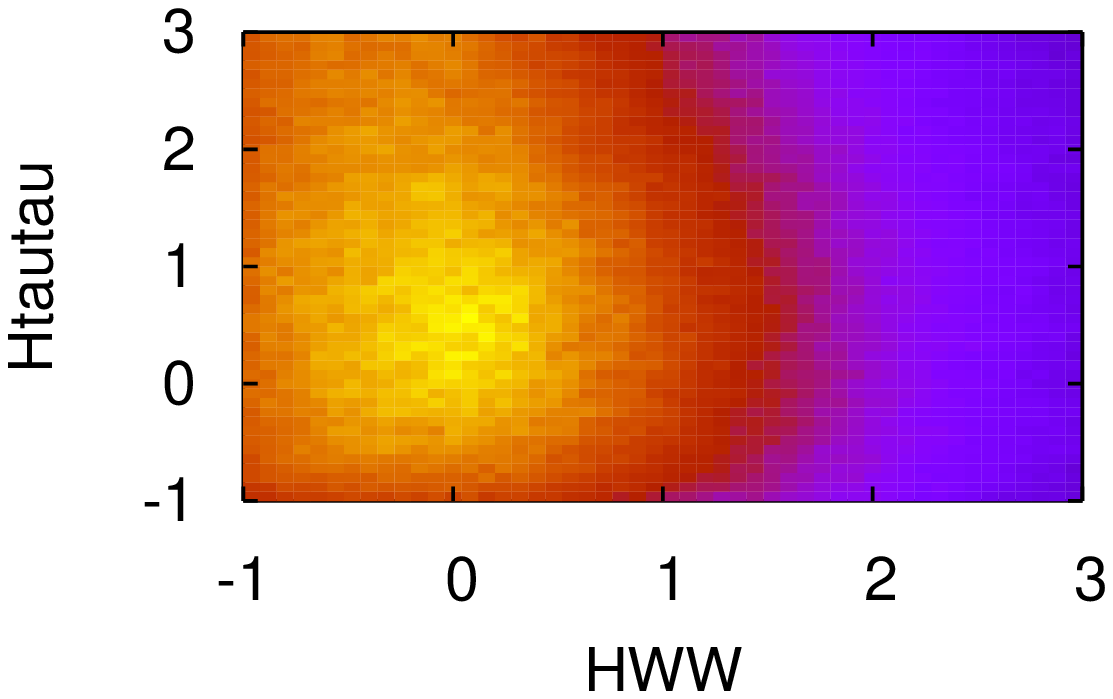} \hspace*{-2ex}
 \includegraphics[width=0.24\textwidth]{%
     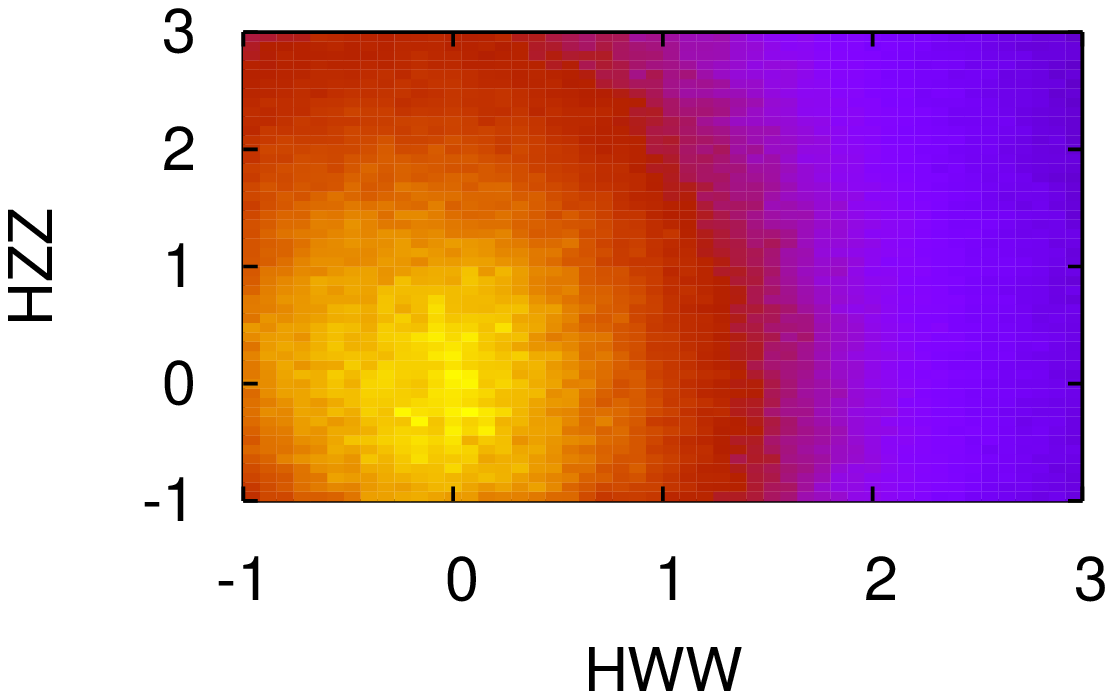} \\
 \includegraphics[width=0.24\textwidth]{%
     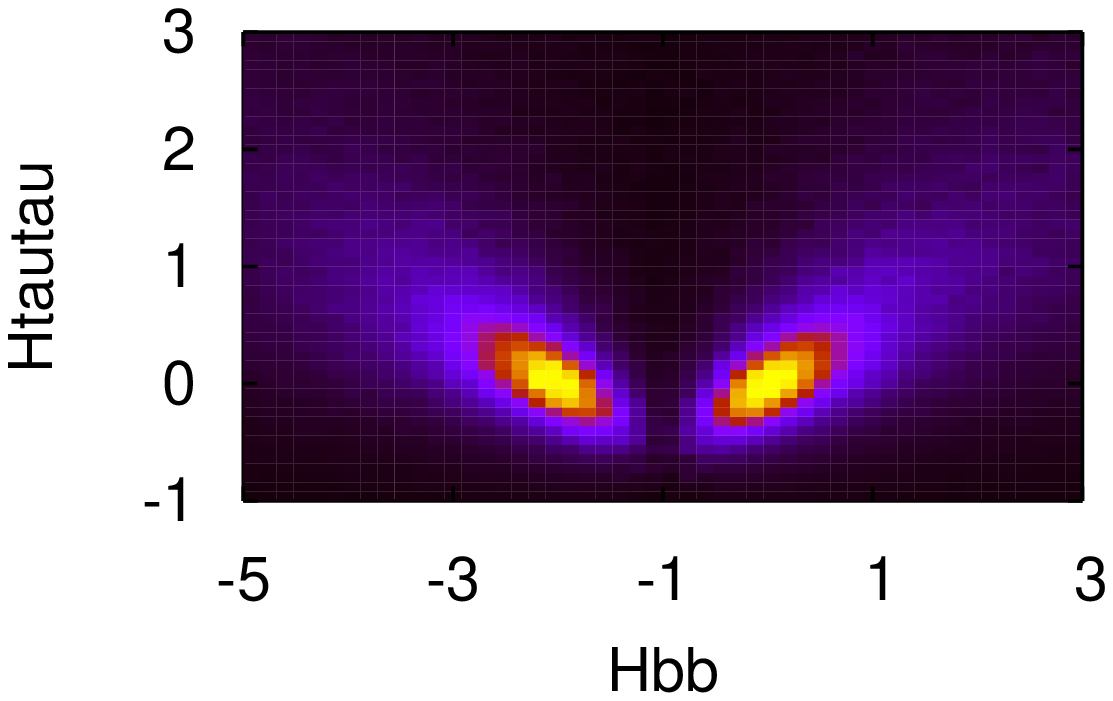} \hspace*{-2ex}
 \includegraphics[width=0.24\textwidth]{%
     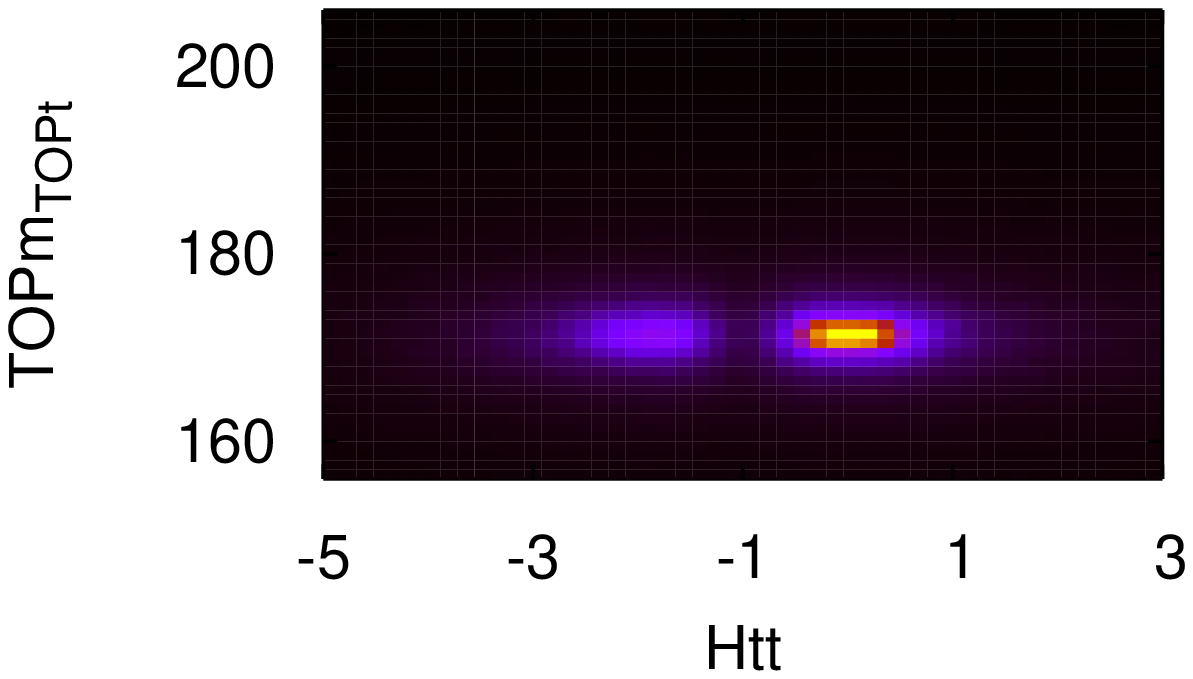}\hspace*{-2ex}
 \includegraphics[width=0.24\textwidth]{%
     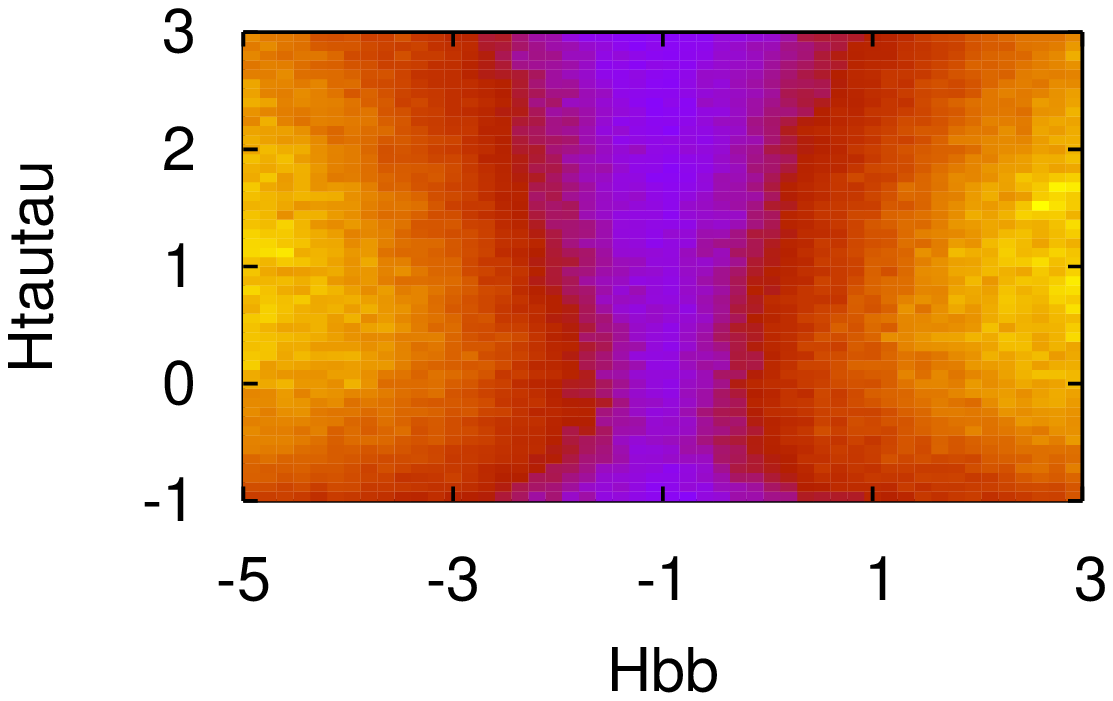} \hspace*{-2ex}
 \includegraphics[width=0.24\textwidth]{%
     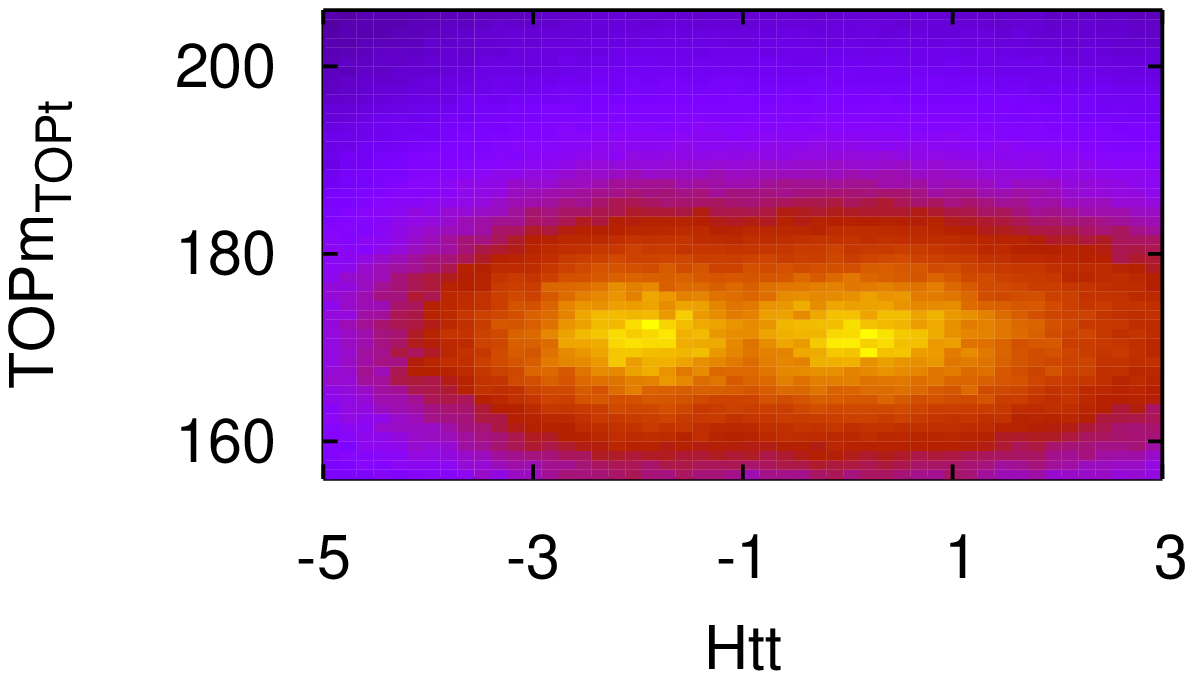} 
 \end{center}
\caption[]{Correlations in profile likelihoods (left) and Bayesian
  probabilities (right), not allowing additional effective
  couplings. All experimental and theory errors included for
  low-luminosity running.}
\label{fig:corr_2}
\end{figure}

To highlight the importance of two-dimensional correlations, we show
an extended set of them in Fig.~\ref{fig:corr_2}. In contrast to the
correlations shown in Fig.~\ref{fig:corr_1} these are not induced by
effective couplings, \ie through quantum effects, but purely based on
the $(\sigma \cdot \br)$ structure our LHC measurements.\bigskip

First of all, there will be a general correlation between all
couplings. Similar to scaling the total width as described in
Section~\ref{sec:higgs} we can shift the bottom Yukawa couplings which
dominates the total Higgs width in the denominator of all rate
measurements. We can compensate for such a shift by simultaneously
scaling all relevant couplings, which gives a general tendency of a
positive correlation of all couplings in this analysis.\bigskip

The correlation between $g_{WWH}$ and $g_{\tau\tau H}$ is induced by
the two weak-boson-fusion signatures, where from the same production
process we can observe both decays at the LHC. This means, that the
ratio $g_{WWH}/g_{\tau\tau H}$ is well constrained, without the
possibility of compensating a shift of only one of the two decay
couplings with another shift in the production cross section. This
adds to the positive diagonal correlation $g_{WWH} \propto g_{\tau\tau
  H}$ in the profile likelihood panel in Fig.~\ref{fig:corr_2}. There
will be a slight distortion due to the fact that we can keep the
weak-boson-fusion $H\to \tau\tau$ rate constant with a negative
correlation $g_{WWH} \propto 1/g_{\tau\tau H}$.

The correlation between $g_{WWH}$ and $g_{ZZH}$ is less pronounced,
mostly due to low statistics.  In the weak-boson-fusion production
process, the $ZZ$ fusion channel typically accounts for
$\mathcal{O}(25\%)$ of all signal events. Hence, we expect a fairly
flat negative correlation between the two couplings. In
the gluon-fusion production channel (and to some degree in the
weak-boson-fusion channel) we can observe both decays to heavy gauge
bosons, which means that the ratio of the two couplings protect the
ratio of the two branching ratios. These two effects together with the
general positive correlation described above again pull the
$g_{WWH}-g_{ZZH}$ correlation into opposite directions and largely
cancel.  This is confirmed by Fig.~\ref{fig:bottom} where the quality
of the $g_{bbH}$ extraction is reduced and the positive correlation
becomes dominant.

In the third panel we show the correlation of the Yukawa couplings
$g_{bbH}$ and $g_{\tau\tau H}$, \ie the ratio of the two down-type
Yukawas. In many models, like in type-II 2HDM this ratio will be
invariant under leading modifications for example by an additional
factor $\tan\beta$. For LHC measurements,these two Yukawas are not
directly linked, if we neglect the two numerically marginal
contributions to $g_{\gamma\gamma H}$. However, we see the over-all
scaling of all couplings also between these two.

Last, but not least we show the correlation between the top Yukawa
coupling and the on-shell top mass. The latter we can measure at the
LHC to $\mathcal{O}(1~\gev)$ and is already measured at the Tevatron 
to 1.2~GeV~\cite{topmass_tev}. We see that at this level a possible
correlation of the top Yukawa coupling in the numerator of the
effective gluon-Higgs coupling and the loop suppression by the top
mass plays hardly any role.\bigskip

Based on Figs.~\ref{fig:corr_1} and \ref{fig:corr_2} we can already
conclude that for this kind of analysis Bayesian probabilities are
less useful to understand the correlations around the best-fit
parameter points and that alternative global maxima will likely not
play a big role in our analysis. Therefore, for the rest of this paper
we limit ourselves to (cooling) profile likelihoods.\bigskip

\begin{figure}[t] 
 \begin{center}
 \includegraphics[width=0.24\textwidth]{%
     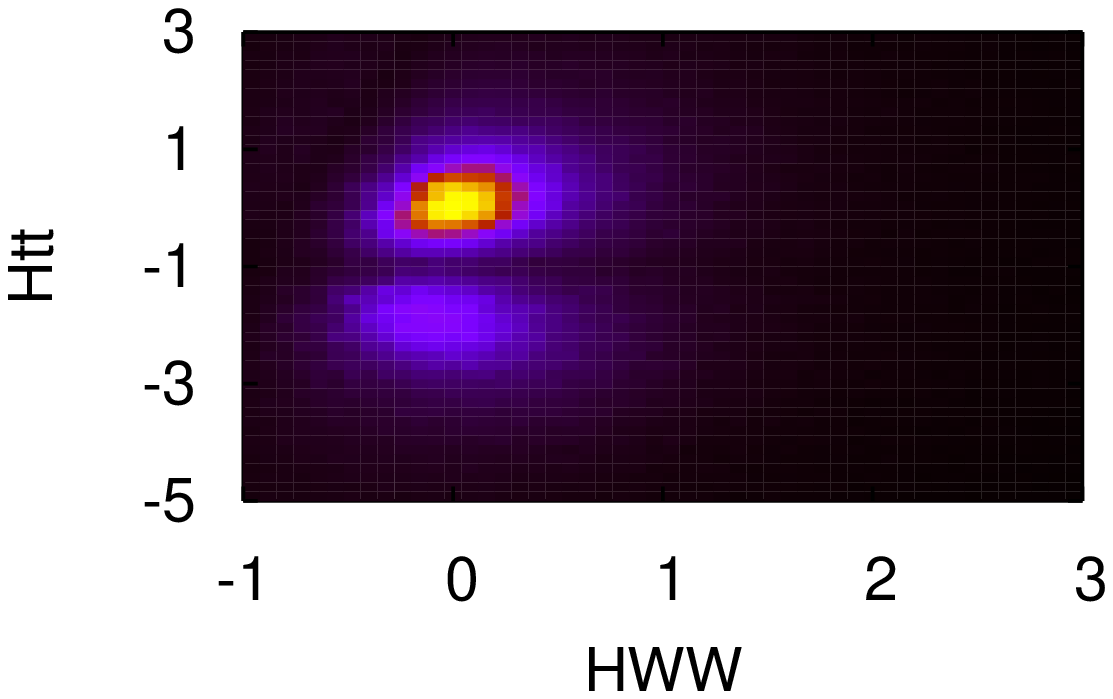}  \hspace*{-2ex}
 \includegraphics[width=0.24\textwidth]{%
     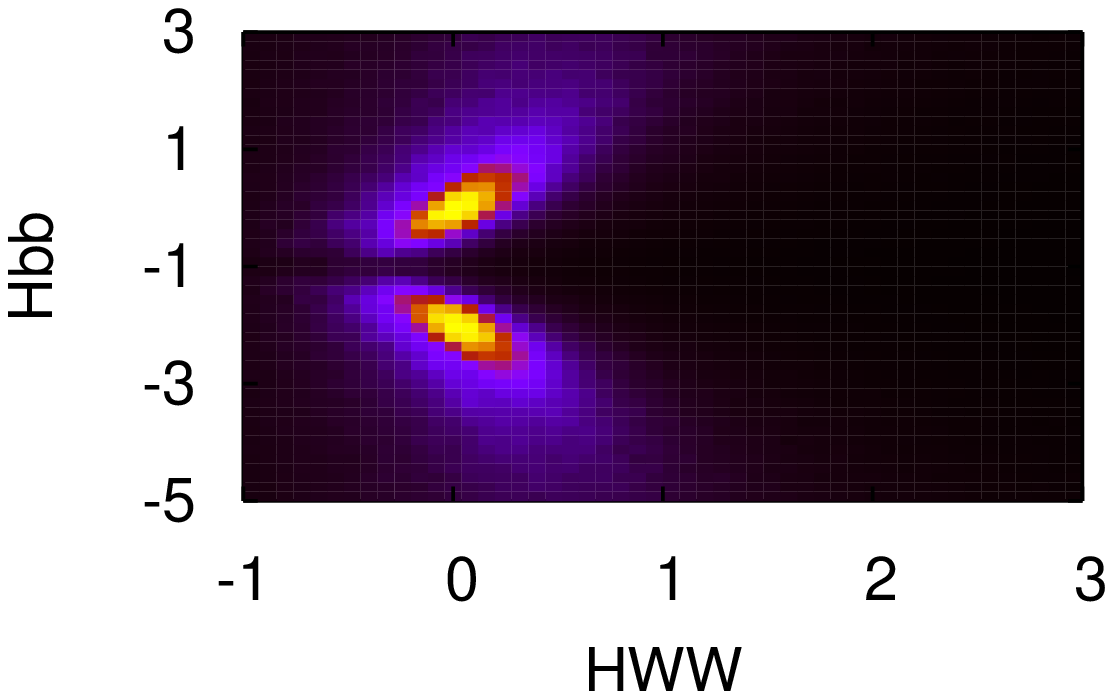}  \hspace*{-2ex}
 \includegraphics[width=0.24\textwidth]{%
     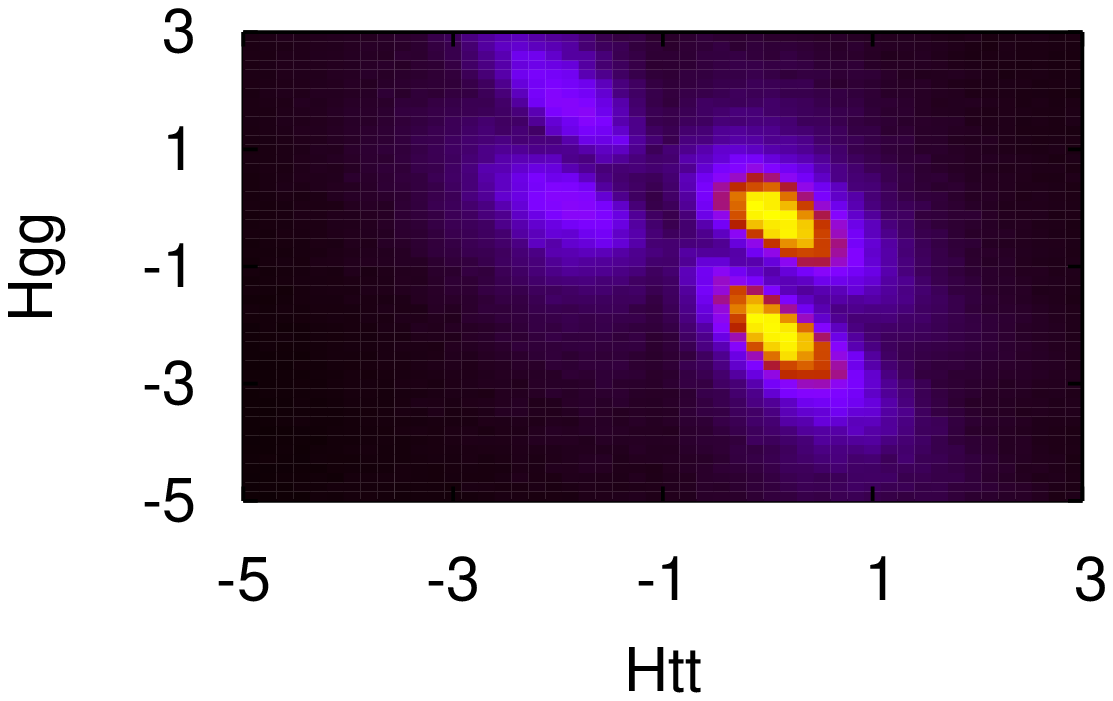}  \hspace*{-2ex}
 \includegraphics[width=0.24\textwidth]{%
     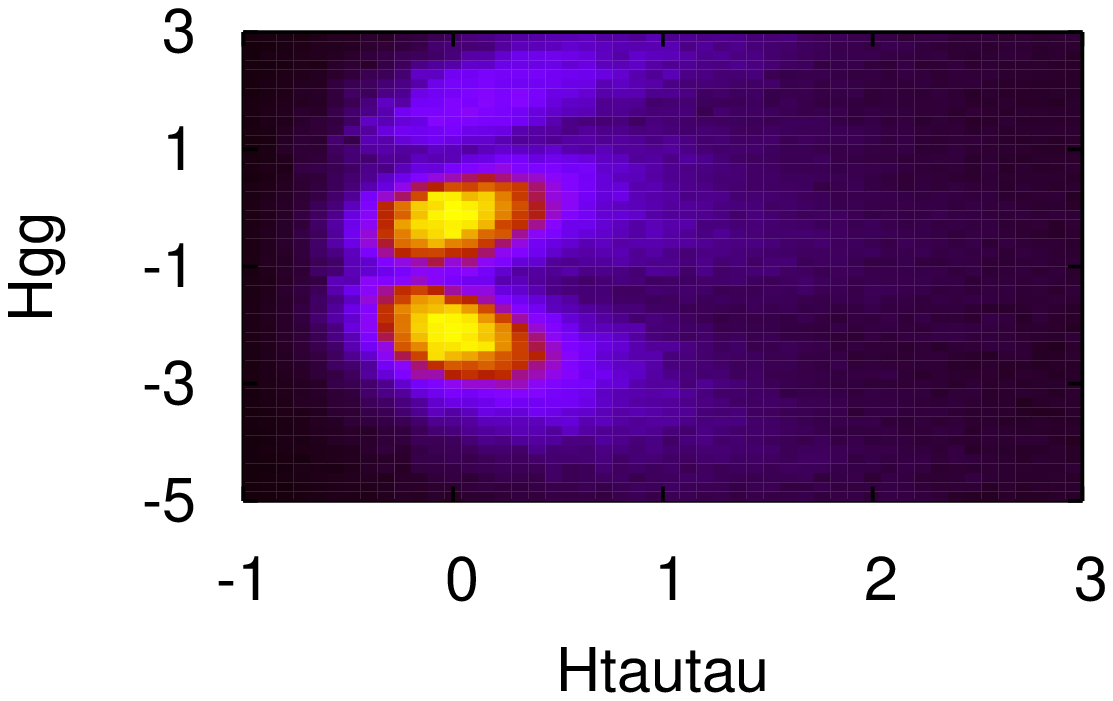} \\
 \includegraphics[width=0.24\textwidth]{%
     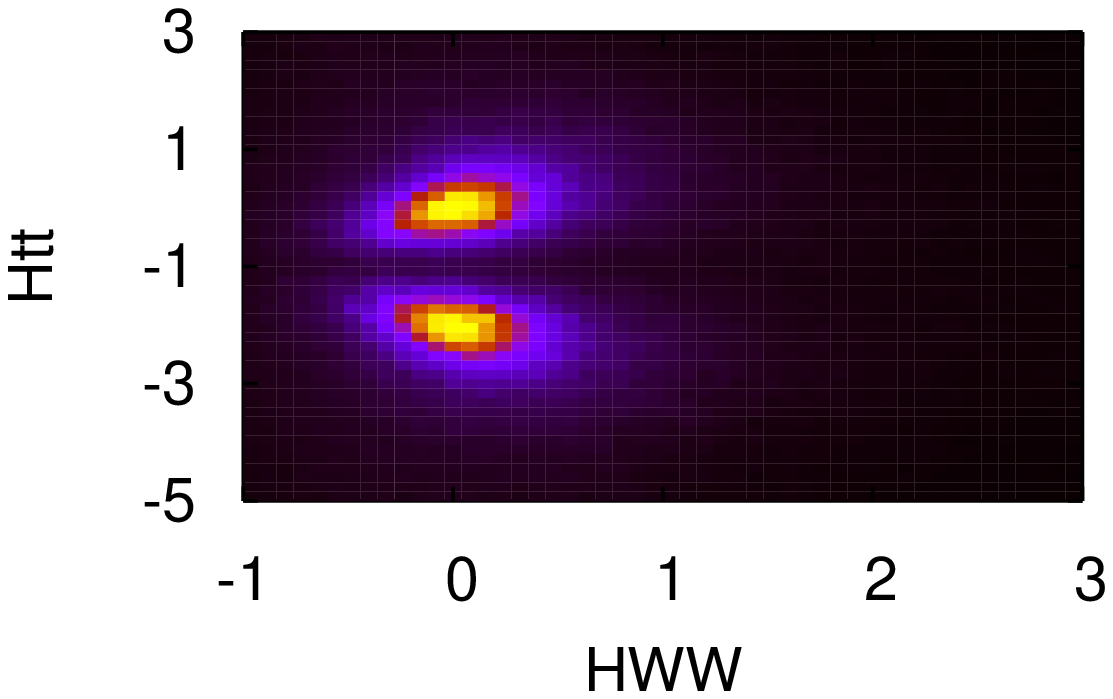}  \hspace*{-2ex}
 \includegraphics[width=0.24\textwidth]{%
     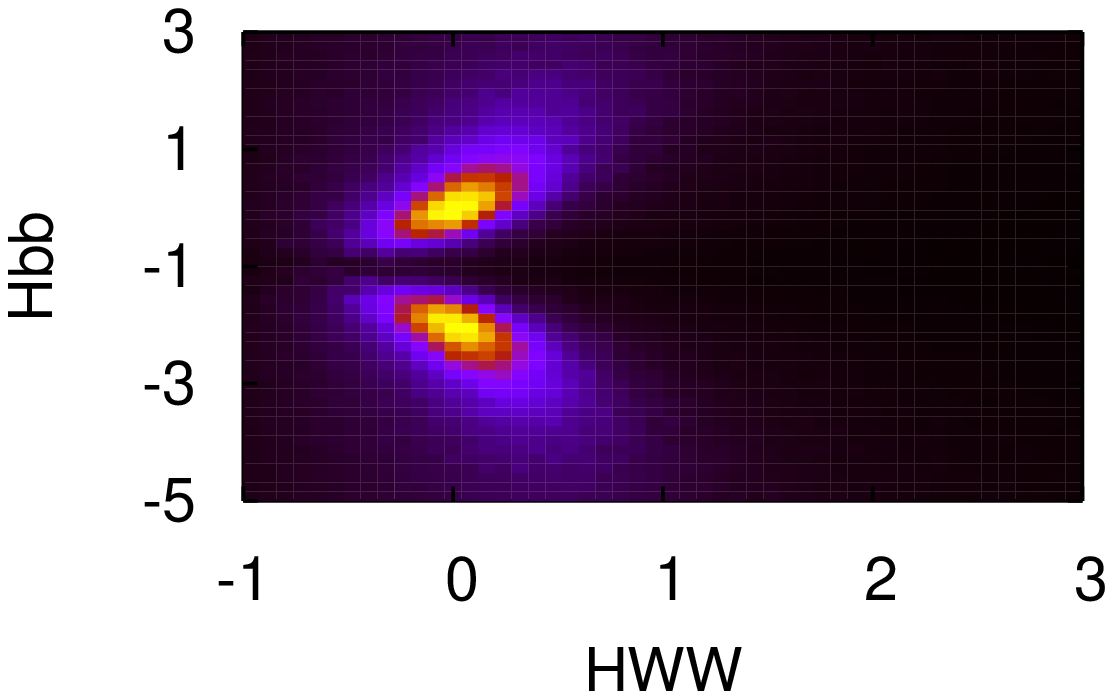}  \hspace*{-2ex}
 \includegraphics[width=0.24\textwidth]{%
     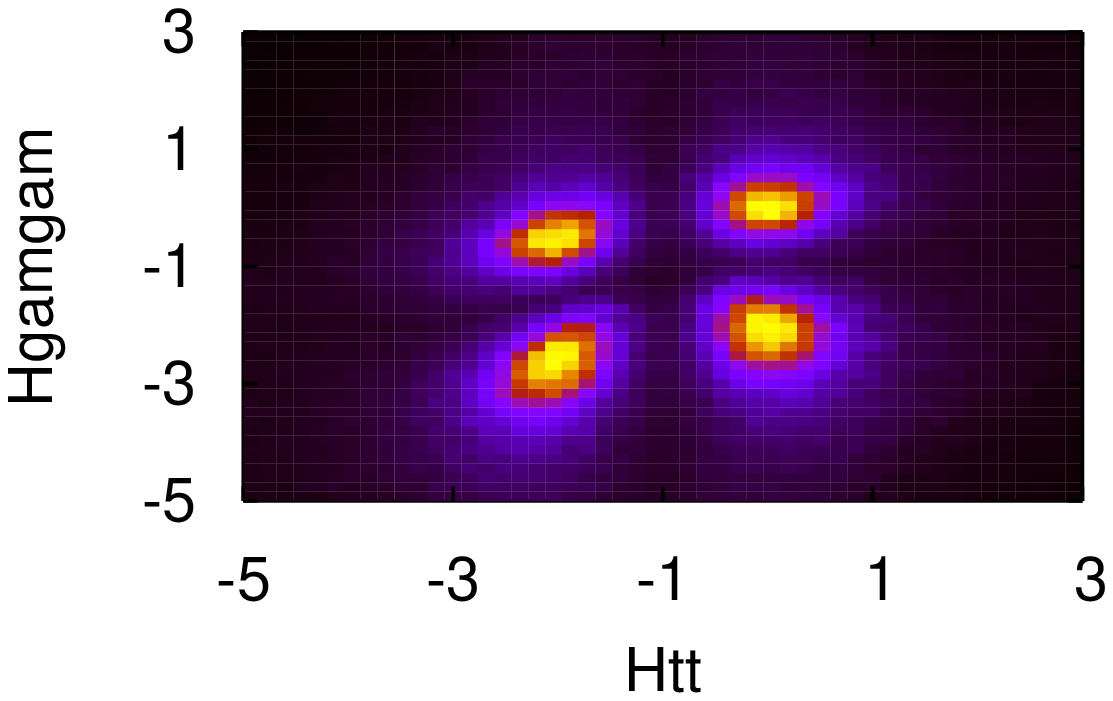} \hspace*{-2ex}
 \includegraphics[width=0.24\textwidth]{%
     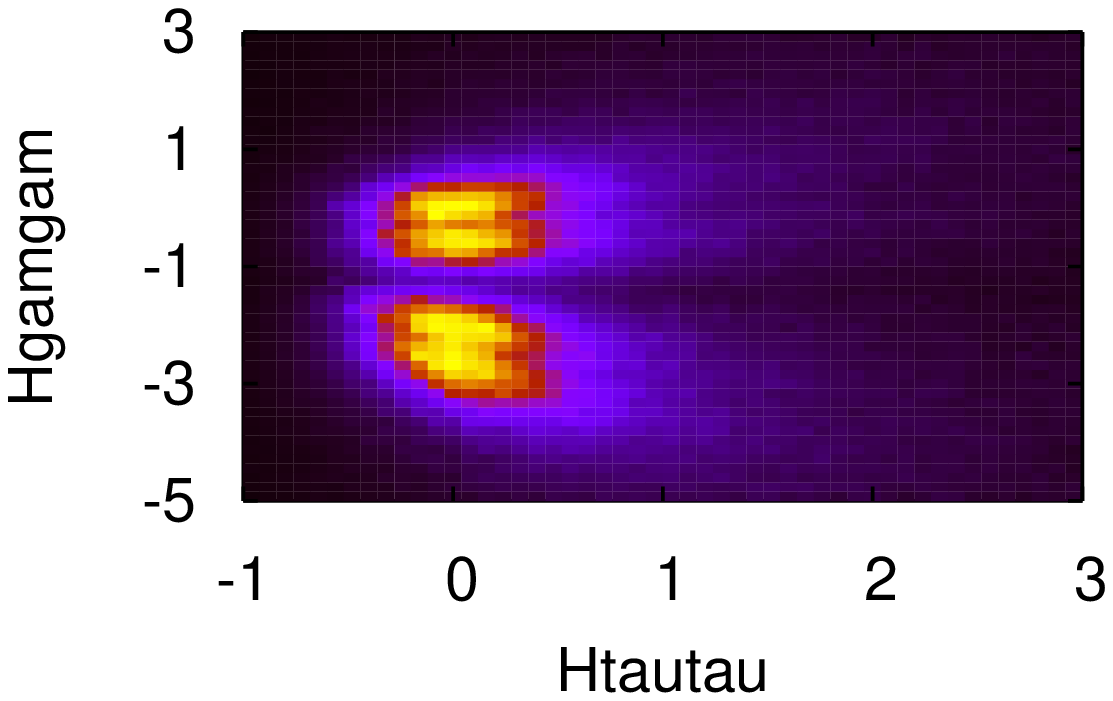} \\
 \includegraphics[width=0.24\textwidth]{%
     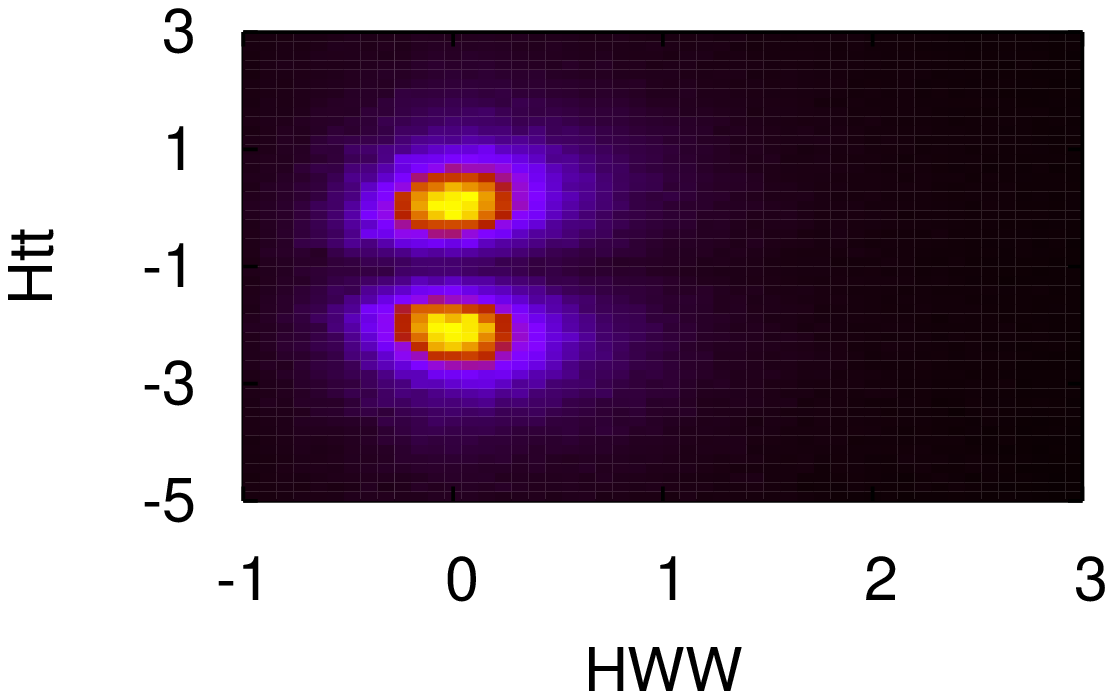} \hspace*{-2ex} 
 \includegraphics[width=0.24\textwidth]{%
     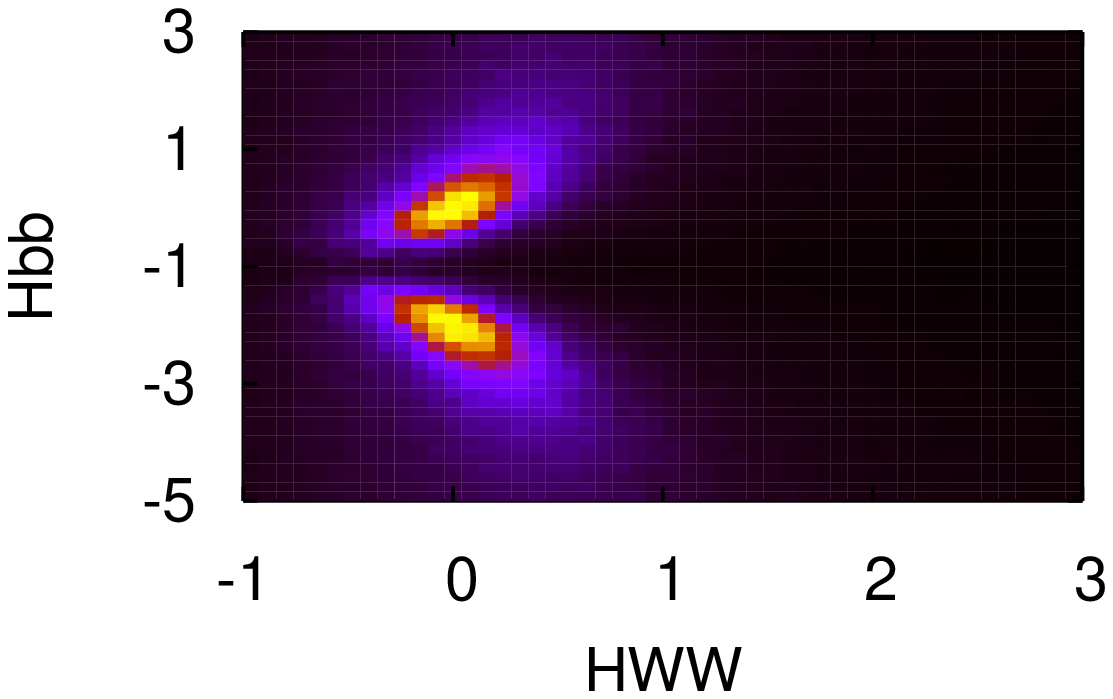} \hspace*{-2ex}
 \includegraphics[width=0.24\textwidth]{%
     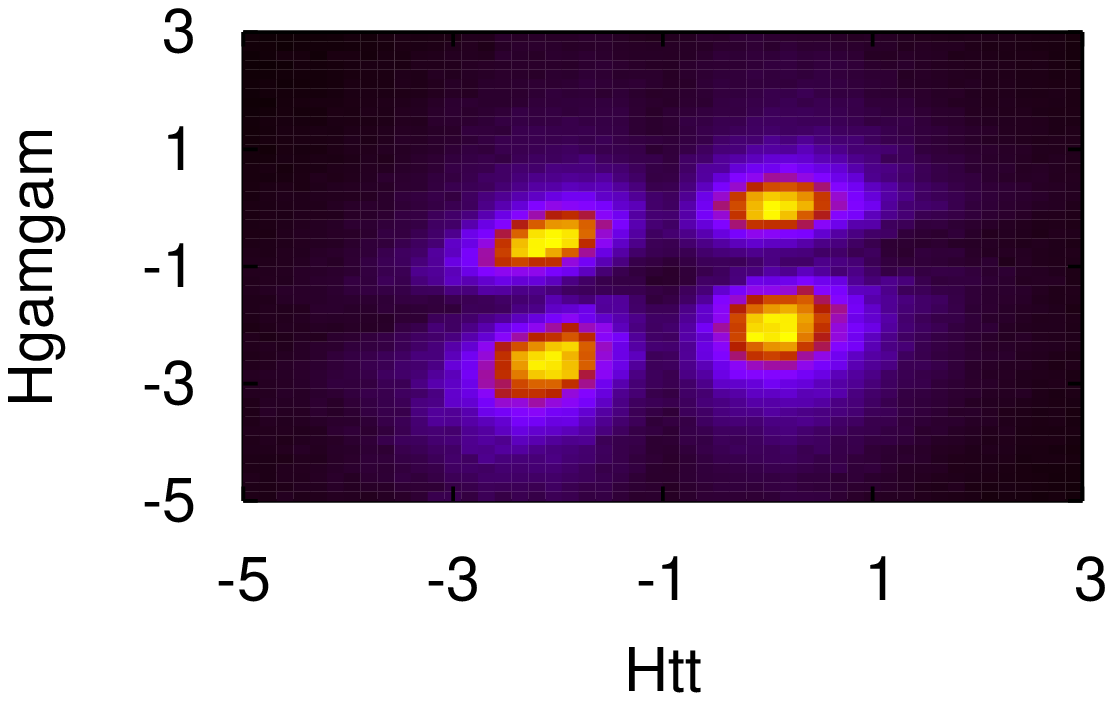} \hspace*{-2ex}
 \includegraphics[width=0.24\textwidth]{%
     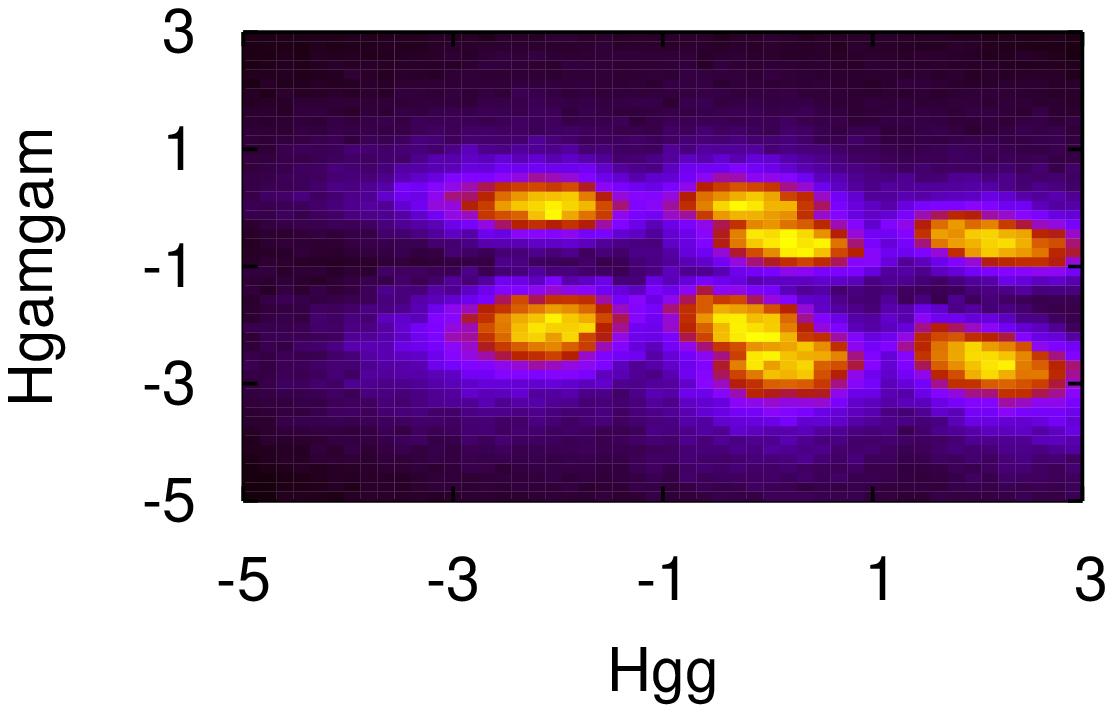} \\
 {\raggedright $1/\Delta \chi^2$\\[2ex]}
 \includegraphics[width=0.135\textwidth]{%
     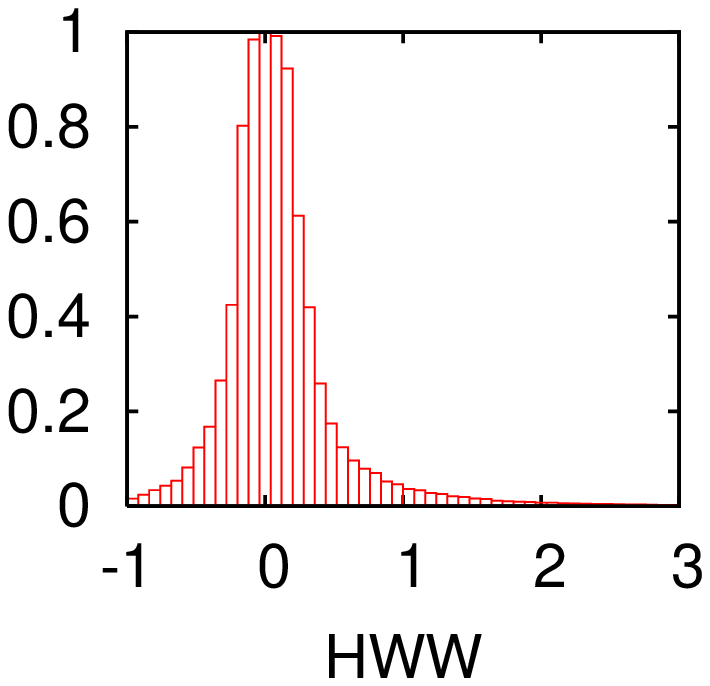} 
 \includegraphics[width=0.135\textwidth]{%
     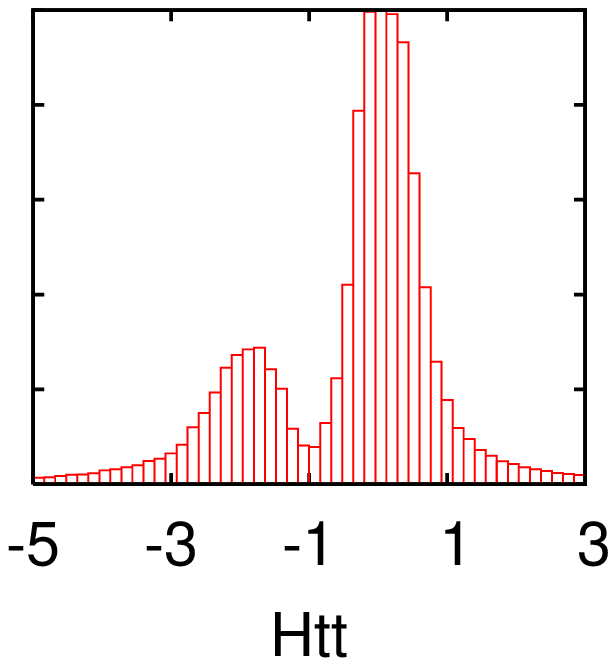} 
 \includegraphics[width=0.135\textwidth]{%
     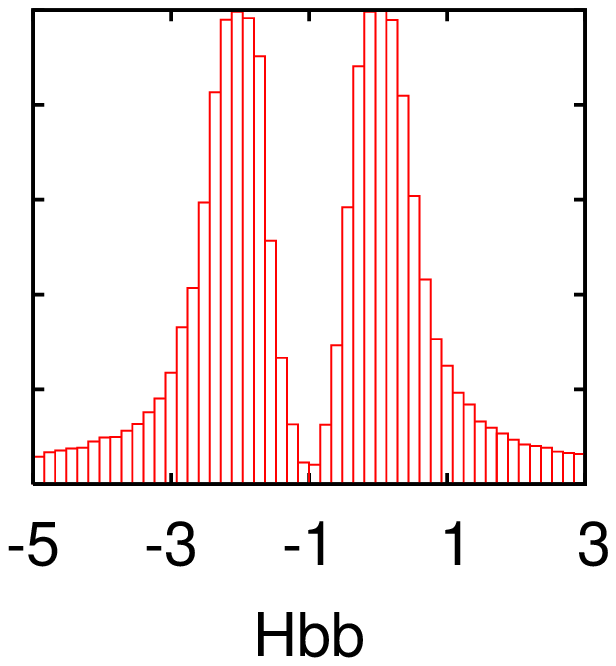} 
 \includegraphics[width=0.135\textwidth]{%
     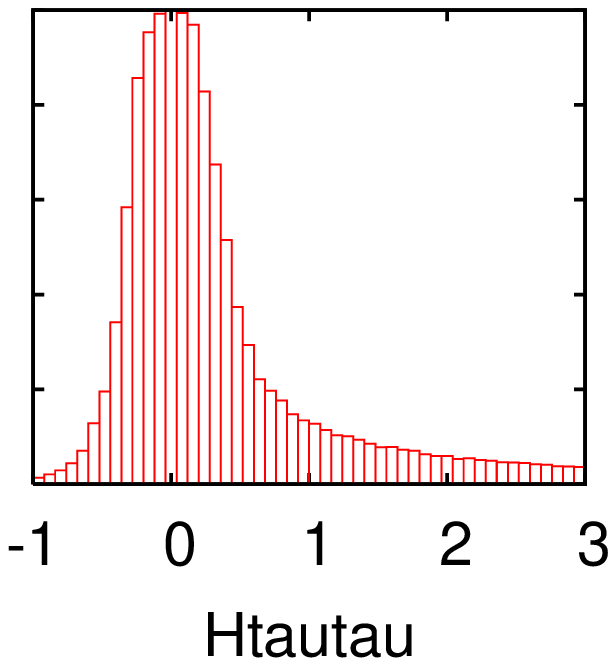} 
 \includegraphics[width=0.135\textwidth]{%
     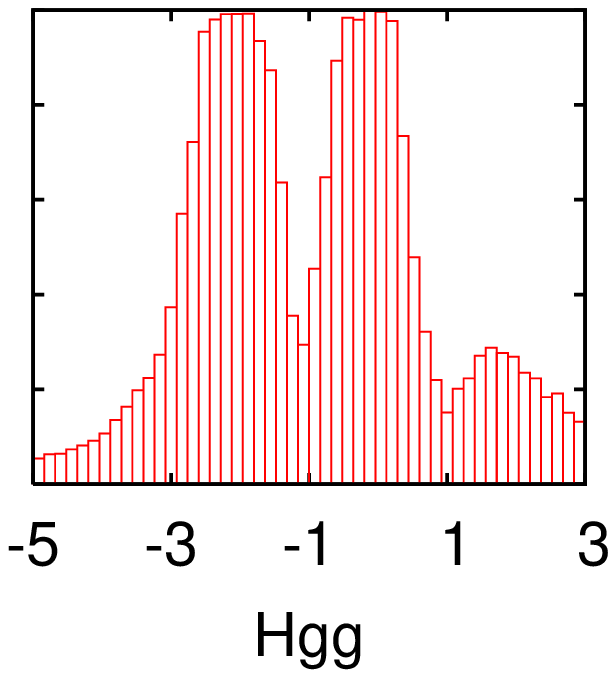} 
 \hspace*{0.135\textwidth} \\[2ex]

 \includegraphics[width=0.135\textwidth]{%
     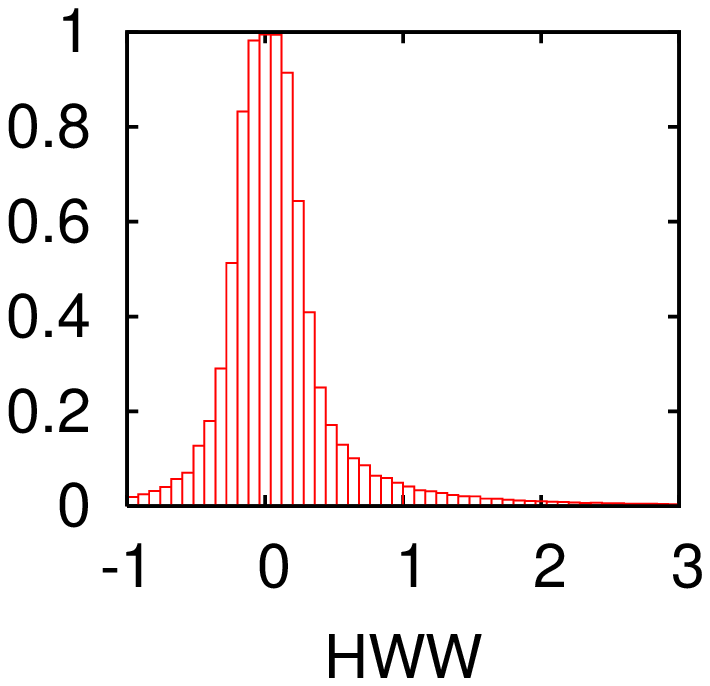} 
 \includegraphics[width=0.135\textwidth]{%
     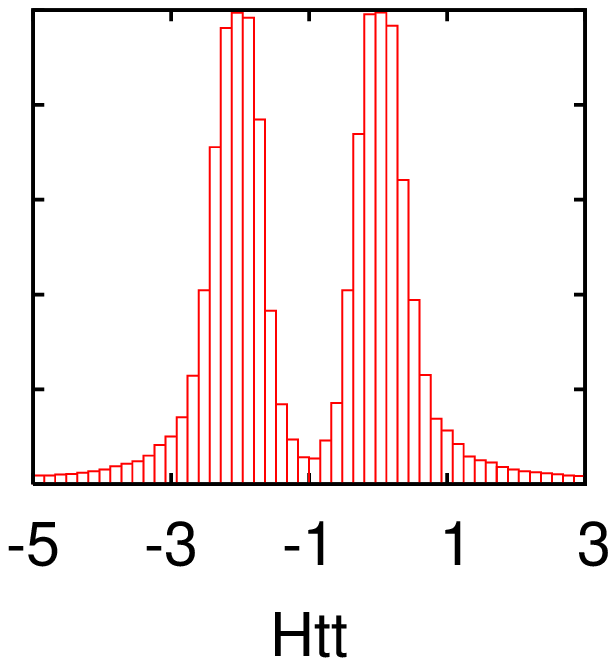} 
 \includegraphics[width=0.135\textwidth]{%
     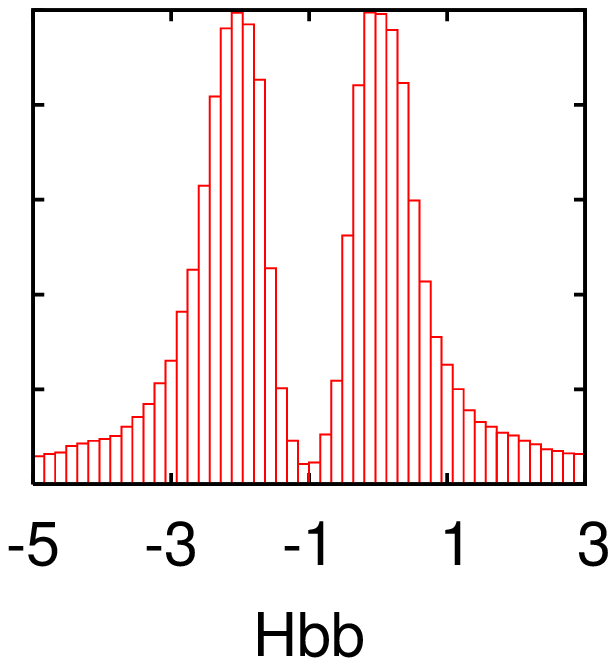} 
 \includegraphics[width=0.135\textwidth]{%
     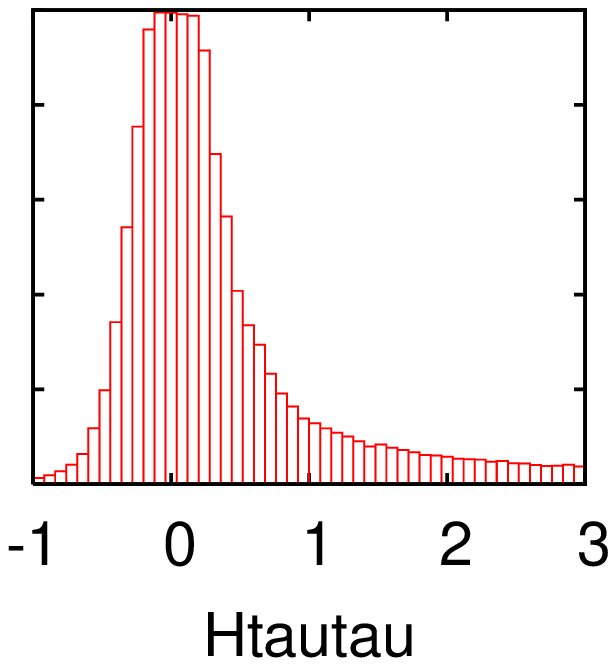} 
 \hspace*{0.135\textwidth} 
 \includegraphics[width=0.135\textwidth]{%
     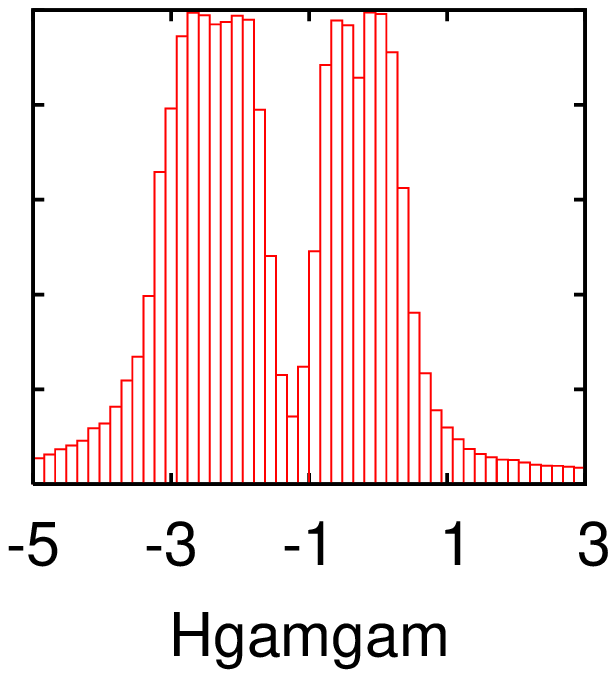} \\[2ex]

 \includegraphics[width=0.135\textwidth]{%
     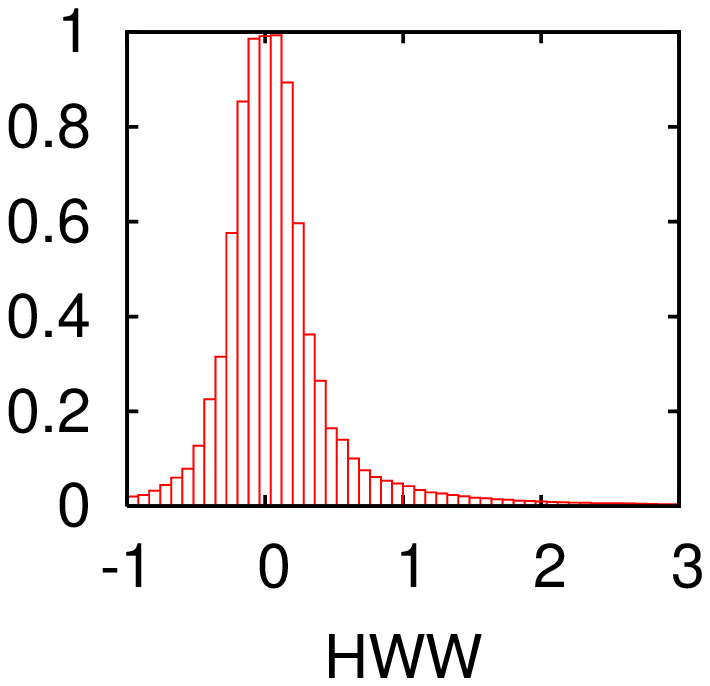} 
 \includegraphics[width=0.135\textwidth]{%
     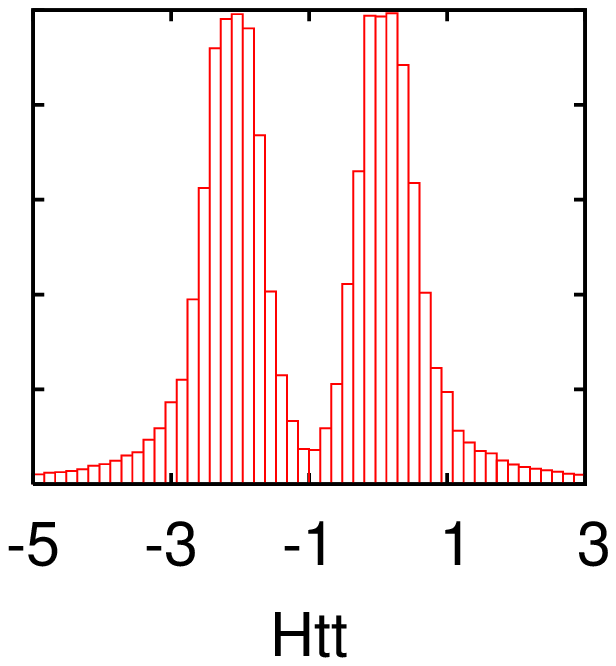} 
 \includegraphics[width=0.135\textwidth]{%
     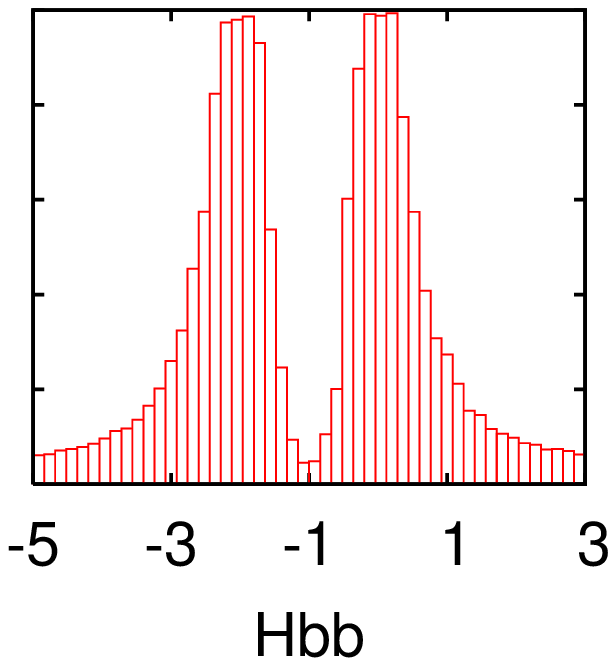} 
 \includegraphics[width=0.135\textwidth]{%
     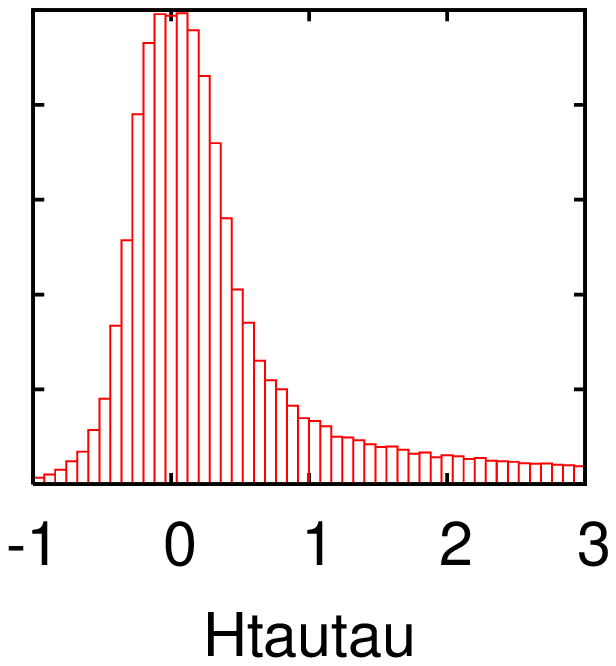} 
 \includegraphics[width=0.135\textwidth]{%
     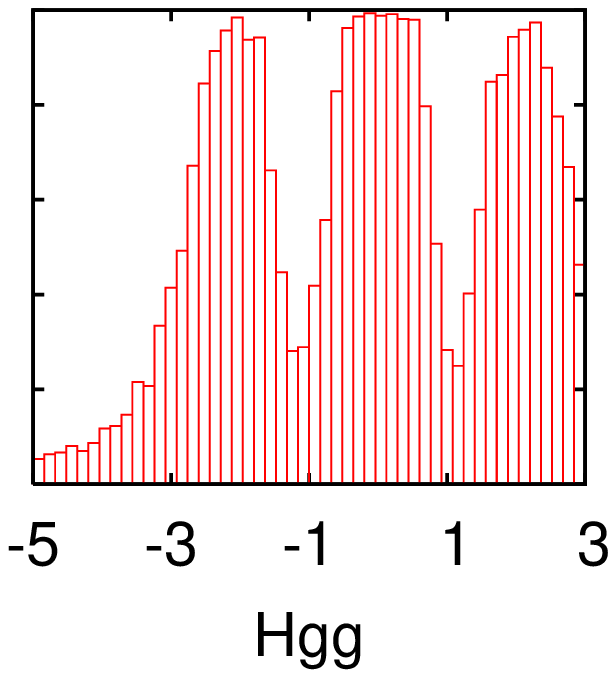} 
 \includegraphics[width=0.135\textwidth]{%
     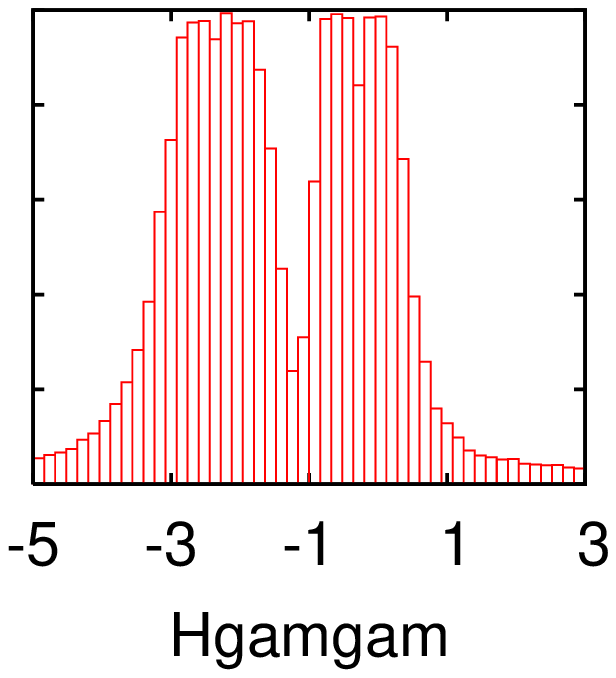}
 \end{center}
\caption[]{Profile likelihoods; the three rows correspond to (1) only
  a $ggH$ effective coupling, (2) only a $\gamma\gamma H$ effective
  coupling, and (3) both of them. The corresponding results without
  any effective coupling are shown in Fig.~\ref{fig:corr_1}.}
\label{fig:corr_f3}
\end{figure}

In Fig.~\ref{fig:corr_f3} we study in detail the impact of the
additional effective $g_{ggH}$ and $g_{\gamma\gamma H}$. Both of them
appear in the Standard Model, but they are completely described by the
tree-level $g_{ttH}$ and $g_{WWH}$ couplings, with minor contributions
from the bottom quark loop. Physics beyond the Standard Model will
always appear in these loops, in weakly interacting as well as in
strongly interacting models.

We start from the discussion of Fig.~\ref{fig:corr_1} which indicates
that without effective couplings there is a clear correlation between
$g_{ttH}$ and $g_{WWH}$, which are the two loops contributing to the
one-loop $H\gamma\gamma$ coupling. The question is what happens if we
allow for an additional effective coupling as shown in
eq.(\ref{eq:coupl_dim5}). In the Standard Model, such small couplings exist, 
and in a general TeV-scale physics
model they can become strong.  In Fig.~\ref{fig:corr_f3} we first add
an additional $g_{ggH}$. This is at the expense of one of the
measurements of the top Yukawa, since the dimension-five $ggH$
operator in the gluon-fusion production cross section is the main
source of the $g_{ttH}$ measurement. This is confirmed by their
correlation or anti-correlation, depending on the sign of the top
Yukawa. Both signs of $g_{ttH}$ are in principle allowed, but their
degeneracy is broken by the $\gamma\gamma H$ coupling. Since the sign
of the sub-leading bottom Yukawa couplings is poorly constrained this
implies that little changes as long as the top Yukawa couplings is
still determined by the $g_{\gamma \gamma H}$. The correlation between
$g_{ggH}$ and $g_{\tau \tau H}$ we show as an example how these
correlations are fed through the parameter space, combining the
effects of loop contributions and production and decay rates.

Adding only an effective $g_{\gamma \gamma H}$ immediately allows for
an equally likely negative top Yukawa coupling. Again, we see a weak
correlation between $g_{\gamma \gamma H}$ and the top Yukawa, but
since the latter is well determined through $g_{ggH}$ it is
quantitatively less relevant. We also see a correlation between
$g_{\gamma \gamma H}$ and $g_{\tau \tau H}$, with four equally likely
branches representing the two possible signs of $g_{ttH}$ combined with
the modulo-two
degeneracy of $\Delta_{\gamma \gamma H} = -2$ or zero.

Finally, we allow for both effective couplings. The correlation
between both effective couplings now shows eight distinct maxima, with
modulo-two steps in $\Delta_{ggH}$ and $\Delta_{\gamma \gamma H}$ and
a sign switch of the top Yukawa coupling. The left pair of solutions
in the $g_{ggH}$--$g_{\gamma \gamma H}$ correlation has a (correct)
positive top Yukawa coupling and a shifted $\Delta_{ggH} \sim -2$. The
central two pairs include the Standard Model solution with positive
top Yukawa and $\Delta_{ggH} \sim 0$ as well as a negative top Yukawa
coupling combined with $\Delta_{ggH} \sim 0$.  In the right pair the
negative top Yukawa coupling is compensated by $\Delta_{ggH} \sim 2$.
The same features we see in the one-dimensional likelihood for the
effective Higgs-gluon coupling. Note that these solutions are only
degenerate because the general effective couplings remove any
sensitivity on the signs of the different Higgs couplings ---
confirming that the sign-flip symmetry is only broken by interferences
in the loop couplings.

\subsection{Theory errors}
\label{sec:theory}

\begin{figure}[t]
 \begin{center}
 \includegraphics[width=0.24\textwidth]{%
     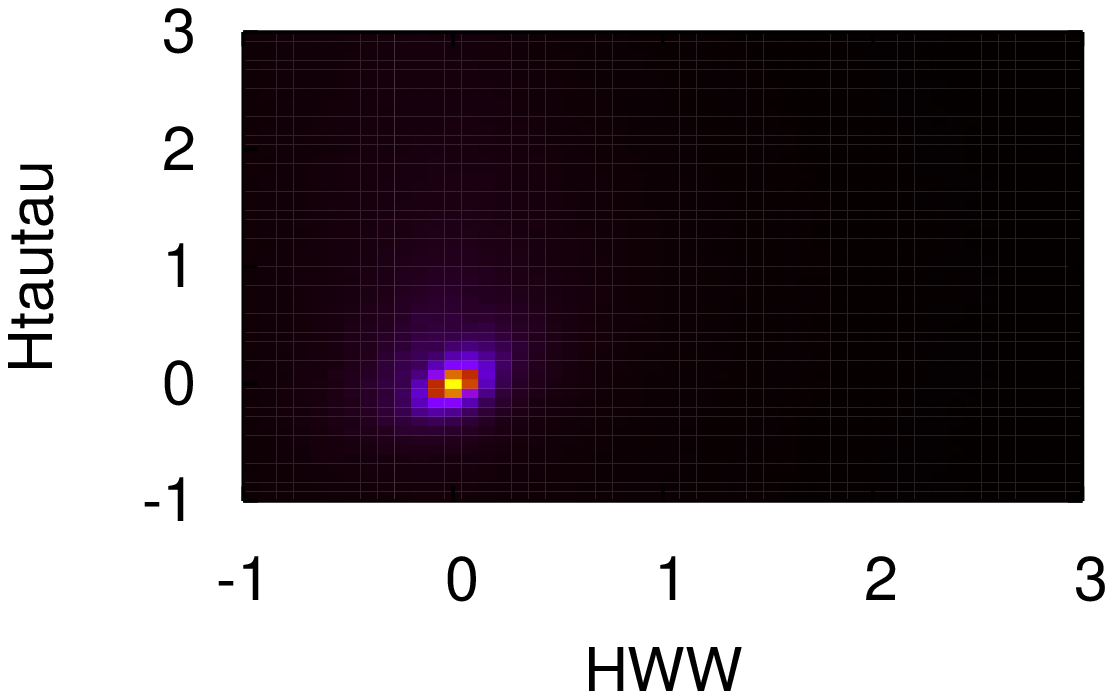} 
 \hspace*{-2ex}
 \includegraphics[width=0.24\textwidth]{%
     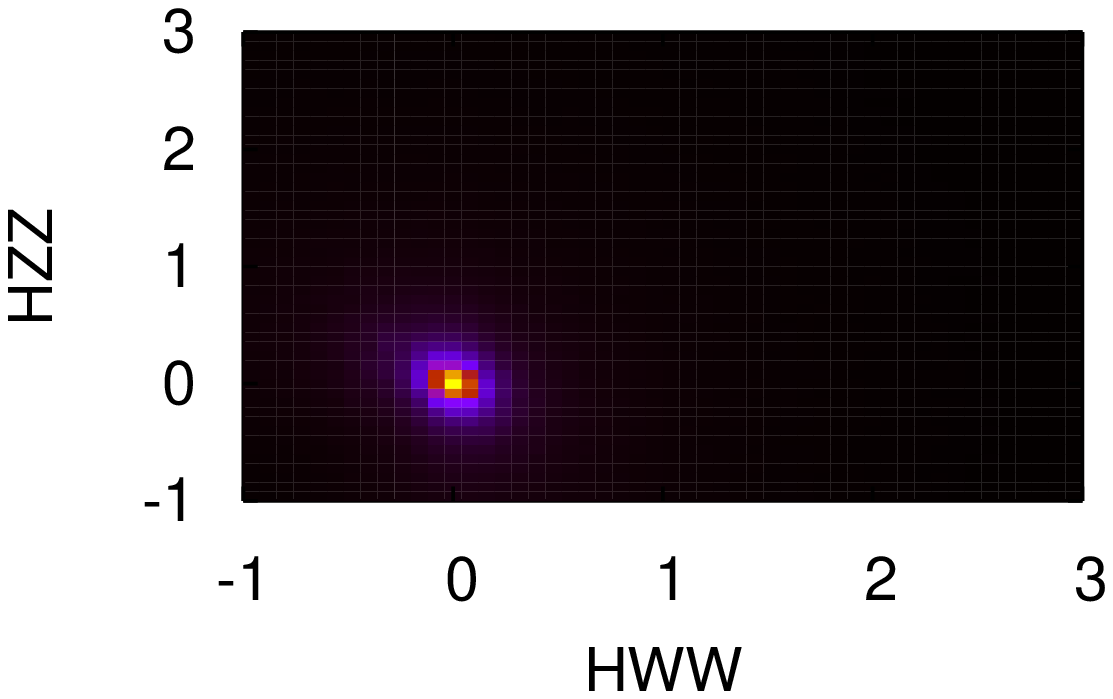} 
 \hspace*{-2ex}
 \includegraphics[width=0.24\textwidth]{%
     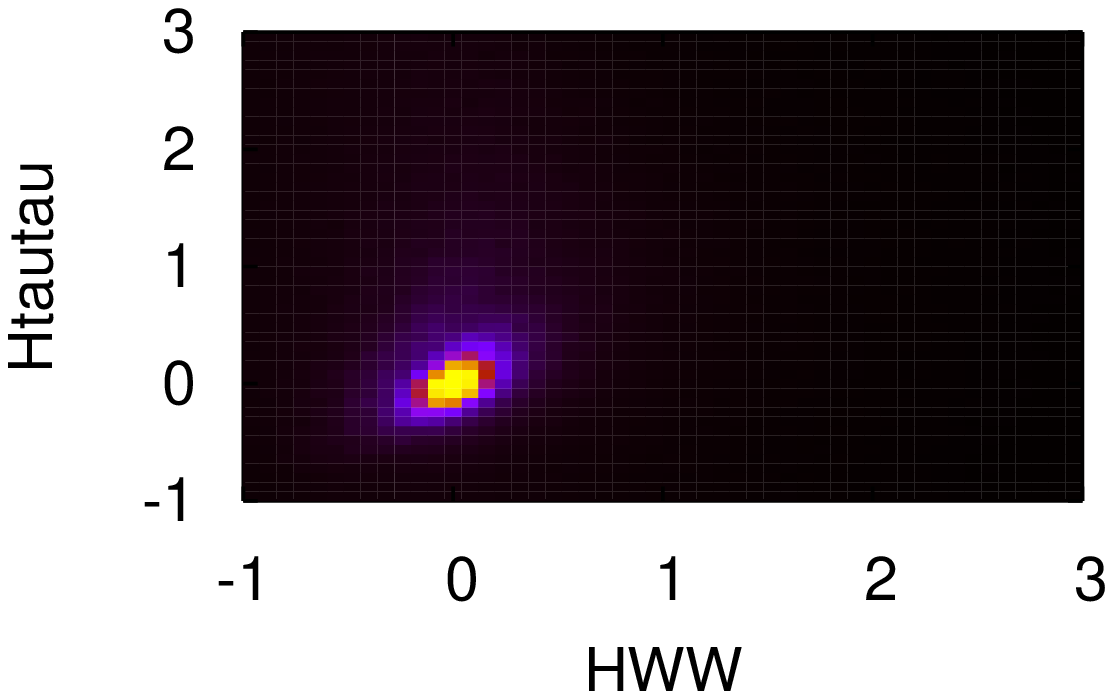} 
 \hspace*{-2ex}
 \includegraphics[width=0.24\textwidth]{%
     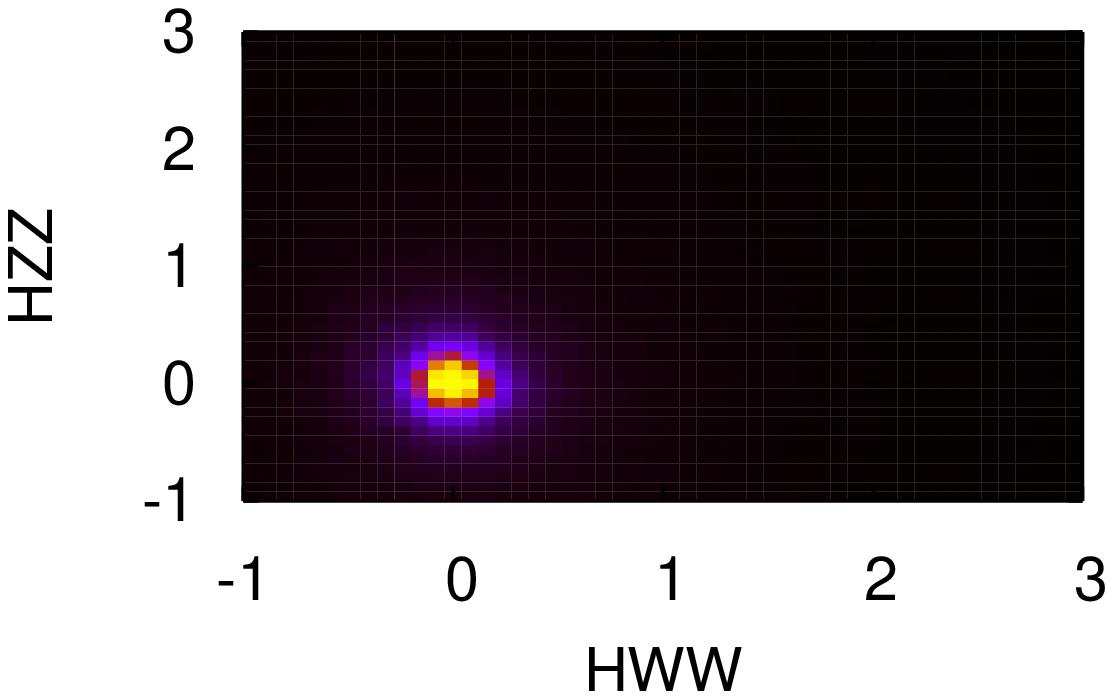} \\

 {\raggedright $1/\Delta \chi^2$\\[2ex]}
 \includegraphics[width=0.135\textwidth]{%
     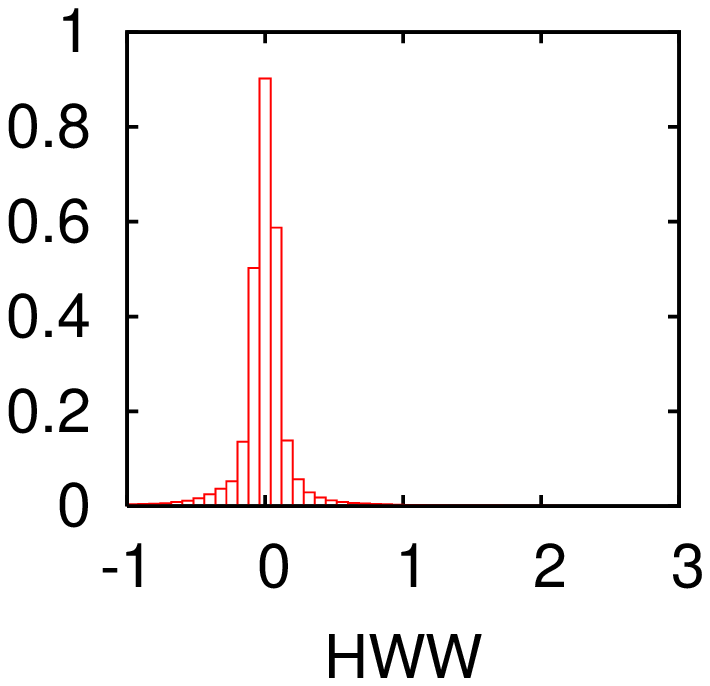} 
 \includegraphics[width=0.135\textwidth]{%
     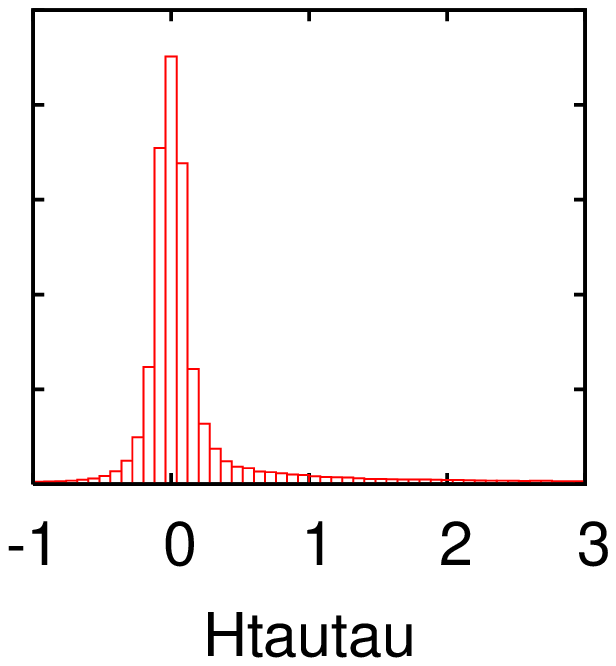} 
 \includegraphics[width=0.135\textwidth]{%
     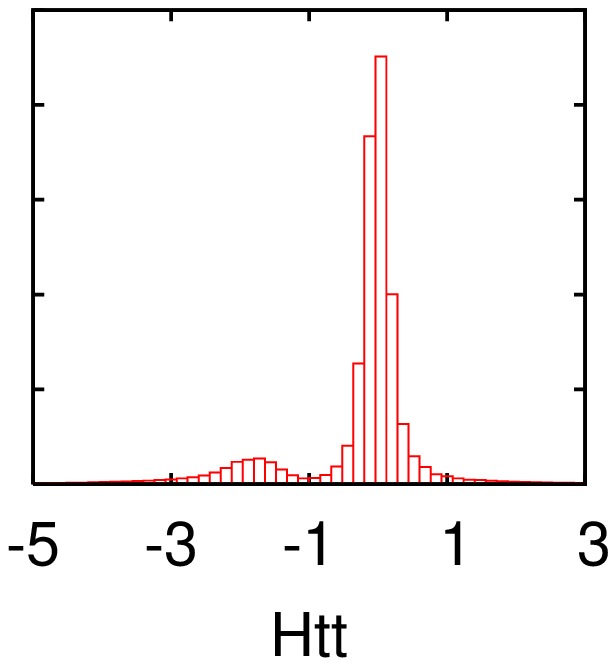} 
 \hspace*{6mm}
 \includegraphics[width=0.135\textwidth]{%
     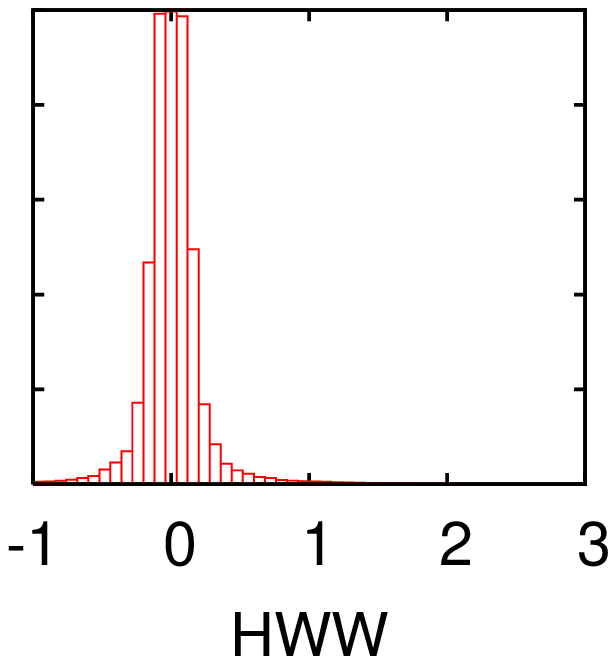} 
 \includegraphics[width=0.135\textwidth]{%
     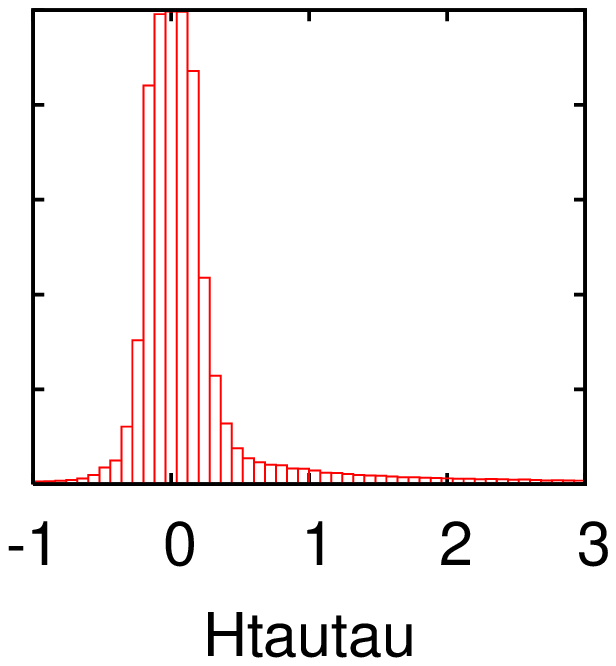} 
 \includegraphics[width=0.135\textwidth]{%
     sfitter_markovbins3.Htt.phaseW.truenoeffsubjet300.eps} \\

 \includegraphics[width=0.24\textwidth]{%
     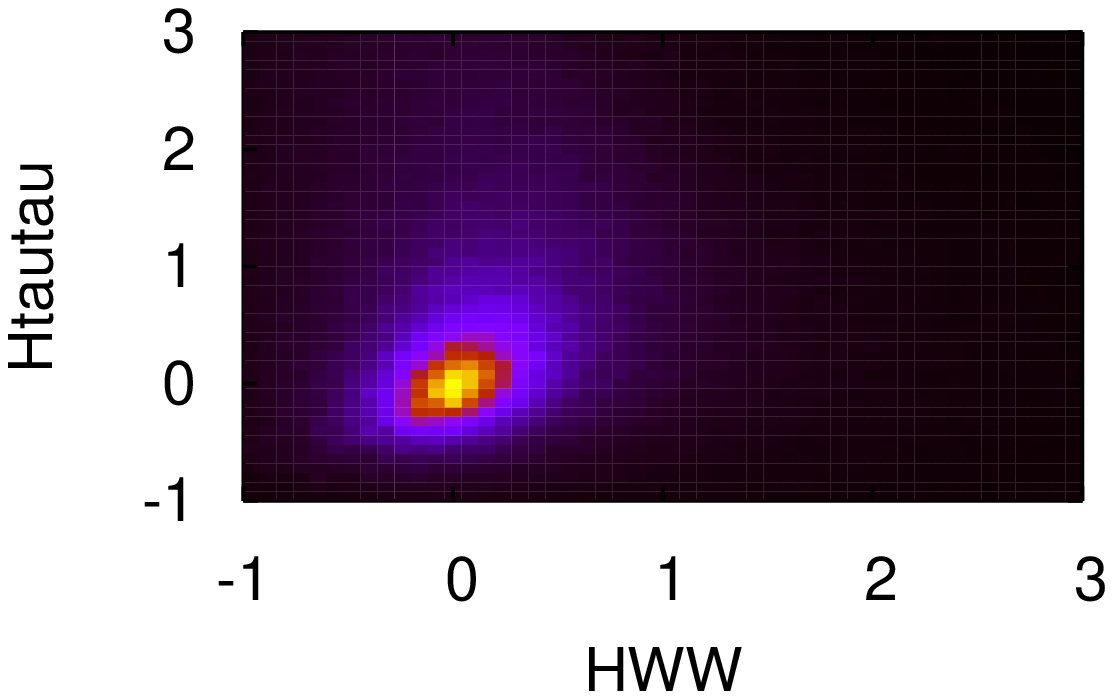} 
 \hspace*{-2ex}
 \includegraphics[width=0.24\textwidth]{%
     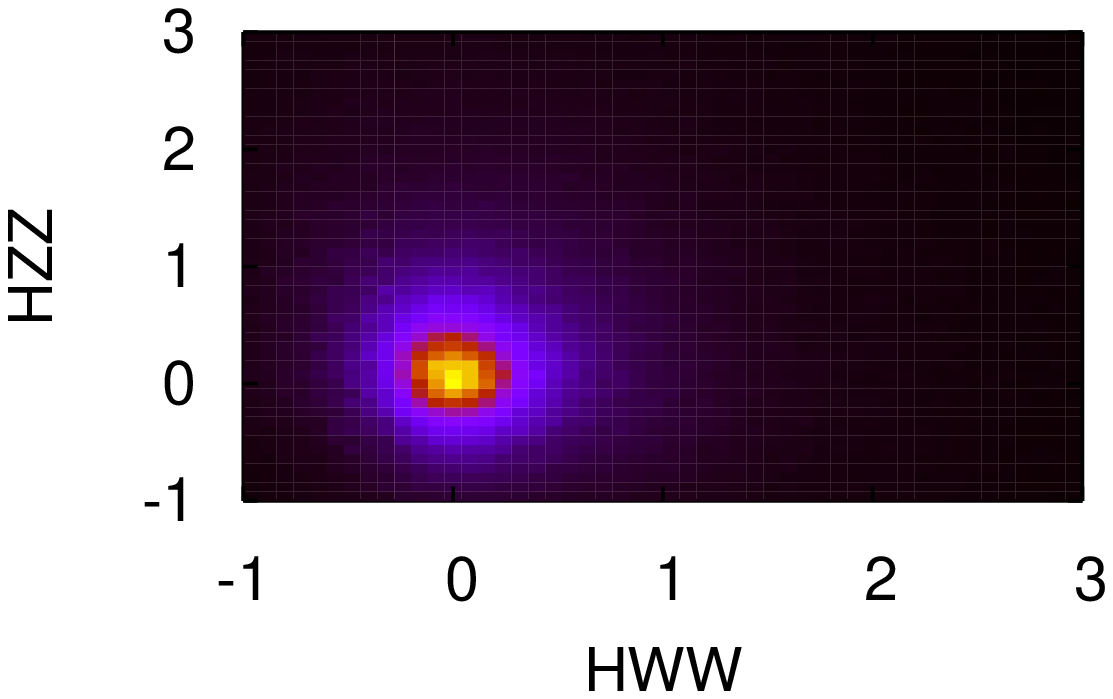} 
 \hspace*{-2ex}
 \includegraphics[width=0.24\textwidth]{%
     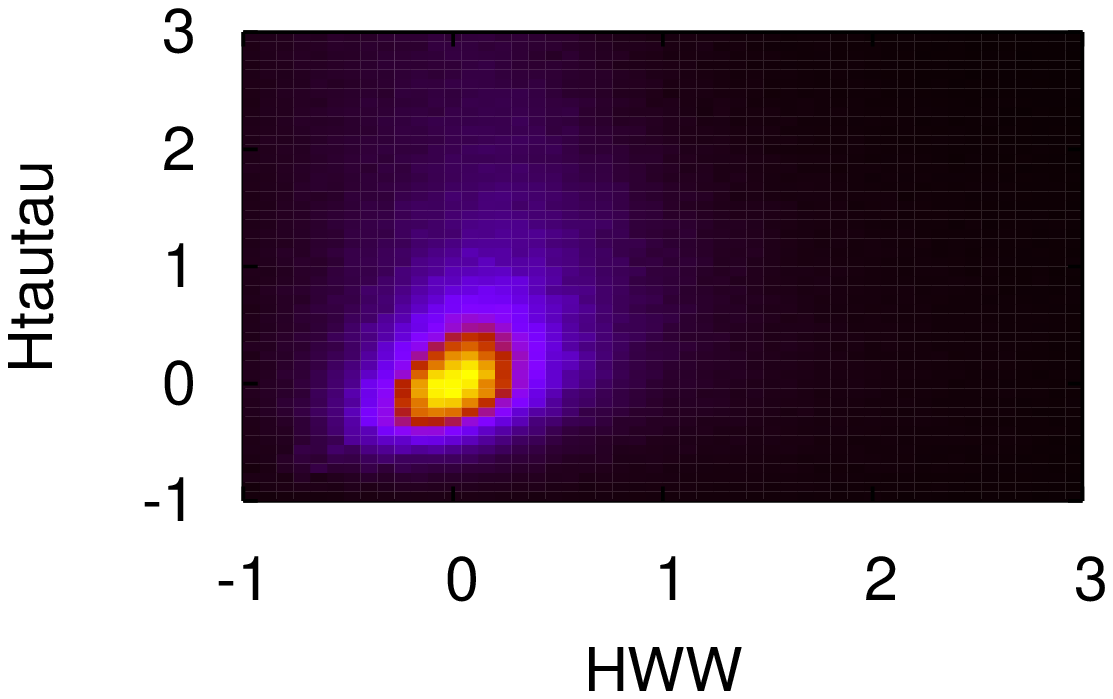} 
 \hspace*{-2ex}
 \includegraphics[width=0.24\textwidth]{%
     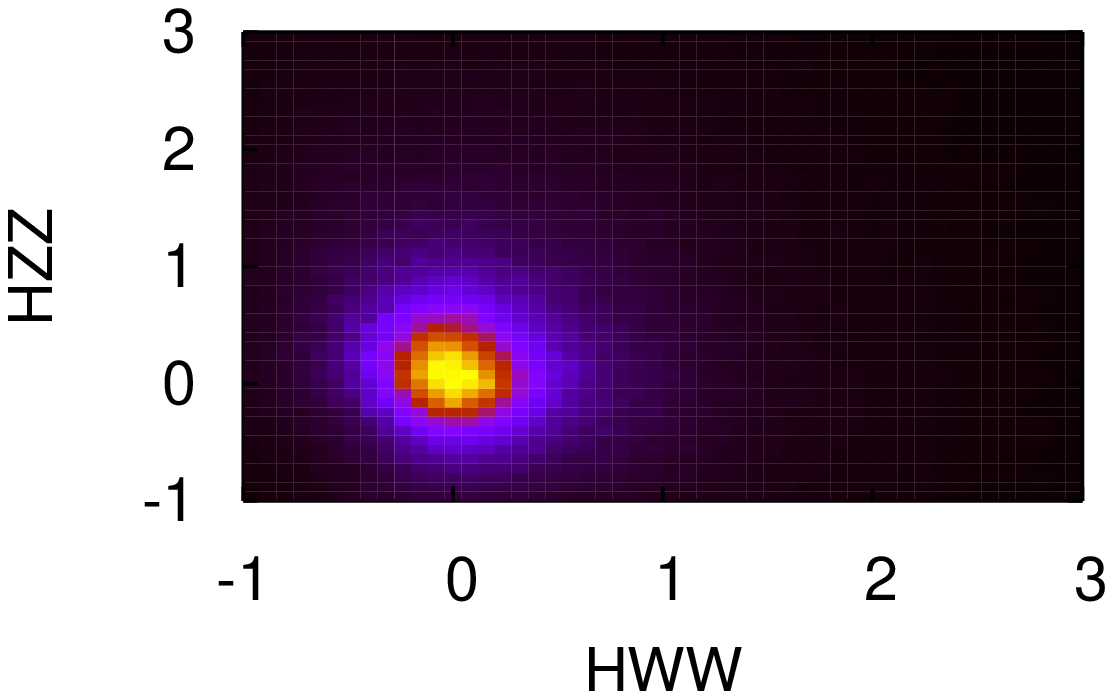} \\

 {\raggedright $1/\Delta \chi^2$\\[2ex]}
 \includegraphics[width=0.135\textwidth]{%
     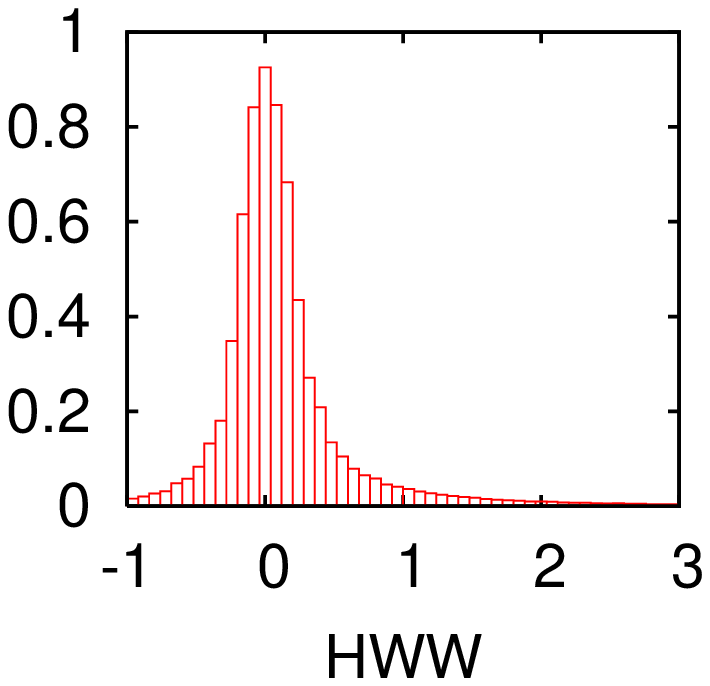} 
 \includegraphics[width=0.135\textwidth]{%
     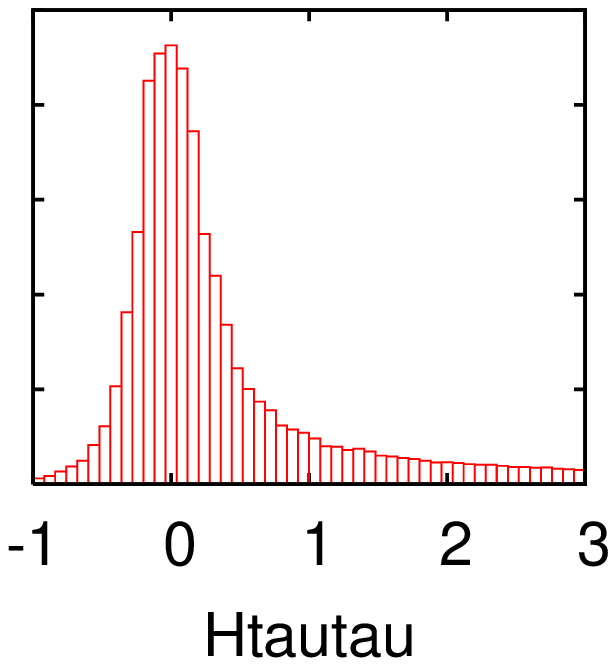} 
 \includegraphics[width=0.135\textwidth]{%
     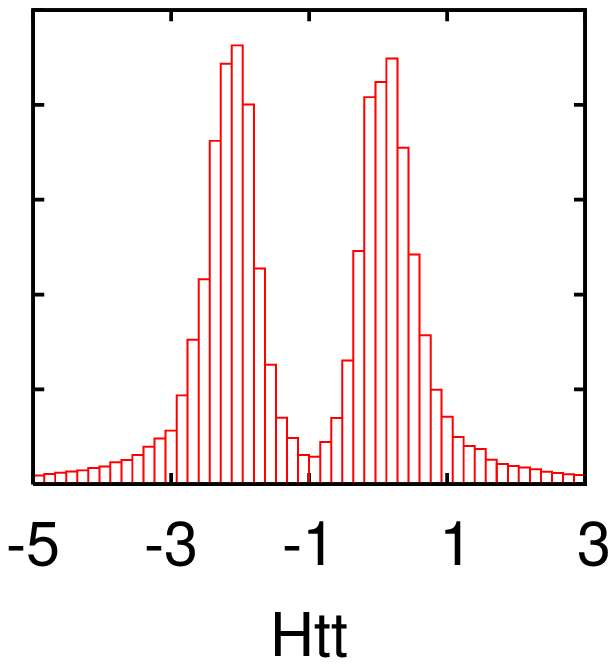} 
 \hspace*{6mm}
 \includegraphics[width=0.135\textwidth]{%
     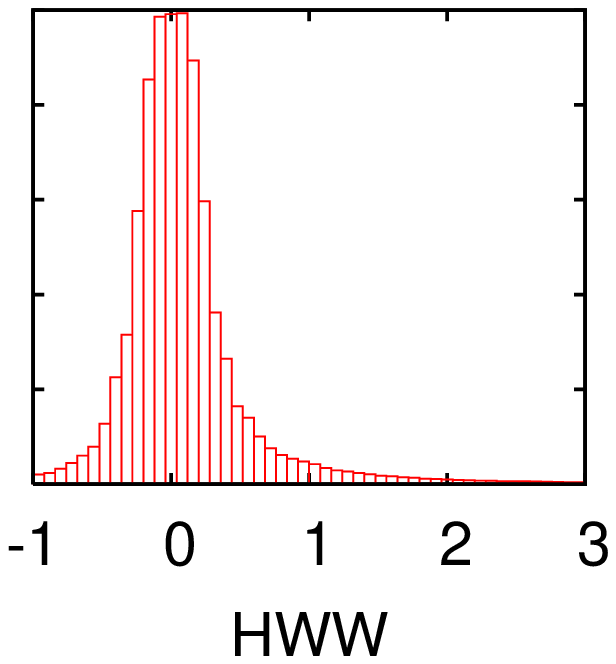} 
 \includegraphics[width=0.135\textwidth]{%
     sfitter_markovbins3.Htautau.phaseW.truesubjet.eps} 
 \includegraphics[width=0.135\textwidth]{%
     sfitter_markovbins3.Htt.phaseW.truesubjet.eps} \\
 \end{center}
\caption[]{Profile likelihoods without (left) and including (right)
  theory errors. The top rows assume $300~\ifb$ and no effective
  couplings, the bottom rows $30~\ifb$ and both effective $ggH$ and
  $\gamma\gamma H$ couplings.}
\label{fig:theory_f1}
\end{figure}

In general, to compare different hypotheses with data the central
values of a parameter fit are useless without appropriate error
bars. Therefore, also in this Higgs sector analysis we have to ensure
that the error treatment is correct and that we understand the impact
of different errors.

While the proper treatment of statistical and (correlated) systematic
experimental errors is well established and standard in all analyses,
the treatment of theory errors deserves a closer look. In
Fig.~\ref{fig:theory_f1} we compare the impact of theory errors on the
(hypothetical) most constraining scenario we can imagine: a parameter
set without effective couplings and assuming a full $300~\ifb$ of
integrated luminosity. Obviously, we expect theory errors to have a
sizable effect on the results.  The second scenario for which we
study the effect of theory errors includes both effective couplings
and a lower luminosity of $30~\ifb$. As discussed in
Section~\ref{sec:corr}, including the two additional effective
couplings has a strong effect on correlations and on alternative
minima. 

Including theory errors on the right-hand side we clearly see the flat
region with maximum log-likelihood around the correct central value
spanning a few bins. This is a direct consequence of the Rfit scheme.
For each measurement there is a flat region around the central value
with the size of the corresponding theory error and likelihood
one. This multi-dimensional box translates into a region in parameter
space, \ie we can slightly vary each parameter around its central
value without any penalty in the likelihood. Outside this flat region
the curves should not drop equally fast in both cases. This is indeed
visible, in particular for the $30~\ifb$ case where the fall-off is
not as steep as for $300~\ifb$. For example the tail for large
$g_{\tau\tau H}$ looks exactly the same. It is interesting to note
that theory errors also increase the likelihood of the alternative
solution $\Delta_{ttH}=-2$ for forbidden effective couplings.  This
alters the effective coupling of the Higgs to photons which within
theory errors we can slightly shift without increasing $\chi^2$.

Theory errors can affect background predictions in addition to signal
rates. However, most of the LHC Higgs searches rely on control regions
or side bins.  The classical example is the two-sided side bins in the
$H \to \gamma \gamma$ search where the theory error does not have to
account for higher-order corrections of the background rate. Therefore,
we neglect the effect of background theory errors.  The systematic
errors on background rates are listed in
Table~\ref{tab:syst_error}. 

\subsection{Unobserved or invisible?}
\label{sec:cch}

\begin{figure}[t]
 \begin{center}
 {\raggedright $1/\Delta \chi^2$\\[2ex]}
 \includegraphics[width=0.135\textwidth]{%
     sfitter_markovbins3.HWWlabel.phaseW.truesubjet.eps} 
 \includegraphics[width=0.135\textwidth]{%
     sfitter_markovbins3.Htt.phaseW.truesubjet.eps} 
 \includegraphics[width=0.135\textwidth]{%
     sfitter_markovbins3.Hgg.phaseW.truesubjet.eps} 
 \includegraphics[width=0.135\textwidth]{%
     sfitter_markovbins3.Htautau.phaseW.truesubjet.eps} 
 \includegraphics[width=0.135\textwidth]{%
     sfitter_markovbins3.Hbb.phaseW.truesubjet.eps} 
 \hspace*{0.135\textwidth} \\[2ex]

 \includegraphics[width=0.135\textwidth]{%
     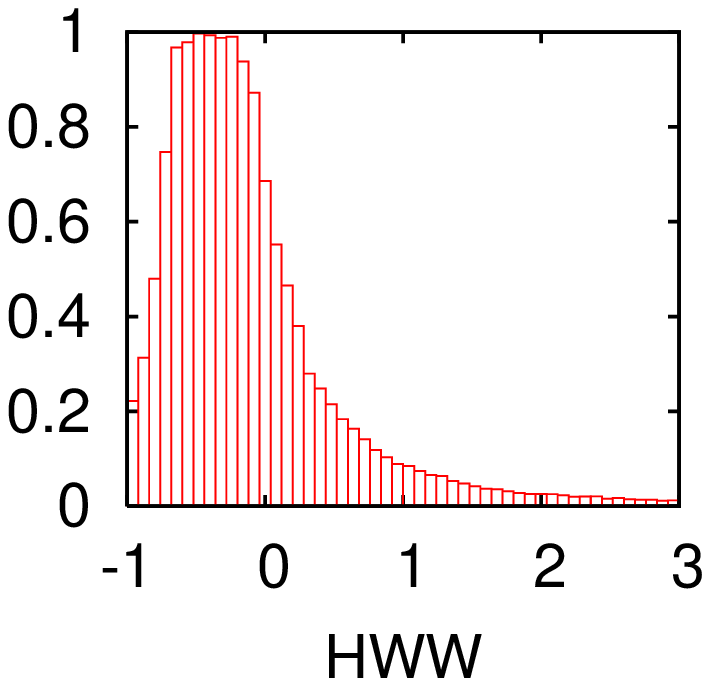} 
 \includegraphics[width=0.135\textwidth]{%
     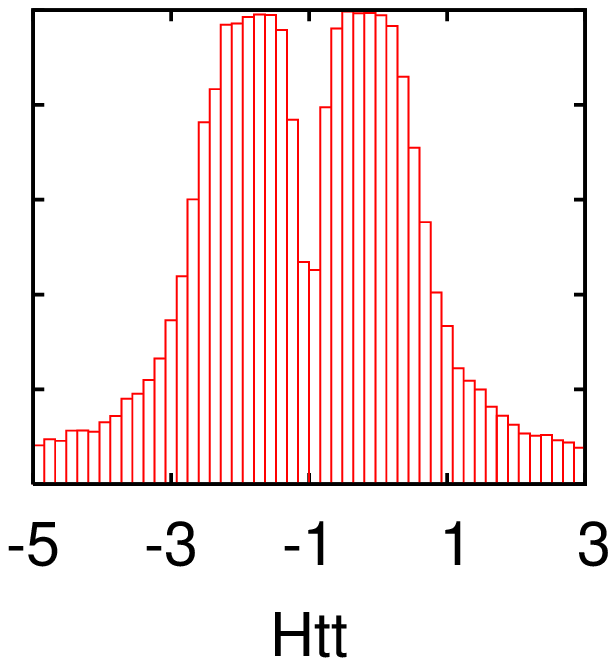} 
 \includegraphics[width=0.135\textwidth]{%
     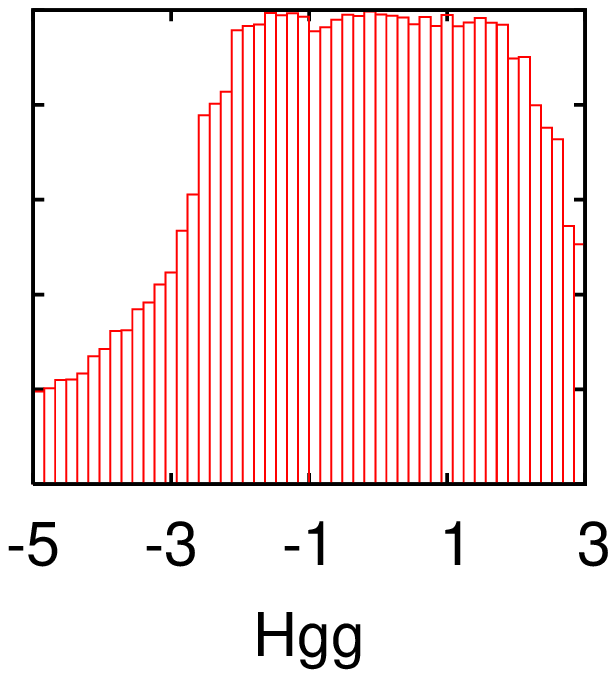} 
 \includegraphics[width=0.135\textwidth]{%
     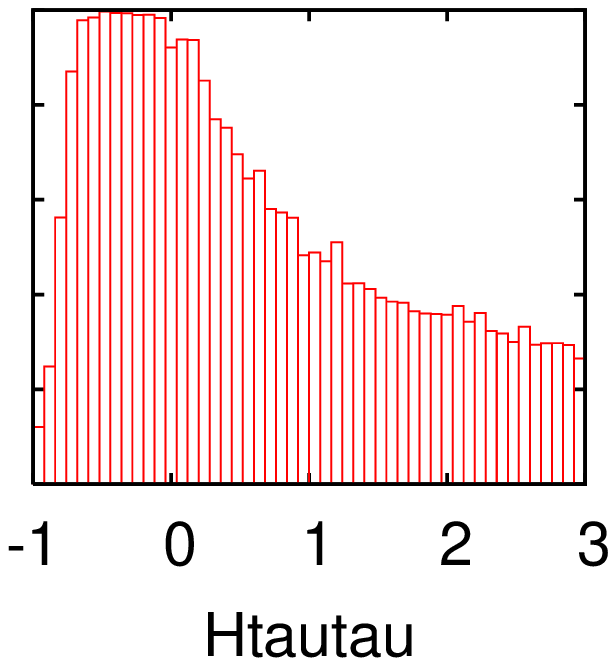} 
 \includegraphics[width=0.135\textwidth]{%
     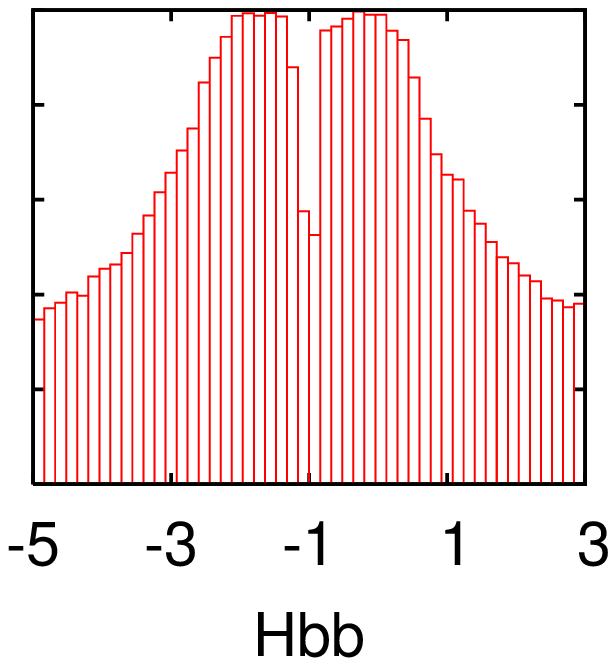} 
 \hspace*{0.135\textwidth} \\[2ex]

 \includegraphics[width=0.135\textwidth]{%
     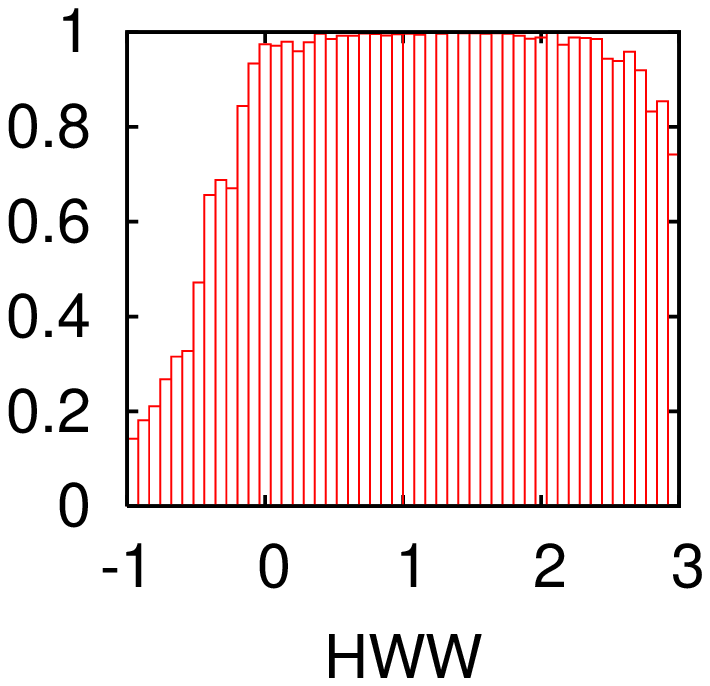} 
 \includegraphics[width=0.135\textwidth]{%
     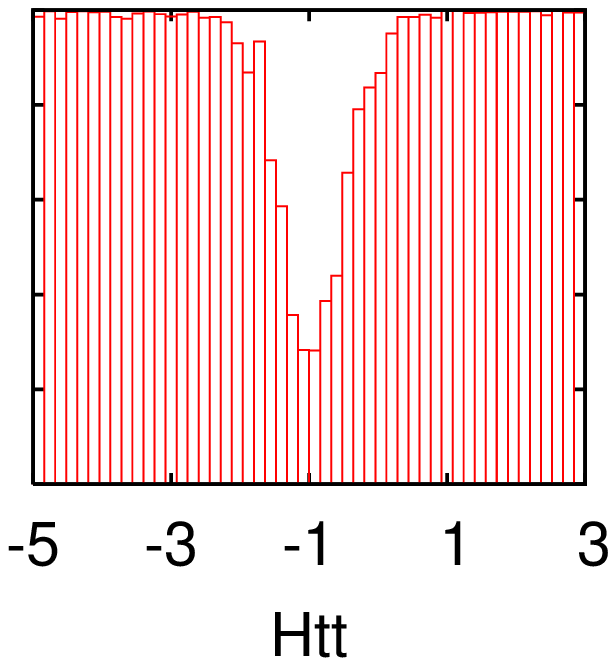} 
 \includegraphics[width=0.135\textwidth]{%
     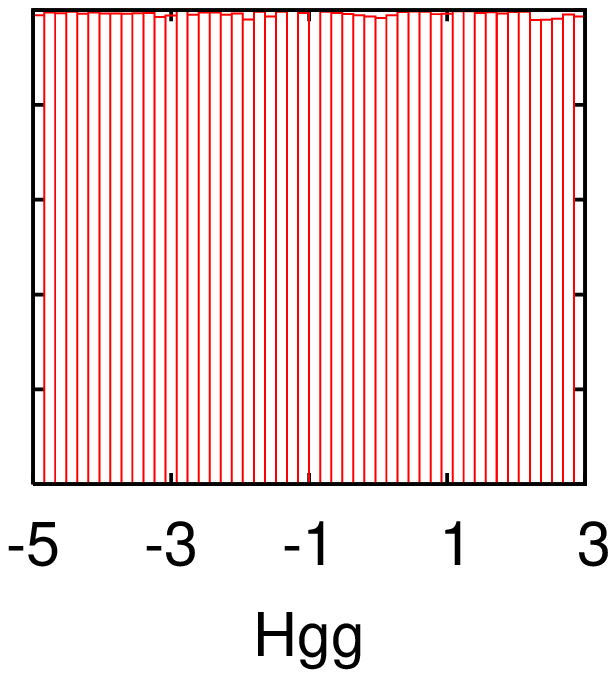} 
 \includegraphics[width=0.135\textwidth]{%
     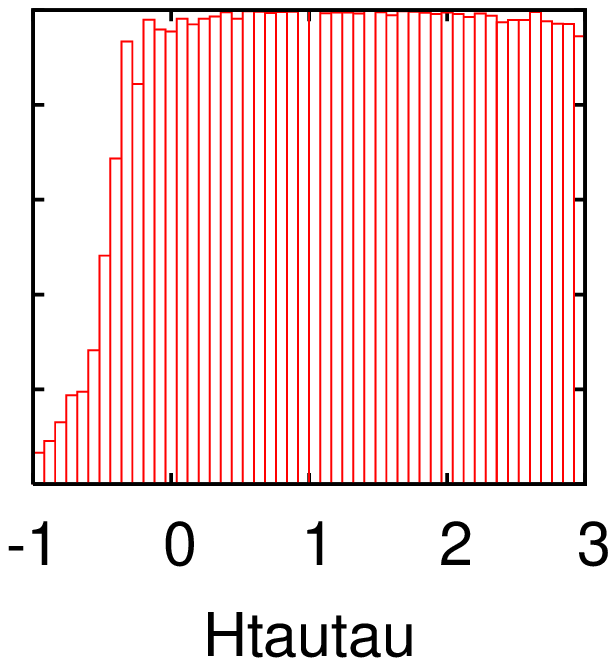} 
 \includegraphics[width=0.135\textwidth]{%
     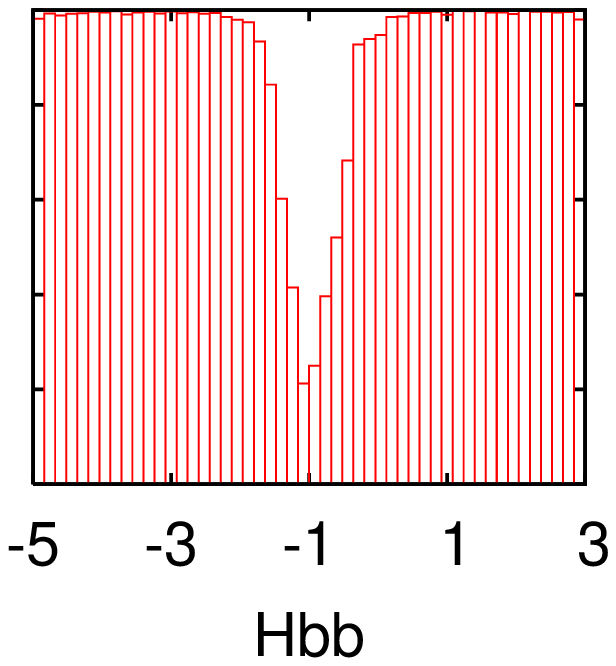} 
 \includegraphics[width=0.135\textwidth]{%
     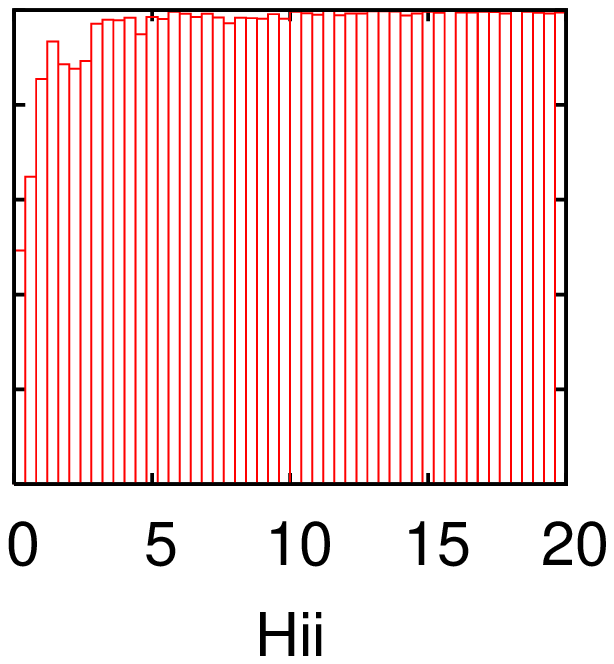} \\[6ex]

 \hspace*{0.135\textwidth} 
 \hspace*{0.135\textwidth} 
 \hspace*{0.135\textwidth} 
 \hspace*{0.135\textwidth} 
 \hspace*{0.135\textwidth} 
 \psfrag{Yc}[c][c][1][0]{$Y_c [\gev]$}
 \psfrag{SM}[c][c][1][0]{SM\hspace*{3ex}}
 \psfrag{15.4}[c][c][1][0]{\hspace*{1.5ex}15.4}
 \psfrag{50}[c][c][1][0]{\hspace*{1ex}50}
 \psfrag{100}[c][c][1][0]{100}
 \includegraphics[width=0.135\textwidth]{%
     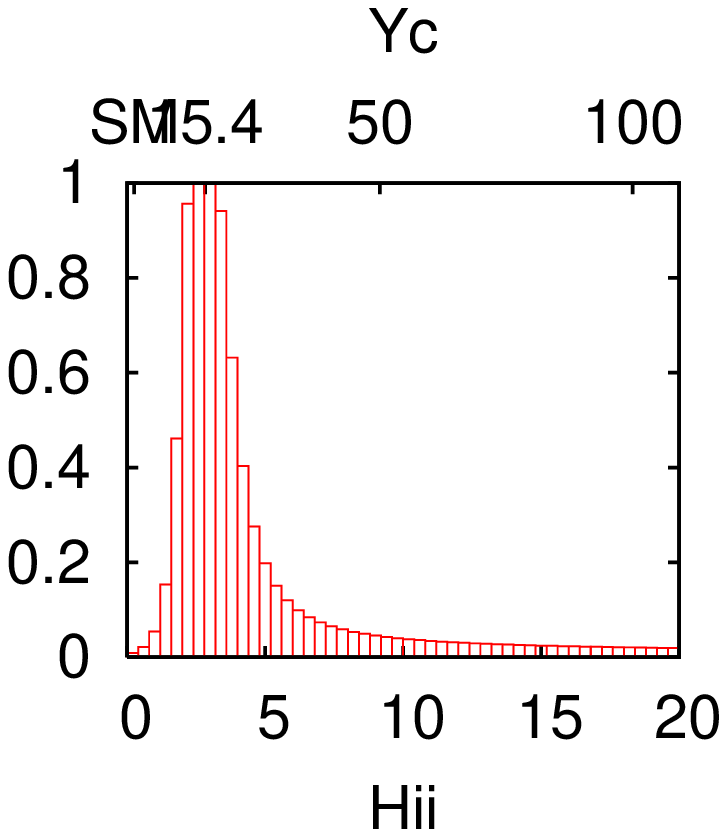} \\
 \end{center}
\caption[]{Profile likelihoods for the standard analysis (top row,
  copied from Fig.~\ref{fig:corr_1}), an increased $ccH$ coupling with
  no additional contribution to the Higgs width (second row), an
  increased $ccH$ coupling and a scaling factor for the total width
  (third row), and finally this scaling factor only (single bottom
  panel).  All experimental and theory errors included for
  low-luminosity running.}
\label{fig:cch}
\end{figure}

There are two kinds of Higgs decays which we have not yet considered
in this analysis. First, the Higgs boson can decay into invisible
particles, like dark-matter agents, which would appear as missing
transverse energy in the detector.  As discussed in the introduction
we skip this kind of analysis.\bigskip

What is more interesting is an actually unobservable Higgs decay, \ie
a Higgs decay which at the LHC we could not see. Looking at the
particle content of the Standard Model, any decay into light leptons
would be immediately visible. Shifted couplings to third-generation
fermions are already part of our parameter set, as is the coupling to
gluons and the electroweak gauge bosons. One thing that could still
happen are Higgs decays to light quarks, which would completely vanish
in the QCD continuum.\bigskip

As an example, we consider an increase of the charm Yukawa coupling to
$15.4~\gev$, which corresponds to a branching ratio of roughly
80\%. To be consistent, we also include this coupling in the rate for
inclusive Higgs production. Identifying the light-jet coupling with
the charm Yukawa is conservative in the sense that it minimizes the
production rate for a given light-jet partial width because of the
small charm parton densities.

In the parameter extraction, the first effect of this additional
coupling is the increase of the inclusive production rate. This effect
is stronger than suggested by just comparing $g_{ggH}$ and $g_{ccH}$,
because the latter induces a tree-level decay. This rate enhancement
can be counter balanced by reducing the branching ratios, \ie by
reducing all relevant Higgs couplings to final-state particles. This
implies very small rates from weak boson fusion, driving them deep
into the Poisson region for $30~\ifb$. The effect on the parameter
extraction we see in Fig.~\ref{fig:cch}: in the second row the most
likely points of all couplings are shifted by roughly the same
amount. In addition, the errors are increased significantly, in
particular for $g_{\tau\tau H}$. The separate peaks for $g_{ggH}$,
corresponding to the different signs of the individual couplings, get
smeared out into a large band of possible solutions.

The consequence of such a result should be a more careful analysis,
now including a free Higgs width. Such a shift shown in
eq.(\ref{eq:deltagamma}) can be implemented as a decay to an
unobservable new-physics particle which does not appear inside the
proton.  Known examples for such effects are unobservable four-body
Higgs decays~\cite{four_body}: The result we show in the third row of
Fig.~\ref{fig:cch}. We always find a best-fitting parameter point and
the error bars are hugely inflated, in particular to large couplings.
For the Standard Model couplings only the region around $\Delta=-1$,
where the coupling vanishes, is excluded. The smallest possible solution
we obtain when the additional contribution to the width is zero. In
that case we reproduce the slope from the no-additional-contribution
scenario, modulo statistics effects. The region of large couplings is
similar to the scaling of a free total Higgs width discussed in
Section~\ref{sec:higgs}.  Any increase can be compensated by a
corresponding increase of the invisible decay mode. The predictivity
of such an analysis is therefore rather limited.\bigskip

There is a second scenario we can test at this stage: if there appears
to be a problem with the total width we can fix all Standard Model
couplings to their respective nominal values and only allow for an
additional experimentally unobservable decay mode. This leads to a very
simple one-dimensional analysis, where only $\Delta_\Gamma$ is varied.
Technically, we identify this increase with a large $ccH$ coupling, but
without including its contribution to the Higgs production rate. Under
this assumption we can determine the size of the additional contribution
to the width, as shown in the bottom line of Fig.~\ref{fig:cch}.\bigskip

\begin{figure}[t]
 \begin{center}
 {\raggedright $1/\Delta \chi^2$\\[2ex]}
 \includegraphics[width=0.135\textwidth]{%
     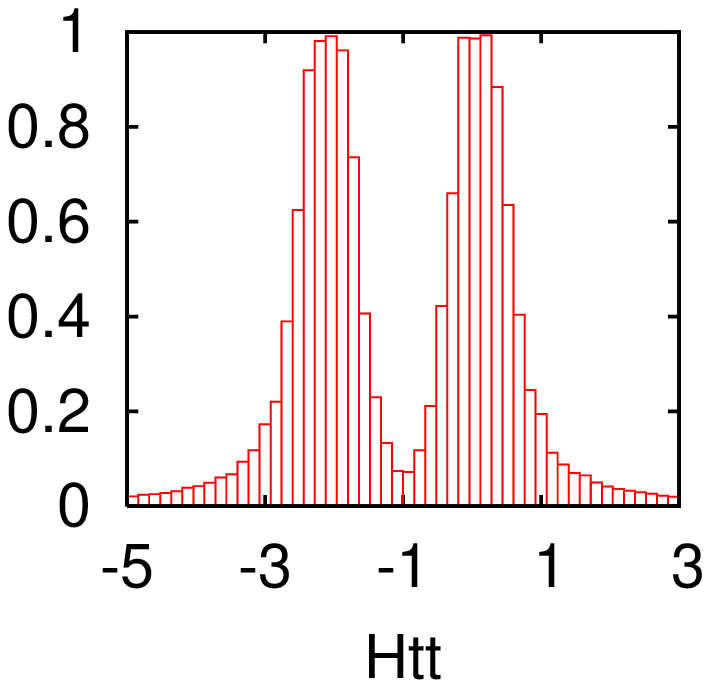} 
 \includegraphics[width=0.135\textwidth]{%
     sfitter_markovbins3.Hgg.phaseW.truesubjet.eps} 
 \includegraphics[width=0.135\textwidth]{%
     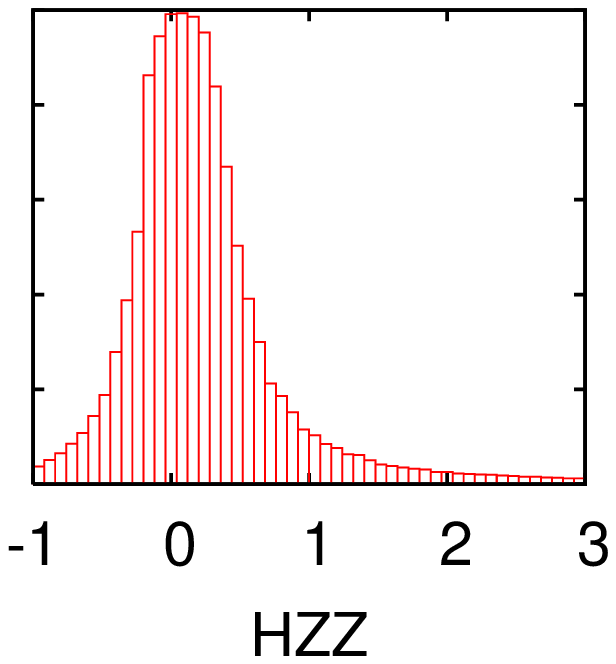} 
 \includegraphics[width=0.135\textwidth]{%
     sfitter_markovbins3.Htautau.phaseW.truesubjet.eps} 
 \includegraphics[width=0.135\textwidth]{%
     sfitter_markovbins3.Hbb.phaseW.truesubjet.eps} 
 \hspace*{0.135\textwidth} \\[2ex]
 \includegraphics[width=0.135\textwidth]{%
     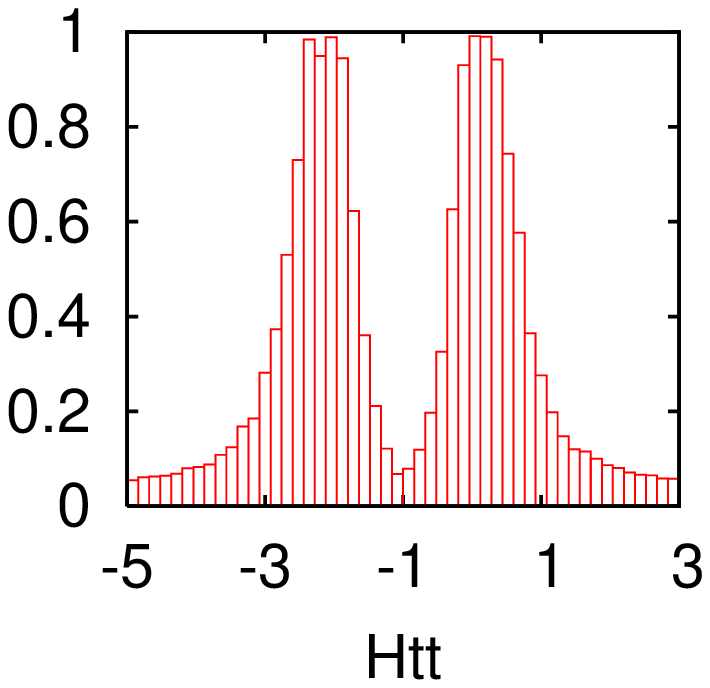} 
 \includegraphics[width=0.135\textwidth]{%
     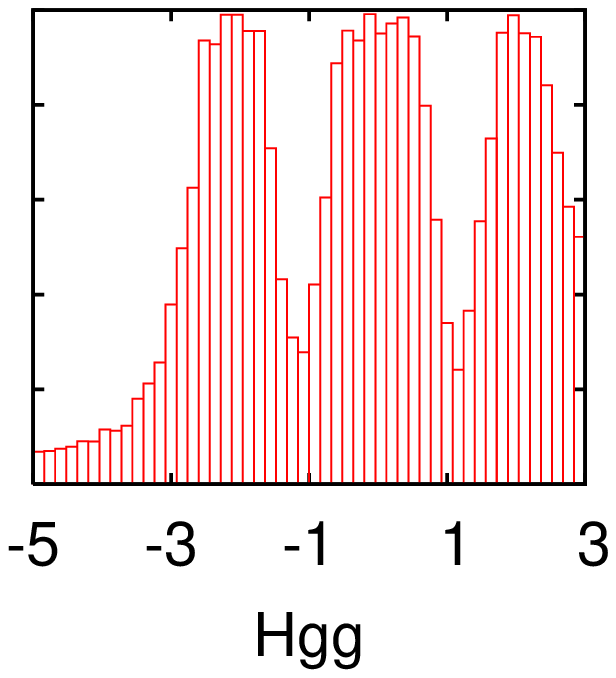} 
 \includegraphics[width=0.135\textwidth]{%
     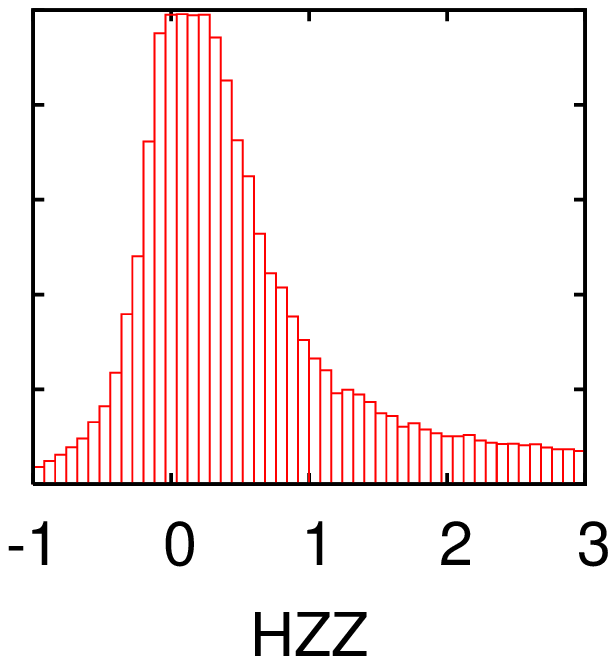} 
 \includegraphics[width=0.135\textwidth]{%
     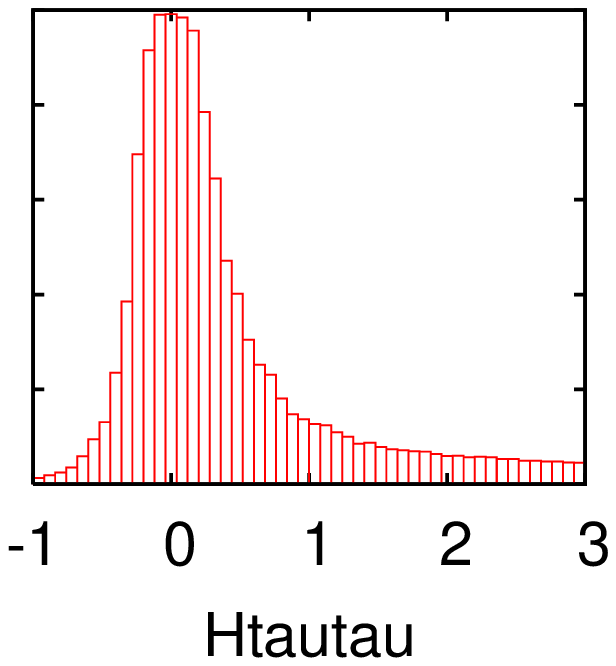} 
 \includegraphics[width=0.135\textwidth]{%
     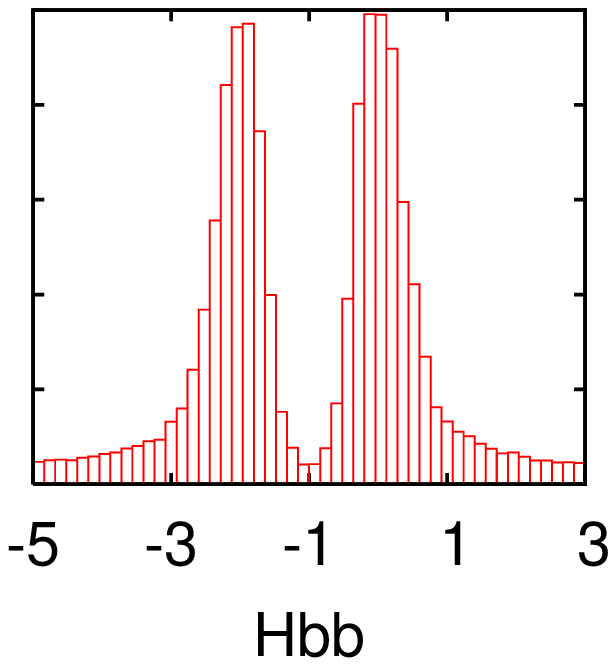}
 \includegraphics[width=0.135\textwidth]{%
     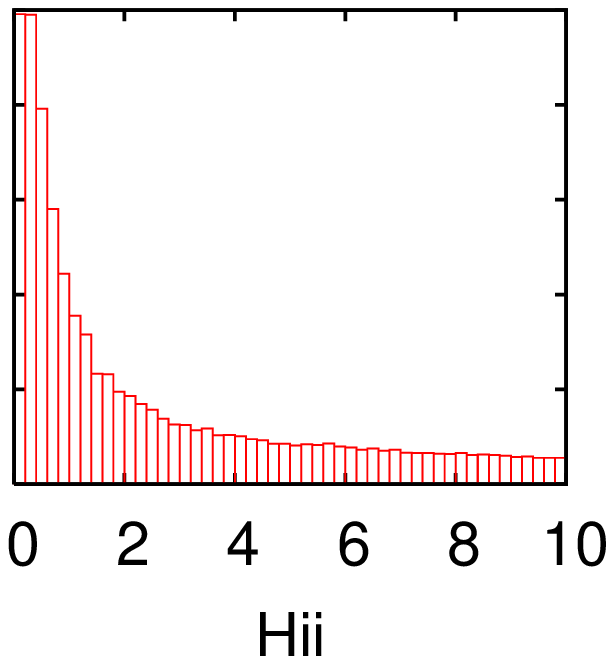} 
 \end{center}
\caption[]{Profile likelihoods for the standard parameter analysis
  (top row, copied from Fig.~\ref{fig:corr_1}), and an additional
  contribution to the total width while at the same time keeping
  $g_{WWH}$ fixed to its Standard model value (bottom rows).  All
  experimental and theory errors included for low-luminosity running.}
\label{fig:inv}
\end{figure}

Finally, there is a third way to establish an upper bound on
additional contributions to the Higgs width. Following the previous
subsection we cannot simply add an additional contribution to the
total width, because this would eliminate the possibility of
determining the absolute scale of the couplings. On the other hand,
following the argument in Section~\ref{sec:higgs} we can trade the
assumption on the Higgs width for a limit in any other coupling, for
example fixing $g_{WWH}$ to its Standard Model value. As we will see
in Section~\ref{sec:err} this is the best-determined coupling of the
parameter set. In models with more than one doublet this means that we
either assume a rapid decoupling of the additional heavy states or
close enough Higgs modes that we cannot resolve them in the transverse
mass reconstruction of the $WW$ final state~\cite{wbf_broad}.

In Fig.~\ref{fig:inv} we see that this indeed retains well-defined
solutions. The errors on the different parameters stay at a similar
level as the original analysis, with $g_{bbH}$ and $g_{\tau\tau H}$
being measured mildly more precisely. Both of them are determined in
channels with a production-side $g_{WWH}$.  The impact of the
$Z$ boson replacing the $W$ in these channels is always
sub-leading. The additional coupling ranges around zero with a
standard deviation of roughly twice the Standard-Model width. This
error corresponds to the error in the bottom Yukawa coupling, which is
the main observable partial width and well determined.\bigskip

The results for these different scenarios give us some confidence in
the robustness of our weak-scale analysis. While the LHC cannot make
any direct statement about the total Higgs width (as discussed in
Section~\ref{sec:higgs}), effects from unobserved decays or
shortcomings of our effective Lagrangian hypothesis will surface as
unexplained effects in the parameter extraction. These effects then
trigger for example modifications of our ansatz and can be studied.

\subsection{Coupling ratios}
\label{sec:ratios}

Given the sizable correlations in the Higgs-sector parameter space
we could try to define better-suited parameters than individual Higgs
couplings. In a way, this is in analogy to the extraction of
supersymmetric masses from cascade decays, where we can extract mass
differences much more precisely than the actual masses. In the Higgs
sector the hope is that ratios of couplings are better suited for an
extraction from LHC measurements, for example coming from different
decays but the same production process (or vice
versa)~\cite{couplings_first,duehrssen}. In this situation some of the
systematic and theory errors will cancel.  The same trick allows us to
extract flavor-physics limits from data in the presence of large
uncertainties in low-energy QCD predictions or form factors.\bigskip

\begin{figure}[t]
 \begin{center}
\psfrag{log\(HZZ\)}[c][][1][0]{%
 \raisebox{-3ex}{$\log(1+\Delta_{ZZH})$}}
\psfrag{log\(Htt\)}[c][][1][0]{%
 \raisebox{-3ex}{$\log(1+\Delta_{ttH})$}}
\psfrag{log\(Htautau\)}[c][][1][0]{%
 \raisebox{-3ex}{$\log(1+\Delta_{\tau\tau H})$}}
\psfrag{log\(HZZ/HWW\)}[c][][1][0]{%
 \raisebox{-3ex}{$\log(\frac{1+\Delta_{ZZH}}{1+\Delta_{WWH}})$}}
\psfrag{log\(Htt/HWW\)}[c][][1][0]{%
 \raisebox{-3ex}{$\log(\frac{1+\Delta_{ttH}}{1+\Delta_{WWH}})$}}
\psfrag{log\(Htautau/HWW\)}[c][][1][0]{%
 \raisebox{-3ex}{$\log(\frac{1+\Delta_{\tau\tau H}}{1+\Delta_{WWH}})$}}
\psfrag{log\(Hbb/HWW\)}[c][][1][0]{%
 \raisebox{-3ex}{$\log(\frac{1+\Delta_{bbH}}{1+\Delta_{WWH}})$}}
\psfrag{log\(Hgg/HWW\)}[c][][1][0]{%
 \raisebox{-3ex}{$\log(\frac{1+\Delta_{ggH}}{1+\Delta_{WWH}})$}}
\psfrag{log\(Hgamgam/HWW\)}[c][][1][0]{%
 \raisebox{-3ex}{$\log(\frac{1+\Delta_{\gamma\gamma H}}{1+\Delta_{WWH}})$}}
 {\raggedright $\qquad 1/\Delta \chi^2$\\}
 \includegraphics[width=0.3\textwidth]{%
     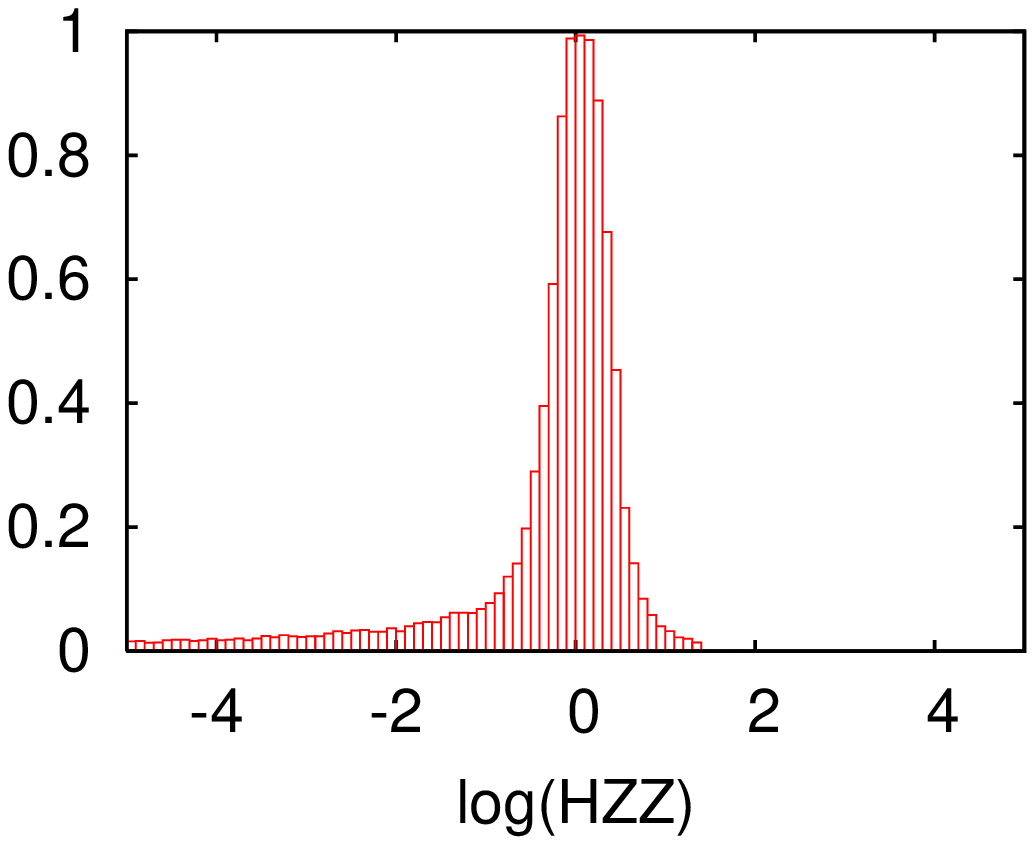} 
 \includegraphics[width=0.3\textwidth]{%
     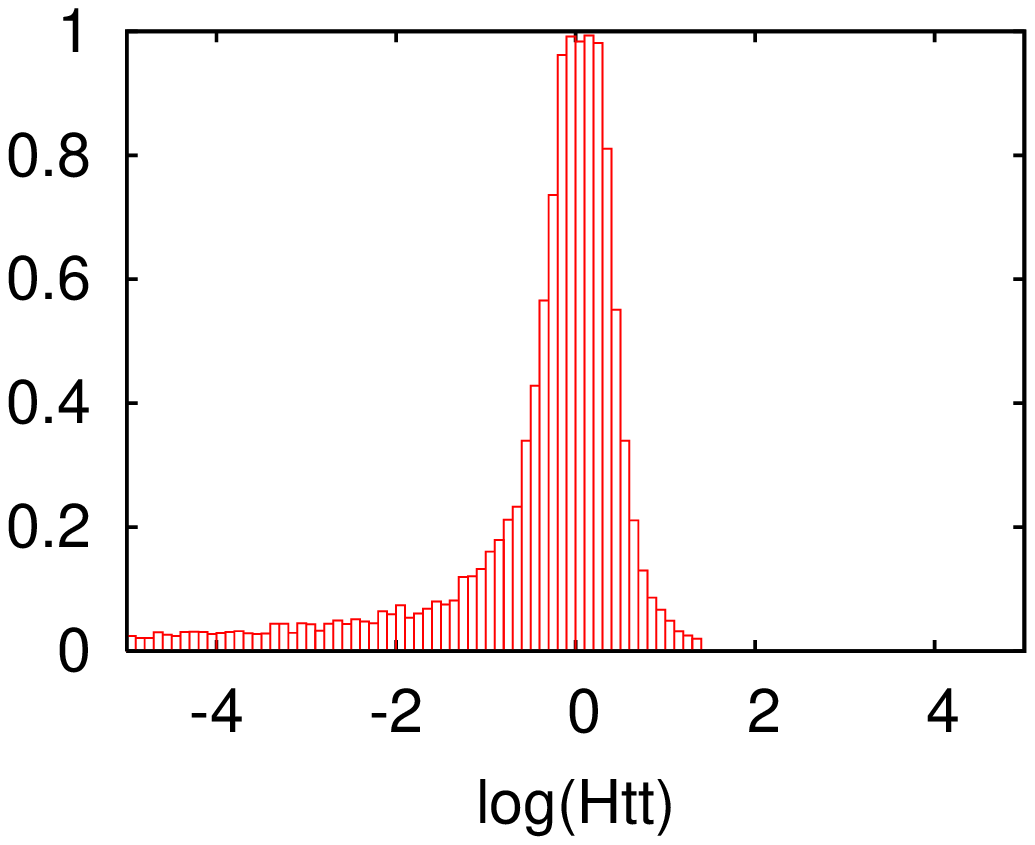} 
 \includegraphics[width=0.3\textwidth]{%
     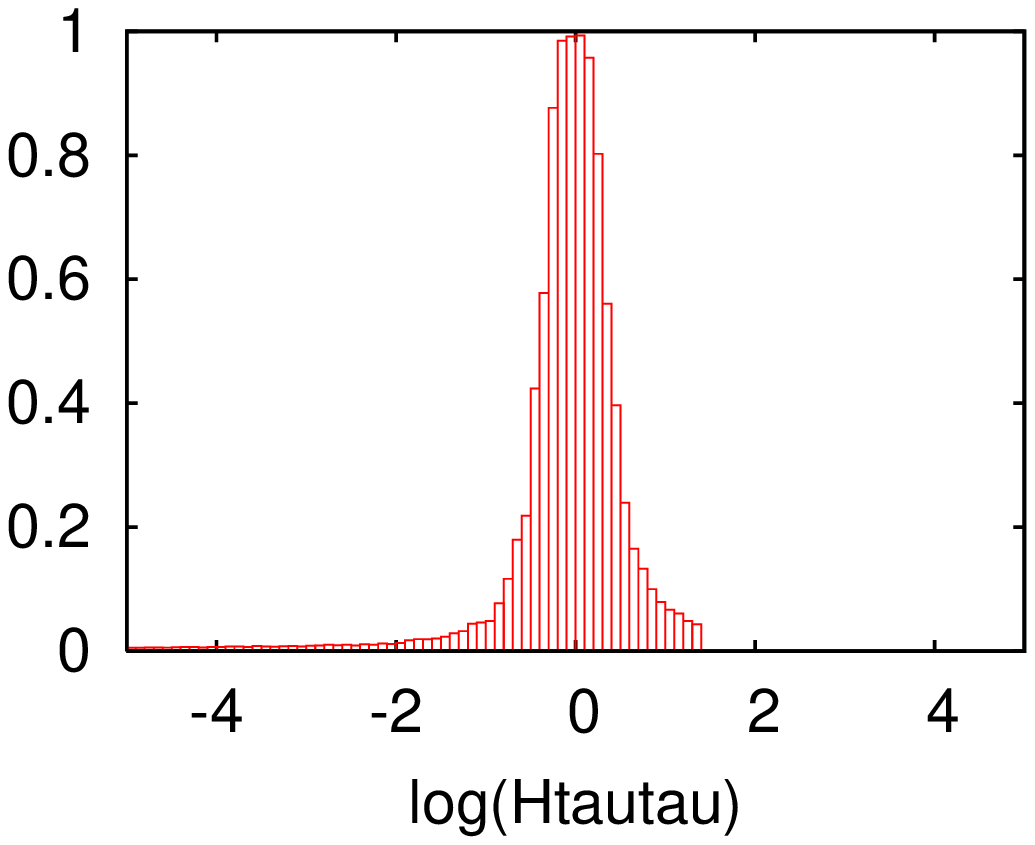} \\[2ex]
 \includegraphics[width=0.3\textwidth]{%
     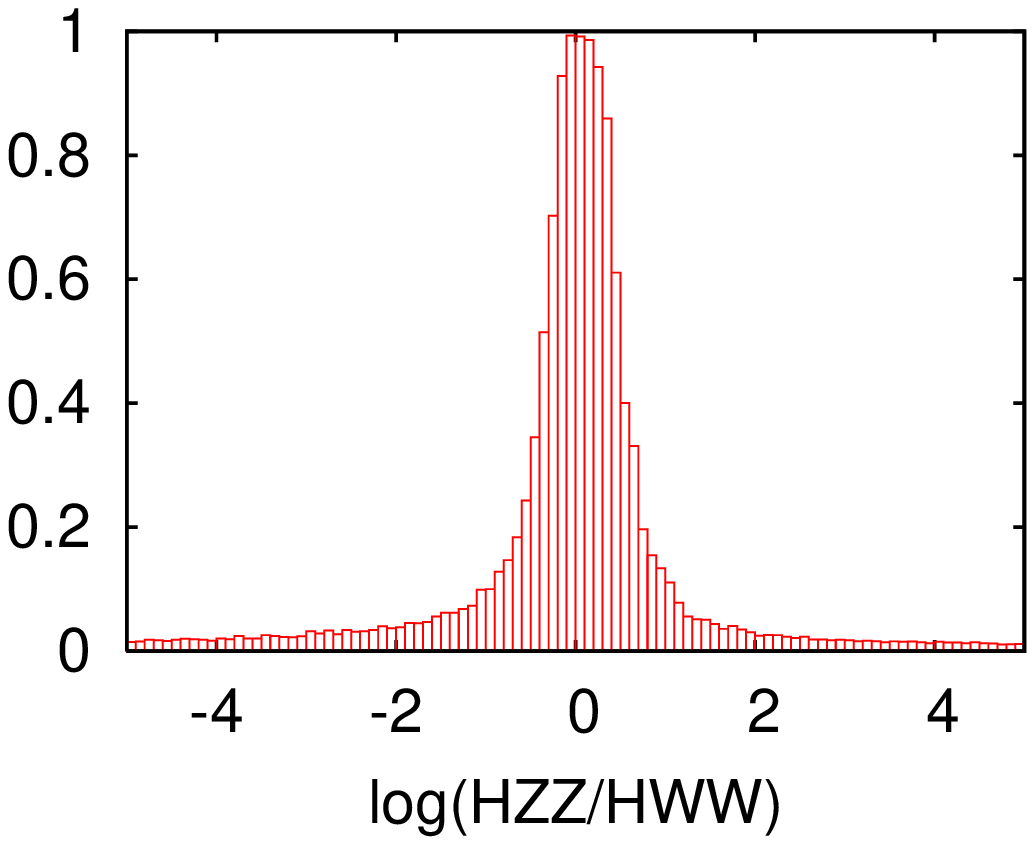} 
 \includegraphics[width=0.3\textwidth]{%
     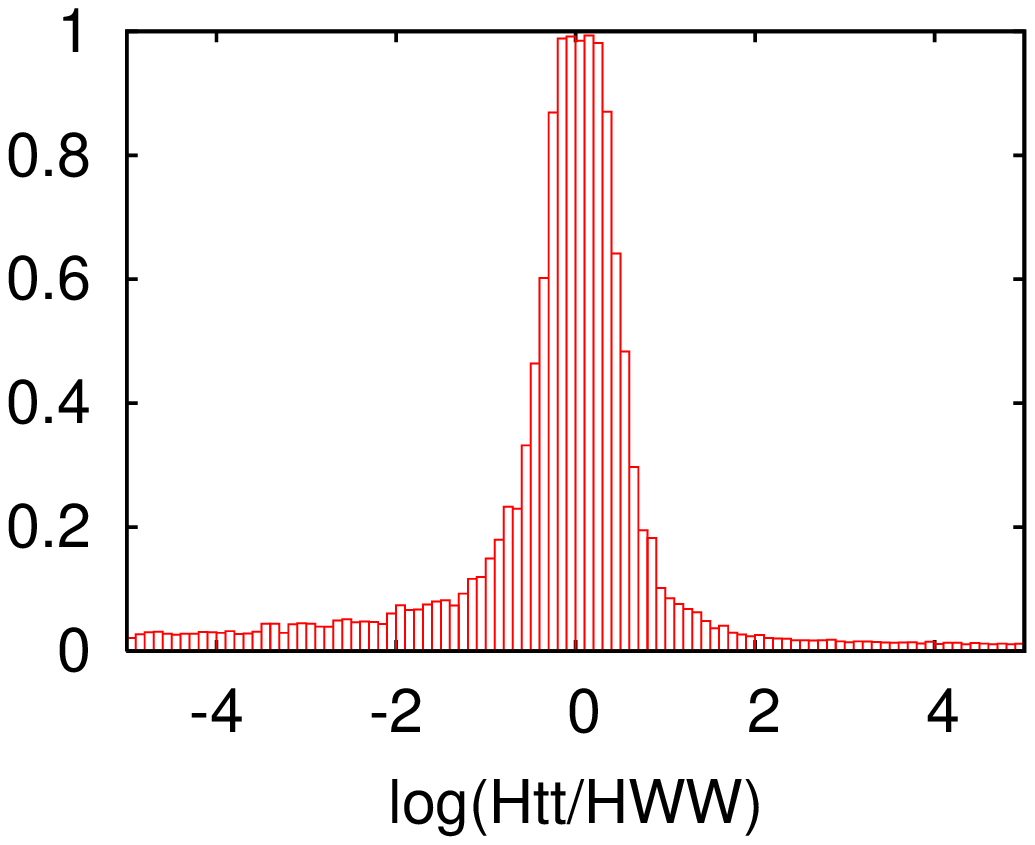} 
 \includegraphics[width=0.3\textwidth]{%
     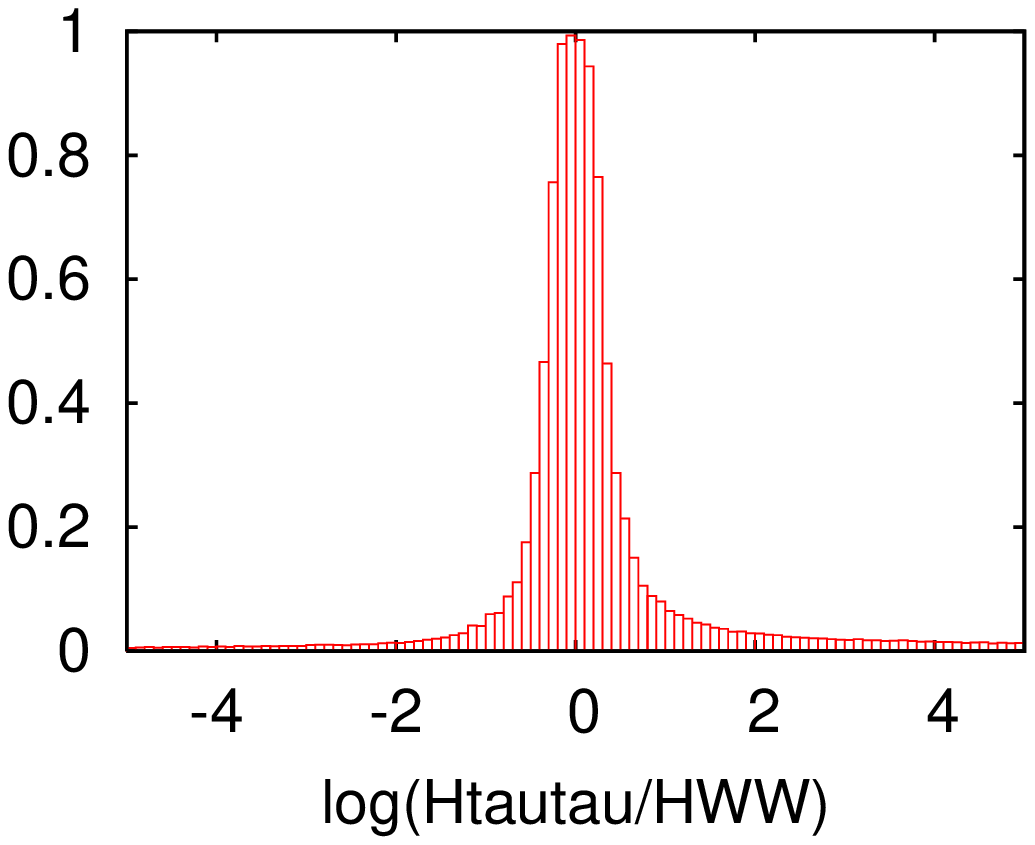} \\[2ex]
 \includegraphics[width=0.3\textwidth]{%
     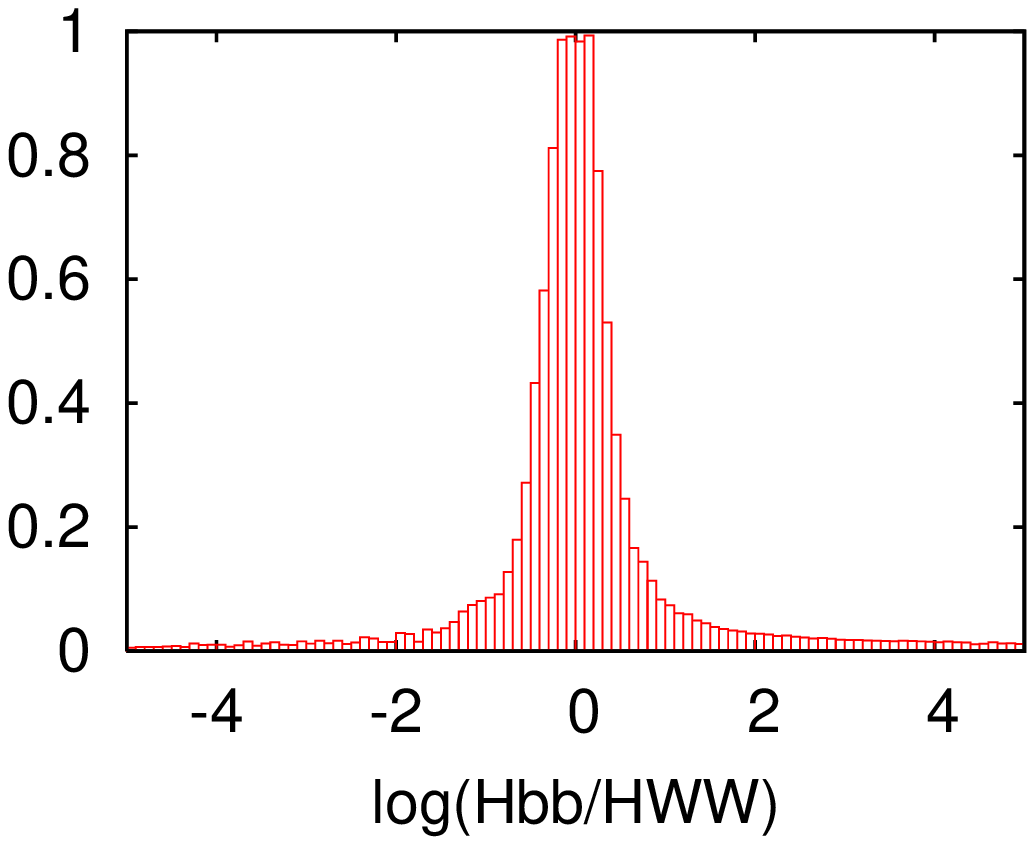} 
 \includegraphics[width=0.3\textwidth]{%
     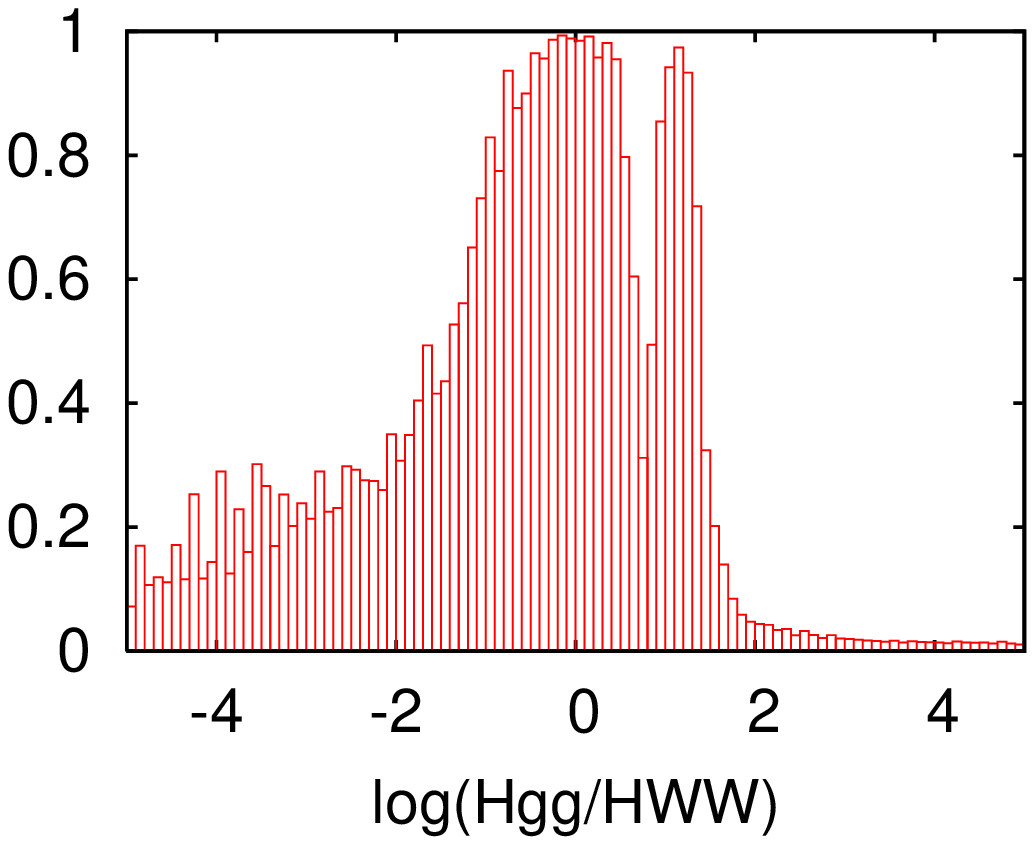} 
 \includegraphics[width=0.3\textwidth]{%
     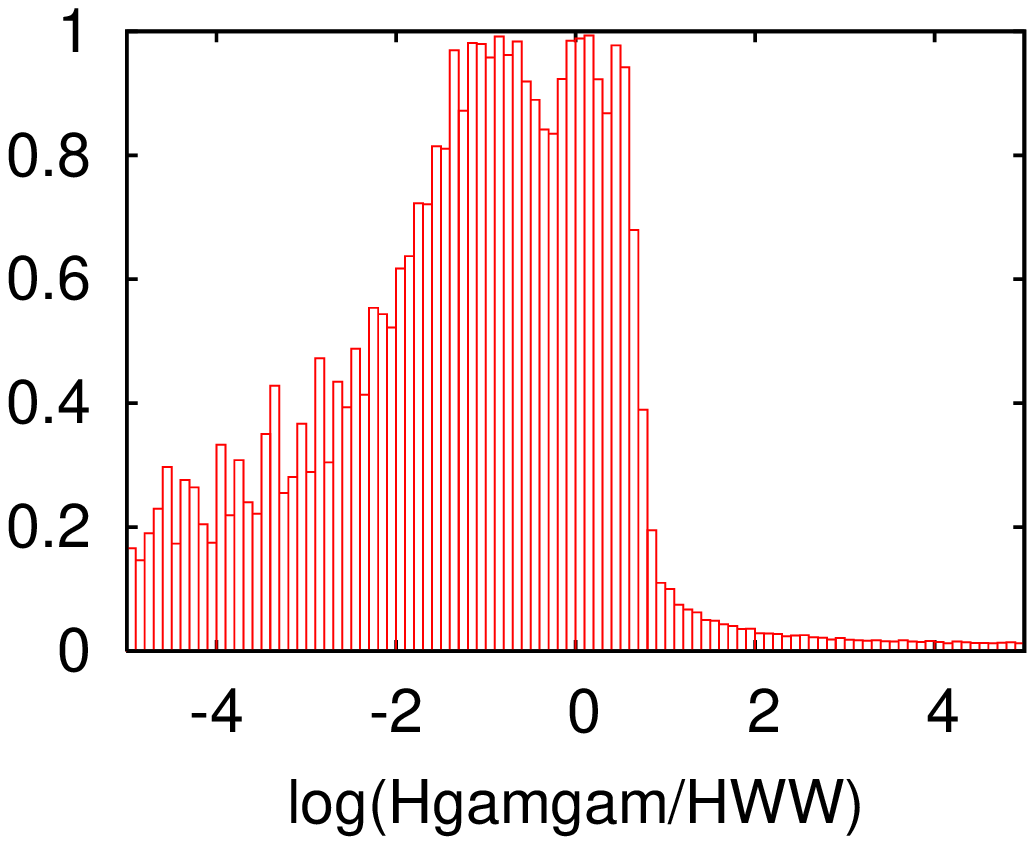} 
 \end{center}
\caption[]{Profile likelihoods for the absolute couplings (top row,
  same input as in Fig~\ref{fig:corr_f3}), for the logarithm of the ratio of the
  $g_{ZZH}$, $g_{ttH}$, $g_{\tau \tau H}$ (center row, left
  to right), $g_{bbH}$, $g_{ggH}$ and
  $g_{\gamma\gamma H}$ (bottom row, left to right) couplings and the $g_{WWH}$
  coupling. All couplings/ratios are normalized to their Standard Model value. 
We include all experimental and theory errors for a running of $30~\ifb$.}
\label{fig:ratios}
\end{figure}

In general, we might expect for example coupling ratios of $1/2$ and $2$ to be
symmetric and equally likely. However, naively showing the coupling ratio would tilt this
distribution, so in Fig.~\ref{fig:ratios} we instead show the logarithms of
the absolute values of coupling ratios, normalized to their respective
Standard Model values; $g_{WWH}$ serves as the general
normalization. For the couplings present in the Standard Model we see
the expected peak around zero. The secondary solution of $\Delta_{ttH}
= -2$ also gets mapped to zero by our formulas for the ratios. For the
effective couplings $g_{ggH}$ and $g_{\gamma\gamma H}$ we see
secondary peaks, originating from the alternative solutions in
Fig.~\ref{fig:corr_f3}. For the $ggH$ coupling both the peaks at
$\Delta_{ggH}=0$ and $-2$, where the latter corresponds to a top
coupling with the correct sign and over-compensation by the effective
coupling, appear at zero. The peak at $\Delta_{ggH}=2$, compensating
for a negative top coupling, appears just above one for the
logarithmic ratio. In the $\gamma\gamma H$ case we see in principle
the same effects. As the top-quark contribution is sub-leading we now
have four solutions.  They appear as three different peaks in the
logarithmic ratio, because again $\Delta_{\gamma\gamma H}=0$ and $-2$
coincide at zero deviation.

Both effective couplings also show a softer shoulder below the peak
than above.  The same effect we see for $g_{ZZH}$ and $g_{ttH}$,
albeit less pronounced. This is due to the fact that reducing the
couplings gives us a well-defined worst case of zero coupling, \ie no
signal events.  In contrast, increasing a coupling we can generate
arbitrarily large numbers of events with an arbitrarily small inverse
likelihood.\bigskip

To answer the crucial question, we show the coupling constants itself
in the top row of Fig.~\ref{fig:ratios}.  For $g_{ZZH}$ we see hardly
any difference between the coupling and the coupling ratio.  The main
measurement of this coupling arises from the gluon-fusion initial
state where $W$ bosons do not play any role. For $g_{ttH}$ the
situation is similar; both, $g_{ttH}$ and $g_{WWH}$ are determined
from a multitude of measurements so there is no particular benefit in
looking at this ratio. In contrast, for $g_{\tau\tau H}$ we can
determine the ratio more precisely than the absolute coupling, because
of the two final states in weak-boson-fusion production. The most
striking improvement we find for $g_{bbH}$, where we exploit the
correlation between the different occurrences of $g_{WWH}$ and the
total width, dominated by bottom decays. The detailed discussion of errors on the ratios we defer
to Section~\ref{sec:err}; it will give us an improvement on the
extraction of typically roughly $\sim 10 \%$.

\section{Higgs couplings}
\label{sec:couplings}

Before we can start determining the errors on the Higgs-sector
parameters given a complete simulated data set at the LHC, we have one
final step to implement. Until now, we have always tested our model on
a set of perfect measurements, \ie these measurements sit on their
central value as predicted by the Standard Model without any
smearing. As long as we do not extract error bars from such a data set
this assumption is safe. However, for the final word on errors we need
to move the signal and background rates away from the central value
simulating an expected statistical fluctuation.

\subsection{Smearing the data}
\label{sec:smear}

\begin{figure}[t]
 \begin{center}
 \includegraphics[width=0.24\textwidth]{%
     sfitter_markovbins3.HWW_Htt.phaseW.truesubjet.eps} \hspace*{-2ex}
 \includegraphics[width=0.24\textwidth]{%
     sfitter_markovbins3.HWW_Hbb.phaseW.truesubjet.eps} \hspace*{-2ex}
 \includegraphics[width=0.24\textwidth]{%
     sfitter_markovbins3.Hgamgam_Htt.phaseW.truesubjet.eps} \hspace*{-2ex}
 \includegraphics[width=0.24\textwidth]{%
     sfitter_markovbins3.Hgamgam_Hgg.phaseW.truesubjet.eps} \\

 \includegraphics[width=0.24\textwidth]{%
     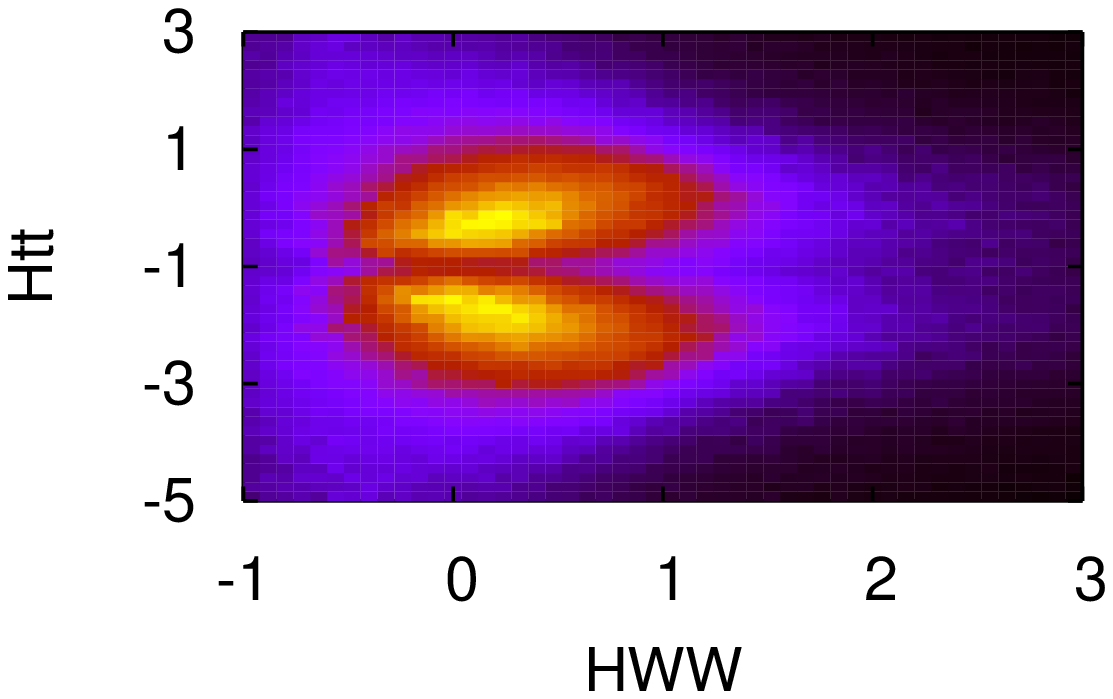} \hspace*{-2ex}
 \includegraphics[width=0.24\textwidth]{%
     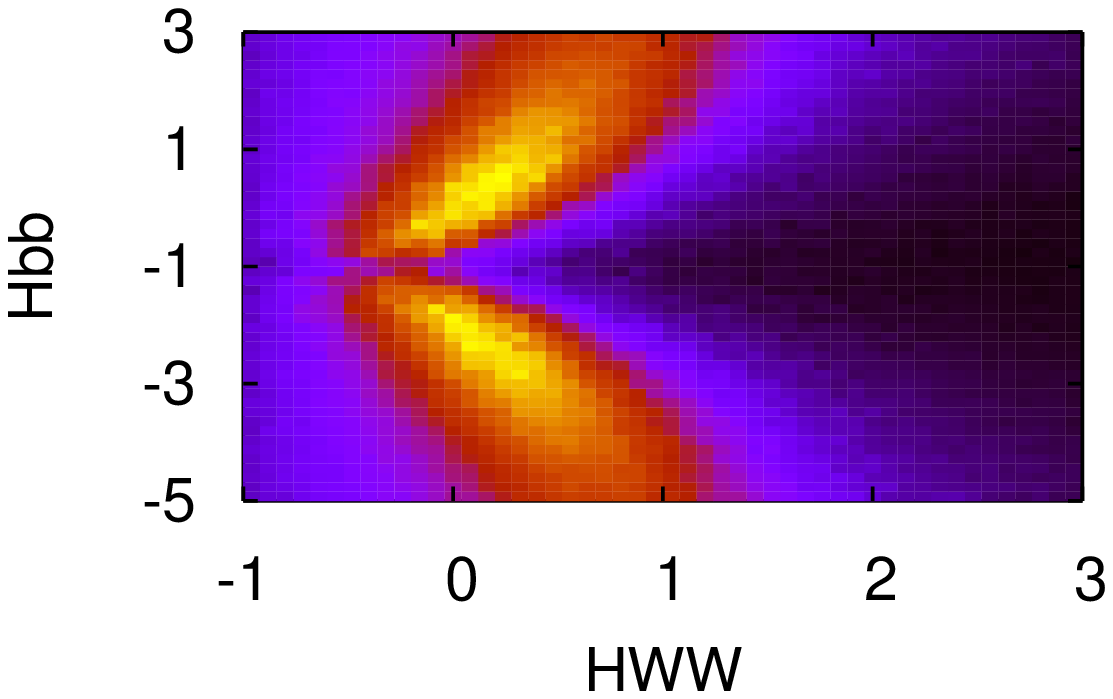} \hspace*{-2ex}
 \includegraphics[width=0.24\textwidth]{%
     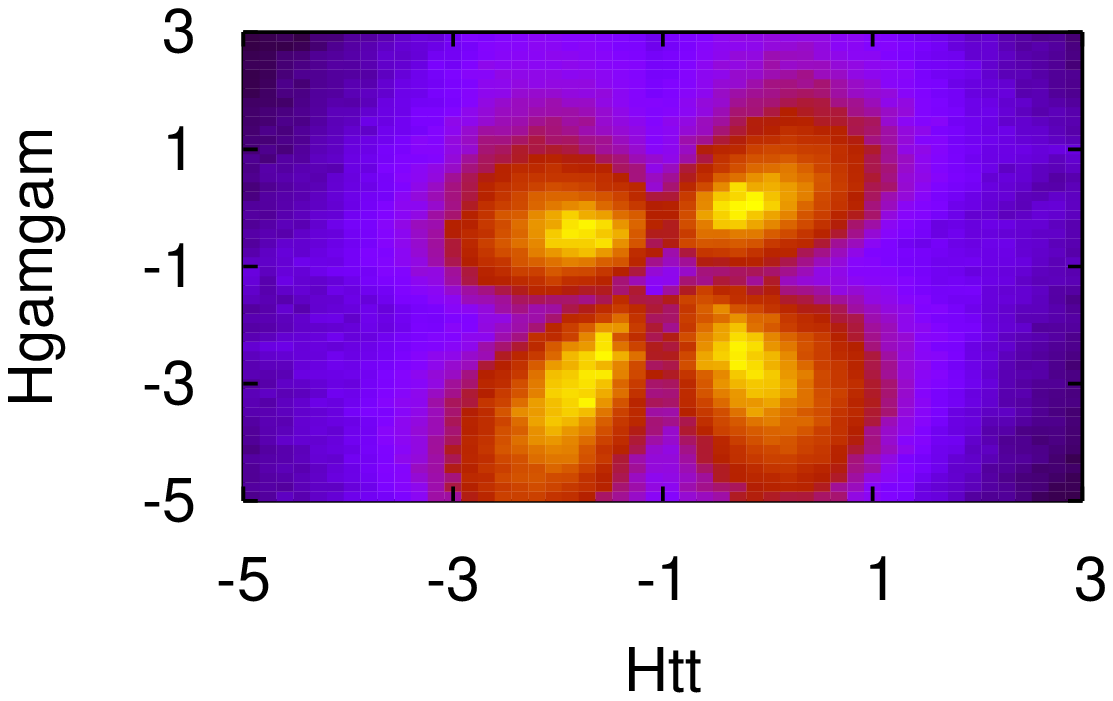} \hspace*{-2ex}
 \includegraphics[width=0.24\textwidth]{%
     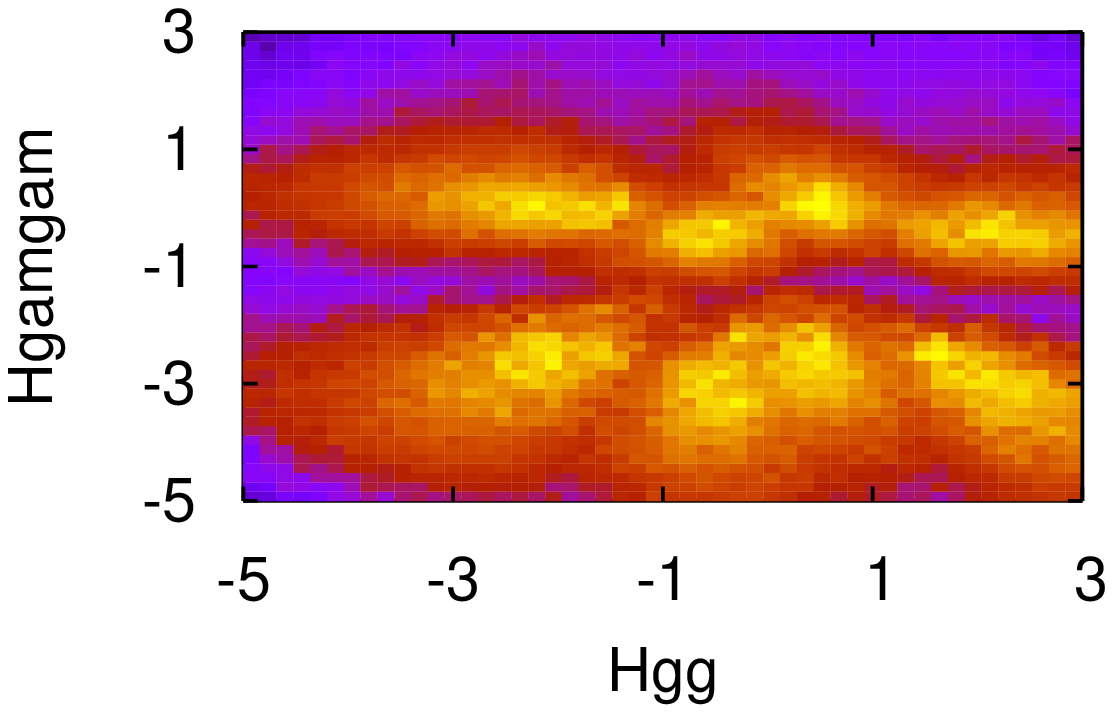} \\

 {\raggedright $1/\Delta \chi^2$\\[2ex]}
 \includegraphics[width=0.135\textwidth]{%
     sfitter_markovbins3.HWWlabel.phaseW.truesubjet.eps} 
 \includegraphics[width=0.135\textwidth]{%
     sfitter_markovbins3.Htt.phaseW.truesubjet.eps} 
 \includegraphics[width=0.135\textwidth]{%
     sfitter_markovbins3.Hbb.phaseW.truesubjet.eps} 
 \includegraphics[width=0.135\textwidth]{%
     sfitter_markovbins3.Htautau.phaseW.truesubjet.eps} 
 \includegraphics[width=0.135\textwidth]{%
     sfitter_markovbins3.Hgg.phaseW.truesubjet.eps} 
 \includegraphics[width=0.135\textwidth]{%
     sfitter_markovbins3.Hgamgam.phaseW.truesubjet.eps} \\[2ex]

 \includegraphics[width=0.135\textwidth]{%
     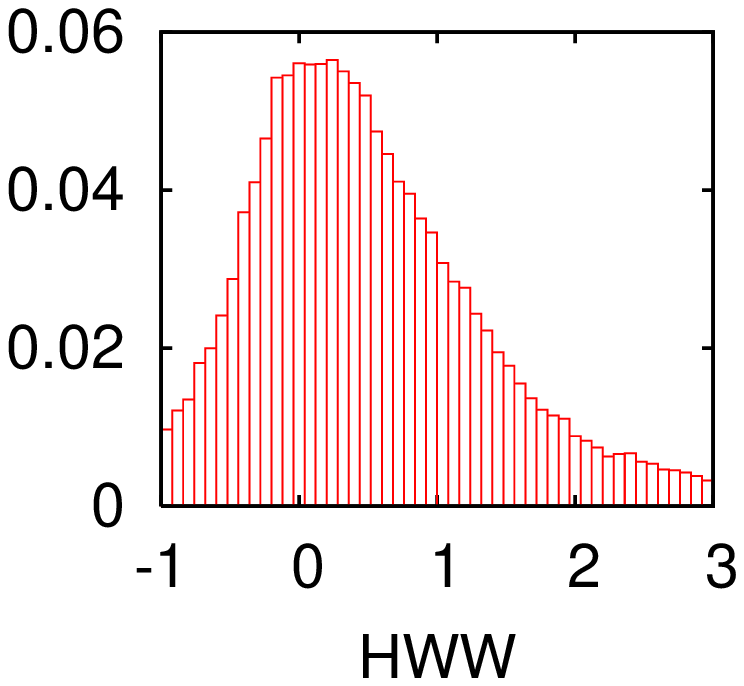} 
 \includegraphics[width=0.135\textwidth]{%
     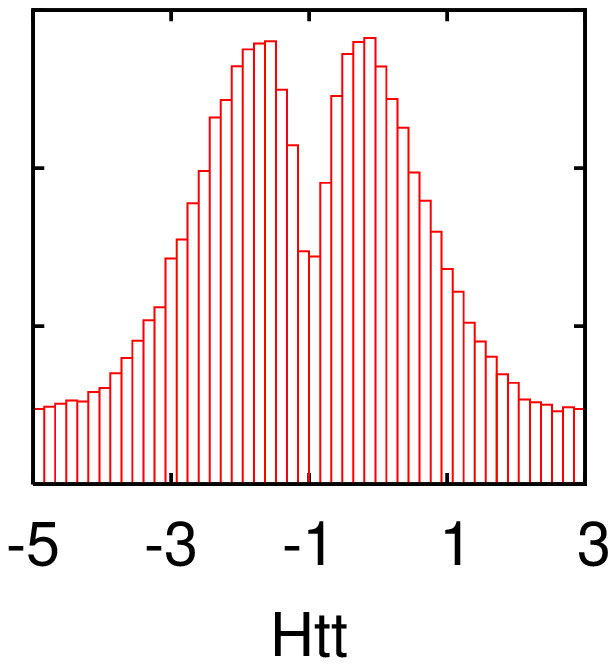} 
 \includegraphics[width=0.135\textwidth]{%
     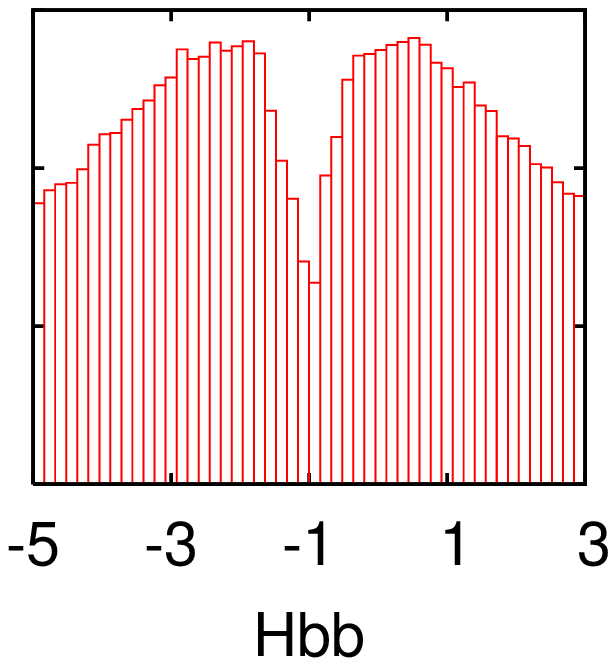} 
 \includegraphics[width=0.135\textwidth]{%
     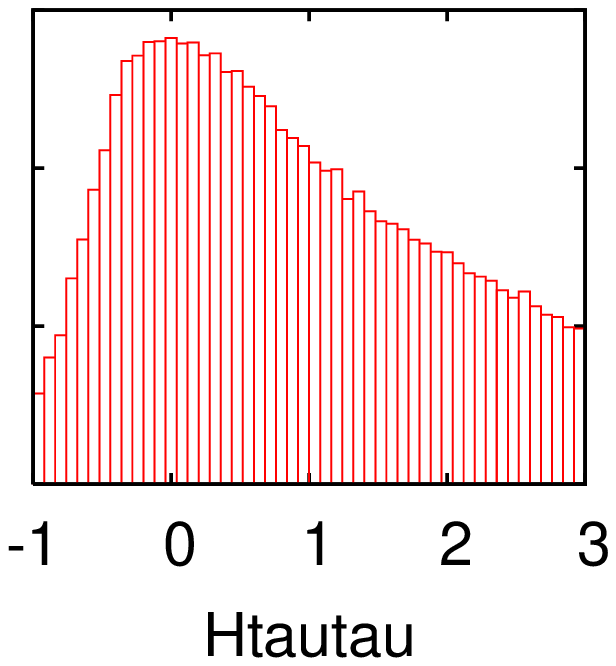} 
 \includegraphics[width=0.135\textwidth]{%
     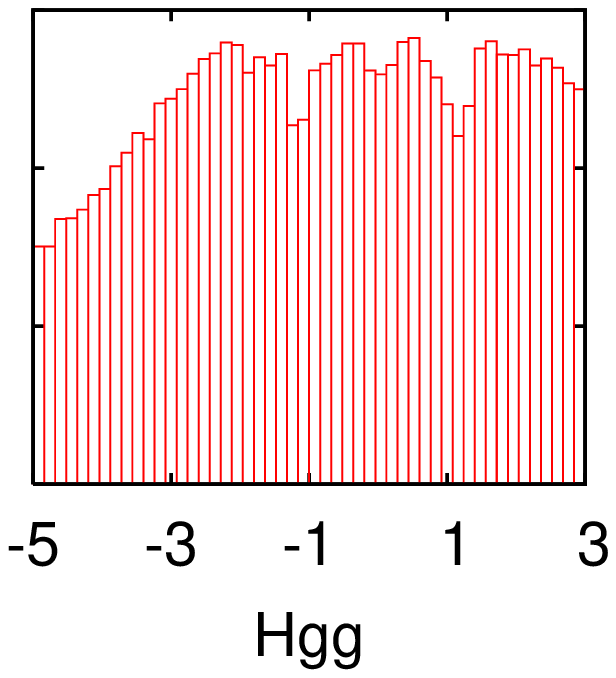} 
 \includegraphics[width=0.135\textwidth]{%
     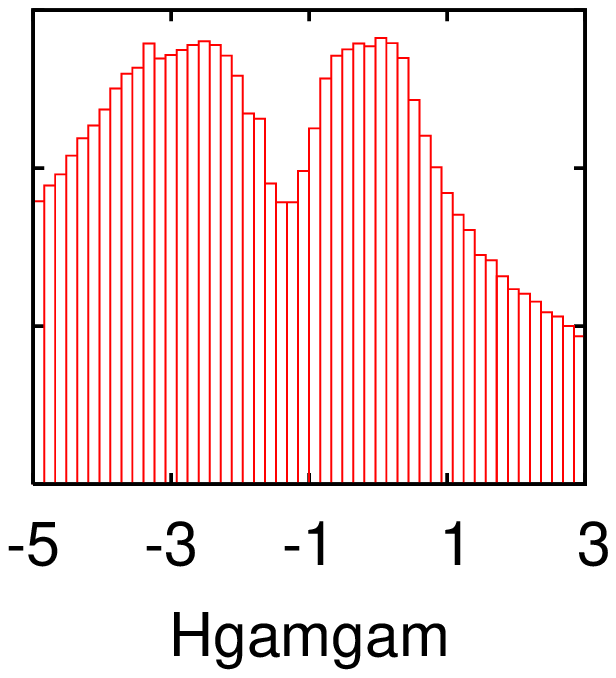}
 \end{center}
\caption[]{Profile likelihoods for unsmeared (top, taken from
  Fig.~\ref{fig:corr_f3}) and smeared (bottom) measurements assuming
  $30~\ifb$.  We include both experimental and theory errors in our
  analysis.}
\label{fig:smeared}
\end{figure}

One- and two-dimensional profile likelihoods for the Higgs parameter
extracted from a smeared set of measurements we show in
Fig.~\ref{fig:smeared}. This smearing of the one pseudo-measurement 
includes all experimental but no theory errors. 
We see that the behavior of our results does not change
significantly whether we use the true or smeared data set. The
best-fitting points are slightly shifted away from the correct
Standard-Model values. Moreover, the peak structures get 
broader, as the different measurements now attempt to pull the
parameters into different directions.  This indicates that any kind of
error estimate from these likelihood distributions has to be taken
with a grain of salt and we better rely on toy
experiments~\cite{sfitter} to extract reliable error bars for the
model parameters.\bigskip

\TABLE[b]{
\begin{small}
\begin{tabular}{l|r|r|r|r|r}
       & \multicolumn{5}{c}{best solutions} \\\hline
$\Delta_{WWH}$           &$  0.12$&$  0.11$&$  0.11$&$  0.09$&$  0.09$ \\
$\Delta_{ZZH}$           &$  0.12$&$  0.13$&$  0.11$&$  0.10$&$  0.09$ \\
$\Delta_{ttH}$           &$ -0.29$&$ -0.29$&$ -0.30$&$ -1.70$&$ -1.70$ \\
$\Delta_{bbH}$           &$  0.17$&$  0.16$&$ -2.15$&$ -2.12$&$ -2.11$ \\
$\Delta_{\tau\tau H}$    &$  0.02$&$  0.03$&$  0.03$&$ -0.10$&$ -0.10$ \\
$\Delta_{\gamma\gamma H}$&$ -0.01$&$ -2.43$&$ -2.42$&$ -0.38$&$ -2.76$ \\
$\Delta_{ggH}$           &$  0.47$&$ -1.85$&$  0.34$&$ -0.48$&$ -0.47$ \\
\end{tabular}
\end{small}
\caption[]{Best solutions for one (experimentally) smeared data set with
  $30~\ifb$. The Higgs mass is 120~GeV.
  Both experimental and theory errors included.}
\label{tab:best}
}

In Table~\ref{tab:best} we list the five best-fitting parameter
points, obtained from an additional Minuit run starting from the best
points of our (weighted and cooling) Markov chains.  The first point
is the closest to the Standard Model, where all couplings have the
correct sign.  All alternative solutions appear after flipping the
sign of one or several of the couplings via $\Delta=-2$.  Only in the
$\gamma\gamma H$ effective coupling, where the $W$ and the top
interfere destructively does the sign play a role. In the second
column the sign of the two effective couplings is changed. The
additional contributions to the effective couplings just
over-compensate the Standard-Model parts. In the third column the
bottom Yukawa coupling has a negative sign and this is reflected in a
slight decrease in the effective gluon coupling. When we flip the
signs from both the top and bottom quark in column four the $ggH$
contribution is back to its original absolute value.  In addition, we
see the shift in the effective photon coupling, which has to
compensate for the now constructive interference of the $W$ and $t$
loops. The same shift appears in the fifth column where the effective
photon coupling over-compensates the Standard-Model
contributions.\bigskip

Note that none of these solutions are `true' alternative solutions,
\ie minima in $\chi^2$ not correlated to the best-fit values and with
a distinctly different underlying set of model parameters. This by
hindsight justifies our choice of profile likelihoods for this
analysis, as compared to Bayesian probabilities. Practically it also
means that for the remaining part of this paper we focus on the `most
physical' parameter point, \ie we ignore secondary points with
$\Delta_{jjH} \sim -2$.

\subsection{Best fit with error bars}
\label{sec:err}

\TABLE[b]{
\begin{small}
\begin{tabular}{l|l|l|ll|l|l|ll}
 &         \multicolumn{4}{c|}{no effective couplings} &
           \multicolumn{4}{c}{with effective couplings} \\
 & 
  RMS & $\sigma_\text{symm}$ & $\sigma_\text{neg}$ & $\sigma_\text{pos}$ & 
  RMS & $\sigma_\text{symm}$ & $\sigma_\text{neg}$ & $\sigma_\text{pos}$ \\\hline 
$m_H$                      
 & $\pm\,0.36$ & $\pm\,0.26$ & $-\,0.26$ & $+\,0.26$ 
 & $\pm\,0.38$ & $\pm\,0.25$ & $-\,0.26$ & $+\,0.25$ \\
$\Delta_{WWH}$             
 & $\pm\,0.31$ & $\pm\,0.23$ & $-\,0.21$ & $+\,0.26$ 
 & $\pm\,0.29$ & $\pm\,0.24$ & $-\,0.21$ & $+\,0.27$ \\
$\Delta_{ZZH}$             
 & $\pm\,0.49$ & $\pm\,0.36$ & $-\,0.40$ & $+\,0.35$ 
 & $\pm\,0.46$ & $\pm\,0.31$ & $-\,0.35$ & $+\,0.29$ \\
$\Delta_{ttH}$       
 & $\pm\,0.58$ & $\pm\,0.41$ & $-\,0.37$ & $+\,0.45$ 
 & $\pm\,0.59$ & $\pm\,0.53$ & $-\,0.65$ & $+\,0.43$ \\
$\Delta_{bbH}$       
 & $\pm\,0.53$ & $\pm\,0.45$ & $-\,0.33$ & $+\,0.56$ 
 & $\pm\,0.64$ & $\pm\,0.44$ & $-\,0.30$ & $+\,0.59$ \\
$\Delta_{\tau\tau{}H}$ 
 & $\pm\,0.47$ & $\pm\,0.33$ & $-\,0.21$ & $+\,0.46$ 
 & $\pm\,0.57$ & $\pm\,0.31$ & $-\,0.19$ & $+\,0.46$ \\
$\Delta_{\gamma\gamma{}H}$ & 
 \phantom{$\pm$} --- &\phantom{$\pm$} --- &\phantom{$-$} --- &\phantom{$+$} --- 
 & $\pm\,0.55$ & $\pm\,0.31$ & $-\,0.30$ & $+\,0.33$ \\
$\Delta_{ggH}$             & 
 \phantom{$\pm$} --- &\phantom{$\pm$} --- &\phantom{$-$} --- &\phantom{$+$} --- 
 & $\pm\,0.80$ & $\pm\,0.61$ & $-\,0.59$ & $+\,0.62$ \\
$m_b$                      
 & $\pm\,0.073$& $\pm\,0.071$& $-\,0.071$& $+\,0.071$
 & $\pm\,0.070$& $\pm\,0.071$& $-\,0.071$& $+\,0.072$\\
$m_t$                      
 & $\pm\,1.99$ & $\pm\,1.00$ & $-\,1.03$ & $+\,0.98$ 
 & $\pm\,1.99$ & $\pm\,0.99$ & $-\,1.00$ & $+\,0.98$ 
\end{tabular}
\end{small}
\caption[]{Errors on the measurements from 10000 toy experiments. We
  quote errors for Standard Model couplings only and including
  effective $ggH$ and $\gamma\gamma H$ couplings using $30\ \ifb$ of
  integrated luminosity. The different $\sigma$ measures are defined in the text.}
\label{tab:toyerrors}
}

\TABLE[b]{
\begin{small}
\begin{tabular}{l|l|ll|l|ll}
         & \multicolumn{3}{c|}{no effective couplings} &
           \multicolumn{3}{c}{with effective couplings} \\
 & 
  $\sigma_\text{symm}$ & $\sigma_\text{neg}$ & $\sigma_\text{pos}$ & 
  $\sigma_\text{symm}$ & $\sigma_\text{neg}$ & $\sigma_\text{pos}$ \\\hline 
$\Delta_{ZZH/WWH}$             & $\pm\,0.46$ & $-\,0.36$ & $+\,0.53$ 
 & $\pm\,0.41$ & $-\,0.40$ & $+\,0.41$ \\
$\Delta_{ttH/WWH}$       & $\pm\,0.30$ & $-\,0.27$ & $+\,0.32$ 
 & $\pm\,0.51$ & $-\,0.54$ & $+\,0.48$ \\
$\Delta_{bbH/WWH}$       & $\pm\,0.28$ & $-\,0.24$ & $+\,0.32$ 
 & $\pm\,0.31$ & $-\,0.24$ & $+\,0.38$ \\
$\Delta_{\tau\tau{}H/WWH}$ & $\pm\,0.25$ & $-\,0.18$ & $+\,0.33$ 
 & $\pm\,0.28$ & $-\,0.16$ & $+\,0.40$ \\
$\Delta_{\gamma\gamma{}H/WWH}$ & 
  \phantom{$\pm$} --- & \phantom{$-$} --- & \phantom{$+$} --- 
 & $\pm\,0.30$ & $-\,0.27$ & $+\,0.33$ \\
$\Delta_{ggH/WWH}$             & 
  \phantom{$\pm$} --- & \phantom{$-$} --- & \phantom{$+$} --- 
 & $\pm\,0.61$ & $-\,0.71$ & $+\,0.46$ 
\end{tabular}
\end{small}
\caption[]{Errors on the ratio of couplings for a 120~GeV Higgs, corresponding to
  Table~\ref{tab:toyerrors}. The coupling ratio is normalized to the 
Standard Model value according to eq.~\ref{Eq:ratios}. For the Standard Model dataset
used in this study the central value of $\Delta_{jjH/WWH}$ is zero, therefore 
only the errors are quoted.}
\label{tab:toyerrors_ratio}
}

In Table~\ref{tab:toyerrors} we show the errors on the
extraction of Higgs coupling parameters. These errors we obtain from
10000 toy experiments, each smeared around the true data point
including all experimental and
theory errors. Besides calculating a root-mean-square error, we
histogram the best fits for each parameter and extract
$\sigma_\text{symm}$ using a Gaussian fit.  As we do not expect the
errors to be symmetric, we also fit a combination of two Gaussians
with the same maximum and the same value at the maximum, but different
widths. Depending on whether we are below or above the maximum we use
the first ($\sigma_\text{neg}$) or the second Gaussian
($\sigma_\text{pos}$). For $\Delta_{ZZH}$ we see that as we approach
$-1$, or vanishing coupling, the histogram goes to a constant value
and is not well-fitted by a Gaussian. Therefore, for the lower branch
of this coupling we fit only the central part within one standard
deviation. Using a different measure for the shape of the error
distribution we also show the standard RMS values for all
couplings. They are systematically larger, owed to individual toy
experiments far from the best-fit points. Only for symmetric Gaussian
behavior we can expect these three error measures to coincide.\bigskip

First, we see that a correlation between the mass measurements and the
couplings plays hardly any role. The errors on the Higgs mass are
symmetric and the values correspond to our input data. 

A bit surprisingly, $g_{WWH}$ is basically unchanged whether
we allow effective couplings or not. This coupling is the main
contribution to the effective $g_{\gamma\gamma H}$, so we would
expect that allowing for additional effective coupling should remove
the decay to photons from the $g_{WWH}$ determination. The results
indicate that the accuracy from the remaining measurements is
sufficient to determine $g_{WWH}$.

For the $ttH$ coupling the situation is different. Besides its small
tree-level impact it is the dominating contribution to $g_{ggH}$ and a
sub-leading contribution to $g_{\gamma\gamma H}$.  Hence, allowing
effective couplings increases the errors.  With and without effective
couplings a Gaussian does not describe the errors accurately. In fact,
including effective couplings we see a flat part around the Standard
Model value, originating from our flat theory errors, and then a
slightly steeper exponential fall-off.

The $\tau\tau H$ and $bbH$ couplings are both strongly linked to
$g_{WWH}$, because of their respective production modes.  The
$t\bar{t}$-associated production channel has too large errors to play
a significant role.  Consequently, we do not see any dependence on the
existence of effective couplings.

The $ZZH$ coupling shows a particular effect when we include the
effective couplings; its error is decreased. Such a behavior is
counter-intuitive, because naively we would expect that such an
additional coupling can always be considered a nuisance parameter
which we need to remove. No matter if we integrate it out or project
it away, the resulting error should form some kind of
envelope. However, the situation changes once we include correlations.
Let us consider two couplings $g_1$ and $g_2$ (with a strong positive
correlation) contributing to the same observable as $g_1^2
g_2^2$. With a combined shift the relative errors $\delta_1$ and
$\delta_2$ on the couplings yield a total shift $g_1^2 g_2^2 (\delta_1
+ \delta_2)$ on the observable, limited by the measurement. For
example, if due to the positive correlation both terms in the
parenthesis are strictly positive, the upper limit on the observable
now splits into two contributions, one for each parameter. In
contrast, if we fix $\delta g_1=0$ $\delta g_2$ alone accounts for the
complete deviation in the observable.

The main source for determining $g_{ZZH}$ is gluon-fusion production
with subsequent Higgs decay into two $Z$. Without any effective
couplings only the top Yukawa coupling determines the production rate,
which is constrained by many other measurements. To account for the
measured event rate we need to adjust $g_{ZZH}$ accordingly. With
additional contributions to $g_{ggH}$ the top Yukawa coupling becomes
largely irrelevant and the production and decay couplings of the $gg
\to H \to ZZ$ channel form exactly the combination discussed
above. From Section~\ref{sec:corr} we know that they are positively
correlated, so indeed the error on $\Delta_{ZZH}$ can
decrease.\bigskip

In Table~\ref{tab:toyerrors_ratio} we show the errors on $\Delta_{jjH/WWH}$,
defined in the following equation as the deviation from~$1$ of the ratios of
the coupling constants to the ${WWH}$ coupling
normalized to the Standard Model value:
\begin{equation}
\frac{g_{jjH}}{g_{WWH}} \longrightarrow
\left( \frac{g_{jjH}}{g_{WWH}} \right)^\text{SM} \;
\left( 1 + \Delta_{jjH/WWH} \right)
\label{Eq:ratios}
\end{equation}
A non-zero central value for $\Delta_{jjH/WWH}$ would arise in new physics scenarios. 
Thus $\Delta_{jjH/WWH}$ provides information on a possible shift of
the coupling ratios as well as the error on the coupling ratio.
The central value of $\Delta_{jjH/WWH}$ is zero for the Standard Model
dataset used in this study, therefore we discuss only the 
errors on $\Delta_{jjH/WWH}$. 

Table~\ref{tab:toyerrors_ratio} confirms the
qualitative results from the profile likelihoods in
Section~\ref{sec:ratios}. The error on $g_{ZZH}$ even increases once we
form the ratio. The main determination mode is via the gluon-fusion
initial state, which is independent of $g_{WWH}$. An additional
constraint enters via the subleading contribution to weak boson fusion,
where $g_{WWH}$ and $g_{ZZH}$ occur additively, so forming the ratio
exacerbates deviations instead of decreasing them. For the top-quark
coupling the effect of the effective photon coupling becomes clearly
visible. Without this additional parameter $g_{ttH}$ and $g_{WWH}$ are
linked and consequently the error on the ratio is significantly
smaller than the error on the coupling itself. The additional
$g_{\gamma\gamma H}$ breaks this correlation and we no longer gain
anything by using the ratio. For all other couplings we observe more
or less significantly smaller errors for ratios of couplings, because
the production-side $g_{WWH}$ enters the determination of the
decay-side couplings. Only the correlation between the well-measured
$g_{WWH}$ and $g_{bbH}$ sticks out; it is due to the appearance of the
total width in all rate predictions, which leads to strong
correlations as discussed in Section~\ref{sec:corr}.

A note of caution should be added: the error on the error determination of
the ratios of the coupling constants is non--zero. This is due to the fact that the
coupling ratios have been determined from the coupling central
values for a given toy experiment. Then for each toy experiment the ratio of
the couplings calculated and
the distribution of these ratios analyzed. The flat theory errors
do not ensure a cancellation of correlated theory errors in these ratios. As
the errors for the integrated luminosity of $30~\ifb$ are dominated by the
expected experimental error 
(78\% to 94\% of the total error as shown in Table~\ref{tab:channels}), 
we defer the development
a more sophisticated treatment ensuring the full cancellation to a later
date. We have checked however that by setting the theory errors to
zero (implying a full cancellation of all errors) the error
on the coupling ratios is improved by at most 10\%.   

\subsection{Bottom Yukawa coupling}
\label{sec:bottom}

\begin{figure}[t]
 \begin{center}
 {\raggedright $1/\Delta \chi^2$\\[2ex]}
 \includegraphics[width=0.135\textwidth]{%
     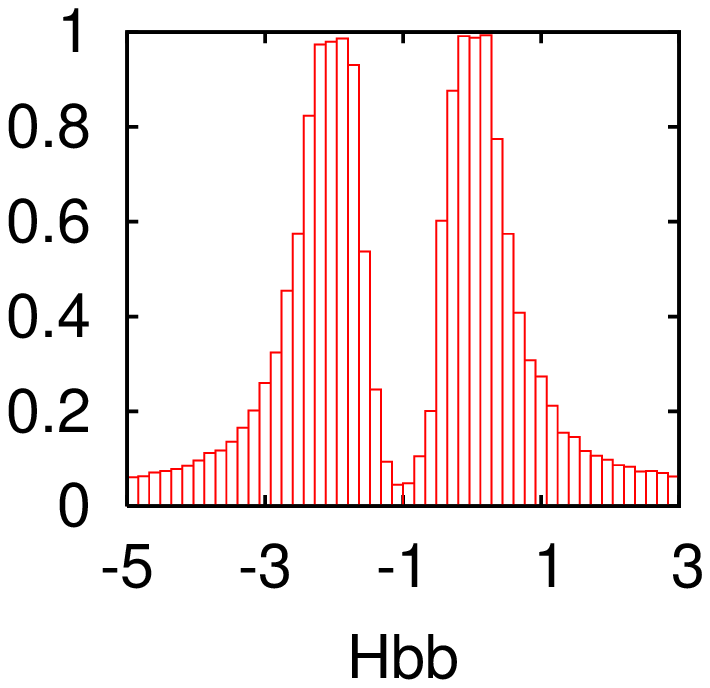} 
 \includegraphics[width=0.135\textwidth]{%
     sfitter_markovbins3.HWW.phaseW.truesubjet.eps} 
 \includegraphics[width=0.135\textwidth]{%
     sfitter_markovbins3.Htautau.phaseW.truesubjet.eps} 
 \includegraphics[width=0.135\textwidth]{%
     sfitter_markovbins3.Htt.phaseW.truesubjet.eps} 
 \includegraphics[width=0.135\textwidth]{%
     sfitter_markovbins3.Hgg.phaseW.truesubjet.eps} 
 \includegraphics[width=0.135\textwidth]{%
     sfitter_markovbins3.Hgamgam.phaseW.truesubjet.eps} \\[2ex]
 \includegraphics[width=0.135\textwidth]{%
     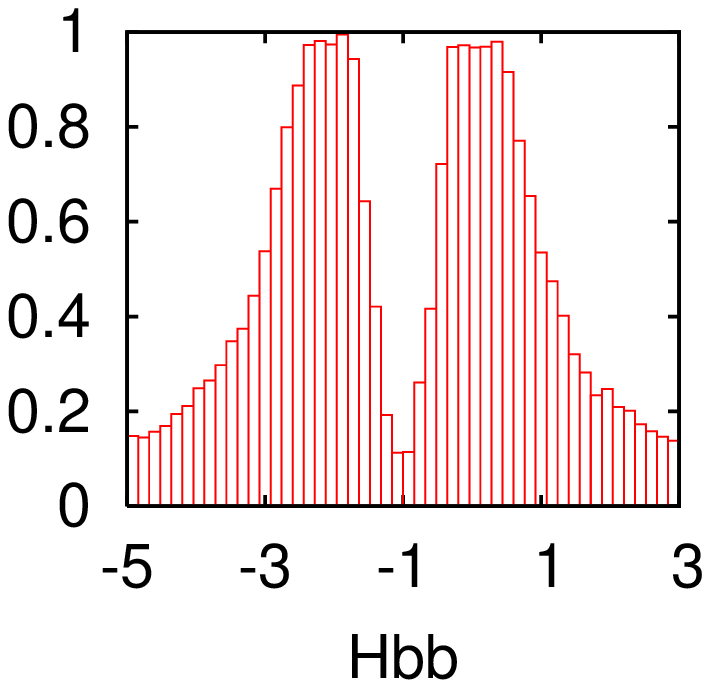} 
 \includegraphics[width=0.135\textwidth]{%
     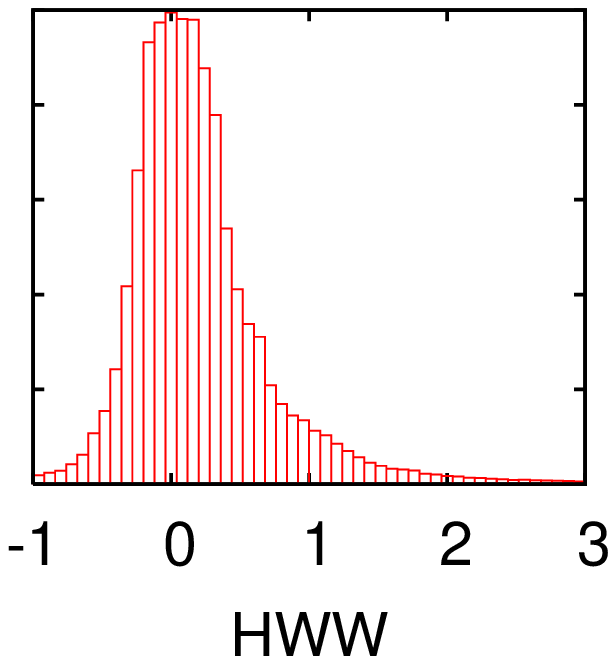} 
 \includegraphics[width=0.135\textwidth]{%
     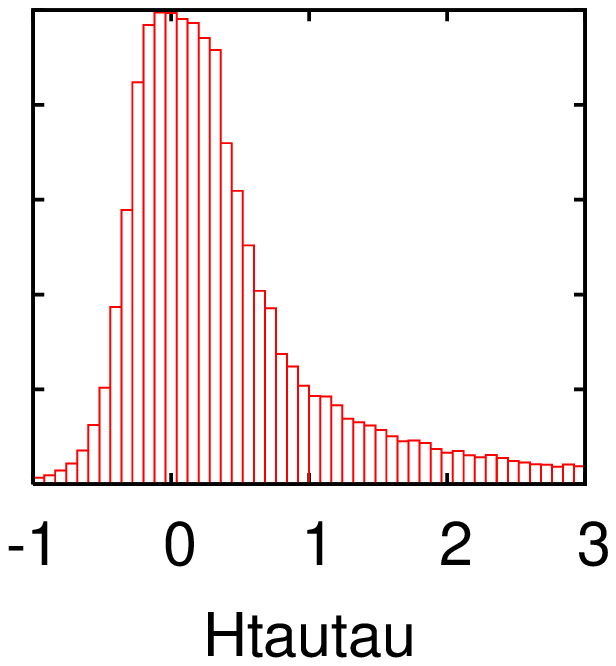} 
 \includegraphics[width=0.135\textwidth]{%
     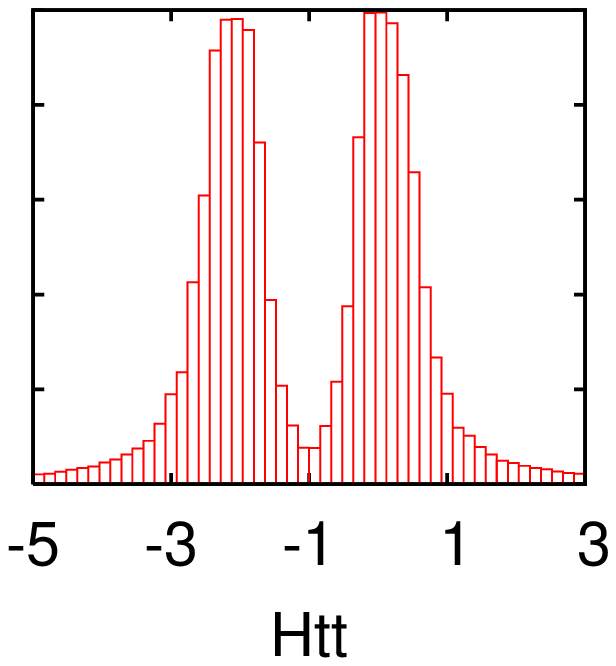} 
 \includegraphics[width=0.135\textwidth]{%
     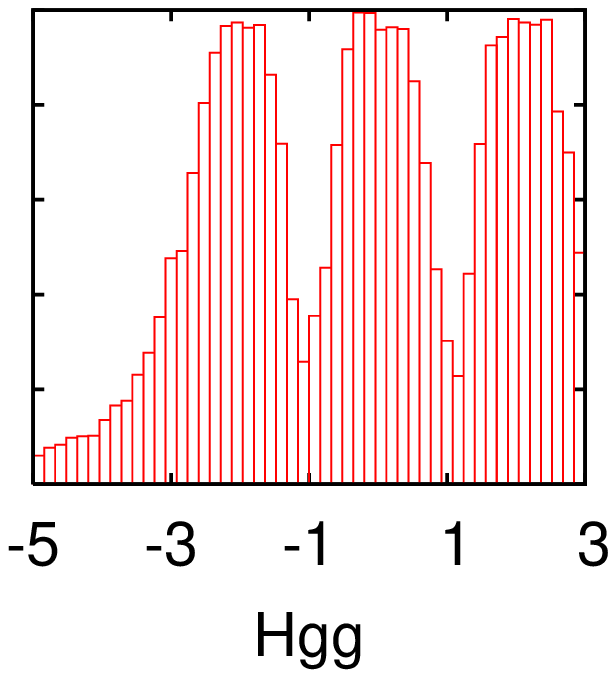} 
 \includegraphics[width=0.135\textwidth]{%
     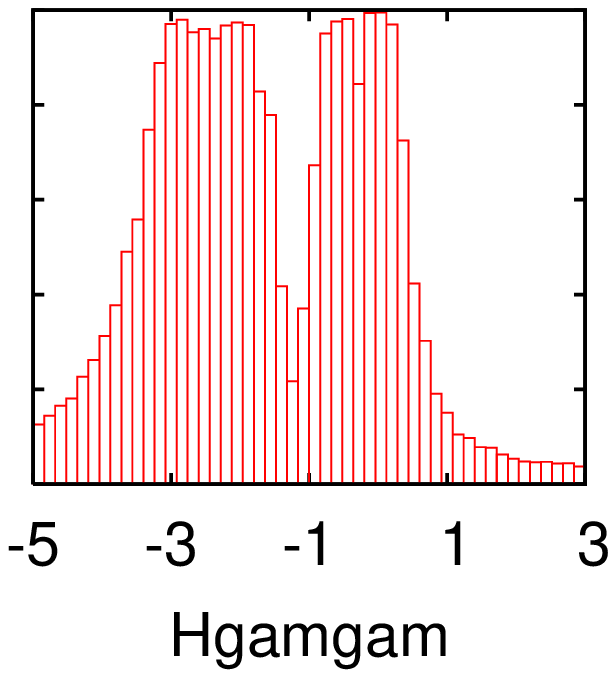} \\[2ex]
 \includegraphics[width=0.135\textwidth]{%
     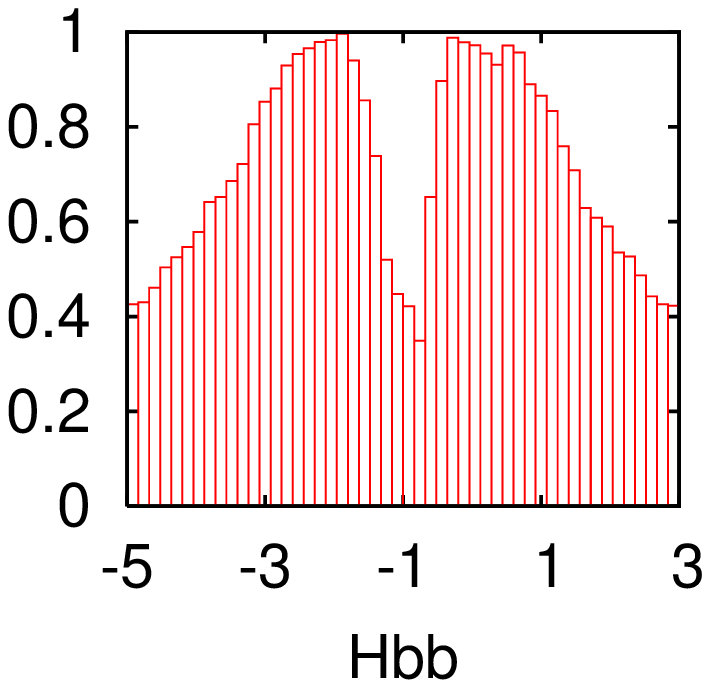} 
 \includegraphics[width=0.135\textwidth]{%
     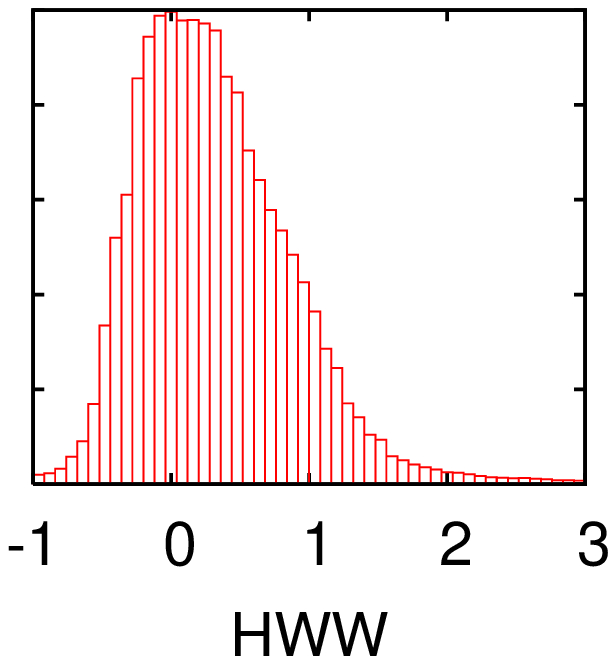} 
 \includegraphics[width=0.135\textwidth]{%
     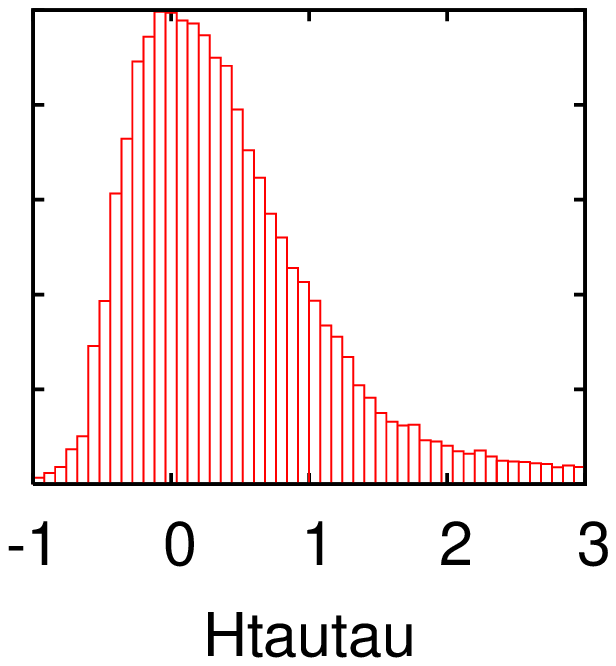} 
 \includegraphics[width=0.135\textwidth]{%
     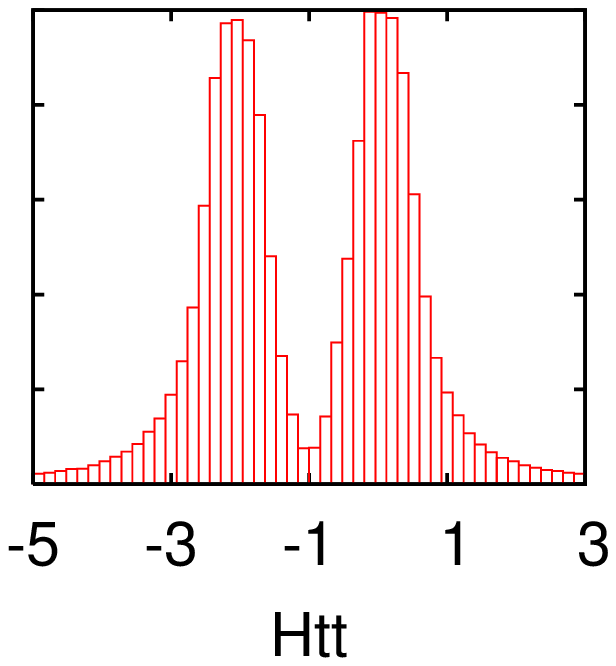} 
 \includegraphics[width=0.135\textwidth]{%
     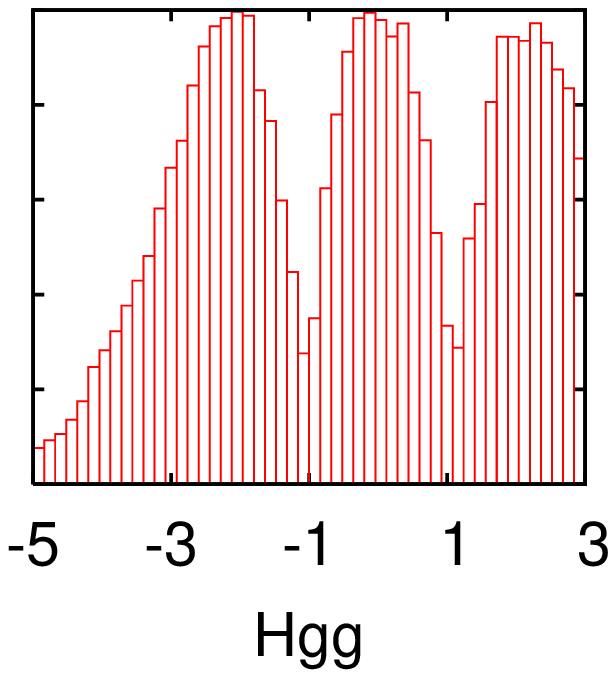} 
 \includegraphics[width=0.135\textwidth]{%
     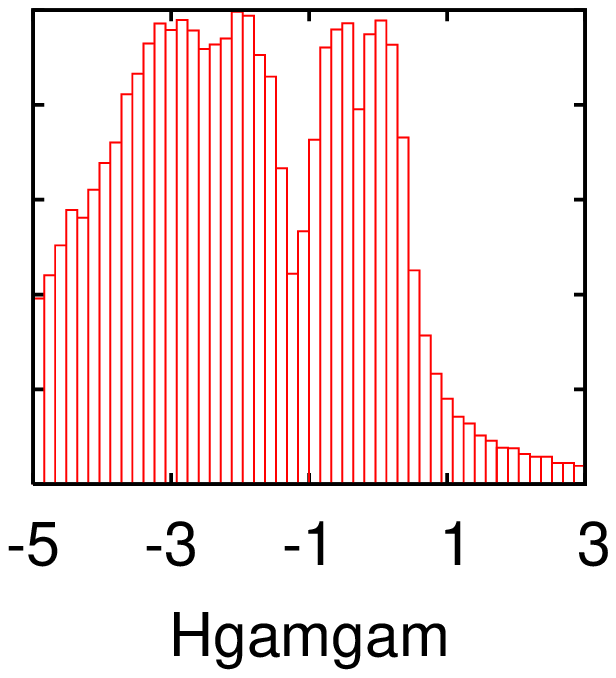} \\
 \includegraphics[width=0.24\textwidth]{%
     sfitter_markovbins3.HWW_HZZ.phaseW.truesubjet.eps} \hspace*{-2ex}
 \includegraphics[width=0.24\textwidth]{%
     sfitter_markovbins3.HWW_Htautau.phaseW.truesubjet.eps} \hspace*{-2ex}
 \includegraphics[width=0.24\textwidth]{%
     sfitter_markovbins3.HWW_Hbb.phaseW.truesubjet.eps} \hspace*{-2ex}
 \includegraphics[width=0.24\textwidth]{%
     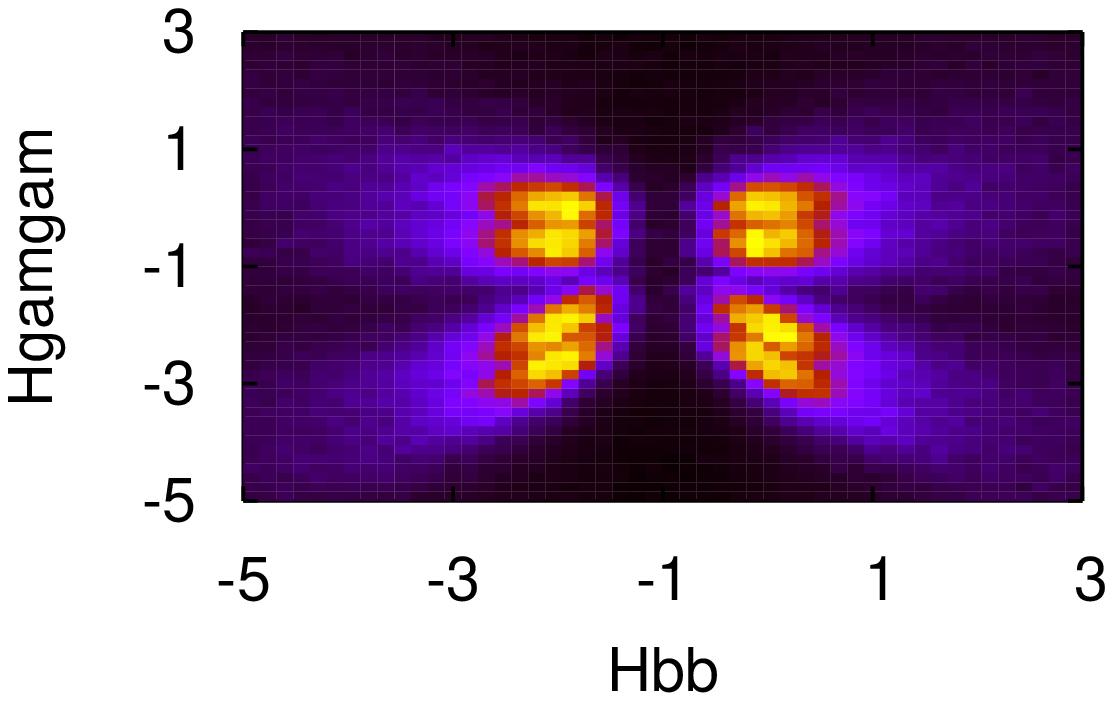} \\
 \includegraphics[width=0.24\textwidth]{%
     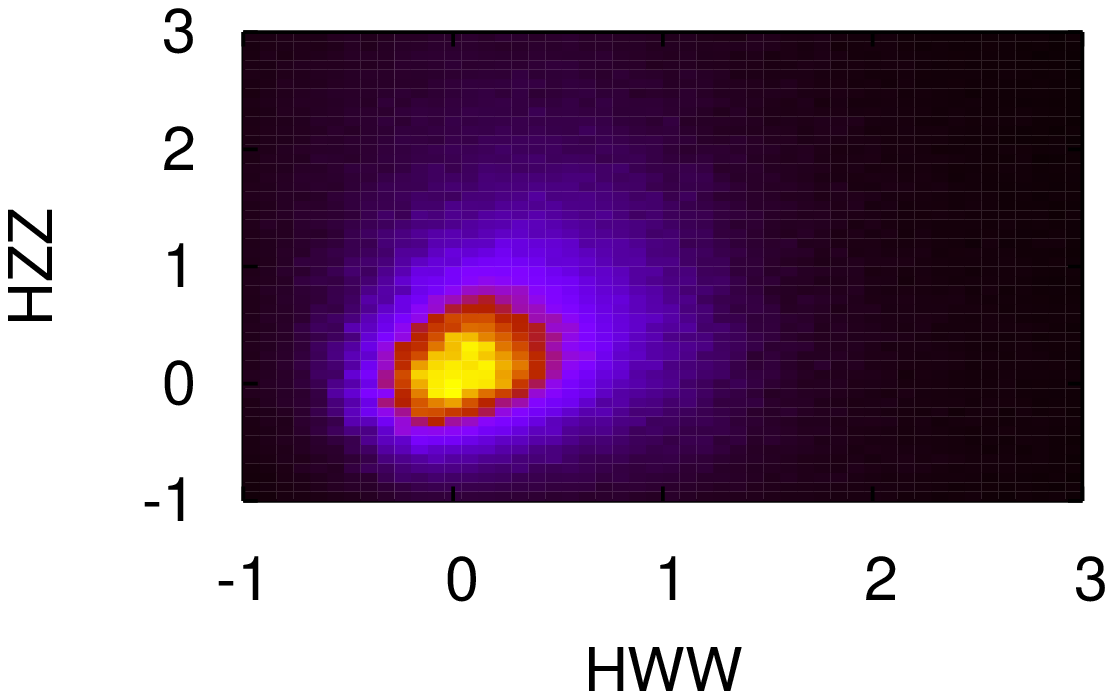} \hspace*{-2ex}
 \includegraphics[width=0.24\textwidth]{%
     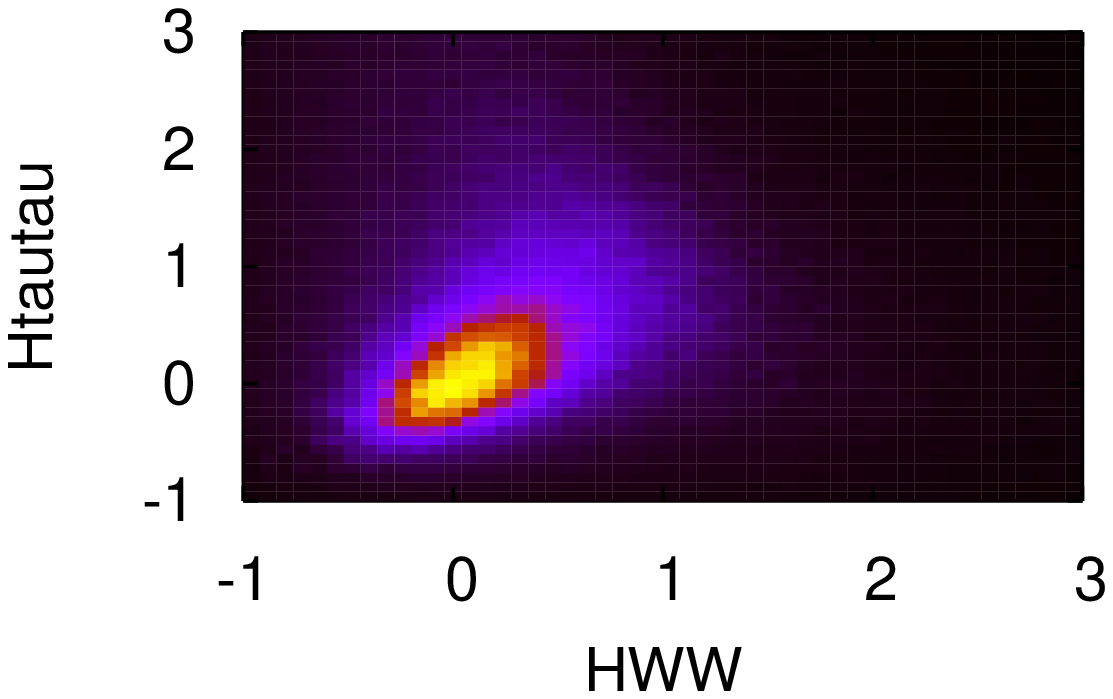} \hspace*{-2ex}
 \includegraphics[width=0.24\textwidth]{%
     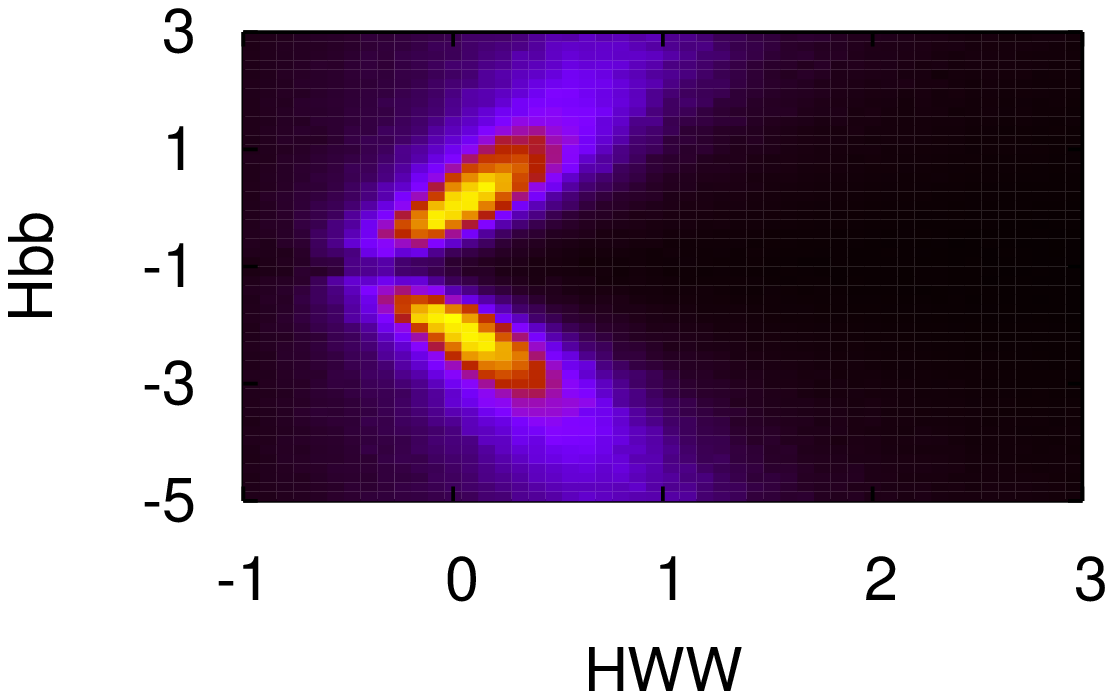} \hspace*{-2ex}
 \includegraphics[width=0.24\textwidth]{%
     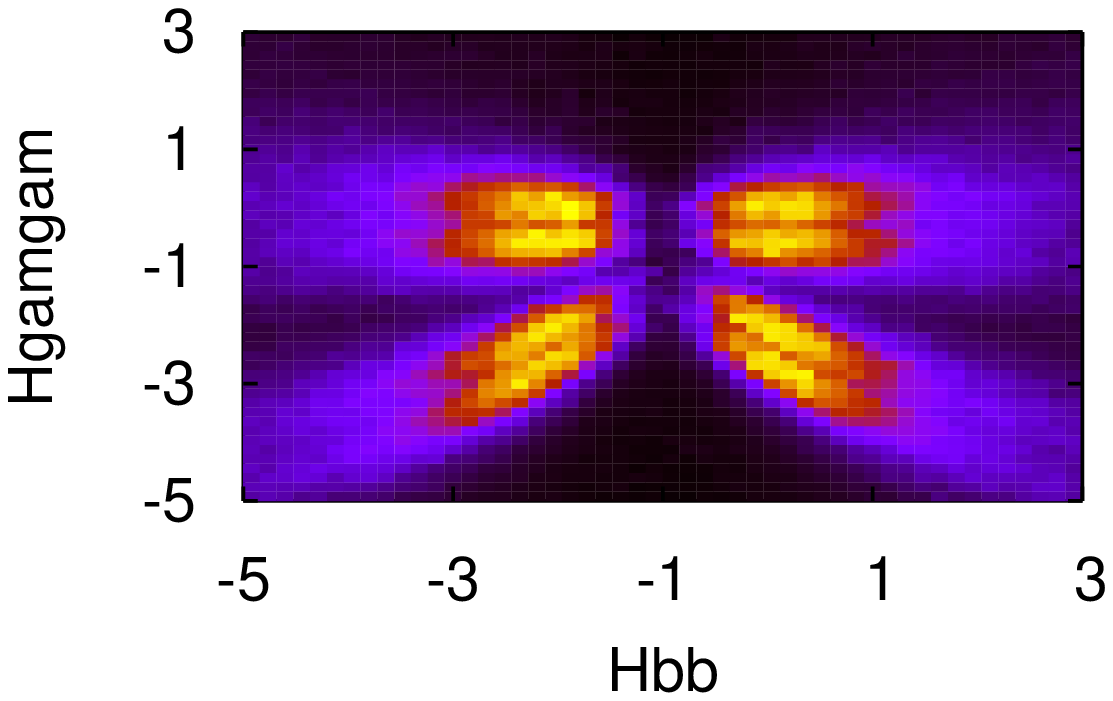} \\
 \includegraphics[width=0.24\textwidth]{%
     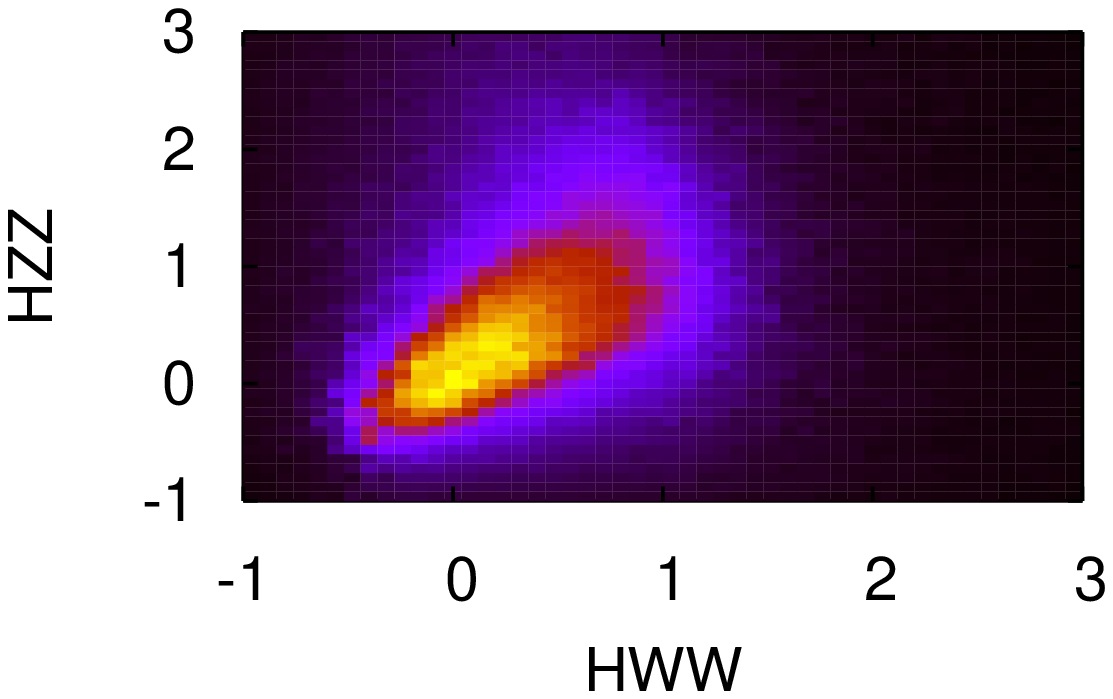} \hspace*{-2ex}
 \includegraphics[width=0.24\textwidth]{%
     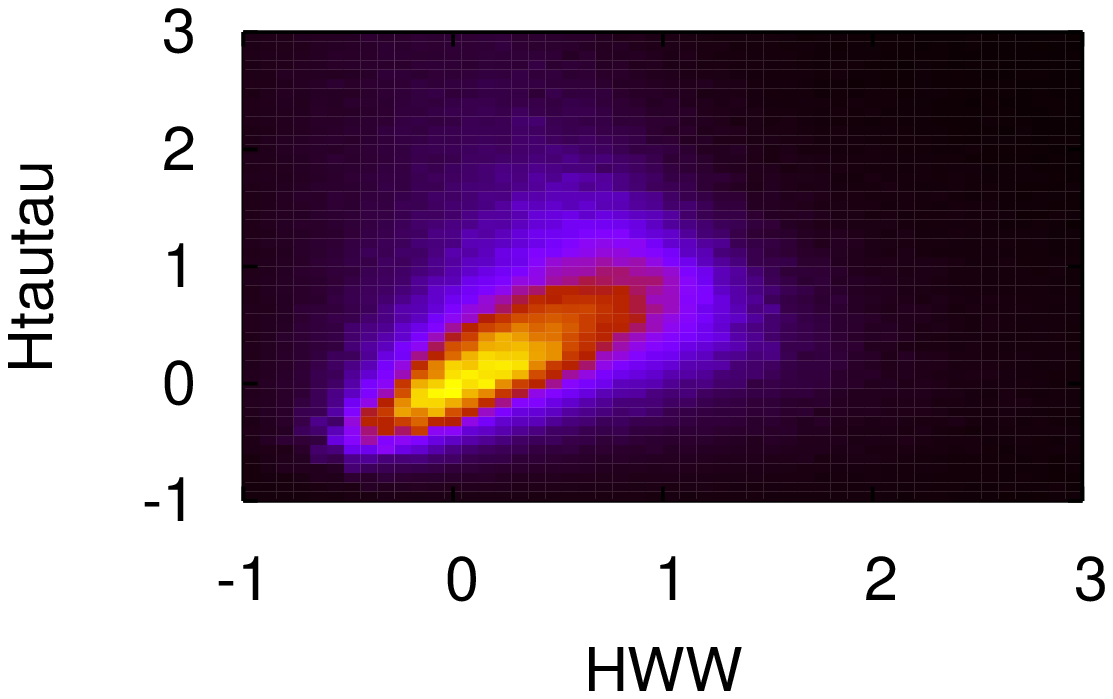} \hspace*{-2ex}
 \includegraphics[width=0.24\textwidth]{%
     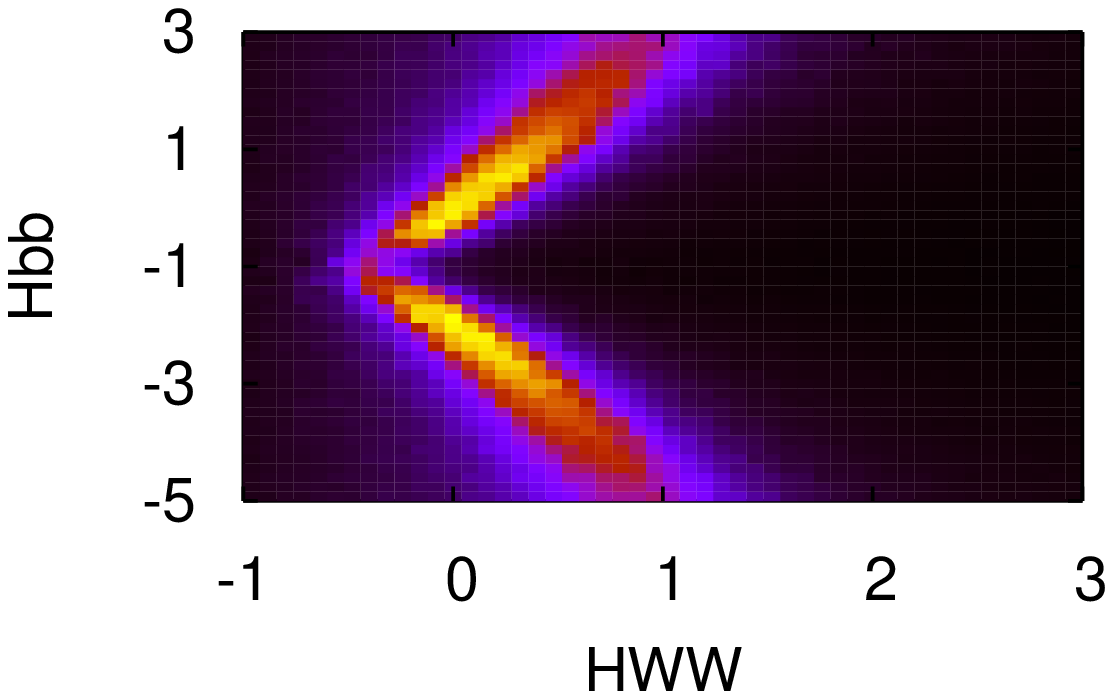} \hspace*{-2ex}
 \includegraphics[width=0.24\textwidth]{%
     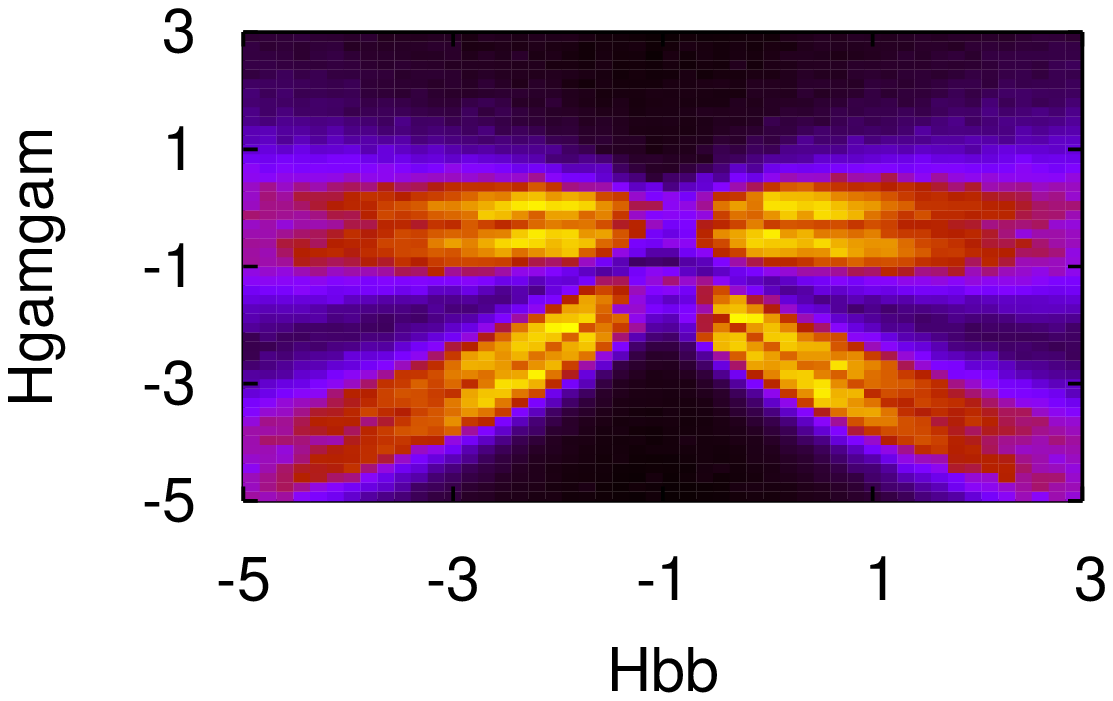} 
 \end{center}
\caption[]{Profile likelihoods including the $WH/ZH$ channel (upper
  row), $WH/ZH$ with reduced sensitivity (center) and only $t\bar{t}H$
  (lower) with a decay $H \to b\bar{b}$. All experimental and theory
  errors included for low-luminosity running (true data points).}
\label{fig:bottom}
\end{figure}

As mentioned in Section~\ref{sec:higgs} the measurement of the bottom
Yukawa coupling is crucial to the success of this parameter analysis;
without it we would hugely underestimate the total width of the Higgs
boson and scale down all couplings correspondingly to obtain the
measured production and decay rates. For a long time the LHC
experiments expected the $t \bar{t} H$ production process to be best
suited for Higgs decaying to bottoms. Such a complex signature is more
promising than simply looking for inclusive $b \bar{b}$ production or
weak-boson-fusion production, because it includes more information to
reduce backgrounds.  On the other hand, several bottom jets lead to
combinatorial errors, which turn out to largely kill this
signature. Recently, there has been progress in understanding the
structure of bottom jets coming from a Higgs resonance as compared to
continuum QCD production. These two sources can be distinguished using
information from the merging of subjets inside the jet algorithm
used. Triggering requires for example additional leptons in the final
state, which makes $VH$ the preferred production channel. While these
results have not yet been confirmed by detailed simulations, we
nevertheless use them to illustrate the impact of the $H \to b\bar{b}$
channels on our analysis.\bigskip

In Fig.~\ref{fig:bottom} we first show results using the published
$WH, H \to b\bar{b}$ results at face value. This is the signature we
base our entire analysis on.  To be consistent with the other
channels, we use the rates as in Ref.~\cite{subjet} but included our
own systematic and theory errors.

In the second line we show results with a $50\%$ reduced signal
rate. As expected, all error bars are slightly increased, but the
features of the distributions are virtually unchanged. Focusing on
$g_{bbH}$ we see that even with the reduced sensitivity a precise
measurement is possible. The slight degradation of this coupling (and
with it of the total width) then propagates into all remaining
couplings.\bigskip

Once we remove the $WH$/$ZH$ subjet channel from our analysis a
determination of the bottom Yukawa coupling becomes impossible. Its
error bar extends to a continuum $\mathcal{O} (150 \%)$ around the
input value, and only a zero coupling is significantly less likely.
The one measurement to determine $g_{bbH}$ is top-quark-associated
Higgs production. However, for this channel a measurement of zero
events only requires a fluctuation of 0.7 standard deviations.  An
additional constraint enters via the total width, but we already know
that there is a scaling symmetry. And finally, for the bottom
contributions to the effective $g_{ggH}$ and $g_{\gamma\gamma H}$ to
play any role we need an enhancement of the bottom-Higgs coupling far
beyond those considered here.

Aside from a broader peak structure of $\Delta_{bbH}$ we also observe
considerably higher shoulders towards large values of the bottom
Yukawa coupling. In this regime the branching ratio into bottom quarks
gradually approaches one, so any increase in the coupling hardly
shifts the number of $H \to b\bar{b}$ events. In the center of the
$\Delta_{bbH}$ distribution the $\Delta \chi^2$ value for vanishing
coupling ranges around $1.5$ in the absence of the subjet analysis.
The $ttH$ channel cannot give such a large value, which means there
is an additional indirect contribution from the total width.

Precisely due to the increased uncertainty on the total width the
otherwise well determined $g_{WWH}$ suffers visibly.  Not
surprisingly, this increased error bar immediately propagates into
$g_{\tau \tau H}$ because of the correlated production process.  In
contrast, the top Yukawa coupling is virtually unchanged: first, the
$t \bar{t} H$ production channel of bottom quarks 
contributes nothing to the top Yukawa
measurement, and secondly the bottom contribution to $g_{ggH}$ is
small enough to still vanish in the error. The same is true for the
effective $g_{ggH}$ shown in Fig.~\ref{fig:bottom}.  Both these
couplings appear only on the production side of our measurements, so
the bigger uncertainty on $g_{bbH}$ will already be absorbed by the
respective decay channel.  The situation is different for the
effective $g_{\gamma \gamma H}$, where the broader range of the
$\Delta_{WWH}$ requires a larger variation of $\Delta_{\gamma\gamma
  H}$ to fulfill the constraints in particular from the well-measured
inclusive $H\to \gamma\gamma$ channel.

As already alluded to in Section~\ref{sec:corr}, with most
measurements significantly degraded we now observe the proper
correlation between $g_{WWH}$, $g_{ZZH}$ and $g_{\tau \tau H}$ from
the weak-boson-fusion channels. Moreover, we see the positive
correlation from the two observed decays in gluon fusion.\bigskip

\TABLE[b]{
\begin{small}
\begin{tabular}{l|l|ll|l|ll|l|ll}
 &         \multicolumn{3}{c}{full measurements} 
 &         \multicolumn{3}{|c}{reduced sensitivity} 
 &         \multicolumn{3}{|c}{only $t\bar{t}H$, $H\to b\bar{b}$} \\
 & $\sigma_\text{symm}$ & $\sigma_\text{neg}$ & $\sigma_\text{pos}$ 
 & $\sigma_\text{symm}$ & $\sigma_\text{neg}$ & $\sigma_\text{pos}$ 
 & $\sigma_\text{symm}$ & $\sigma_\text{neg}$ & $\sigma_\text{pos}$ \\\hline 
$\Delta_{WWH}$             
 & $\pm\,0.24$ & $-\,0.21$ & $+\,0.27$ 
 & $\pm\,0.32$ & $-\,0.25$ & $+\,0.40$ 
 & $\pm\,0.33$ & $-\,0.24$ & $+\,0.43$ \\
$\Delta_{ZZH}$             
 & $\pm\,0.31$ & $-\,0.35$ & $+\,0.29$ 
 & $\pm\,0.46$ & $-\,0.49$ & $+\,0.45$ 
 & $\pm\,0.59$ & $-\,0.33$ & $+\,0.64$ \\
$\Delta_{ttH}$       
 & $\pm\,0.53$ & $-\,0.65$ & $+\,0.43$ 
 & $\pm\,0.54$ & $-\,0.60$ & $+\,0.50$ 
 & $\pm\,0.48$ & $-\,0.56$ & $+\,0.41$ \\
$\Delta_{bbH}$       
 & $\pm\,0.44$ & $-\,0.30$ & $+\,0.59$ 
 & $\pm\,0.75$ & $-\,0.64$ & $+\,0.80$ 
 & $\pm\,0.78$ & $-\,0.43$ & $+\,0.84$ \\
$\Delta_{\tau\tau{}H}$ 
 & $\pm\,0.31$ & $-\,0.19$ & $+\,0.46$ 
 & $\pm\,0.36$ & $-\,0.18$ & $+\,0.56$ 
 & $\pm\,0.39$ & $-\,0.20$ & $+\,0.60$ \\
$\Delta_{\gamma\gamma{}H}$ 
 & $\pm\,0.31$ & $-\,0.30$ & $+\,0.33$ 
 & $\pm\,0.31$ & $-\,0.31$ & $+\,0.30$ 
 & $\pm\,0.33$ & $-\,0.33$ & $+\,0.33$ \\
$\Delta_{ggH}$             
 & $\pm\,0.61$ & $-\,0.59$ & $+\,0.62$ 
 & $\pm\,0.59$ & $-\,0.54$ & $+\,0.64$ 
 & $\pm\,0.66$ & $-\,0.48$ & $+\,0.82$ 
\end{tabular}
\end{small}
\caption[]{Errors on the measurements including the $WH/ZH$ channel
  (left, from Table~\ref{tab:toyerrors}), $WH/ZH$ with reduced
  sensitivity (center) and only $t\bar{t}H, H \to b\bar{b}$
  (right). We assume $30~\ifb$ of integrated luminosity.}
\label{tab:toyerrors_red}
}

In Table~\ref{tab:toyerrors_red} we show errors on the parameters for
the three different scenarios. Again we fit only the central peak
around the Standard Model solution. These numbers confirm the findings
from Fig.~\ref{fig:bottom}. A very prominent shift we see of course in
$g_{bbH}$, as this is directly affected by changes in the subjet
analysis. There we also see a large fraction of solutions where the
coupling vanishes. Other couplings are extracted with a reduced
precision of $50 - 100\%$. Only the error on $g_{ttH}$ is largely
unchanged with remaining differences in part due to statistical
fluctuations.

\subsection{Minijet veto}
\label{sec:mjv}

\begin{figure}[t]
 \begin{center}
 {\raggedright 
 \hspace*{0.135\textwidth} 
 $1/\Delta \chi^2$\\[2ex]}
 \includegraphics[width=0.15\textwidth]{%
     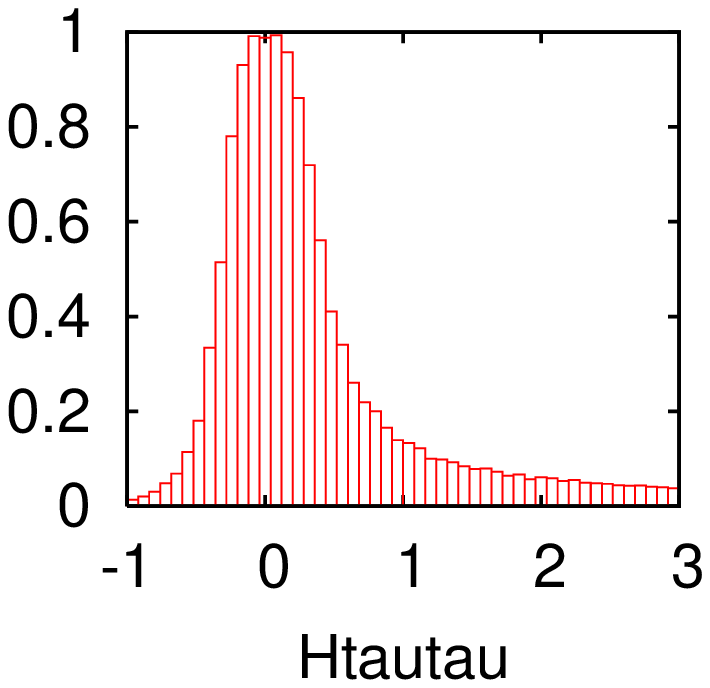} 
 \hspace*{0.25\textwidth} 
 \includegraphics[width=0.15\textwidth]{%
     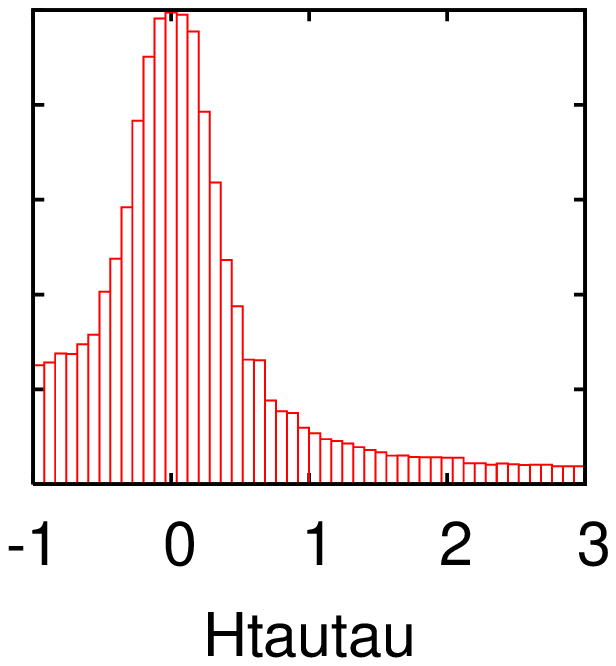} 
 \end{center}
\caption[]{Profile likelihood for $\Delta_{\tau\tau H}$ 
  including the standard error of $5\%$
  (left) and an increased error of $100 \%$ (right) on the central
  mini-jet veto. We show results for true data points and $30~\ifb$.}
\label{fig:mjv}
\end{figure}
\TABLE[b]{
\begin{small}
\begin{tabular}{l|l|ll|l|ll}
 &         \multicolumn{3}{c}{$5\%$ error on minijet veto} 
 &         \multicolumn{3}{|c}{$100\%$ error on minijet veto} \\
 & $\sigma_\text{symm}$ & $\sigma_\text{neg}$ & $\sigma_\text{pos}$ 
 & $\sigma_\text{symm}$ & $\sigma_\text{neg}$ & $\sigma_\text{pos}$ \\\hline 
$\Delta_{WWH}$             
 & $\pm\,0.24$ & $-\,0.21$ & $+\,0.27$ 
 & $\pm\,0.26$ & $-\,0.29$ & $+\,0.23$ \\
$\Delta_{ZZH}$             
 & $\pm\,0.31$ & $-\,0.35$ & $+\,0.29$ 
 & $\pm\,0.32$ & $-\,0.37$ & $+\,0.30$ \\
$\Delta_{ttH}$       
 & $\pm\,0.53$ & $-\,0.65$ & $+\,0.43$ 
 & $\pm\,0.55$ & $-\,0.63$ & $+\,0.49$ \\
$\Delta_{bbH}$       
 & $\pm\,0.44$ & $-\,0.30$ & $+\,0.59$ 
 & $\pm\,0.47$ & $-\,0.31$ & $+\,0.64$ \\
$\Delta_{\tau\tau{}H}$ 
 & $\pm\,0.31$ & $-\,0.19$ & $+\,0.46$ 
 & $\pm\,0.35$ & $-\,0.33$ & $+\,0.37$ \\
$\Delta_{\gamma\gamma{}H}$ 
 & $\pm\,0.31$ & $-\,0.30$ & $+\,0.33$ 
 & $\pm\,0.37$ & $-\,0.31$ & $+\,0.43$ \\
$\Delta_{ggH}$             
 & $\pm\,0.61$ & $-\,0.59$ & $+\,0.62$ 
 & $\pm\,0.61$ & $-\,0.44$ & $+\,0.75$ 
\end{tabular}
\end{small}
\caption[]{Errors assuming the $5\%$ standard error on the mini-jet veto
  (left, from Table~\ref{tab:toyerrors}), and assuming an increased
  error of $100\%$ (right), using $30~\ifb$ of integrated luminosity.}
\label{tab:vbferrors}
}

One of the crucial ingredients to the weak-boson fusion production
analyses is the central mini-jet veto~\cite{mjv}. The two tagging jets
are emitted into the forward regions of the detector, while the
central region stays free of jets. This behavior is reflected in the
small next-to-leading order QCD corrections to this
process~\cite{wbf_nlo,wbf_susy}.  Typical backgrounds as well as
gluon-fusion production involving two jets do not share this
feature. How well Monte-Carlo simulations describe the mini-jet veto
probability is not yet established. Without
any claim of a useful range we increase the uncertainty on the
mini-jet veto probability from $5\%$ to clearly unrealistic $100\%$.

In Table~\ref{tab:vbferrors} we see that even with this huge theory
error the numbers hardly change at all.  The fit of a symmetric
Gaussian to the toy experiments consistently enlarges the errors,
although the effect is mostly minimal. The varying behavior for the
asymmetric errors can be traced back to statistical fluctuations. Only
for $g_{\tau\tau H}$ we see a significant change compared to our
normal analysis.  This coupling is determined exclusively via
weak-boson-fusion channels, so an increase in the error directly
affects this coupling.  In Fig.~\ref{fig:mjv} we see a clearly finite
likelihood for vanishing coupling. In the complete scan we even see a
slight peak in this place, because due to the large errors it is now
possible that the rates for the weak-boson-fusion channels fluctuate
below the background. To illustrate a non-negligible effect of the
increased error: we see that while the zero-signal solution is clearly
excluded in the standard case, we now cannot even reach a $95 \%$
confidence level assuming an integrated luminosity $30~\ifb$.

While this discussion seems to indicate that the exact knowledge of
the minijet veto probabilities is not crucial to our analysis, there
are two caveats: first of all, we only study the effect of the minijet
veto probability for the signal, while we assume that it can be
measured precisely for the backgrounds. This assumption might or might
not be valid once we face real data. In particular in tough analyses
like invisible Higgs searches (which we do not use in this analysis)
this assumption needs to be reconsidered. Secondly, we limit ourselves
to low-luminosity running, which hurts the statistical pull of the
weak-boson-fusion channels. The moment we consider for example
$100~\ifb$ of data this situation will likely change.

\section{Beyond the Standard Model}
\label{sec:bsm}

The main aim of the kind of Higgs sector analysis presented in this
paper is to probe the nature of the Higgs sector in case that we only
see a light scalar Higgs boson at the LHC. In this section we study
the application of the Higgs parameter analysis and the ansatz
described in Section~\ref{sec:higgs} on a supersymmetric Higgs sector.
We restrict the analysis to the standard without 
additional observables~\cite{Arnesen:2008fb}.

This ultraviolet completion of the Standard Model has the advantage or
disadvantage that it contains a the decoupling limit in which the LHC
might quite realistically be left with a single light Higgs boson with
very similar properties to its Standard--Model counter part. Such an
outcome is predicted for fairly small values of $\tan\beta \lesssim
10$, where the Yukawa couplings of the heavy Higgs states are not
sufficiently enhanced to warrant a direct observation. In this regime
the only hope to see a heavy Higgs at the LHC might be an on-shell
decay $H \to hh$ with a subsequent decay for example to
$b\bar{b}\gamma\gamma$~\cite{self_rare}. On the other hand, for small
$\tan\beta$ the decoupling limit is approached far less rapidly, so
that a careful analysis of all couplings of an observed light Higgs
state might reveal significant deviations from their Standard--Model
values.

In this section we study two different scenarios: first, we choose a
fairly low mass for the heavy CP-odd Higgs boson $m_A$, so that we
are not too far in the decoupling regime.  In a second part we consider a
gluophobic Standard-Model like Higgs, \ie the top loop in the
effective $ggH$ coupling is largely canceled by a corresponding stop
loop~\cite{gluophobic}. Just like in Section~\ref{sec:likelihood} we
use true data points to disentangle the effects of the models from
smearing. In addition, we use the same effective theory as described
in Section~\ref{sec:higgs} without including new Higgs search channels
in the decays of new states and without including loop effects from
particles seen elsewhere in Atlas or CMS.

\subsection{Supersymmetric Higgs}
\label{sec:susy}

\TABLE[b]{
\begin{small}
\begin{tabular}{l|r|r|r|r|r|r|r|r}
 &
$\Delta_{WWH}$           &
$\Delta_{ZZH}$           &
$\Delta_{ttH}$           &
$\Delta_{bbH}$           &
$\Delta_{\tau\tau H}$    &
$\Delta_{\gamma\gamma H}$&
$\Delta_{ggH}$           &
$m_H$                    \\\hline
true &
$-0.13$ & $-0.13$ & $-0.19$ & $3.27$ & $3.29$ & $0.19$ & $-0.28$ &
$120.0$ \\
errors &
$\pm 0.45$ &
$\pm 0.61$ &
$\pm 0.63$ &
$\pm 2.34$ &
$\pm 3.35$ &
$\pm 0.99$ &
$\pm 1.12$ &
$\pm 0.29$ \\
 & 
$- 0.43$   &
$- 0.99$   &
$- 0.60$   &
$- 3.68$   &
$- 3.23$   &
$- 0.70$   &
$- 0.69$   &
$- 0.29$   \\
 & 
$+ 0.48$   & 
$+ 0.52$   & 
$+ 0.65$   & 
$+ 1.52$   & 
$+ 3.58$   & 
$+ 1.30$   & 
$+ 1.46$   & 
$+ 0.30$    
\end{tabular}
\end{small}
\caption[]{Couplings for the SPS1a-inspired scenario. We give the true
  values of our input and the error bars assuming 10000 toy
  experiments and $30~\ifb$ of data.}
\label{tab:sps1alow}
}

As a first step of this application we need to replace all Standard
Model measurements with the respective rates for a supersymmetric
parameter point based on SPS1a~\cite{sps}. After evolving the original
SPS1a parameters from the high scale to the low scale using
SuSpect~\cite{suspect}, we change $\tan\beta$ to 7, $A_t$ to
$-1100~\gev$ and $m_A$ to $151~\gev$. For this parameter choice
$g_{WWH}$ and $g_{ZZH}$ are significantly smaller than in the Standard
Model. It also predicts a light CP-even Higgs mass of $120~\gev$. The
true values for all Higgs couplings we give in the first line of
Table~\ref{tab:sps1alow}. Comparing these deviations with the Standard
Model error bands in Table~\ref{tab:toyerrors} we can guess that only
$g_{bb H}$ and $g_{\tau\tau H}$ will be useful individually, provided
we can extract them in this scenario. While the couplings to gauge
bosons are different in our supersymmetric scenario, their shifts with
respect to the Standard Model values are all within their respective
errors.

For the fairly small value of $\tan\beta$ the heavy Higgs bosons $H$,
$A$ and $H^+$ will not be observed at the LHC, and certainly not using
$30~\ifb$ of data. The question is: can we distinguish such a scenario
from a Standard--Model Higgs sector just looking for a light Higgs
boson~\cite{wbf_nolose}?\bigskip

Strictly speaking, we should account for the extended particle content
of supersymmetry with an increased theory error on the perturbative
predictions of the production and decay rates. However, in most
production channels the quantum effects of new states are mass
suppressed and hence within the errors quoted in
Table.~\ref{tab:theo_error}~\cite{wbf_susy,gf_susy}.  With real LHC
data this procedure would be an iterative process: first we would try
to establish a deviation assuming the Standard Model particle content
as the correct hypothesis. The general new-physics mass scale
accounting for deviations from the Standard Model couplings will then
feed back into our new physics analysis. In turn, taking information
from other sectors into account we can refine the Higgs sector
analysis.\bigskip

\begin{figure}[t]
 \begin{center}
 \includegraphics[width=0.24\textwidth]{%
     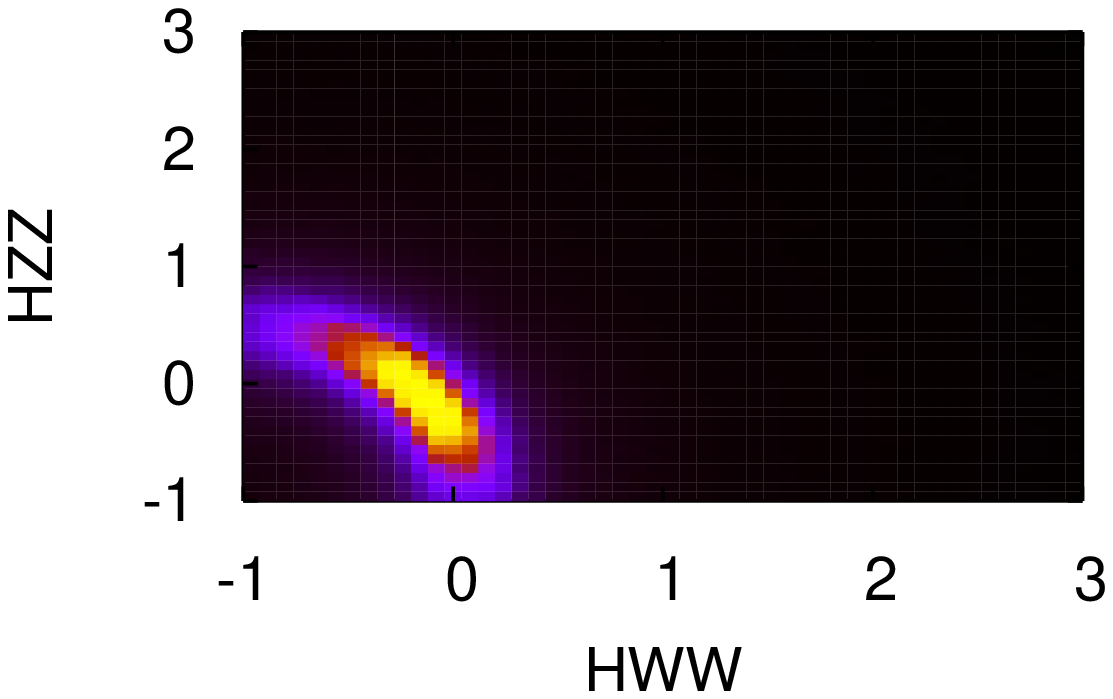} \hspace*{-2ex}
 \includegraphics[width=0.24\textwidth]{%
     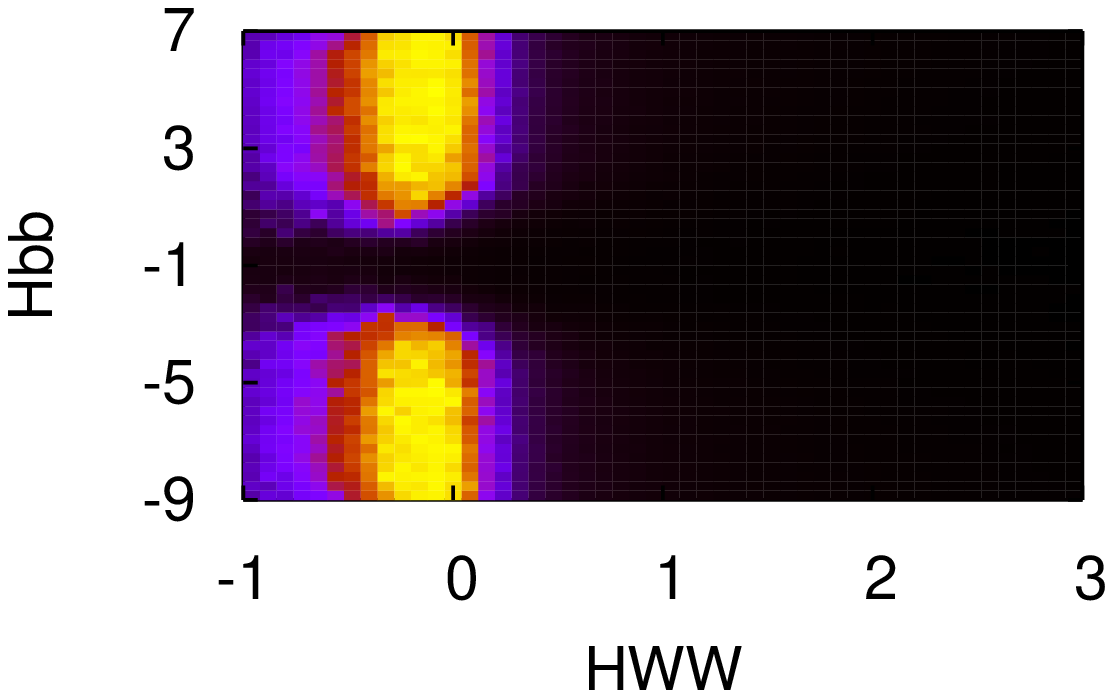} \hspace*{-2ex}
 \includegraphics[width=0.24\textwidth]{%
     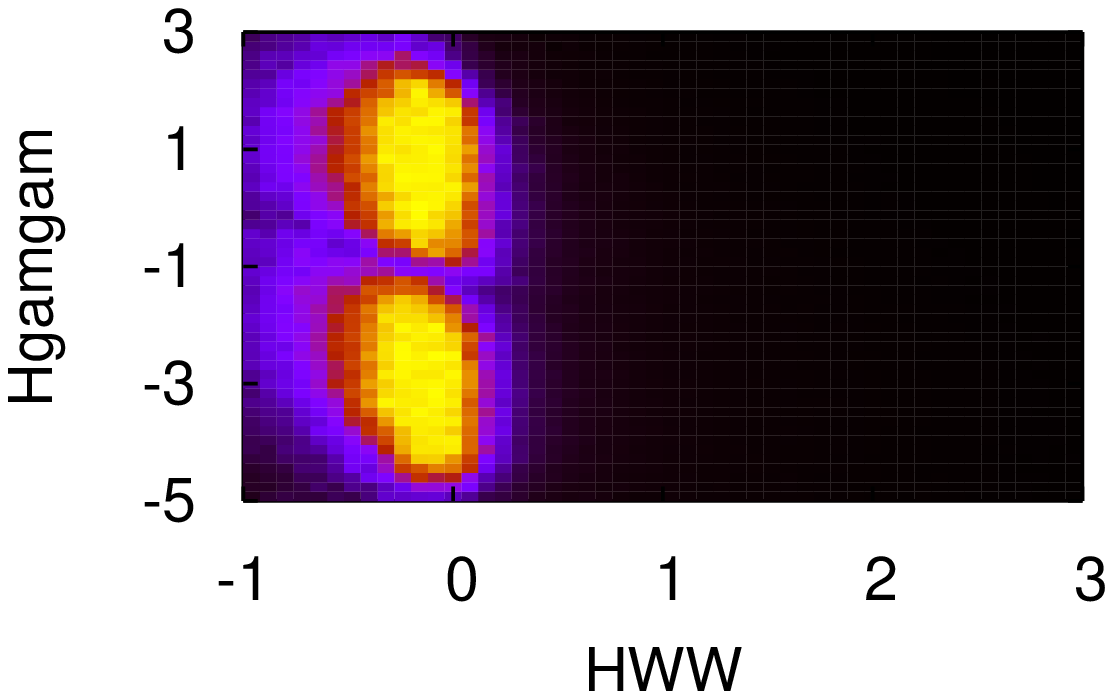} \hspace*{-2ex}
 \includegraphics[width=0.24\textwidth]{%
     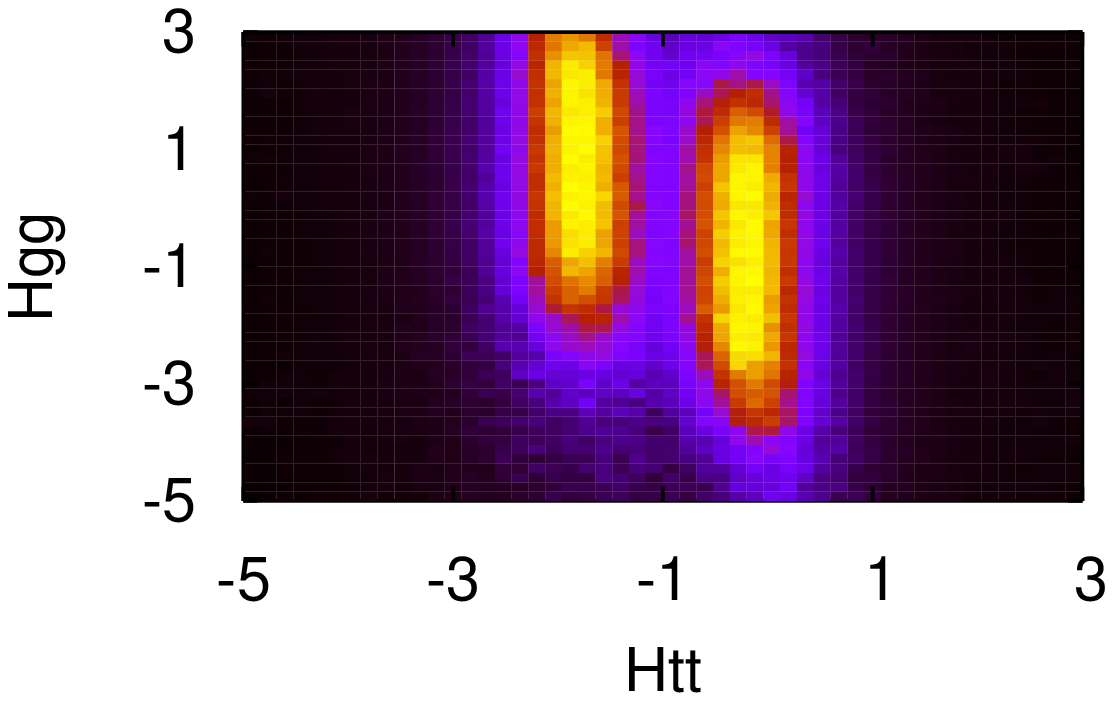} \\
 \includegraphics[width=0.24\textwidth]{%
     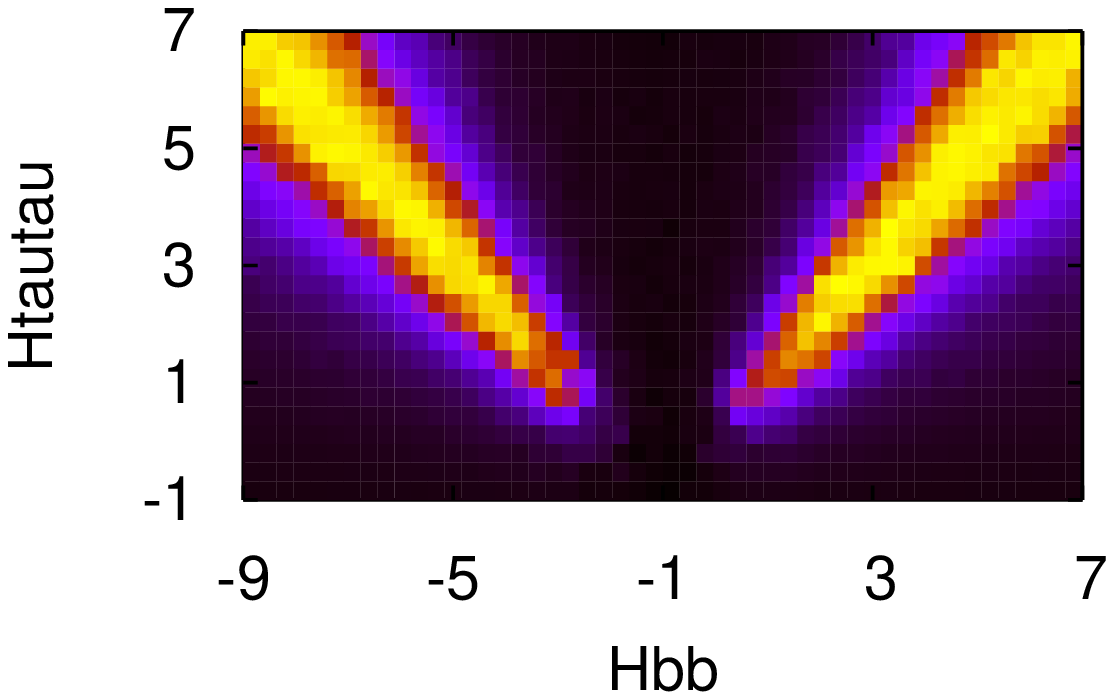} \hspace*{-2ex}
 \includegraphics[width=0.24\textwidth]{%
     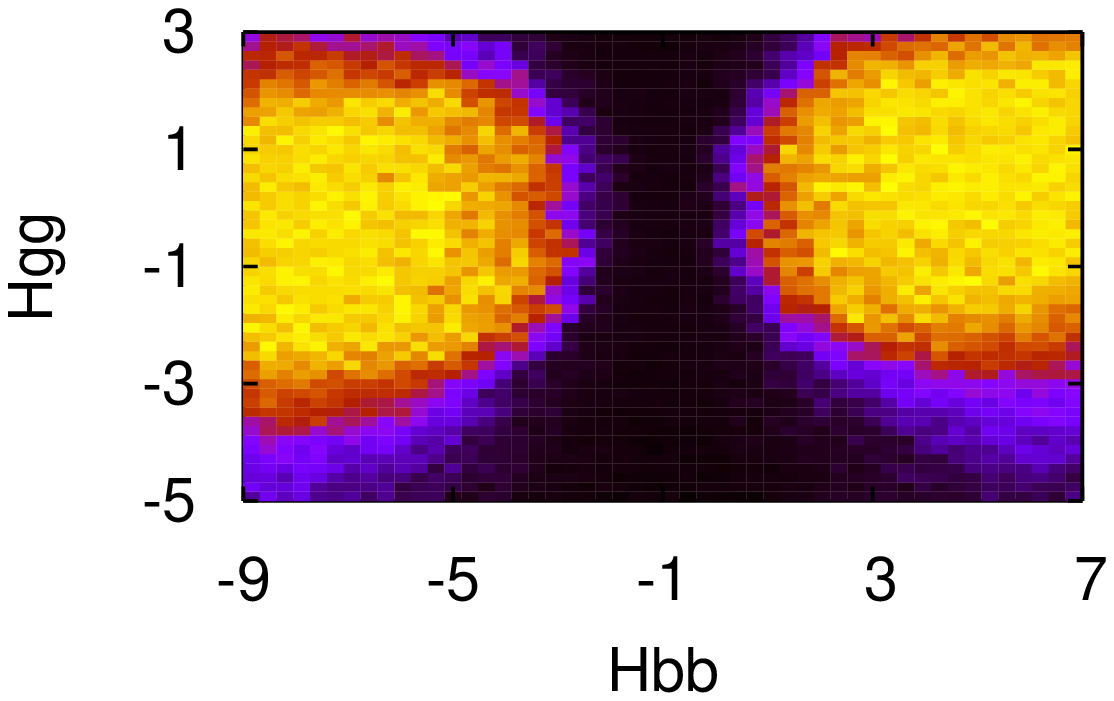} \hspace*{-2ex}
 \includegraphics[width=0.24\textwidth]{%
     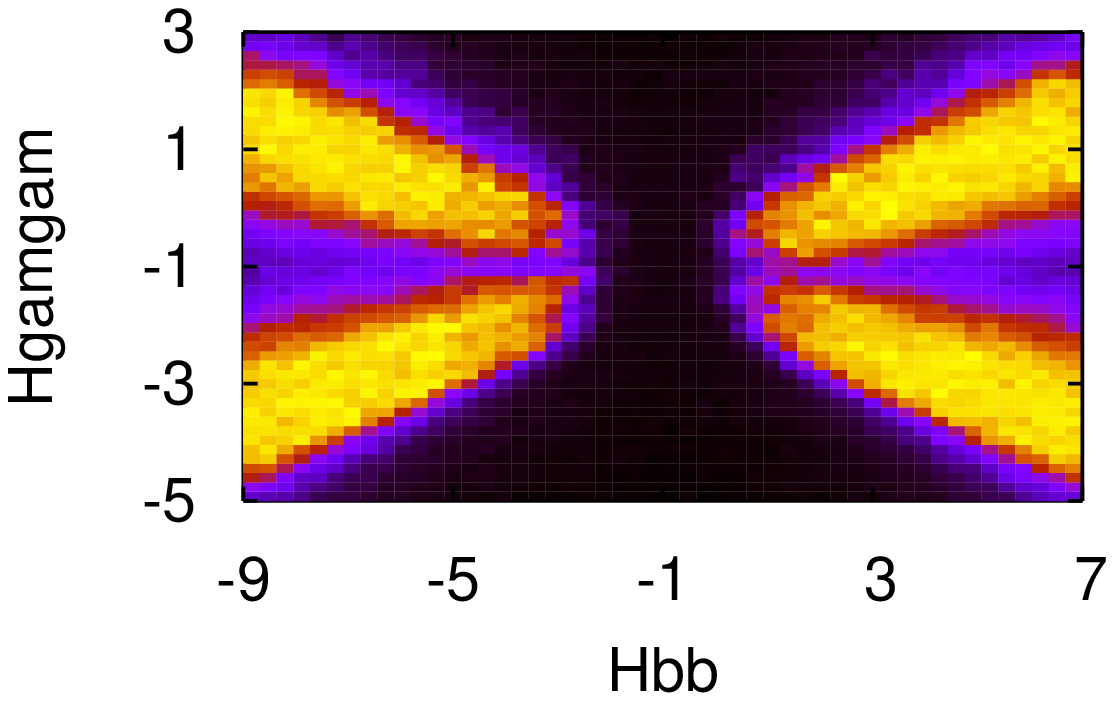} \hspace*{-2ex}
 \includegraphics[width=0.24\textwidth]{%
     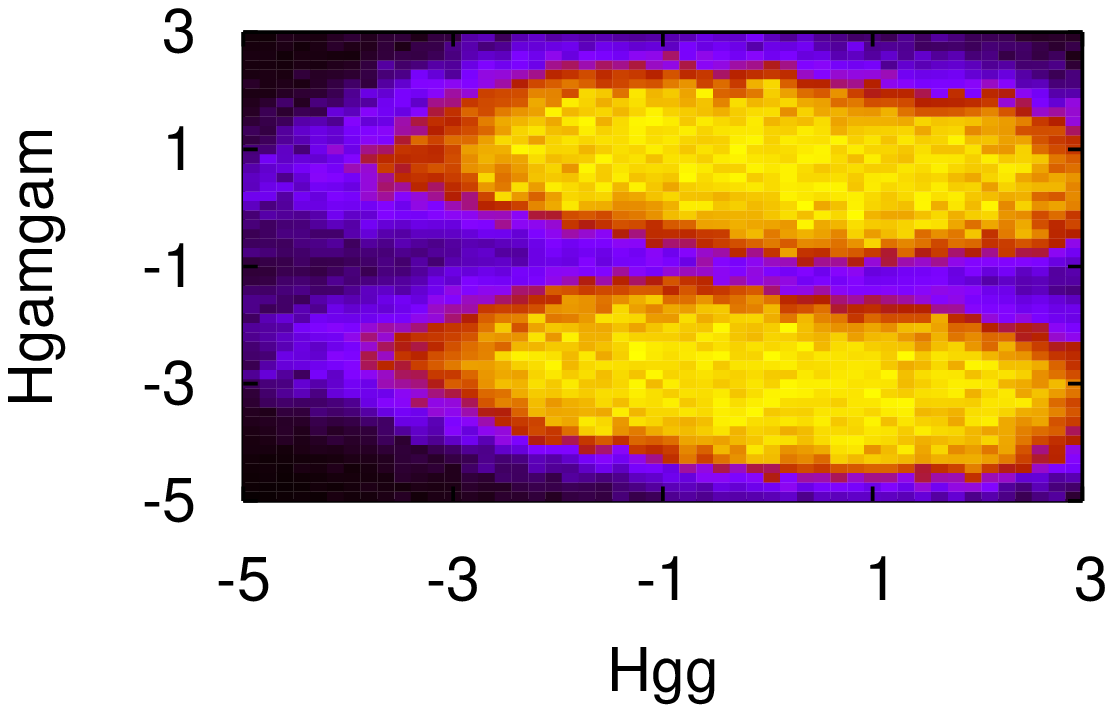} \\

 {\raggedright $1/\Delta \chi^2$\\[2ex]}
 \includegraphics[width=0.135\textwidth]{%
     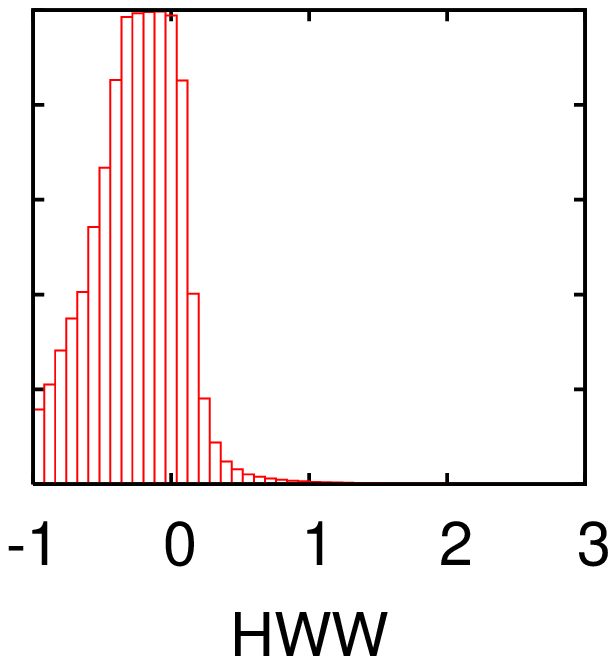} 
 \includegraphics[width=0.135\textwidth]{%
     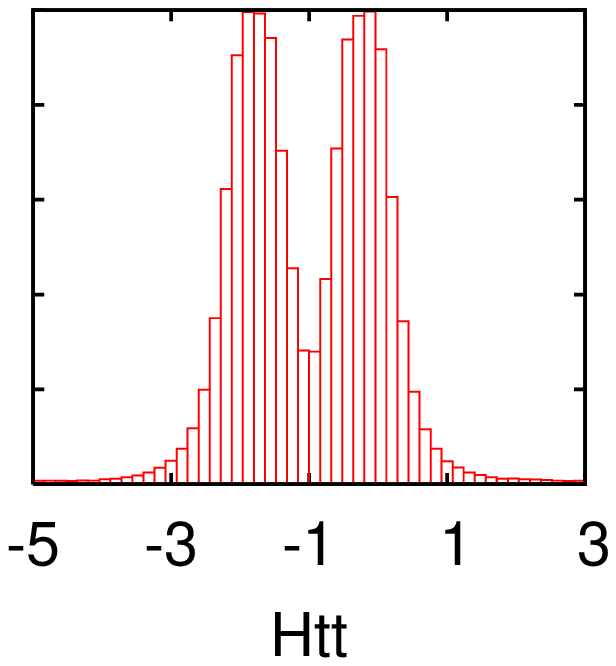} 
 \includegraphics[width=0.135\textwidth]{%
     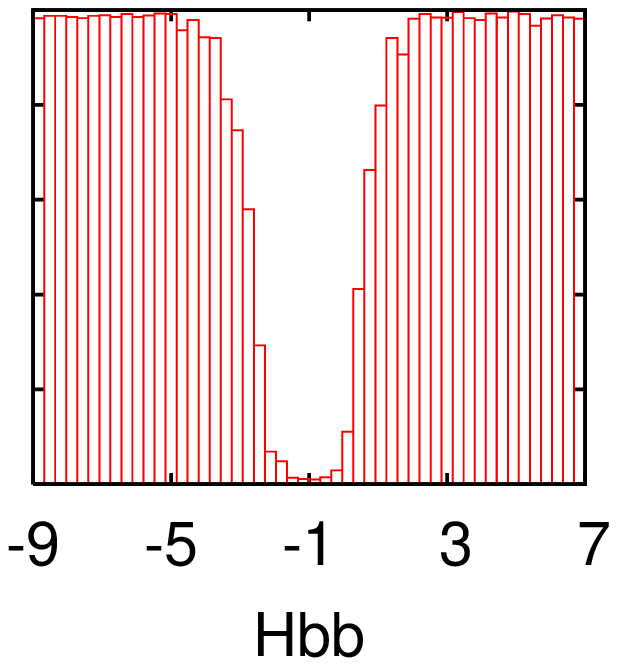} 
 \includegraphics[width=0.135\textwidth]{%
     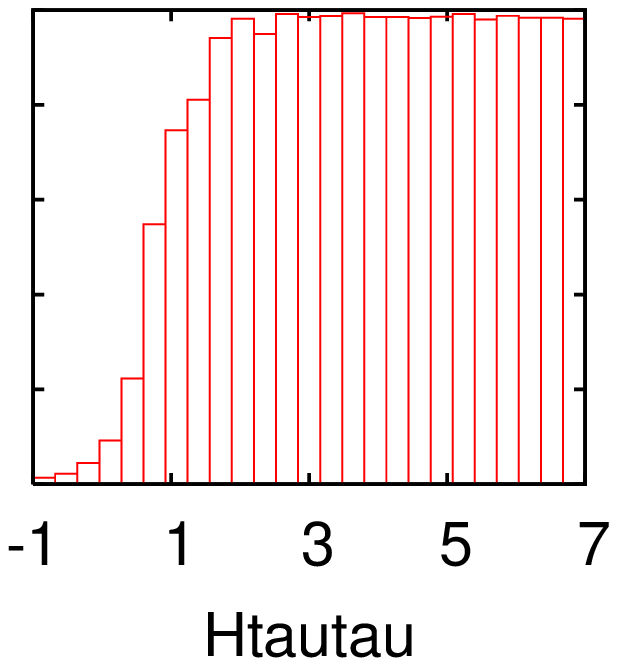} 
 \includegraphics[width=0.135\textwidth]{%
     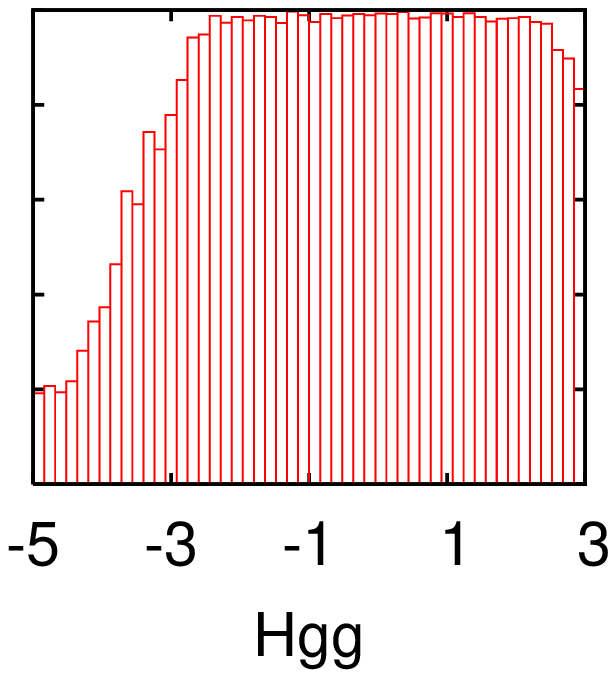} 
 \includegraphics[width=0.135\textwidth]{%
     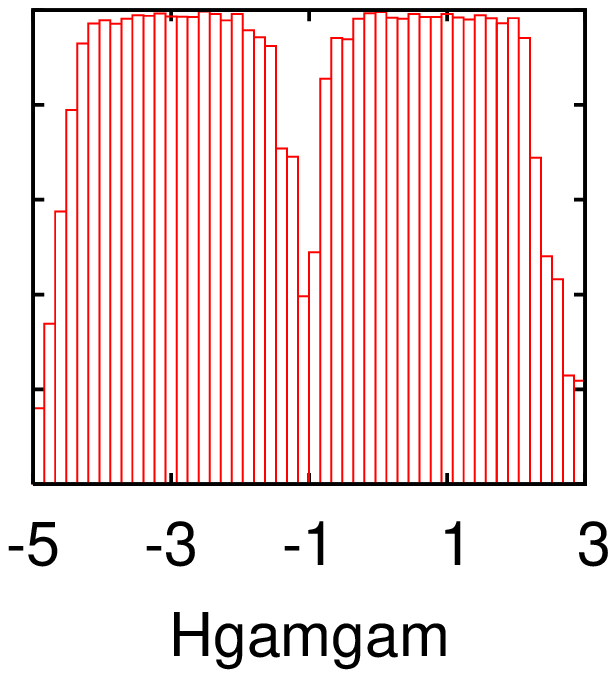} \\
 \end{center}
\caption[]{Profile likelihoods for the parameter point SPS1a with
  light $m_A$ and modified $A_t$ and $\tan\beta$ for $30~\ifb$. 
  We include both experimental and theory errors in our analysis.}
\label{fig:sps1alow}
\end{figure}

In the first line of Table~\ref{tab:sps1alow} we can guess that
essentially the four measurements with $\tau$ leptons or bottom quarks
in the final state will contribute to our likelihood. All other
measurements have too low production rates, even consistent with a
null result with $30~\ifb$.  We nevertheless include all measurements
in our analysis for consistency. For the dominant measurements
$g_{ttH}$ $g_{WWH}$ and $g_{ZZH}$ appear on the production side. The
latter is a sub-leading contribution to both weak boson fusion and the
subjet analysis.  For these four parameters we have four high-quality
measurements, so we can hope to obtain a unique solution in this
subspace when considering the complete set of measurements.

In the one-dimensional distributions in Fig.~\ref{fig:sps1alow} we
immediately see that the Standard Model hypothesis is not correct. For
$g_{bbH}$ we do not see an upper limit on the coupling strength. This
is a consequence of the branching ratio into bottom quarks which event
at unity would be compatible with the data. This limit removes any
kind of rate measurement involving $H \to b\bar{b}$ from our
measurements. The question remains if this leads to the free scaling
of the total width with all couplings, as we have seen before.

In the subjet analysis channel we cannot make the branching ratio into
bottom quarks larger than one, which induces a lower limit on
$g_{WWH}$. On the other hand, we cannot decrease the branching ratio
significantly, because this would require a large value of
$g_{WWH}$. Weak-boson-fusion production with subsequent decays into a
pair of $W$ bosons prohibits this solution, giving us an upper limit
on $g_{WWH}$ and inducing the lower limit on $g_{bbH}$ in
Fig.~\ref{fig:sps1alow}.

The same argument we can now apply to $g_{ttH}$ and
$t\bar{t}$-associated Higgs production.  The lower limit on $g_{bbH}$
induces an upper limit on $g_{ttH}$, which we also see in our
results. Additional information we obtain from the non-observation of
other channels with top-quark initial states. As this measurement is
much less precise than the subjet analysis, the allowed values of the
top-quark coupling extend further to positive values. The lower limit
again reflects the fact that the bottom branching ratio cannot be
larger than one.

The fourth parameter $\Delta_{\tau\tau H}$ is determined via weak
boson fusion. The range in $g_{WWH}$ implies an allowed range of
branching ratios into a $\tau$ pair, and therefore of the ratio
$g_{\tau\tau H}/g_{bbH}$ via the total width. Accordingly we see a
strong positive correlation between these two couplings. This way,
$g_{\tau\tau H}$ can extend to arbitrarily large values, while we have
a definite lower limit and the Standard-Model value is clearly
excluded.\bigskip

For the remaining couplings we can extract limits from a
non-observation in Higgs search channels. In $g_{\gamma\gamma H}$ we
again see a scaling with the total width and therefore a strong
correlation with $g_{bbH}$. The boundaries of $\Delta_{\gamma\gamma
  H}$ are simply given by our (arbitrary) limits on $g_{bbH}$ in this
scan, as we see in the correlation plot. The excluded region around
$-1$ is a normalization effect: the $WWH$ and $ttH$ couplings
contributing to the effective coupling are smaller than in the
Standard Model, while our additional contribution is normalized to the
Standard Model value. Setting $\Delta_{\gamma\gamma H}=-1$ would not
give a vanishing total coupling but something too large. The bands we
see are the regions where the additional contribution cancels the
loop-induced coupling.

For the $g_{ggH}$ coupling we do not see a scaling with $g_{bbH}$,
because any effect of varying this coupling will already be compensated
by the associated decay coupling. Including the $\tan \beta$ enhancement
the bottom loop can now become large enough to give a relevant
contribution to the effective coupling, so we then need to dial the
additional parameter accordingly to cancel the combinations from both
heavy quarks. Therefore, we do not see a sharp peak in $\Delta_{ggH}$
but a wide band of best-fitting points.\bigskip

Given such a scenario where we see clear deviations from the Standard
Model, the question arises at which level can we exclude
it. Obviously, in the presence of correlations and non-Gaussian errors
we cannot simply add the individual log-likelihoods.  Instead, we
determine the exclusion limits using the log-likelihood $q$ as an
estimator.

The first question is how well we can rule out the Standard Model
prediction given a data set $d_\text{SUSY}$ consistent with the
superysmmetric Higgs model predictions $m_\text{SUSY}$.  The purely
Standard Model log-likelihood distribution
$q(d_\text{SM}|m_\text{SM})$ usually gives us the desired confidence
level as the integral over $q$ from minus infinity to the central
value of the $m_\text{SUSY}$. However, we also have to take into
account that the new-physics data can fluctuate within its
experimental and theoretical errors. Basing the statement on one
single supersymmetric toy experiment would be incomplete.  Therefore,
we first compute a 90\% confidence level for the Standard Model
hypothesis given Standard Model data $\bar{q}$. In a second step we
evaluate the log-likelihood distribution for data consistent with the
supersymmetric scenario, but still based on the Standard Model couplings
$q(d_\text{SUSY}|m_\text{SM})$.  The percentage of toy experiments
giving $q(d_\text{SUSY}|m_\text{SM}) < \bar{q}$ we find to be
77\%. This means that of all our toy experiments assuming the
supersymmetric Higgs sector described in Table IX, 77\% are not
described by the Standard Model within the 90\% confidence level and
can be ruled out accordingly.

The second question is whether the new physics model is a better
description of our new physics data ($d_\text{SUSY}$) than the
Standard Model. The relevant information is the distribution of the
difference of the two log-likelihood values for each toy experiment:
$\Delta q = q(d_\text{SUSY}|m_\text{SUSY}) -
q(d_\text{SUSY}|m_\text{SM})$. Note that for most cases we expect this
difference to be positive. Because of the non-standard distribution of
our log-likelihood the usual definition corresponding to $\Delta
\chi^2 > s^2$ (with $s$ a number of standard deviations) does not
hold. A detailed discussion on testing different hypotheses and
confidence levels we postpone to a later paper, while at this stage we
simply request $\Delta q > q_0$ where $q_0$ is chosen of the order of
the quality of the consistently supersymmetric fit.  Of all our toy
experiments we find that only 4\% are better described by the
supersymmetric hypothesis than by the Standard Model given this naive
choice.

\subsection{Gluophobic Higgs}
\label{sec:gluophobic}

\TABLE[b]{
\begin{small}
\begin{tabular}{l|r|r|r|r|r|r|r|r}
 &
$\Delta_{WWH}$           &
$\Delta_{ZZH}$           &
$\Delta_{ttH}$           &
$\Delta_{bbH}$           &
$\Delta_{\tau\tau H}$    &
$\Delta_{\gamma\gamma H}$&
$\Delta_{ggH}$           &
$m_H$                    \\\hline
true &
$-0.01$ & $ 0.00$ & $ 0.00$ & $0.14$ & $0.13$ & $0.28$ & $-0.76$ &
$112.36$ \\
errors &
$\pm 0.37$ &
$\pm 0.53$ &
$\pm 0.50$ &
$\pm 0.67$ &
$\pm 0.62$ &
$\pm 0.68$ &
$\pm 0.69$ &
$\pm 6.83$ \\
 & 
$- 0.41$   &
$- 1.21$   &
$- 0.59$   &
$- 0.39$   &
$- 0.24$   &
$- 0.48$   &
$- 0.59$   &
$- 1.84$   \\
 & 
$+ 0.32$   & 
$+ 0.29$   & 
$+ 0.42$   & 
$+ 1.10$   & 
$+ 0.99$   & 
$+ 0.86$   & 
$+ 0.79$   &
$+10.21$
\end{tabular}
\end{small}
\caption[]{Couplings for the gluophobic Higgs scenario. We give the true
  values of our input and the error bars assuming 10000 toy
  experiments and $30~\ifb$ of data.}
\label{tab:badmh}
}

The determination of the Higgs mass plays an important role in the
$120 - 140$~GeV mass range because the different branching ratios
are steep functions of the mass. For a low-mass Higgs the best mass
measurement comes from the decay into two photons, which is usually
extracted from overwhelming backgrounds using two side bands. The
excellent invariant mass resolution of the detectors then allows for a
Higgs mass determination with a dominant experimental error around
$\mathcal{O} (100~\mev)$. However, in models where the loop-induced
Higgs couplings to gluons is strongly reduced this mass measurement
gets degraded by lower statistics. A supersymmetric Higgs can be a
good example for such a possible challenge.

Naively, we would first expect problems with the effective coupling to
photons including superpartners.  Aside from the usual $W$ and top
loop, charged Higgs bosons, sfermions and charginos contribute to the
effective coupling of the photons to the Higgs boson. However, the
effective $g_{\gamma \gamma H}$ is unlikely to decrease significantly
after taking into account current mass limits.

This is different for the inclusive Higgs production rate from gluon
fusion. Here, the top loop naturally cancels with the stop loop due to
the opposite spin of the two states.  Moreover, the current
experimental bounds on the stop mass is weak enough to allow for such
a scenario, commonly called a gluophobic Higgs.  The effect of this
supersymmetric scenario on the Higgs parameters we show in the first
line of Table.~\ref{tab:badmh}. Similarly to the supersymmetric model
shown in Table~\ref{tab:sps1alow} individual deviations from the
Standard Model couplings will not be conclusive, and the sizable
$\Delta_{ggH}$ will turn out to be dangerous, because it reduces the
number of Higgs bosons produced altogether and with that the precision
on all model parameters.\bigskip

The reduced rate for the inclusive $H \to \gamma\gamma$ signal means
that this channel is no longer suited for a precise determination of the
Higgs mass. An alternative is a Higgs decay to $ZZ$, but for low Higgs
masses and strongly reduced inclusive production rates it is even less
promising.  For an integrated luminosity of $30~\ifb$ the
weak-boson-fusion production process with a subsequent decay to
photons predicts 13.5~signal events with a background uncertainty of
close to~20 events, which means it will only contribute to a Higgs
mass measurement at higher luminosities. The last method for
determining the Higgs mass is weak-boson-fusion production with a
decay into a $\tau$ pair.  The $\tau \tau$ invariant mass can be
reconstructed in the collinear limit, with a mass resolution around
$8~\gev$~\cite{wbf_tau,coll_taus}, which we assume as the error on the
Higgs mass in this scenario.\bigskip

\begin{figure}[t]
 \begin{center}
 \includegraphics[width=0.24\textwidth]{%
     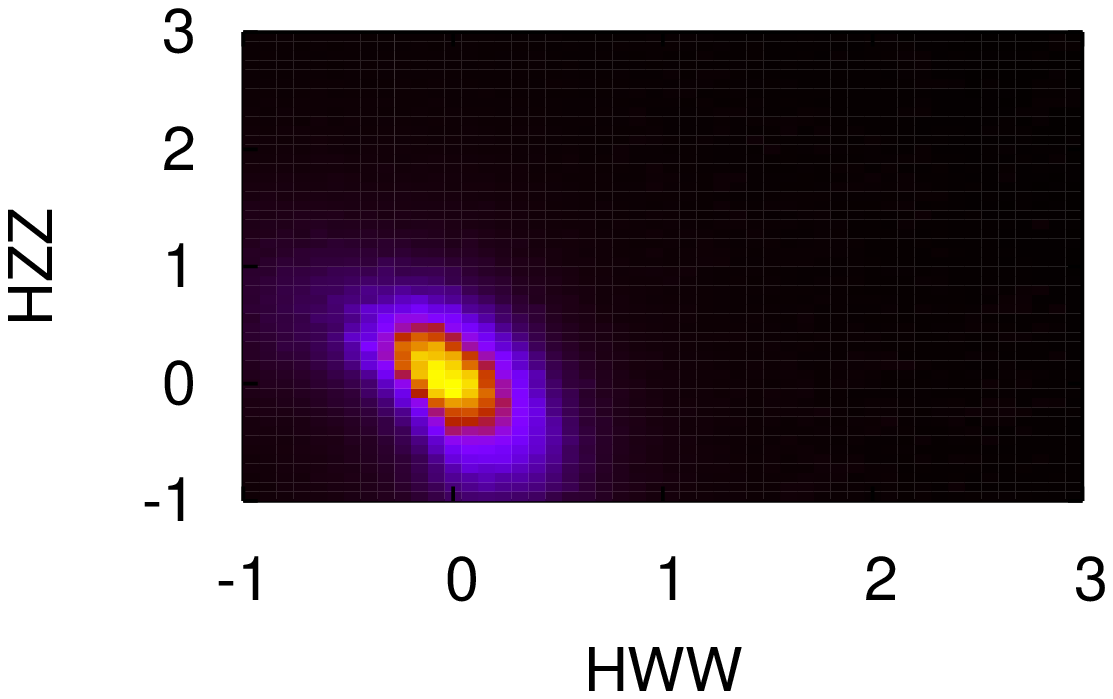} \hspace*{-2ex}
 \includegraphics[width=0.24\textwidth]{%
     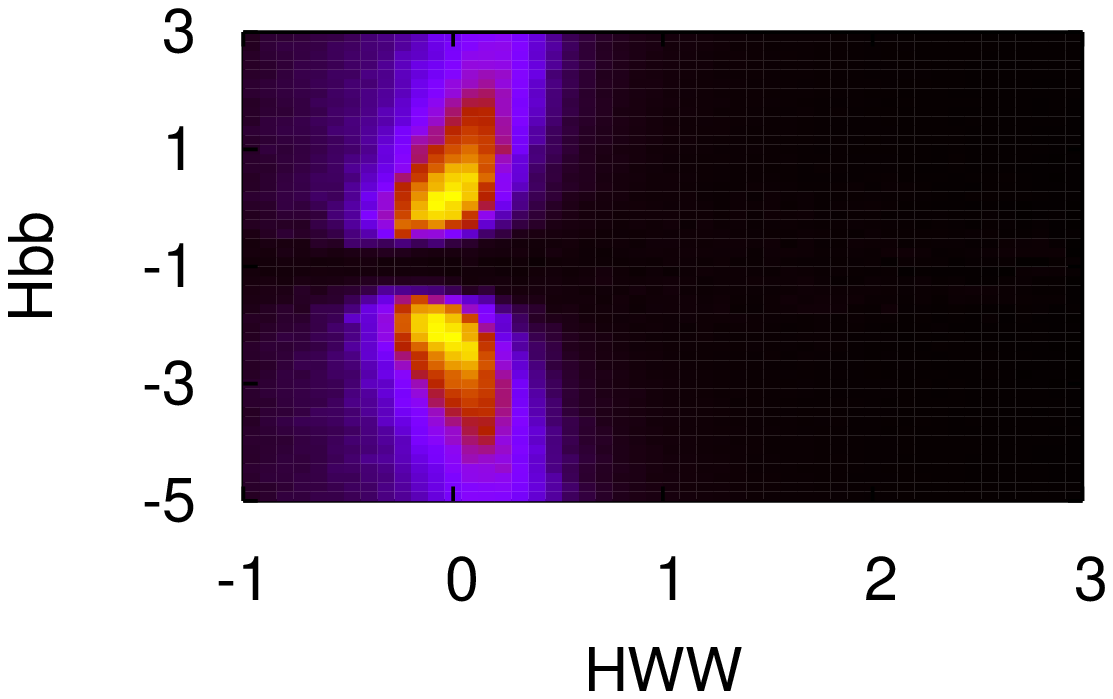} \hspace*{-2ex}
 \includegraphics[width=0.24\textwidth]{%
     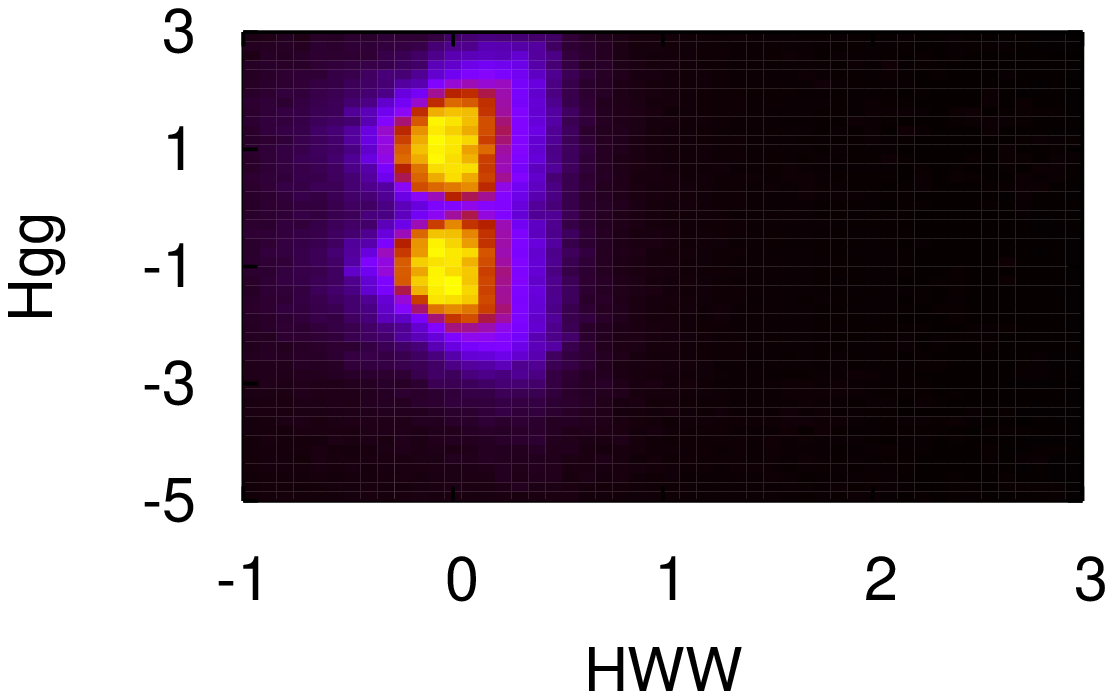} \hspace*{-2ex}
 \includegraphics[width=0.24\textwidth]{%
     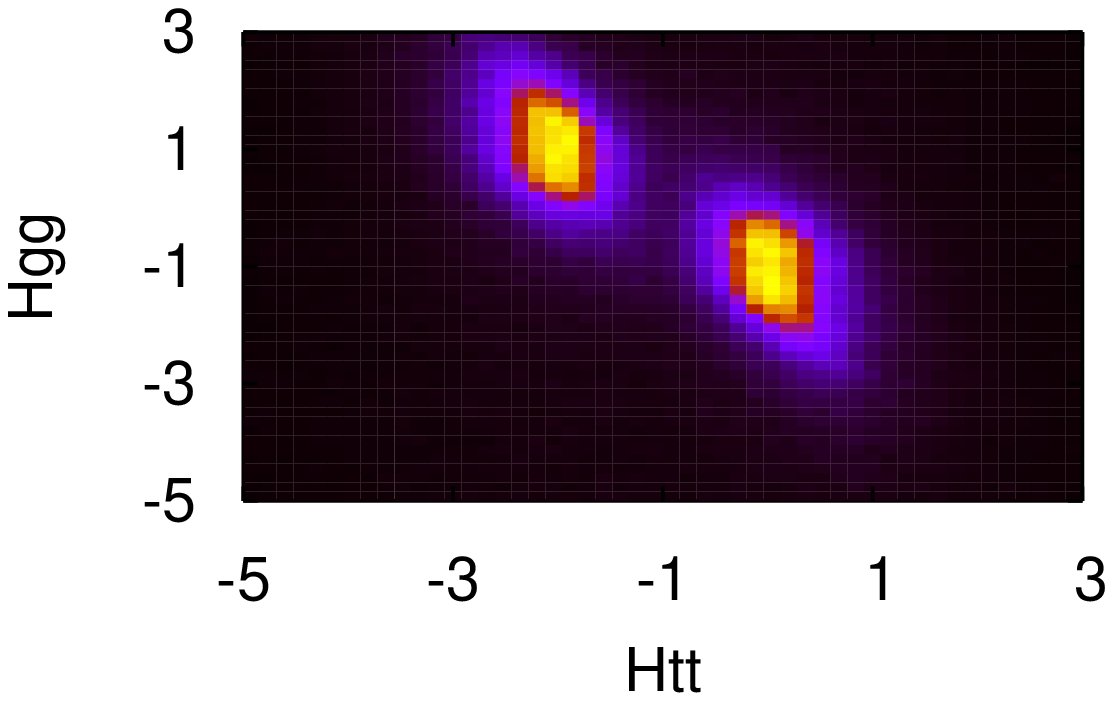} \\
 \includegraphics[width=0.24\textwidth]{%
     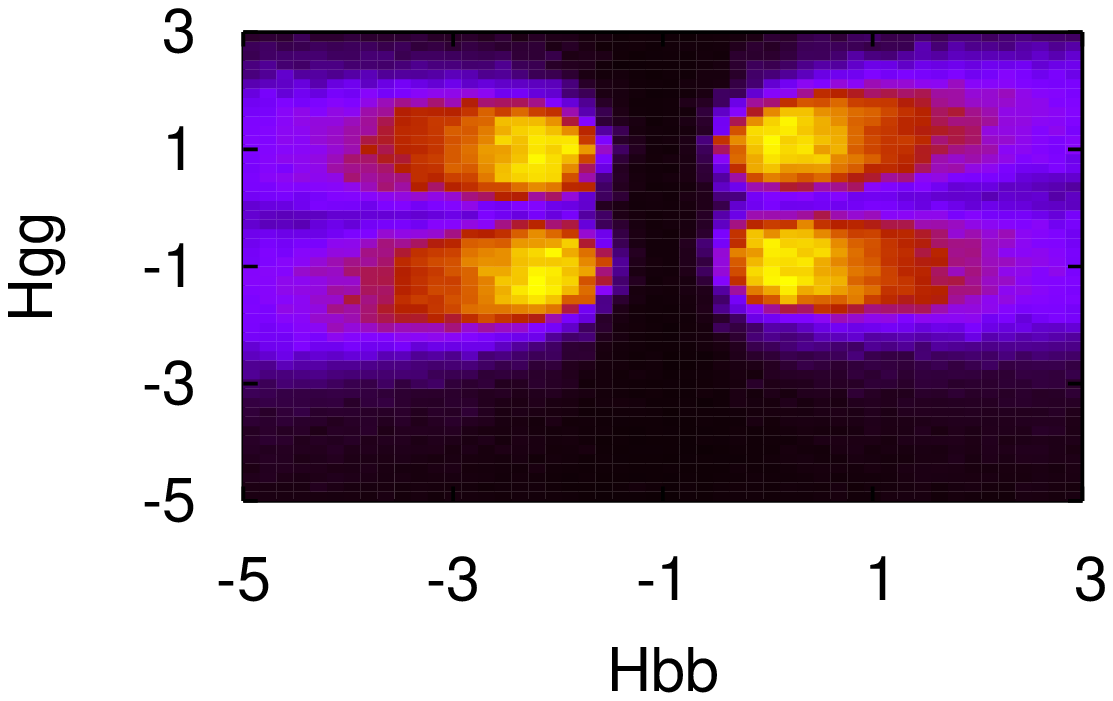} \hspace*{-2ex}
 \includegraphics[width=0.24\textwidth]{%
     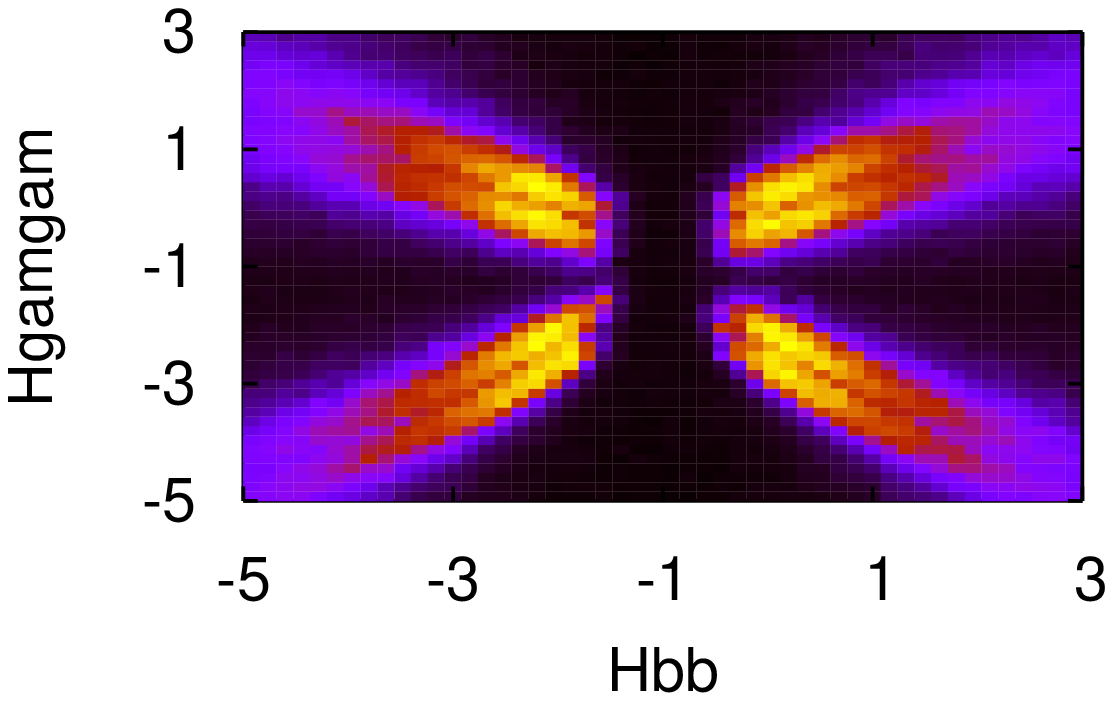} \hspace*{-2ex}
 \includegraphics[width=0.24\textwidth]{%
     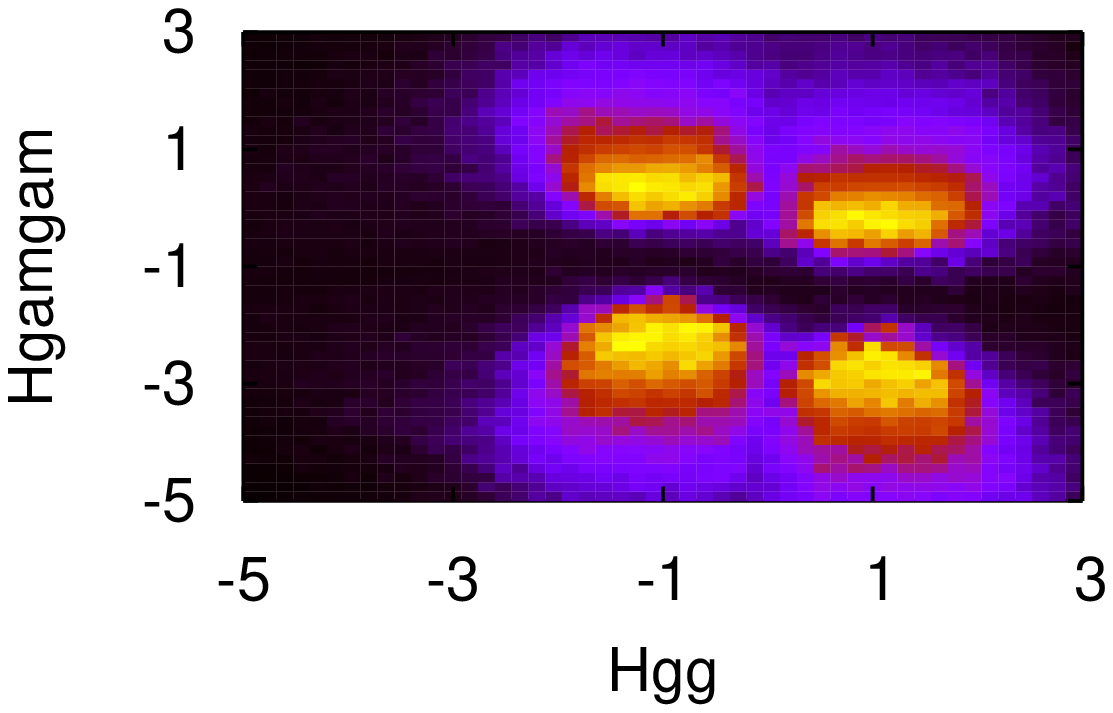} \hspace*{-4ex}
 \psfrag{mH}[c][][1][0]{$m_H$}
 \psfrag{ 90}[c][c][1][0]{90}
 \psfrag{ 105}[c][c][1][0]{105}
 \psfrag{ 120}[c][c][1][0]{120}
 \psfrag{ 135}[c][c][1][0]{135}
 \raisebox{0.16\textwidth}{$1/\Delta \chi^2$ }
 \includegraphics[width=0.135\textwidth]{%
     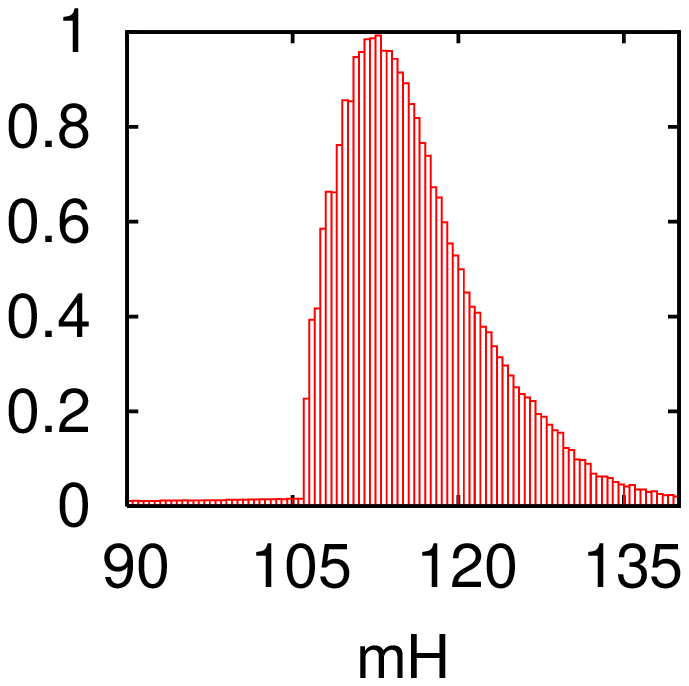} \hspace*{2ex} \\

 {\raggedright $1/\Delta \chi^2$\\[2ex]}
 \includegraphics[width=0.135\textwidth]{%
     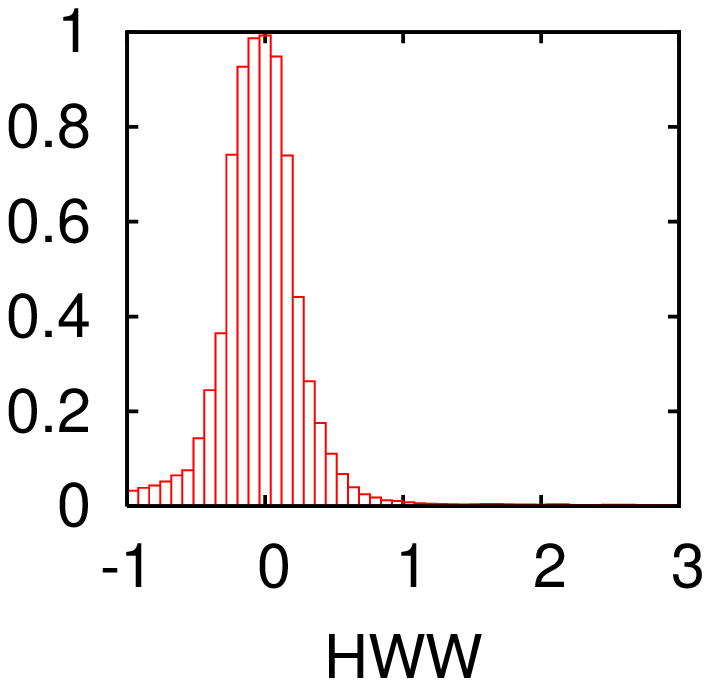} 
 \includegraphics[width=0.135\textwidth]{%
     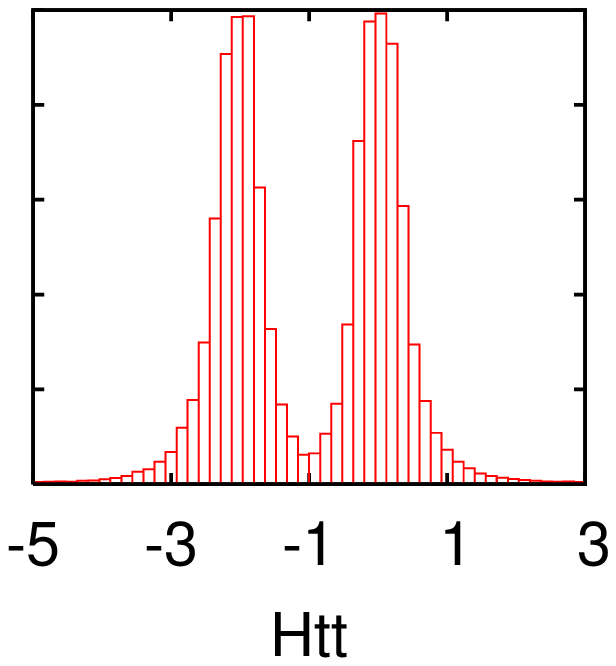} 
 \includegraphics[width=0.135\textwidth]{%
     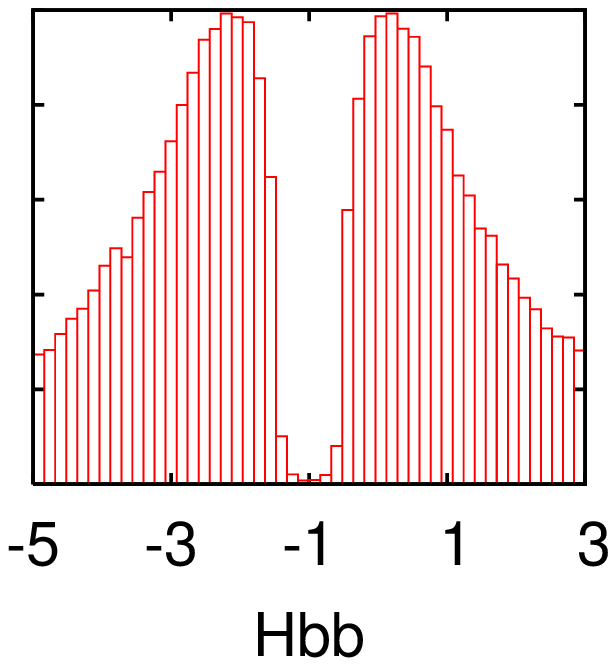} 
 \includegraphics[width=0.135\textwidth]{%
     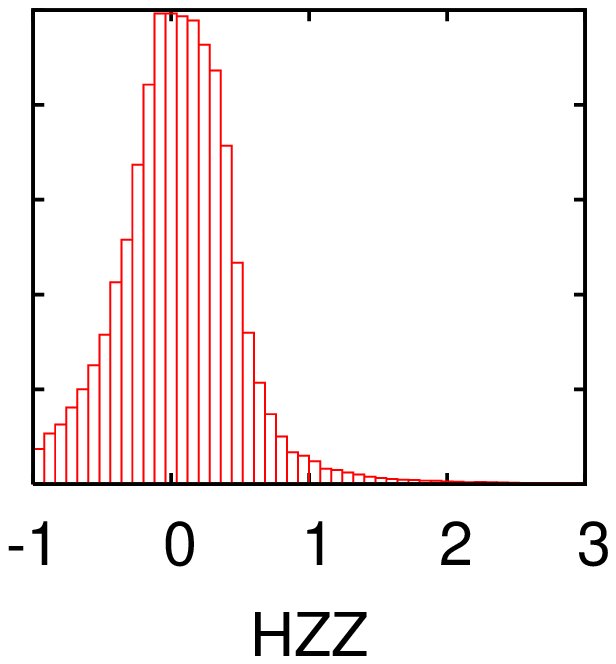} 
 \includegraphics[width=0.135\textwidth]{%
     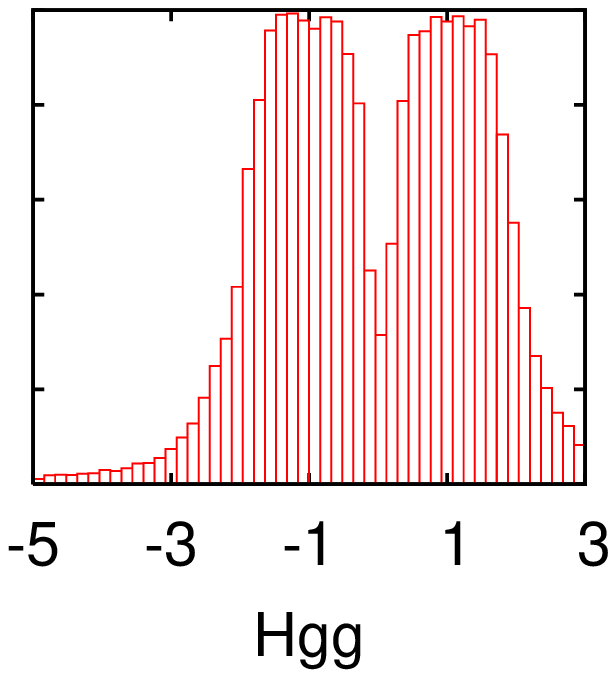} 
 \includegraphics[width=0.135\textwidth]{%
     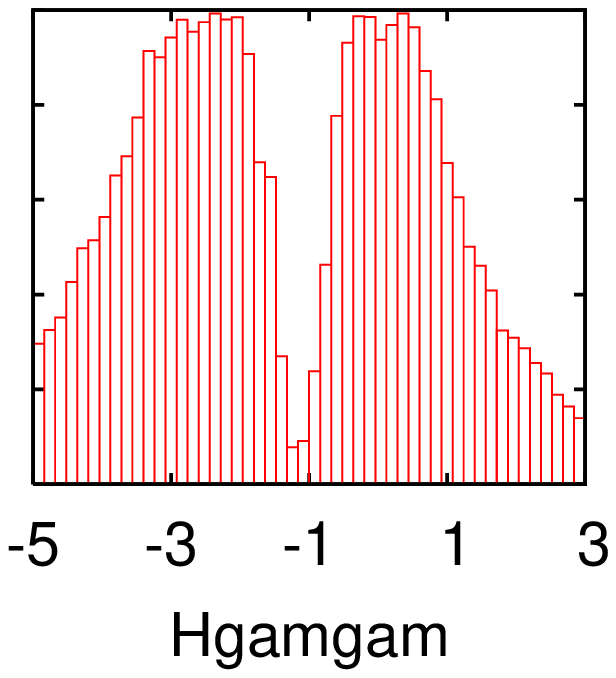} \\
 \end{center}
\caption[]{Profile likelihoods for a gluophobic Higgs scenario at 
  $30~\ifb$. 
  We include both experimental and theory errors in our analysis.}
\label{fig:badmh}
\end{figure}

In Fig.~\ref{fig:badmh} we show a set of profile likelihoods.  The
parameter determination suffers from the effective absence of all
gluon-fusion production channels and the larger uncertainty in the
Higgs boson mass. On the other hand, in contrast to the supersymmetric
scenario discussed in Section~\ref{sec:susy} we do not observe any
severe holes in the analysis.  All parameters which do not depend,
directly or indirectly, on $g_{ggH}$ are determined well, with an
increased error in part from the Higgs mass measurement.  This
includes $g_{WWH}$, $g_{ZZH}$, $g_{ttH}$, and $g_{\tau\tau H}$. Even
for the $\Delta_{ggH}$ itself the Standard Model value of zero is
excluded. The four solutions for correct and flipped sign of the
top-quark coupling and zero or double contribution from the additional
effective coupling are now grouped in two pairs, each forming a broad
peak. This we also observe in the two-dimensional likelihood.

The $bbH$ coupling depends on $g_{ggH}$ via the total width. In the
Standard Model loop-induced decays of the Higgs into gluons have a
sizable branching ratio. A wide variation in $\Delta_{ggH}$
significantly shifts the total width and needs to be compensated by
adjusting $g_{bbH}$, so the decays into bottoms account for the
correct rates. The production-side $g_{ggH}$ directly impacts the
measurement of $g_{\gamma\gamma H}$. Both, the large range for the
gluon coupling and the increased statistical error on the inclusive
photon channel lead to a poorer determination of this
coupling.\bigskip

Again, we compute the significance with which we can exclude the
Standard Model hypothesis. Follow the prescription outlined in the
previous section we find that for 46\% of our toy experiments we can
rule out the Standard Model at a 90\% confidence level. For data
consistent with a gluophobic Higgs, the new-physics hypothesis gives a
better description than the Standard Model in only $2\%$ of the cases.

\section{Outlook}

In this paper we have studied quantitatively how well LHC measurements
can determine the parameters in the effective weak-scale Higgs
sector~\cite{duehrssen}. SFitter~\cite{sfitter} uses Markov chains to
map the multi-dimensional Higgs parameter space onto a set of
measurements which we expect for a 120~GeV Higgs boson at the LHC. The
resulting exclusive likelihood map allows us to show correlations and
to determine the error bars for individual Higgs couplings. For an
appropriate description of these errors we allow for general
correlated statistical, systematic and theory errors. Since we do not
find distinct alternative maxima in the exclusive likelihood map, we
rely on profile likelihoods (as compared to Bayesian probabilities) to
describe likelihood distributions in one or two dimensions.\bigskip

We find that Higgs couplings can be extracted with typical errors
around $20 - 40\%$ using an integrated luminosity of $30~\ifb$
and properly simulating all errors involved. The different parameters
are strongly correlated, and sign ambiguities in particular in the
presence of dimension-five operators occur. One main correlation
between all channels is due to the total width, which also means that
a reliable measurement of the bottom Yukawa coupling for example using
subjet analyses~\cite{subjet} is vital for our analysis. Coupling
ratios instead of individual couplings can have somewhat reduced
errors, also depending on the treatment of the total Higgs
width.\bigskip

Unobservable (not `invisible') Higgs decays at the LHC can include a
coupling to light quarks which also affects the inclusive production
cross section.  We discuss strategies of dealing with different
scenarios of this kind.  While the LHC is clearly not going to measure
the total Higgs width directly, unexpected effects not contributing to
the observed decay channels will be visible in our fit. 

Given our results for the error on the couplings it is unlikely that
we will be able to use a general Higgs-sector analysis to distinguish
between different decoupling model hypotheses (like the MSSM vs
Standard Model), but drastic modifications like low-$m_A$
supersymmetry or a gluophobic Higgs boson will be clearly visible. We
quantify the confidence levels of distinguishing the Standard Model
hypothesis from the respective new physics hypotheses in a
two-dimensional plane, including a possible fluctation of the data as
well as a variation of the model predictions, as predicted by the
complete error structure.

\acknowledgments

We are particularly grateful to Dave Rainwater for many fruitful
discussions in the early phase of this project and to Peter Zerwas for his comment on the final version. 
Moreover, we are
grateful to the Fittino and CKMfitter groups for the constructive
interaction over many years of studying high--dimensional parameter
spaces. We are also grateful to the GDR Supersym\'etrie (CNRS), the Les
Houches Workshops as well as the Aspen center of physics. In particular,
we would like to thank the University of of Washington for organizing a
Higgs workshop with may enlightening discussions on our at that time
preliminary results.
This work was supported in part by the DOE under Task TeV of contract
DE-FGO3-96-ER40956.
MR acknowledges support by the Deutsche Forschungsgemeinschaft via the
Sonderforschungsbereich/Transregio SFB/TR-9 “Computational Particle
Physics” and the Initiative and Networking Fund of the Helmholtz
Association, contract HA-101 (“Physics at the Terascale”). 

\begin{appendix}

\section{Combining log-likelihoods}
\label{app:combine}

For each experimental channel in our analysis we need to combine
different sources of errors. The number of events is given by a
Poisson distribution which for non-integer event numbers we compute as
$P(m,d) = e^{-m} m^d/\Gamma(d+1)$, with $m$ the predicted number of
events and $d$ the number of measured events.  Systematic errors
follow a Gaussian distribution, since they are measured using large
background samples. We assume Higgs signals and background
measurements of the systematics to be independent. Therefore, their
combined probability is the convolution of the individual parts. For
two Gaussians this can be done analytically, leading to another
Gaussian with the errors added in quadrature. The convolution of a
Gaussian and a Poisson distribution, on the other hand, can only be
evaluated numerically.  This is practically not feasible for each step
in our Markov chain, so instead we use an approximate analytic
formula.\bigskip

The basic quantity for SFitter is a generalized $\chi^2$, related to
the log-likelihood by $\chi^2 = -2 \log L$. First, we transform the
Poisson probability into a likelihood $L_P$. In analogy to the
Gaussian case we normalize it to $L(d=m)=1$. Since the peak of the
Poisson distribution is close to $d=m-0.5$ our likelihood would exceed
unity for $d \, \epsilon \, [m-1, m]$. Therefore we fix it to unity in
this interval. This way the maximum of a combined Gaussian and Poisson
likelihood is well defined and unique at $d=m$. Compared to an exact
calculation this method slightly overestimates the likelihood over the
whole range, thereby erring on the side of caution.
This leads us to
\begin{equation}
-2 \log L_P = 
-2 \log \frac{P(m,d)}{P(m,m)} =
-2 \left[ (d-m) \log m 
         + \log \Gamma(m+1) 
         - \log \Gamma(d+1)
   \right]
\end{equation}
which in the Gaussian limit becomes $-2 \log L_G = (d-m)^2/\sigma^2$
with the combined error $\sigma$.

\begin{figure}
 \begin{center}
\psfrag{exact}[r][r][1][0]{exact}
\psfrag{approximation}[r][r][1][0]{approximation}
\psfrag{chi2}[r][r][1][0]{$\chi^2$}
\psfrag{Deltachi2}[r][r][1][0]{$\Delta \chi^2$}
\psfrag{m}[r][r][1][0]{$m$}
\psfrag{ 0.002}[c][c][1][0]{0.002}
\psfrag{ 0.004}[c][c][1][0]{0.004}
\psfrag{ 0.006}[c][c][1][0]{0.006}
\psfrag{ 0.008}[c][c][1][0]{0.008}
\psfrag{ -0.02}[c][c][1][0]{-0.02}
\psfrag{ 0.02}[c][c][1][0]{0.02}
\psfrag{ 0.04}[c][c][1][0]{0.04}
\psfrag{ 0.06}[c][c][1][0]{0.06}
\psfrag{ 0.08}[c][c][1][0]{0.08}
\psfrag{ 0.1}[c][c][1][0]{0.1}
\psfrag{ 0.12}[c][c][1][0]{0.12}
\psfrag{ 0.2}[c][c][1][0]{0.2}
\psfrag{ 0.4}[c][c][1][0]{0.4}
\psfrag{ 0.6}[c][c][1][0]{0.6}
\psfrag{ 0.8}[c][c][1][0]{0.8}
\psfrag{ 0}[c][c][1][0]{0}
\psfrag{ 1}[c][c][1][0]{1}
\psfrag{ 2}[c][c][1][0]{2}
\psfrag{ 3}[c][c][1][0]{3}
\psfrag{ 4}[c][c][1][0]{4}
\psfrag{ 5}[c][c][1][0]{5}
\psfrag{ 6}[c][c][1][0]{6}
\psfrag{ 7}[c][c][1][0]{7}
\psfrag{ 8}[c][c][1][0]{8}
\psfrag{ 9}[c][c][1][0]{9}
\psfrag{ 10}[c][c][1][0]{10}
\psfrag{ 15}[c][c][1][0]{15}
\psfrag{ 20}[c][c][1][0]{20}
\psfrag{ 160}[c][c][1][0]{160 }
\psfrag{ 180}[c][c][1][0]{180 }
\psfrag{ 200}[c][c][1][0]{200 }
\includegraphics[width=0.40\textwidth]{%
     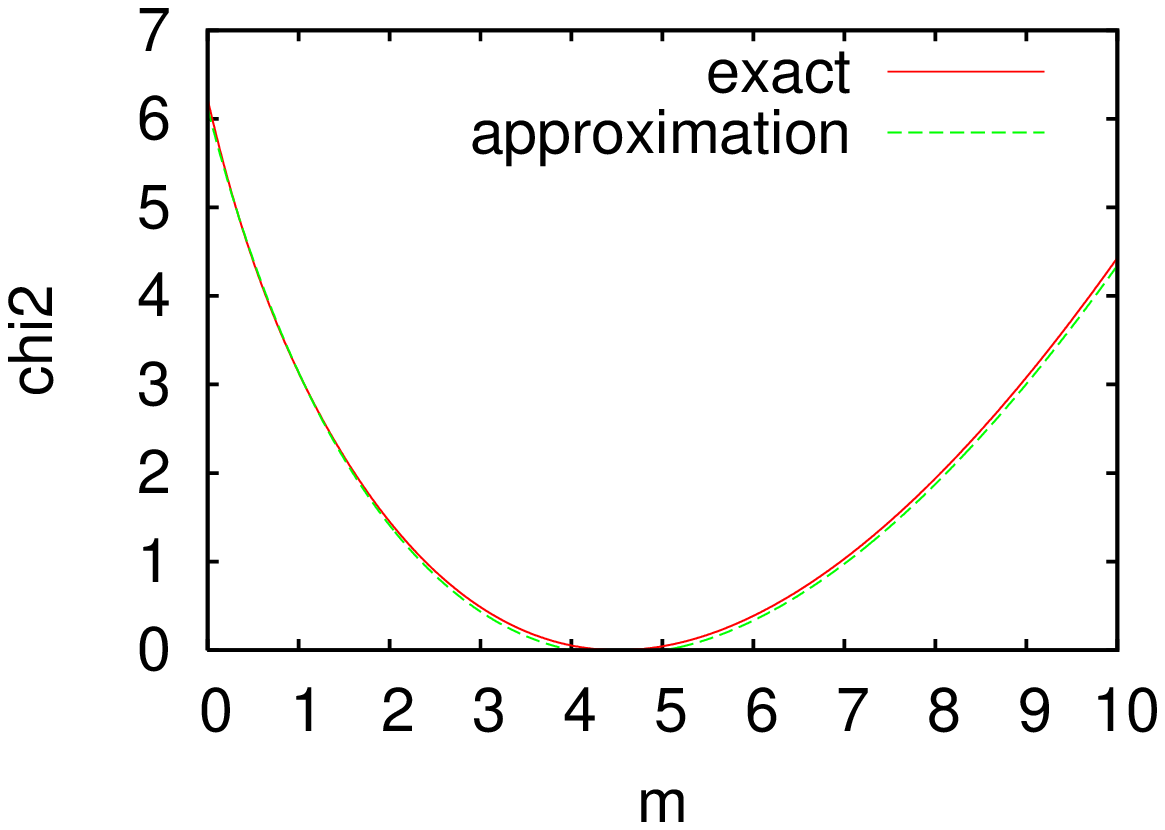} \hspace*{5ex}
\includegraphics[width=0.40\textwidth]{%
     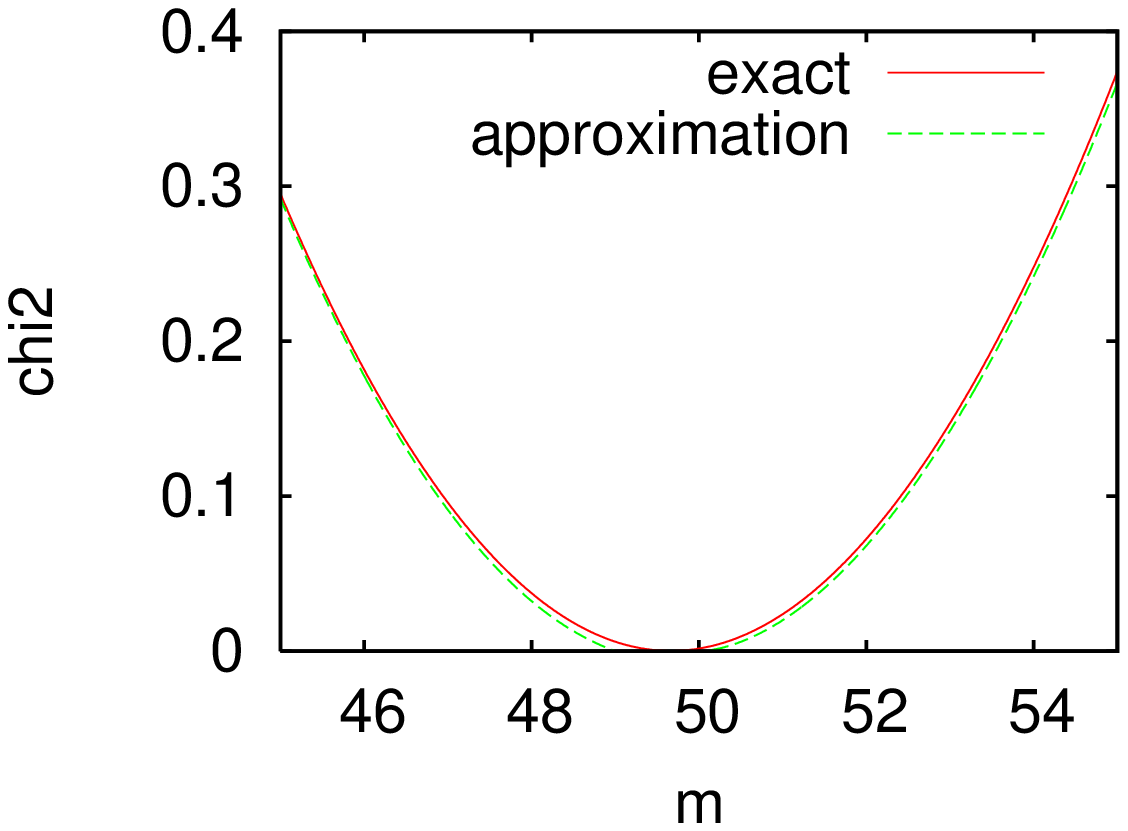}\\
\includegraphics[width=0.40\textwidth]{%
     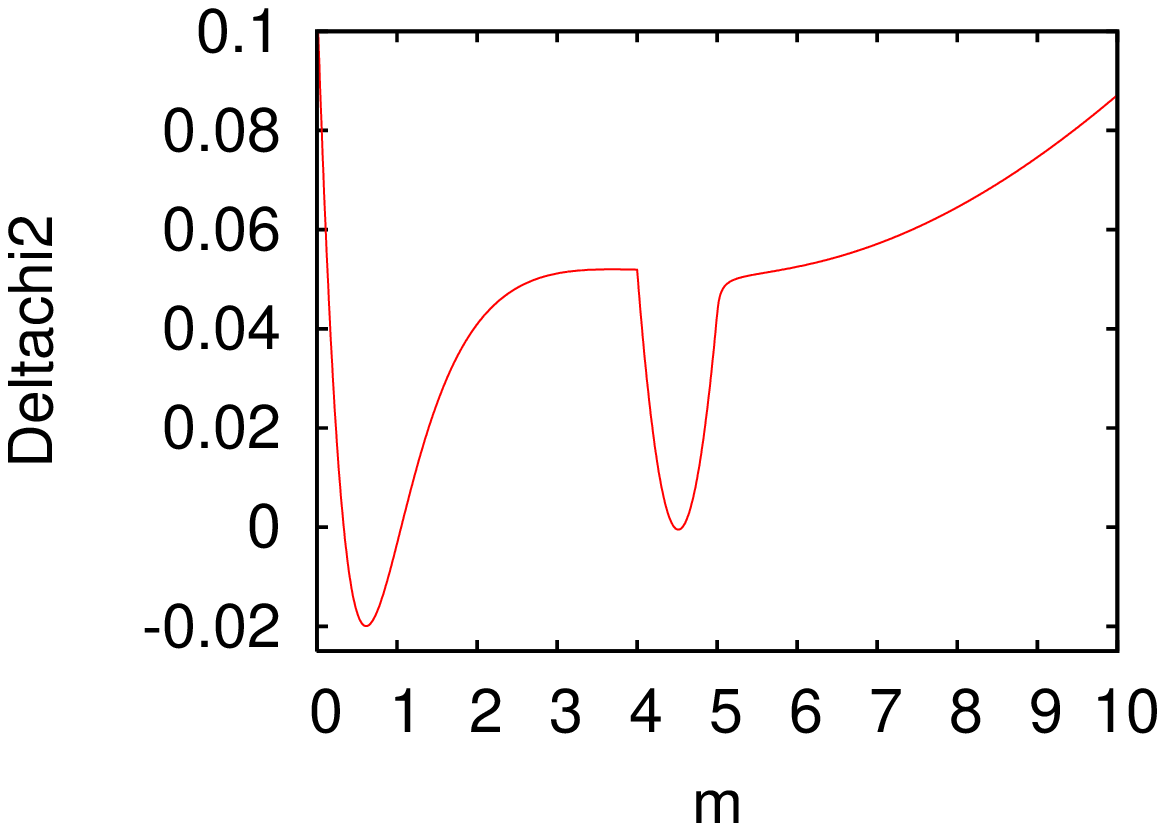} \hspace*{5ex}
\includegraphics[width=0.40\textwidth]{%
     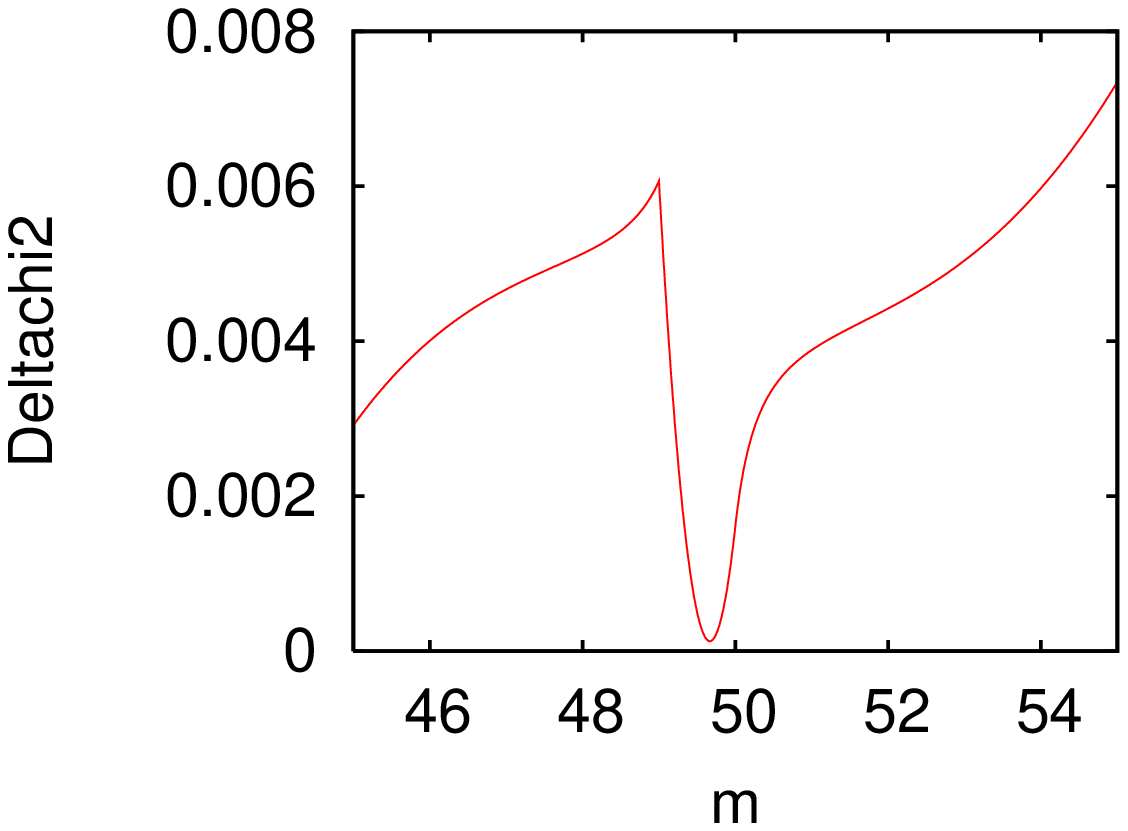}
 \end{center}
\caption[]{Comparison of our approximate (dashed green) with the exactly
  convoluted (solid red) log-likelihood. We show the absolute and the
  relative difference in $\chi^2$ for 5 (left) and 50 (right) events
  with a Gaussian error of $10\%$.}
\label{fig:app_comp}
\end{figure}

Next, we combine the two log-likelihoods into a single form, guided by
two limiting cases: the combination of two Gaussians should return the
exact formula for the combined Gaussian. Secondly, if one of the
log-likelihoods becomes very large, the result should approach the
other log-likelihood.  The form
\begin{equation}
\frac1{\log L} = \frac1{\log L_P} + \frac1{\log L_G}
\qquad \qquad \text{or} \qquad \quad 
\frac1{\chi^2} = \frac{-1}{2 \log \frac{\displaystyle P(m,d)}
                                      {\displaystyle P(m,m)}} +
                  \frac{\sigma_G^2}{(d-m)^2}
\end{equation}
fulfills both conditions. To test this setup we can compare our
result with a numerical convolution for typical cases as they appear
in our data. In Fig.~\ref{fig:app_comp} we show results for $5$ and
$50$ events. The Gaussian error is $10\%$, \ie $0.5$ and $5$
respectively. In both cases the difference of our approximation to the
exact result is small. Furthermore, we very slightly underestimate
$\chi^2$, and consequently slightly overestimate the errors, erring on
the safe side.\bigskip

Note that our results can differ from the naive approach of just
multiplying the individual likelihoods arising from different errors
for the same channel. This approximation is only valid for different
channels in the limit of vanishing correlations between these.  We
illustrate this for two errors in one channel: let us look at a
measurement with two Gaussian distributions with a difference between
predicted and measured values $x$ and an error of 1 each. Adding the
log-likelihoods gives us $\chi^2 = 2 x^2/1^2 = 2 x^2$.  To compute the
standard deviation we set $\chi^2=1$ and solve for $x=1/\sqrt{2}$. In
contrast, simply adding both errors in quadrature yields $x=\sqrt{2}$,
which means adding the log-likelihoods underestimates the error. This
tendency becomes more pronounced when the two errors are not
equal.\bigskip

\section{Cooling Markov chains}
\label{app:coolmc}

The Markov Chain algorithm is optimized for creating Bayesian
marginalized log-likelihood distributions. This is analogous to an
integration, \ie not only the height of a peak is relevant for its
contribution to the final result, but also its volume in parameter
space. Correspondingly, the Metropolis-Hastings algorithm involves two
steps: first, it suggests the next point with a probability
proportional to the parameter-space volume of a peak
structure. Secondly, it accepts or rejects it based on the height of the
peak. The total probability is a combination of the two.

In a (frequentist) likelihood approach we are only interested in the
height of the peak. Only the value at the tip determines the
profile-likelihood value.  Following the above argument sharp and high
peaks are not very efficiently probed by Markov-Chain algorithms.  
Consequently, we amend the algorithm so that it better scans the
vicinity of points with large function values. A well known approach
which does this is simulated annealing~\cite{simann}. Combining this
with the Metropolis-Hastings algorithm~\cite{metropolis_hastings} 
we modify the accept or reject condition to
\begin{equation}
\frac{f_{\text{suggested}}}{f_{\text{previous}}} \ge r^{100/(j \cdot c)} \quad ,
\end{equation}
where $f_{\text{suggested}}$ and $f_{\text{previous}}$ are the
function values of the newly suggested and the last point in the
Markov chain, respectively. The complete Markov chain we divide into
100 equally long partitions numbered by $j$. $r$ is a random number
between $0$ and $1$, and $c$ is the cooling factor. We find that $c \sim
10$ gives well-converging results for this analysis. As in the original
algorithm a better point is unconditionally accepted.

In the beginning $r$ is almost entirely mapped to values close to
$r^{100/c} \sim 0$, \ie we accept many points and quickly scan the
parameter space. As $j$ increases we turn via the standard algorithm
into a region where we accept almost only better points. This way we
carefully scan the vicinity of the peaks and we obtain reliable values
for the profile-likelihood distribution. If there are several peaks of
comparable quality, different Markov chains will focus on different
peaks. We are guaranteed to see these alternative maxima because we
combine 30 individual Markov chains for the final result.

\begin{figure}
 \begin{center}
 {\raggedright \hspace*{10ex} $1/\Delta \chi^2$\\[2ex]}
\includegraphics[width=0.25\textwidth]{%
  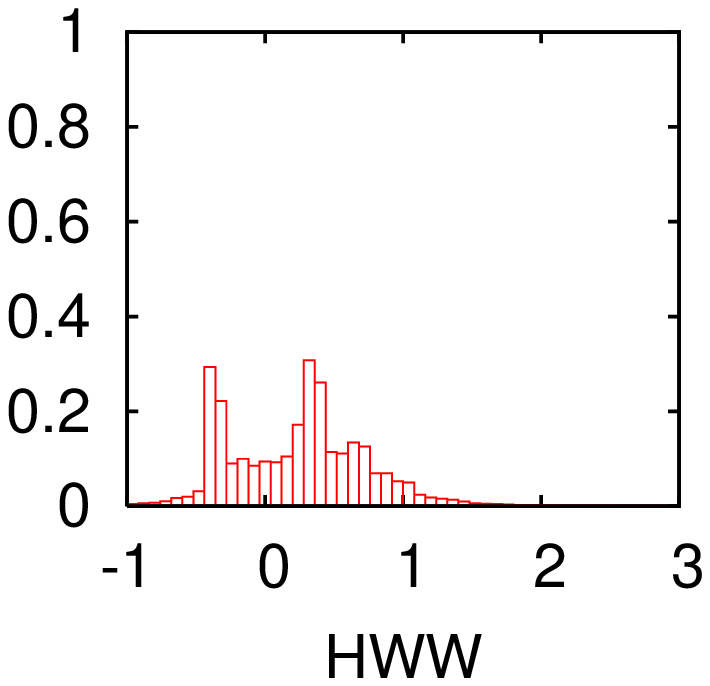} \hspace*{10ex}
\includegraphics[width=0.25\textwidth]{%
  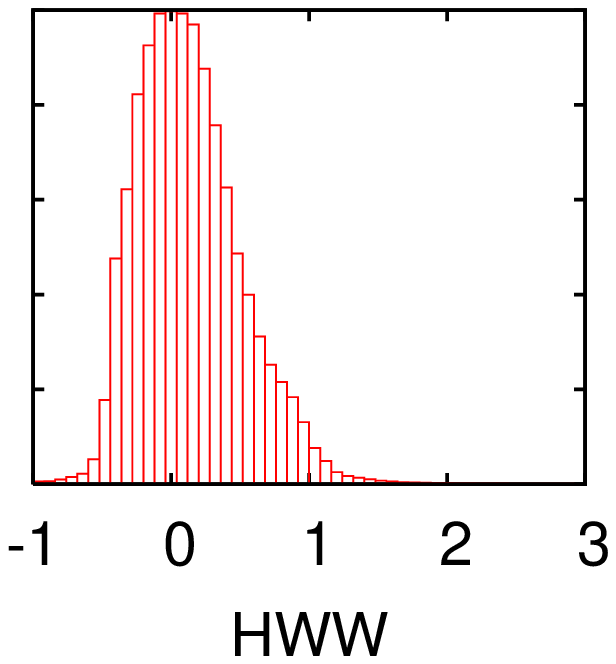}
 \end{center}
\caption[]{Comparison of profile likelihoods obtained from standard (left)
and cooling (right) Markov chains.}
\label{fig:app_cooling}
\end{figure}

To show the power of this modification we show the one-dimensional
profile likelihood for the $WWH$ coupling in
Fig.~\ref{fig:app_cooling}. It includes 30 Markov chains with
200000 points each. We see how the coarse standard algorithm places
the peaks at a completely different location while the cooling Markov
chain gets both the position and the height of the maximum right
(compared to the input).

\end{appendix}


\end{document}



\bibitem{topmass} 
 V.M Abazov et. al., [D0 Collaboration], Nature 429, 638 (2004)

\bibitem{ellis_olive}
 J.~R.~Ellis, K.~A.~Olive, Y.~Santoso and V.~C.~Spanos,
  Phys.\ Rev.\ D {\bf 69}, 095004 (2004);
 O.~Buchmueller {\it et al.},
  arXiv:0707.3447 [hep-ph].

\bibitem{ferrenberg_swendsen} 
 A.~M.~Ferrenberg, R.~H.~Swendsen,
  Phys.\ Rev.\ Lett. {\bf 61}, 2635 (1988).

\bibitem{minuit} 
 F.~James, M.~Roos, 
  Comp.~Phys.~Commun. {\bf 10}, 343 (1975).

\bibitem{ridders} 
 C.~J.~F.~Ridders,
  Advances in Engineering Software, {\bf 4}, 75 (1982).

\bibitem{lepewwg}
 [LEP Collaborations],
  arXiv:hep-ex/0412015.

\bibitem{higher_dim}
 V.~Barger, T.~Han, P.~Langacker, B.~McElrath and P.~Zerwas,
  Phys.\ Rev.\  D {\bf 67}, 115001 (2003).

\bibitem{wh_nlo}
 O.~Brein, A.~Djouadi and R.~Harlander,
  Phys.\ Lett.\  B {\bf 579}, 149 (2004);
 O.~Brein, M.~Ciccolini, S.~Dittmaier, A.~Djouadi, R.~Harlander and M.~Kr\"amer,
  arXiv:hep-ph/0402003.

\bibitem{slhc}
 F.~Gianotti {\it et al.},
  Eur.\ Phys.\ J.\  C {\bf 39}, 293 (2005).